\documentclass[10pt,twoside]{report}

\usepackage[utf8]{inputenc}
\usepackage[T1]{fontenc}
\usepackage[english]{babel}		
\usepackage{times}				
\usepackage{anyfontsize}

\usepackage[a4paper,margin=3cm]{geometry}
\usepackage{eso-pic}					
\usepackage{setspace}

\usepackage{fancyhdr}					

\usepackage[numbers]{natbib}
\usepackage[nottoc,notlot,notlof]{tocbibind}
\usepackage{tocloft}		 			

\usepackage[parfill]{parskip}			

\usepackage{hyperref}

\usepackage{sectsty}
\usepackage{abstract}

\usepackage{longtable}

\usepackage{amsmath}		
\usepackage{amsfonts}		
\usepackage{amssymb}		
\usepackage{amsthm}			

\usepackage{mathrsfs}		
\usepackage{accents}		
\usepackage{bbm}			
\usepackage{bm}      		
\usepackage{bigints}		

\usepackage{latexsym}		

\usepackage{graphicx}		
\usepackage{graphics}		
\usepackage{wrapfig}		
\usepackage{pict2e}			
\usepackage{tikz}			
\usepackage{grffile}        
\usepackage{pdfpages}		

\DeclareGraphicsExtensions{.pdf,.jpeg,.jpg,.png}

\usepackage{algorithm}		
\usepackage{algpseudocode}

\tocloftpagestyle{fancy}
\usepackage{etoc} 						
\usepackage{booktabs}					

\chapterfont{\sffamily\fontsize{20}{24}}
\sectionfont{\sffamily\fontsize{14}{15}}
\subsectionfont{\sffamily\fontsize{13}{15}}
\subsubsectionfont{\sffamily\fontsize{12}{15}}

\expandafter\def\expandafter\normalsize\expandafter{%
    \normalsize
    \setlength\abovedisplayskip{5pt}
    \setlength\belowdisplayskip{3pt}
    \setlength\abovedisplayshortskip{1pt}
    \setlength\belowdisplayshortskip{3pt}
}
\newcommand{\customtoc}{
        \etocdepthtag.toc{mtchapter}
        \etocsettagdepth{mtchapter}{subsection}
        \etocsettagdepth{mtappendix}{none}
        \newpage
        \thispagestyle{fancy}
        \tableofcontents
        \thispagestyle{fancy} 
        \newpage
}

\hypersetup{
	breaklinks=true,				
	colorlinks=true,				
	urlcolor=blue,					
	citecolor=blue,					
	linkcolor=blue,					
	anchorcolor=blue,				
	bookmarksopen=false,			
	linktocpage=false,				
	pdftitle={HDR Report},
	pdfauthor={Julien Garaud},
	pdfsubject={},
	pdfkeywords = {Topological defects, multicomponent superconductors, vortex},
	}

\makeatletter
\def\namedlabel#1#2{\begingroup
    JG#2%
    \def\@currentlabel{JG#2}%
    \phantomsection\label{#1}\endgroup
}
\makeatother
\newcommand{\CVcite}[1]{[\ref{CV:#1}]}

\makeatletter
\newcommand*{\Equation}{\@ifstar\sEquation\oEquation}
\newcommand{\sEquation}[1]{\begin{equation*}#1\end{equation*}}
\newcommand{\oEquation}[2]{\begin{equation}\label{#1}#2\end{equation}}
\makeatother
\makeatletter
\newcommand*{\Align}{\@ifstar\sAlign\oAlign}
\newcommand{\sAlign}[1]{\begin{align*}#1\end{align*}}
\newcommand{\oAlign}[2]{\begin{align}\label{#1}#2\end{align}}
\makeatother
\makeatletter
\newcommand*{\SubAlign}{\@ifstar\sSubAlign\oSubAlign}
\newcommand{\sSubAlign}[1]{\begin{subequations}\begin{align}#1\end{align}\end{subequations}}
\newcommand{\oSubAlign}[2]{\begin{subequations}\label{#1}\begin{align}#2\end{align}\end{subequations}}
\makeatother
\newcommand{\Itemize}[1]{\begin{itemize}#1\end{itemize}}

\newcommand{\Figref}[1]{Fig.~\ref{#1}}
\newcommand{\Tabref}[1]{Table.~\ref{#1}}
\newcommand{\Eqref}[1]{\eqref{#1}}

\newcommand{\Quote}[1]{\guillemotleft\,{\it #1}\,\guillemotright}

\newcommand{\bs}{\boldsymbol}

\newcommand{\ie}{\emph{i.e.}\,}
\newcommand{\eg}{\emph{e.g.}\,}

\newcommand{\rhs}{\emph{r.h.s.}\,}

\newcommand{\Relative}{\mathbbm{Z}}

\newcommand{\Real}{\mathbbm{R}}
\newcommand{\Complex}{\mathbbm{C}}

\newcommand{\groupU}[1]{\mathrm{U}(#1)}  			
\newcommand{\groupSU}[1]{\mathrm{SU}(#1)}			
\newcommand{\groupS}[1]{\mathbb{S}^{#1}} 			
\newcommand{\groupZ}[1]{\mathbb{Z}_{#1}} 			
\newcommand{\groupCP}[1]{\mathbb{C}\mathrm{P}^{#1}} 
\newcommand{\groupUU}{\groupU{1}\!\times\!\groupU{1}} %
\newcommand{\groupUZ}{\groupU{1}\!\times\!\groupZ{2}} %
\newcommand{\Exp}[1]{\mathrm{e}^{#1}}
\renewcommand\Re{\mathrm{Re}}
\renewcommand\Im{\mathrm{Im}}
\newcommand{\argmin}{\mathrm{argmin}\,}
\newcommand{\Grad}{{\bs\nabla}}
\newcommand{\Div}{{\bs\nabla}\!\cdot\!}
\newcommand{\Curl}{{\bs\nabla}\!\times\!}
\newcommand{\ScalarProd}[2]{\left\langle #1,#2\right\rangle}

\newcommand{\x}{\mathbf{x}}

\newcommand{\ez}{\bs e_z}
\newcommand{\etheta}{\bs e_\theta}

\newcommand{\oo}{{(1)}}
\newcommand{\ot}{{(2)}}
\newcommand{\oa}{{(a)}}

\newcommand{\eps}{\epsilon}
\newcommand{\bvarphi}{{\bar{\varphi}}}

\newcommand{\A}{{\bs A}}
\newcommand{\B}{{\bs B}}
\newcommand{\He}{{\bs H}_e}
\newcommand{\Hc}[1]{H_{c#1}}   			
\newcommand{\Tc}[1]{T_{c#1}}   			

\newcommand{\J}{{\bs J}}

\newcommand{\D}{{\bs D}}

\newcommand{\Q}{\mathcal{Q}}
\newcommand{\I}{\mathcal{I}}

\newcommand{\F}{\mathcal{F}}
\newcommand{\G}{\mathcal{G}}

\newcommand{\SRO}{{Sr$_2$RuO$_4$\ }}
\newcommand{\MGB}{{MgB$_2$\ }}

\newcommand{\sis}{{s\!+\!is}}
\newcommand{\spp}{{s_{++}}}
\newcommand{\spm}{{s_{\pm}}}
\newcommand{\sid}{{s\!+\!id}}
\newcommand{\did}{{d\!+\!id}}
\newcommand{\pip}{{p\!+\!ip}}

\newcommand{\Op}{\mathcal{O}}
\newcommand{\U}{\mathcal{U}}

\graphicspath{{Plots/}}
\newif\ifPRINT
\PRINTtrue
\newcommand{\doPrint}[2]{\ifPRINT {#1}\else {#2}\fi}
\title{Topological defects and other properties of multicomponent superconductors}

\author{Julien Garaud}

\begin{document}
\setstretch{1.1}
\includepdf[pages=-]{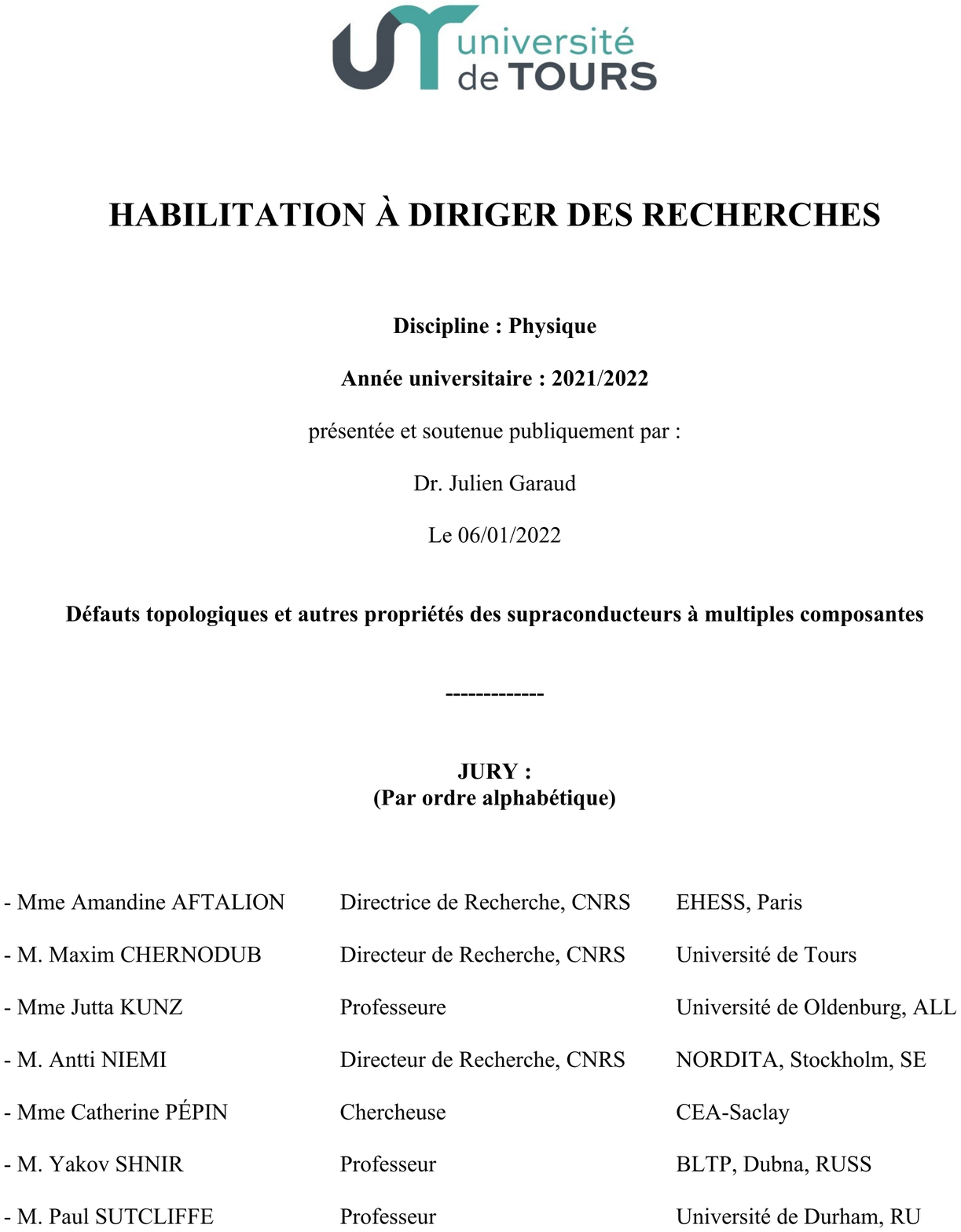} 
\doPrint{ \newpage \thispagestyle{empty}\ \newpage }{ }
%


	\addcontentsline{toc}{chapter}{R\'esum\'e - Abstract}
\section*{R\'esum\'e }

Il y a eu récemment un certain nombre de développements expérimentaux et de découvertes 
de nouveaux matériaux supraconducteurs, dont les degrés de liberté à plusieurs corps 
ont plusieurs composantes. Ces supraconducteurs, qui sont décrits par plusieurs 
condensats supraconducteurs, sont le lieu de nombreux phénomènes nouveaux, qui sont 
absents chez leurs homologues à une seule composante. 
Plusieurs de ces nouveaux aspects de la supraconductivité à plusieurs composantes 
sont présentés dans ce mémoire. 

Ils hébergent d'abord un large éventail de défauts topologiques. En effet, ayant 
plusieurs condensats, les excitations topologiques élémentaires des matériaux à 
plusieurs composantes sont des vortex fractionnaires. Ceux-ci peuvent se combiner 
pour former des états liés d'énergie finie, porteurs de flux. De tels défauts composites 
peuvent être de diverses nature, notamment des vortex, skyrmions, hopfions et murs de 
domaine. De plus, il existe des invariants topologiques supplémentaires qui permettent  
de différencier ces différents types de défauts topologiques. 

Par ailleurs, les supraconducteurs à plusieurs composantes sont généralement décrits 
par des échelles de longueur caractéristiques supplémentaires. Il est donc en général
impossible de construire un unique paramètre de Ginzburg-Landau. Également, l'interaction 
entre vortex peut être différente de soit purement attractive ou soit répulsive. Ainsi, 
il peut exister une phase supraconductrice, qui n'est ni de type-1 ni de type-2, où
les vortex peuvent former des agrégats entourés de régions dans l'état Meissner. 

Enfin, du fait de la compétition entre différents canaux d'appariement, certains 
états peuvent briser spontanément la symétrie d'inversion temporelle. Cela implique non 
seulement de nouvelles excitations topologiques, mais aussi que les modes collectifs 
et les échelles de longueur sont sensibles à cette symétrie brisée. De plus, comme les 
réponses électriques et magnétiques ont des contributions supplémentaires, 
les propriétés thermoélectriques des états supraconducteurs qui brisent la symétrie 
d'inversion temporelle sont modifiées. 


\section*{Abstract}			

In recent years, there were a number of experimental developments and discoveries 
of novel superconducting materials which exhibit multicomponent, many-body degrees 
of freedom. These superconductors, that are described by several superconducting 
condensates, feature many new interesting phenomena that are absent in their 
single-component counterparts.
Several of these new aspects of multicomponent superconductivity are addressed
in this report.

First of all, they feature a rich spectrum of topological defects. Indeed, since 
they have several condensates, the elementary topological excitations in multicomponent 
superconductors are fractional vortices. These can combine to form finite energy, 
flux carrying, bound states. The resulting composite defects can be of various nature 
including vortices, skyrmions, hopfions, and domain-walls. Moreover, there exist 
additional topological invariants that can differentiate between the different kind 
of topological defects.

Also, multicomponent superconductors are typically describe by extra characteristic 
length scales. Thus it is not possible, in general, to construct a single  
Ginzburg-Landau parameter. Moreover since the length scales rule, to some extent, 
the interactions between the quantum vortices, they can be richer than purely attractive 
or purely repulsive. These facts imply that there can exist a new 
superconducting phase, which is neither in the type-1 nor in the type-2, where 
vortices can form aggregates surrounded by Meissner regions.

Finally, because of the competition between different pairing channels, some 
multicomponent superconducting states can spontaneously break the time-reversal 
symmetry. This implies not only that they allow for new topological excitations, 
but also that the collective modes and characteristic length scales are sensitive 
to the new broken symmetry. Moreover, since the electric and magnetic responses feature 
additional contributions, the thermoelectric properties of the superconducting states 
that break the time-reversal symmetry can be substantially altered.


\doPrint{ \newpage \thispagestyle{empty}\ \newpage }{\newpage }
\addcontentsline{toc}{chapter}{Pr\'eface - Preface}

\section*{Pr\'eface}

\paragraph{Nomenclature des citations} les citations comme par exemple 
\CVcite{Rybakov.Garaud.ea:19} désignent des travaux de l'auteur. Les références 
correspondantes sont répertoriées dans la liste des publications, en 
Annexe~\ref{Chap:Papers}. Les autres citations, comme par exemple \cite{Manton.Sutcliffe}, 
sont des citations régulières, qui sont situées dans la bibliographie.

\paragraph{Illustrations:} Les illustrations présentés dans ce rapport sont du nouveau 
matériel qui n'a pas été publié. Elles sont cependant représentatives des résultats 
obtenus et discutés, dans les articles publiés précédemment.

\paragraph{Synth\`ese en fran\c cais:} 
Le réglement de l'Habilitation à Diriger des Recherches requiert que si le mémoire est 
rédigé anglais, il soit accompagné d'un document de synthèse rédigé en fran\c cais. 
Ce document de synthèse en fran\c cais du mémoire d’HDR est reproduit en 
Annexe~\ref{Chap:Synthese}.

\vspace{2cm}
\section*{Preface}


\paragraph{Nomenclature for citations} the citations with for example 
\CVcite{Rybakov.Garaud.ea:19} denote the citations of the author's works. 
The corresponding references are listed in the publication list, given in the Appendix 
\ref{Chap:Papers}. The other citations, as for example \cite{Manton.Sutcliffe} 
are ``regular'' citations, which are located on the bibliography.

\paragraph{Illustrations:} Apart from the introduction,
all the illustrations presented in this report are new material that has not been 
previously published. These are however representative of the results obtained, 
and discussed in the previously published papers.

\paragraph{Summary in French:} 
The regulations for the Habilitation to Supervise Researches require that if the 
dissertation is written in English, it must be accompanied by a summary document 
written in French. This summary in French of the HDR dissertation is reproduced 
in the Appendix~\ref{Chap:Synthese}.

\doPrint{ \newpage \thispagestyle{empty}\ \newpage }{\newpage }

\addcontentsline{toc}{chapter}{Remerciements - Aknowledgements}

\vspace{0.5cm}
\section*{Remerciements}

Avant toute chose, je souhaite remercier les membres du jury de cette habilitation, 
et en particulier les rapporteurs: A.~Niemi, Ya.~Shnir et P.~Sutcliffe. Je remercie 
également M.~Chernodub, référent de l'habilitation, pour son soutient.

Cette mémoire d'habilitation présente les résultats de nombreuses de discussions 
et de collaborations, la plupart d'entre elles ont eu lieu lors de mes années de postdoc.
Après avoir soutenu mon doctorat en physique des hautes énergies, j'ai changé ma thématique 
principale de recherche pour exercer dans le domaine de la physique de la matière condensée.
Je serai toujours reconnaissant envers Egor Babaev, pour m'avoir fait découvrir 
ce domaine de la physique.Je pense souvent à son enthousiasme, lors de nos nombreuses 
discussions stimulantes sur tant de sujets. 
Je pense également à sa fa\c con de penser que la créativité est une qualité essentielle 
pour un chercheur, et que rien ne doit être tenu pour acquis.

Je tiens également à remercier les différents collaborateurs avec qui j'ai eu 
le plaisir de faire des recherches. Parmi eux, j'ai une pensée particulière pour 
Johan Carlstr\"om qui m'a beaucoup aidé pour mon changement de domaine de recherche. 
Je remercie également les étudiants et post-doctorants du groupe à Stockholm, 
Karl Sellin, Mihail Silaev et Filipp Rybakov.
J'ai aussi une pensée pour les étudiants en Master et Doctorat que j'ai (co)encadrés.

Je tiens également à remercier mon directeur de thèse, Mikhail Volkov qui, 
je le crois, m'a communiqué la volonté d'une rédaction soignée et celle d'améliorer 
mes différentes compétences techniques. Je pense aussi que j'ai appris de lui 
l'importance de savoir se remettre en question. Pour tout cela, je lui en suis 
reconnaissant.

Enfin, je tiens à exprimer mes sentiments les plus chaleureux à tous mes proches. 
Ils se reconnaîtront à coup sur. 
Au fil du temps, certains restent, d'autres non. Certains s'en vont. 

\vspace{0.5cm}
\section*{Aknowledgements}

Beforeall, I would like to thank the members of the jury for this habilitation, 
and in particular the referees: A.~Niemi, Ya.~Shnir and P.~Sutcliffe. I also want 
to thank M.~Chernodub, the referent for the habilitation, for his support. 

This thesis presents the results of numerous discussions and collaborations, 
most of them I did as postdoc.
After graduating with the PhD in high energy physics, I changed my main area of 
research and mostly worked in the area of solid state and condensed physics.
%
I will always be grateful to my postdoc advisor, Egor Babaev, for introducing me to 
this area of physics. I often think about his enthusiasm, during all our stimulating 
discussions on so many subjects. 
%
I also think about his way of thinking that creativity is an essential quality 
for a researcher, and that nothing should be taken for granted.

I also want to thank my various collaborators I had the pleasure to do research with. 
Among these, I have a particular thought for Johan Carlstr\"om who greatly helped me 
with changing my research area. I also thank the students and postdocs from the group 
in Stockholm, Karl Sellin, Mihail Silaev and Filipp Rybakov. 
%
I also have a thought for the Master and PhD students I (co)supervised. 

I am also grateful to my PhD advisor Mikhail Volkov whom, I believe, 
communicated me the will of a careful writing, and for improving different 
technical skills. I also think that I learned from him the importance of questioning
oneself. For all this, I am grateful to him.

Finally I want to express my warmest feeling to my close relatives. 
They surely know who they are.
Over the time, some stay, some fade away. Some just go.


\doPrint{ \newpage \thispagestyle{empty}\ \newpage }{ }
\customtoc
\chapter*{Introduction}		\addcontentsline{toc}{chapter}{Introduction}
\graphicspath{{Plots/01-Introduction/}}

\begin{wrapfigure}{R}{0.4\textwidth}
\hbox to \linewidth{ \hss
\includegraphics[width=.975\linewidth]{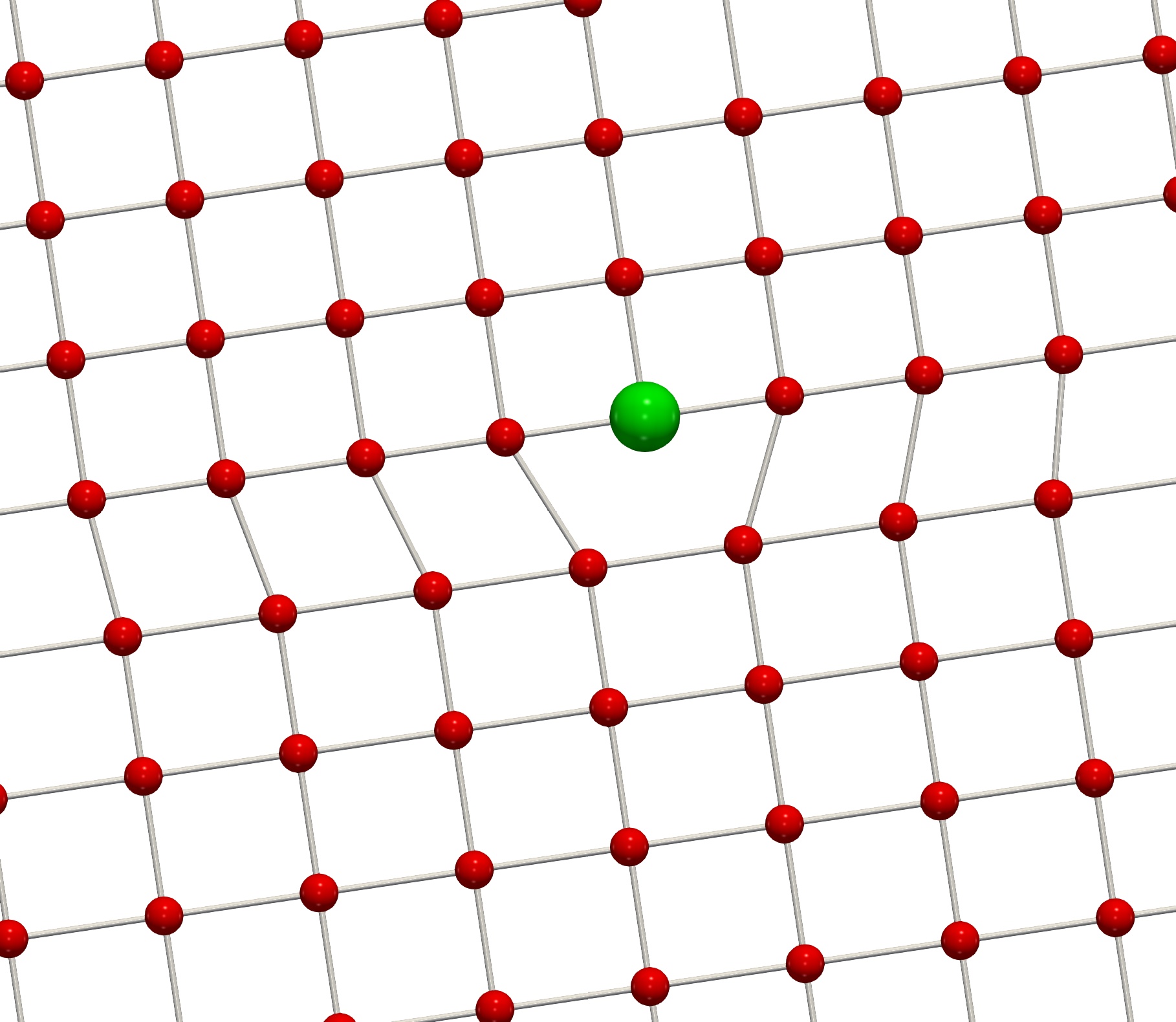}
\hss}
\caption{
A topological defect in a crystal. 
One of the columns of atoms on bottom disappears halfway through the sample. 
The place where it disappears (highlighted in green) is a defect, 
because it doesn't locally look like a piece of the perfect crystal. 
}\label{Fig:Dislocation}
\end{wrapfigure}
The topological defects and their understanding are at the core of modern physics. 
The formalization of their properties, and of the knowledge of their role in a very 
broad range of physical processes is rather recent. However, they have been heuristically 
known by mankind for probably more than three thousand years. Indeed, this is 
approximately as far as the processes of quench hardening of metal by smiths can be 
traced back \cite{Mackenzie:08}. The fast thermal quenches used in metal hardening 
processes creates dislocations of the crystal structures, which are akin to topological 
defects. This is similar to the proliferation of topological defects that occur 
during phase transitions or other kinds of thermal quenches. 
As illustrated in \Figref{Fig:Dislocation}, a dislocation in a crystal is a 
topological defect, because it cannot be removed by any local rearrangement.

Associated with broken symmetries, the topological defects are ubiquitous in physics. 
They indeed arise in a very broad context ranging from early universe cosmology and 
particle physics \cite{Rajaraman,Manton.Sutcliffe,Vachaspati,Vilenkin.Shellard,Shnir,
Volovik,Shnir:18}, to solid state and condensed matter systems \cite{Mermin:79,
Mineev,Volovik}. 
Depending on the underlying theory, the topological defects can be of various kind
including for example dislocation in crystals, monopoles, domain-walls, vortices, 
skyrmions, hopfions, and much more. They are intimately related with phase transitions 
\cite{Kosterlitz:17,Kibble:76,Zurek:85}, and their mere existence can have important 
consequences. For example, the possible formation of topological defects during early 
universe phase transitions could have greatly impacted the structure formation of the 
universe \cite{Vilenkin.Shellard,Kibble:76,Brandenberger:94}. Likewise, they are 
believed to drive certain phase transitions in various physical system, as for example 
the vortex proliferation in superfluids and superconductors \cite{Kosterlitz:17}.
Vortices, which are line-like objects with specific topological properties, 
are probably the most studied topological defects.

\section*{Topological defects -- Superconductors and superfluids}
\addcontentsline{toc}{section}{Topological defects -- Superconductors and superfluids}

The vortex physics have been the subject of an intense scientific activity since the 
second half of the nineteenth century. Shortly after the earlier works of Helmholtz 
\cite{Helmholtz:58} on fluid dynamics, vortices were the cornerstone of the 
``vortex atom" theory of matter conjectured by Kelvin \cite{Thomson:69}. This failed 
attempt to classify the chemical elements as excitations consisting of closed, linked, 
and knotted vortex loops in the luminiferous aether
\footnote{
The luminiferous aether was a postulated ideal fluid, supposed to fill in the whole 
universe, and serving as a medium for the propagation of light waves.
},  
yet led to considerable breakthrough in topology as it motivated the creation of 
the first knot tables by Tait \cite{Tait:78,Tait:84,Tait:86} and to the early knot 
theories works shortly after.

\subsection*{The theory of the vortex atom}

Interestingly, the theory of the ``vortex atom" of Kelvin and Tait still resonates 
with some concepts of modern physics \cite{Kragh:02,Laan:12}, and it inspired 
several works over the years. Hence, it is a story worth being told.

In his 1858 work on fluid dynamics, Helmholtz \cite{Helmholtz:58} demonstrated that
in an ideal fluid (\ie with an incompressible and inviscid flow), the circulation of 
a vortex filament does not vary over the time. He further demonstrated that a vortex 
cannot terminate inside such a fluid, but should either extend to the boundaries of 
the fluid or form closed loops. Also that, in the absence of external rotational 
forces, an initially irrotational flow remains irrotational.

\begin{wrapfigure}{R}{0.4\textwidth}
\hbox to \linewidth{ \hss
\includegraphics[width=.975\linewidth]{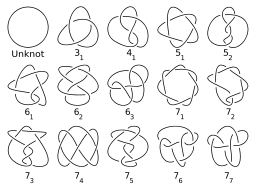}
\hss}
\caption{
A table of inequivalent knots.
}\label{Fig:Kelvin-Knots}
\end{wrapfigure}
Knowing the theorems of Helmholtz, in 1867 Kelvin \cite{Thomson:69} noticed that 
\Quote{(...) this discovery inevitably suggests the idea that Helmholtz's rings 
are the only true atoms.} The general idea was that, because the vortex lines are 
frozen in the flow of an ideal fluid, then their topology should be invariant in time: 
\Quote{
It is to be remarked that two ring atoms linked together, or one knotted in any manner 
with its ends meeting, constitute a system which, however it may be altered in shape, 
can never deviate from its own peculiarity of multiple continuity (...)
}. That ideal fluid would be the luminiferous aether that people believed to fill 
the universe. He then attributed the spectroscopic properties of matter to the topology 
of such vortex lines: 
\Quote{It seems, therefore, probable that the sodium atom may not consist of a single
vortex line; but it may very probably consist of two approximately equal vortex rings 
passing through one another, like two links of a chain.}
\footnote{
More precisely, that the spectroscopic properties of the elements should corresponds 
to vibrational and rotational modes of linked and knotted vortices 
\Quote{It is probable that the vibrations which constitute the
incandescence of sodium vapour are analogous to those which the
smoke-rings had exhibited; and it is therefore probable that the
period of the vortex rotations of the atoms of sodium vapour are
much less than T $1/526$ of the millionth of the millionth of a second,
this being approximately the period of vibration of the yellow
sodium light.}
}. 
Kelvin further noticed that in models of \Quote{(...) knotted or knitted vortex atoms, 
the endless variety of which is infinitely more than sufficient to explain the varieties 
and allotropies of known simple bodies and their mutual affinities.}. In short Kelvin 
conjectured that the different chemical bodies consist in topologically inequivalent 
closed, linked and knotted vortex loops, illustrated in \Figref{Fig:Kelvin-Knots}, 
of luminiferous aether.

Subsequently, Tait started to classify the different inequivalent ways to tie such 
knots \cite{Tait:78,Tait:84,Tait:86}. These works pioneered the field of knots theory 
in algebraic topology. 
Kelvin's theory was eventually falsified, when Michelson and Morley's experiment ruled 
out the existence of aether \cite{Michelson.Morley:87}. Yet the paradigm to associate 
vortices in some underlying field with ``elementary particles" re-emerged on several 
occasions. 
In a way, these knotted vortices can be seen as the first theoretical example of 
topological defects.

\subsection*{Vortices and other topological defects in modern physics}

About 80 years after Kelvin's work, it was realized by Onsager \cite{Onsager:49}, 
and later formalized on solid theoretical grounds by Feynman \cite{Feynman:55}, 
that the vortices occupy an important part in modern physics processes. In his 
work on superfluid $^4$He, Onsager \cite{Onsager:49} observed that the circulation 
of the superfluid velocity is quantized, and he further understood that vortex matter 
basically controls many of the key responses of superfluids. For example, that 
the superfluid to normal state phase transition is a thermal generation, and a 
proliferation of vortex loops and knots \cite{Onsager:49}. Also that vortices 
appear as the rotational response of superfluids.

These ideas somehow partially resonate with Kelvin's theory of the vortex atom.
Indeed, because the circulation of the superfluid velocity is quantized, then 
vortices in superfluids are topological defects. Moreover, the rotation of a 
superfluid results in the formation of a lattice or a liquid of these quantum 
vortices. These lattices could be seen as the vortex-matter realisation 
of crystals and liquids. 
It was also later predicted by Abrikosov \cite{Abrikosov:57a} that the type-II 
superconductors should form magnetic vortices when subjected to an external 
magnetic field, by analogy with the vortices formed as a rotational response of a 
superfluid. Later it was further understood that in three dimensions superfluid 
and superconducting phase transitions are a thermal generation and a proliferation 
of vortex loops \cite{Peskin:78,Dasgupta.Halperin:81}.

Noteworthy, important progresses in modern physics phenomena, where vortices 
occupy a central part, were awarded a Nobel prize. As for example to Abrikosov in 
2003 \cite{Abrikosov:04} for the understanding of their role in superconductors, 
or more recently in 2016 to Haldane, Kosterlitz and Thouless for their role in 
the phase transitions in two-dimensional systems \cite{Haldane:17,Kosterlitz:17}. 
The concept of quantum vortices was later generalized to relativistic theories, 
as for example in the abelian-Higgs model \cite{Nielsen.Olesen:73}; theories that 
might have been relevant in the early universe \cite{Vilenkin.Shellard,Witten:85,
Witten:85a}, and also to the 
bosonic sector of the Weinberg-Salam theory of the electroweak interactions 
\cite{Achucarro.Vachaspati:00}. According to the Kibble-Zurek mechanism \cite{Kibble:76,
Zurek:85}, various kind of topological defects should be produced during possible 
early universe phase transitions. This would imply, among other things, that if 
topological defects were created, they could substantially contribute to the matter 
content of the universe, and have had a nontrivial impact on its history 
\cite{Vilenkin.Shellard,Brandenberger:94}. These interesting ideas are at the origin 
of a lot interest for topological defects. This resulted in many seminal works and 
in a deeper understanding of the mathematical properties of topological defects.

\begin{wrapfigure}{R}{0.4\textwidth}
\hbox to \linewidth{ \hss
\includegraphics[width=.975\linewidth]{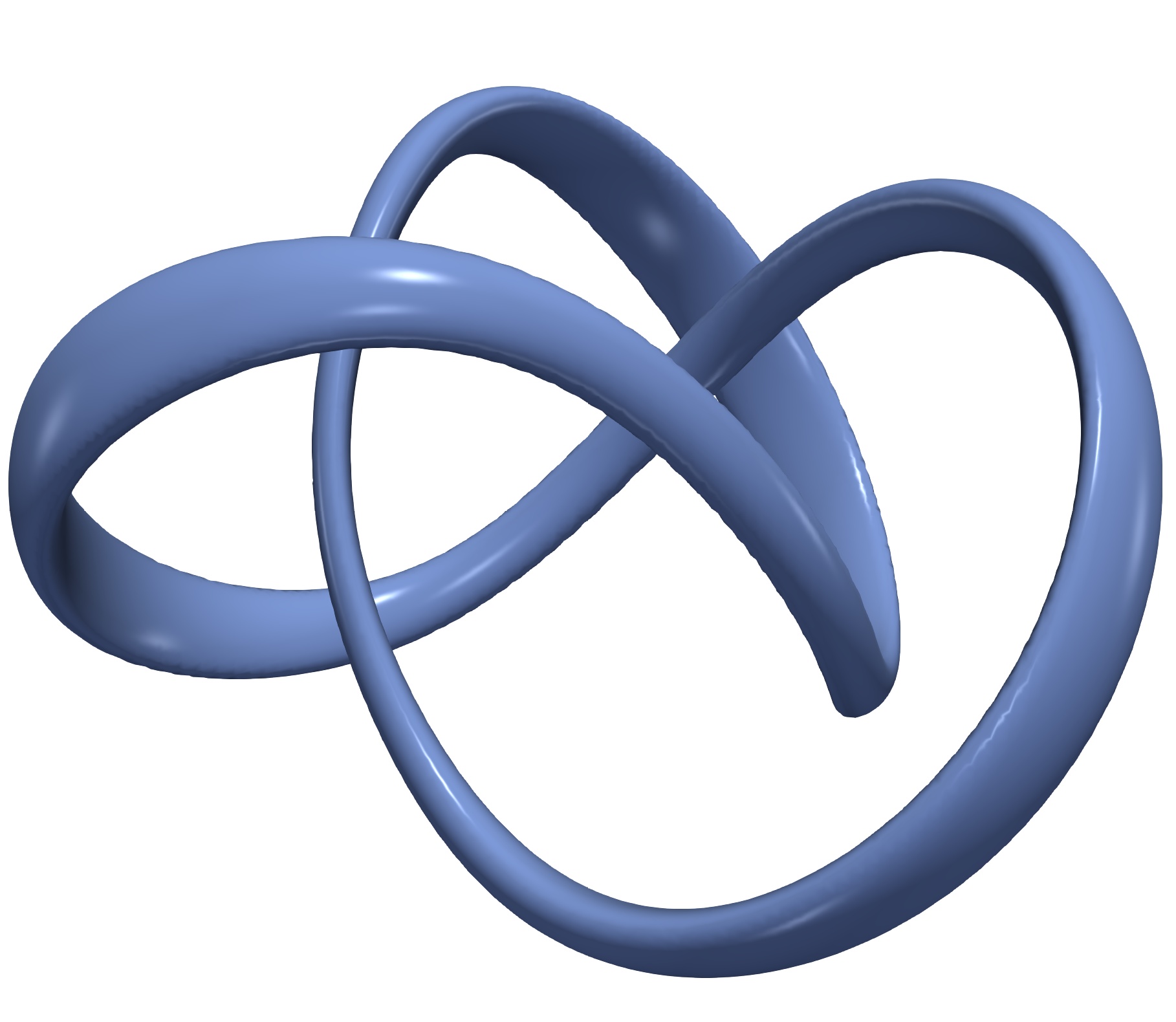}
\hss}
\caption{
A trefoil knot in Skyrme-Faddev model.
}\label{Fig:Knots-SF}
\end{wrapfigure}

Thus, as already emphasized there are plethora of different kind of topological 
defects, in a  broad range of physical systems. It is kind of meaningless to 
exhaustively list all of them. Rather let's mention two particular kinds of 
topological defects that particularly resonate with Kelvin's theory, as they were 
somehow identified with states of matter. 
A first example is that of the topological defects in Skyrme model \cite{Skyrme:61,
Skyrme:62}. The topological defects there are termed \emph{skyrmions}
\footnote{
In the main text, the terminology \emph{skyrmions} is used to characterize slightly 
different kind of topological defects. They are more related to the so-called 
\emph{baby-skyrmions}, but since they share many topological properties they are 
often termed skyrmions as well, with a bit of terminological abuse.
}, and the associated 
topological invariant is interpreted as the baryon number \cite{Brown.Rho}. 
Likewise, the research on models supporting stable knotted topological defects has 
been of great interest in mathematics and physics, after the stability of these 
objects termed \emph{hopfions} was demonstrated in the Skyrme-Faddeev model 
\cite{Faddeev:75,Faddeev.Niemi:97,Gladikowski.Hellmund:97,Hietarinta.Salo:99,
Battye.Sutcliffe:99,Sutcliffe:07} (for a review, see \cite{Radu.Volkov:08}). 
Hopfions in the Skyrme-Faddeev model resemble knots, as illustrated in 
\Figref{Fig:Knots-SF}.

After this general introduction about topological defects, most of the attention 
will be ported on vortices, with a particular focus on those that appear in models 
of superconductivity with multiple components of the order parameter.

\section*{Multi-component superconductors}
\addcontentsline{toc}{section}{Multi-component superconductors}

Superconductivity and superfluidity are states of matter that are characterized by the 
macroscopic coherence of the underlying quantum excitations. The underlying physics 
that describes such systems are quantum fields theories, and these are the many-body 
properties, within those theories, that yield the macroscopic coherence of the quantum 
excitations. The very interesting feature is that such quantum many-body problems 
can be reduced, in the mean field approximation, to nonlinear classical field theories 
describing the macroscopic properties of the coherent state represented by a single 
complex scalar field (the \emph{order parameter}). 
These mean field approximations are known as the Gross-Pitaevskii equations for 
superfluids, and as the Ginzburg-Landau equations for superconductors. Note that 
in the case of superconductors, the complex scalar is supplemented by a real Abelian 
vector field, describing the electromagnetic potential. This gauge field becomes 
massive via the Anderson-Higgs mechanism \cite{Anderson:58,Anderson:63}, which is 
responsible for the Meissner effect \cite{Meissner.Ochsenfeld:33}. While in the simplest 
textbook cases, these order parameters are singlets, they can be scalar multiplets 
in more complicated situations.

In condensed matter systems such as superfluids or Bose-Einstein condensates of 
ultra-cold atoms, theories with order parameters with multiple components 
(\ie described by multiplets or even matrices of complex scalar fields) have been 
considered for a long time.  They have been known to offer an extremely rich zoology 
of topological defects, as for example in superfluid helium \cite{Volovik,Volovik:92}, 
spinor Bose-Einstein condensates \cite{Kawaguchi.Ueda:12,Ueda:14}, 
or in neutron $^3P_2$ superfluids \cite{Fujita.Tsuneto:72,Masuda.Nitta:20}. 
In the context of superconductivity, theories with multiple superconducting gaps 
where considered from the earlier days of the Bardeen-Cooper-Schrieffer theory 
\cite{Moskalenko:59,Suhl.Matthias.ea:59,Tilley:64}. Yet these multiband/multicomponent 
theories where for a long time considered to describe exotic materials.

In the recent years however, there have been an increased interest in such materials, 
as the number of known multiband/multicomponent superconductor is rapidly growing. 
Indeed, in many superconductors, the pairing of electrons is supposed to occur on 
several sheets of a Fermi surface which is formed by the overlapping electronic bands. 
To name a few of them, this is for example the case of MgB$_2$ 
\cite{Nagamatsu.Nakagawa.ea:01,Mazin.Antropov:03}, layer perovskite Sr$_2$RuO$_4$ 
\cite{Maeno.Hashimoto.ea:94,Mackenzie.Maeno:03,Damascelli.Lu.ea:00}, or of the familly 
of iron-based superconductors \cite{Kamihara.Watanabe.ea:08,Mazin.Singh.ea:08,
Kuroki.Onari.ea:08,Chubukov.Efremov.ea:08}. Beyond solid state physics, multicomponent 
theories also apply to more exotic systems, as certain models of nuclear superconductors 
in the interior of neutron stars \cite{Jones:06}, or the superconducting state of Liquid 
Metallic Hydrogen \cite{Ashcroft:68,Ashcroft:00}, Liquid Metallic Deuterium 
\cite{Oliva.Ashcroft:84,Oliva.Ashcroft:84a} and other kind of metallic superfluids 
\cite{Ashcroft:05}.
This opens the possibility of more complicated field theory models where, typically 
due to the existence of multiple broken symmetries, the physics of vortices 
(and of other topological defects) is extremely rich with no counterparts in  
single-component models.

Note that the models where vortices appear, in the context of high-energy physics,
are typically very symmetric because of the underlying properties of the theory.
For example, in the case of the Weinberg-Salam theory of the electroweak interactions, 
the theory is invariant (among other symmetries) under the local $\groupSU{2}$ rotations 
within the scalar doublet (the Higgs field). Models describing condensed matter systems 
are typically much less constrained on symmetry grounds, and thus allow for more 
interaction terms that explicitly break various symmetries. For example, in two-component 
superconductors (described by a scalar doublet) the global $\groupSU{2}$ invariance 
is explicitly broken down to a smaller subgroup (as for example 
$\groupU{1}\times\groupU{1}$). 
The absence of strong symmetry constraints, and thus the existence of various symmetry 
breaking terms is at the origin of many new features. It results, in particular, 
that vortices can acquire new properties and are associated to a broad range of 
new physical phenomena. 
Those new exotic properties can be understood as originating in the new broken 
symmetries. As a results such new phenomena can be used as signatures to trace 
back informations on the actual symmetries of some unknown state.

The crucial importance of the topological excitations in the physics of 
superconductivity made the Ginzburg-Landau vortices one of the most studied example 
of topological defects. Indeed, all the transport properties of superconductors 
crucially depend on the behaviour of magnetic vortices in these materials. 
For example, high critical currents in currently existing commercial superconducting 
transmission lines are only achieved by carefully controlling the vortex motion in 
these materials. 
The theories for multiband/multicomponent superconductors extends the usual 
Ginzburg-Landau theory by considering more than one superconducting order parameters. 
Because of the additional fields and the new broken symmetries, the spectrum of 
topological excitations and the associated signatures are much richer in multicomponent 
systems than in their single-component counterparts. For example multicomponent 
superconductors feature fractional vortices, singular/coreless vortices, skyrmions, 
hopfions, domain-walls, etc. All these topological excitations can be used as 
experimental signatures to probe the multicomponent nature of a superconducting system. 
Their observability can, for example, provide valuable informations about the structure 
of the order parameter and of the underlying pairing symmetry.

The works discussed in this report, deal with various aspects of the theories with 
more than one superconducting condensate. Both for general models of multicomponent 
superconductors and for material specific models.
In particular through the investigation of the properties associated with the 
topological defects that appear therein.

\subsection*{Mean-field Ginzburg-Landau theories}

In the single-component, weak-coupling, mean-field, Bardeen-Cooper-Schrieffer theory 
\cite{Bardeen.Cooper.ea:57}, the superconducting state is described by a classical 
complex field which is proportional to the gap function. Namely, the phenomenologically 
introduced Ginzburg-Landau theory \cite{Ginzburg.Landau:50a} can be derived as the 
classical, mean-field, approximation of the microscopic theory \cite{Gorkov:59}, 
and the modulus of the order parameter is the density of Cooper pairs. 
There are various approaches to characterize the properties of superconducting materials, 
that are different/complementary to the Ginzburg-Landau theory. For example, methods 
such as the Bogoliubov-de Gennes formalism \cite{Gennes,Zhu:16}, the Eilenberger 
\cite{Eilenberger:68} and Usadel \cite{Usadel:70} equations for transport, etc.
However, the rest of this report is restricted only to the classical mean-field 
aspects of superconductivity of multicomponent systems. That is, to the multicomponent 
Ginzburg-Landau theory, and to the topological excitations that occur therein.

Interestingly, the Ginzburg-Landau theory of superconductivity attracted a lot of 
attention from the Numerical Analysts community only since the 1990s, after the report 
of the well-posedness of the problem \cite{Du.Gunzburger.ea:92,Du:94}. Since then, 
there have been quite some activity in understanding the mathematical properties of 
that problem, see for example \cite{Du:05,Frank.Lemm:16}.

\section*{Field theoretical models in the mean field approximation}
\addcontentsline{toc}{section}{Field theoretical models in the mean field approximation}

The details of the multicomponent models may vary, depending on the context of 
the underlying physical problem that is considered. For example, via the dependence of 
the different parameters. Here, we briefly expose the mathematical structure of the 
generic mean field models that describe multicomponent superconductors.
The macroscopic properties of such physical systems are typically described by a 
Ginzburg-Landau (free) energy functional of the form:
\Equation{Eq:General:FreeEnergy}{
\F/\F_0=\int_{\mathbb{R}^3} \frac{1}{2}\big|\Curl\A\big|^2 
+\frac{\kappa_{ab}}{2}(\D\psi_a)^*\D\psi_b^{} 
+ \alpha_{ab} \psi_a^*\psi_b^{}
+ \beta_{abcd} \psi_a^*\psi_b^*\psi_c^{}\psi_d^{}	\,,
}
where $\psi_a$ are the components of the scalar multiplet $\Psi\in\mathbb{C}^N$, 
that accounts for the superconducting degrees of freedom. The scalar multiplet thus reads 
as $\Psi^\dagger=(\psi_1^*,\psi_2^*,\cdots,\psi_N^*)$,  where $a,b,c,d=1,\cdots, N$; 
and the repeated indices are implicitly summed over. The scalar fields are coupled 
to the (Abelian) gauge field $\A$ via the gauge derivative $\D=\Grad+ie\A$, with $e$ 
the gauge coupling (the bold fonts denote the vector quantities).
All the matrix and tensor coefficients $\hat{\kappa}$, $\hat{\alpha}$, $\hat{\beta}$ 
obey some symmetry relations, so that the energy is a real positive definite quantity 
\footnote{The Ginzburg-Landau model \Eqref{Eq:General:FreeEnergy} is isotropic. 
Anisotropies can be incorporated by using more general kinetic term:
$\kappa_{ab;\mu\nu}(D_\mu\psi_a)^*D_\nu\psi_b^{}$.
}.

It might be convenient to collect all the potential terms in \Eqref{Eq:General:FreeEnergy} 
into a single potential term  $V(\Psi,\Psi^\dagger)$ as 
\Equation{Eq:General:Potential}{
V(\Psi,\Psi^\dagger)=\alpha_{ab} \psi_a^*\psi_b^{}
		+ \beta_{abcd} \psi_a^*\psi_b^*\psi_c^{}\psi_d^{} \,.
}
Sometimes, the specific structure of the potential $V(\Psi,\Psi^\dagger)$ will be 
unimportant. On other occasions, the interacting potential will have a central role 
for the definition of the new physical properties. Thus the relevant restriction of 
the most generic potential \Eqref{Eq:General:Potential} will be specified when necessary.

The \emph{ground state} is the state which minimizes the potential energy 
\Eqref{Eq:General:Potential} and that is constant in space: 
$\Psi_0:=\argmin V(\Psi,\Psi^\dagger)$. Moreover the \emph{superconducting ground 
state} is the state that minimizes the energy and that has $\Psi^\dagger\Psi=const.\neq0$. 
The criterion for condensation, that is $\Psi_0\neq0$, is that $\det\hat{\alpha}<0$. 
The superconducting ground state is degenerate in energy and this defines a manifold 
called the \emph{vacuum manifold}. Roughly speaking this is the topology of that vacuum 
manifold that specifies the topological defects that can appear in the theory. For example, 
the ground state energy is invariant under overall phase rotations of the multiplet 
$\Psi$, thus this defines a vacuum manifold that is a circle. The field configurations 
are hence classified by a \emph{winding number} that is an element of the first homotopy 
group $\pi_1(\groupS{1})$ (this can alternatively be understood as a consequence that 
$\Psi$ has to be single-valued). This winding number determines the vortex content of 
the theory.

The functional variation of the free energy with respect to the superconducting condensates 
yields the Euler-Lagrange equations of motion. These, in the framework of superconductivity, 
are the Ginzburg-Landau equations
\Equation{Eq:General:GL}{
\kappa_{ab}\D\D\psi_b=2\frac{\delta V}{\delta\psi_a^*}\,.
}
Similarly, the variation with respect to the gauge field yields the 
Amp\`ere-Maxwell equation
\Equation{Eq:General:Maxwell}{
\Curl\B +e\sum_{a,b}\kappa_{ab}\Im\big( \psi_a^*\D\psi_b \big)=0 	\,,
}
where $\B=\Curl\A$ is the magnetic field. This equation is used to introduce the 
supercurrents 
\Equation{Eq:General:Current}{
\J:=\sum_a \J^\oa \,,~~~~~\text{where}~~~~~
\J^\oa=e\sum_{b}\kappa_{ab}\Im\big( \psi_a^*\D\psi_b \big) 	\,.
}
Here $\J$ is the total supercurrent, while $\J^\oa$ is the partial current associated 
with a given superconducting condensate $\psi_a$.

Depending on the properties of the microscopic model under consideration, there can 
be various additional requirements further constraining the structure of the tensor 
parameters $\hat{\kappa}$, $\hat{\alpha}$, $\hat{\beta}$. This can yield many different 
situations that are useless to be listed here. 
As mentioned above, the vortex content is specified by the winding number of the field 
configuration (more precisely the winding at infinity). The next step is to explicitly 
construct the vortex solutions in a given topological sector specified by this winding 
number. The theory is clearly nonlinear and the explicit construction of a field 
configuration with a given winding number thus has to be addressed numerically. 
In the works that are discussed here, this is done using minimization algorithms on 
the energy within a finite element formulation of the problem (see details in the 
Appendix~\ref{App:numerics}).

\section*{Outline}
\addcontentsline{toc}{section}{Outline}

It is hardly conceivable to disentangle all the aspects related to the new physics 
that appear in multicomponent systems. Hence there will surely be some kind of an 
overlap from time to time. Anyway, the main body of this report is organized as follows: 
First, the Chapter \ref{Chap:Topological-defects} sheds the light on the new properties 
associated with the topology of the phenomenological multicomponent models. Next, 
Chapter \ref{Chap:Semi-Meissner} presents some new physical properties that occur 
because of the existence of additional length scales. Finally, the properties of 
multicomponent superconducting states that spontaneously break the time-reversal 
symmetry are discussed in the Chapter \ref{Chap:TRSB}.

More precisely, the Chapter \ref{Chap:Topological-defects} is focused on the 
nature of the topological excitations that appear in multicomponent superconductors.
It is first demonstrated that the condition for the quantization of the magnetic flux 
implies that the elementary topological excitations there, are \emph{fractional vortices}. 
These are field configurations that carry an arbitrary fraction of the flux quantum, 
but that have a divergent energy per unit length. Yet when the fractional vortices 
combine to form an object that carries an integer amount of flux, they form a topological 
defect which has a finite energy. Depending on the relative position of the ensuing 
fractional vortices, the resulting topological defect is either \emph{singular} or 
\emph{coreless}. In the later case, it can then be demonstrated that there exists an 
additional topological invariant of a different nature than the most common winding number. 
However, the most simple analysis shows that typically the fractional vortices attract 
each other to form a singular defect. It follows that a stabilizing mechanism is necessary 
for the existence of coreless defects. Various occurrence of such stable coreless 
topological defects, termed \emph{skyrmions}, are discussed along that chapter. 
As they have a different core structure, the skyrmions, can interact differently than 
the singular (Abrikosov) vortices, and thus have significantly different observable 
properties.

While the coreless defects feature interesting new properties, the singular defects 
also exhibit a rich new physics. This new physics of the singular defects is discussed 
in the Chapter \ref{Chap:Semi-Meissner}. The properties of the magnetic response of 
superconductors can, to some extent, be seen as the consequence of the interaction 
between vortices. More precisely, the textbook dichotomy that classifies the 
conventional superconductors into type-1 or type-2 can be understood by whether the 
vortices attract or if they repel. The vortex interactions can be determined by the 
analysis of the length-scales of the theory. Vortices attract when the coherence 
length is larger than the penetration depth of the magnetic field (this is the type-1 
regime). On the other hand, if the penetration depth is the largest length scale, 
vortices repel each other (this is the type-2 regime). In multicomponent superconductors 
such a dichotomy is not always possible. Indeed, because they have several superconducting 
condensates, the multicomponent superconductors usually feature additional length scales. 
It thus can happen that the penetration depth is an intermediate length scale, and that 
the vortex interaction is long-range attractive (as in type-1) and short-range repulsive 
(as in type-2). Such a regime with non-monotonic intervortex forces is termed type-1.5. 
In that regime vortices tend to aggregate to form large clusters surrounded by vortex-less 
regions of the Meissner state. The possible formation of such aggregates strongly impact 
the magnetization properties, as compared to the conventional type-1 or type-2 regimes.

The non-monotonic interactions between vortices are (partially) determined by the 
length-scales, and these are determined by the perturbations of the theory around 
the ground state. As discussed in the Chapter \ref{Chap:TRSB}, certain multicomponent 
superconductors feature unusual ground states that spontaneously break the time-reversal 
symmetry. These states are characterized by ground state relative phases between the 
condensates that are neither 0 nor $\pi$. Depending on the pairing symmetries, there 
exists various such superconducting states, \eg termed $\pip$, $\sis$, $\sid$, $\did$, 
etc. Yet, the focus will mostly be about $\sis$ state, which the simplest extension of 
the most abundant $s$-wave state, that break the time-reversal symmetry. The spontaneous 
breakdown of the time-reversal symmetry in the $\sis$ state typically occurs as a 
consequence of the competition between different phase-locking terms. The phase transition 
to the time-reversal symmetry broken states is of the second order, so it is associated 
with a divergent length scale. Notably this transition can occur within the superconducting 
state, where the penetration depth is finite. It follows that, in the vicinity of the 
time-reversal symmetry breaking transition, the penetration depth can be an intermediate 
length scale, thus leading to the non-monotonic vortex interactions mentioned above. 
Moreover, the time-reversal symmetry is a discrete operation, so if it is spontaneously 
broken, then the ground state has a discrete $\groupZ{2}$ degeneracy in addition to the 
usual $\groupU{1}$. This implies that, in addition to the vortices, the theory allows for 
domain-walls excitations. They interact non-trivially with the vortex matter, and this 
results in a new kind of topological excitations with different magnetization properties.
Finally, the superconducting states that break the time-reversal symmetry also feature 
unusual thermoelectric properties. These can be used to induce specific electric 
and magnetic responses, when exposed to an inhomogeneous local heating.

As explain above, the essential of these effects that appear in multicomponent 
superconductors, are discussed here in the framework of the Ginzburg-Landau theory. 
The Appendix \ref{App:Single-Component}, presents the theoretical framework, and the 
textbook properties of the single-component Ginzburg-Landau theory. It indeed might 
be useful, for example in order to compare with the properties of the multicomponent 
Ginzburg-Landau theories. This may provide a better insight to understand the new 
features of the multicomponent theories.

It is emphasized on several occasions that the Ginzburg-Landau theory is a nonlinear 
classical field theory. Being nonlinear implies that except under very special 
circumstances there are no analytic solutions, and the problem has to be addressed 
numerically. This is in particular the case of the results displayed in this work, 
and of the results that are discussed there. Technical aspects of the numerical 
methods are discussed in the Appendix \ref{App:numerics}. This includes a presentation 
of the finite element methods used to handle the spatial discretization of the partial 
differential equations; also the optimization algorithm to handle the nonlinear 
problem. The numerical construction of topological defects also relies on the 
appropriate implementation of the topological properties for the numerical 
algorithm; this is also discussed there.

\paragraph{Remark:}The next chapters will follow the plan presented above. 
They are written, so that they are as self-contained as possible. Yet there 
will occasionally be some overlap. There will also be some redundancy, in particular 
with the general introduction, when developing the introduction part within 
each chapter.


\graphicspath{{Plots/03-Topological-defects/}}
\chapter{Topological defects in multicomponent systems}		
\label{Chap:Topological-defects}

Unlike single-component Ginzburg-Landau theory, where the topological excitations merely 
consists in quantum vortices, theories with multicomponent order parameters feature a much 
richer spectrum of topological excitations. Multicomponent superconductors and superfluids, 
are described by order parameters for which each of the component is commonly described 
by a complex field. Overall, all the superconducting/superfluid degrees of freedom can 
be cast into a multiplet of complex scalar fields.

The following chapter presents results concerning the properties of the topological 
excitations in theories of superconductivity featuring multiple order parameters or 
order parameters with multiple components. In the context of superconductivity, 
theories with multiple superconducting gaps where considered from the earlier days of 
the Bardeen-Cooper-Schrieffer theory \cite{Moskalenko:59,Suhl.Matthias.ea:59,Tilley:64}. 
Yet these multicomponent/multiband theories where for a long time considered to describe 
exotic materials. In the recent years however, there have been an increased interest 
in such materials, as the number of known multicomponent superconductor have been 
rapidly growing. For example materials such as \SRO \cite{Maeno.Hashimoto.ea:94,
Mackenzie.Maeno:03}, \MGB \cite{Nagamatsu.Nakagawa.ea:01}, heavy fermion compounds 
such as UPt$_3$ \cite{Joynt.Taillefer:02}, the family of iron based superconductors 
\cite{Kamihara.Watanabe.ea:08}, are understood to be multicomponent/multiband.

The topological properties of multicomponent systems have been widely investigated
in the context of superfluid $^3$He, see \eg \cite{Mizushima.Tsutsumi.ea:16,
Mizushima.Tsutsumi.ea:15}, and the detailed books \cite{Volovik:92,Volovik}. 
Superfluid $^3$He has been particularly known to host a broad variety of unusual 
topological defects \cite{Anderson.Toulouse:77,Thuneberg:86,Salomaa.Volovik:86,
Salomaa.Volovik:87,Tokuyasu.Hess.ea:90}.
More recently, in the context ultracold atomic gases, the topological properties 
of spinor Bose-Einstein condensates were also investigated in great details 
\cite{Maekelae.Zhang.ea:03,Kawaguchi.Kobayashi.ea:10,Kawaguchi.Ueda:12,Ueda:14}.
This chapter presents the topological properties of phenomenological, multicomponent,
Ginzburg-Landau models of superconductivity. An overview of the properties of the 
conventional, single-component, Ginzburg-Landau models of superconductivity is given 
in the Appendix \ref{App:Single-Component}. This might indeed be useful for the 
comparison with the new properties that appear in multicomponent systems.

In the context of multicomponent superconductors and superfluids, the most elementary 
topological excitations are \emph{fractional vortices}. For superfluids, these objects 
carry a fraction of the circulation of the superfluid velocity, while for superconductors 
they carry a fraction of the flux quantum \cite{Babaev:02,Babaev:04b,Babaev.Ashcroft:07}. 
In short, a \emph{fractional vortex} is a field configuration of a multicomponent 
system for which only a \emph{single} component has a nonzero phase winding.

In some specific models of superconductivity, the fraction carried by the fractional 
vortices is half of a flux quantum (or half of a circulation quantum, in the case of 
superfluids). There, the fractional vortices are rather termed \emph{half-quantum} 
vortices. Half-quantum vortices were originally predicted to exist in $A$-phase of 
superfluid $^3$He \cite{Volovik.Mineev:76a,Mineev:13}. Their existence was relentlessly 
investigated for, and their observation was eventually reported in the polar phase of 
superfluid $^3$He \cite{Autti.Dmitriev.ea:16}.

The search for half-quantum vortices, have also been very active in solid state physics.
In particular for superconductors which have been argued to have a $p$-wave pairing, 
such as \SRO \cite{Chung.Bluhm.ea:07,Chung.Agterberg.ea:09,Chung.Kivelson:10}. The 
observation of half of a flux quantum steps, in the magnetization curves of mesoscopic 
\SRO samples, was claimed to be the hallmark of half-quantum vortices 
\cite{Jang.Ferguson.ea:11}.
The interest in the realization of half-quantum vortices follows from that their 
excitation spectrum contains zero-energy Majorana fermions \cite{Ivanov:01}. It follows 
that the statistics of vortices is non-Abelian \cite{Ivanov:01}, which could potentially 
be used for quantum computations \cite{Kitaev:03}.

As discussed below, the fractional vortices in multicomponent superconductors do not 
have a finite energy (per unit length) and the only finite energy topological excitations 
carry an integer amount of the flux quantum. It follows that, under usual conditions, 
fractional vortices are  thermodynamically unstable in bulk systems. Note however that 
complex setups, such as mesoscopic samples, can allow for fractional vortices to be 
energetically favoured \cite{Chibotaru.Dao:10,Silaev:11}. Despite the non-finiteness 
of their energy, the fractional vortices are crucially important for multicomponent 
superconductors. Indeed, they can form bound states that carry an integer amount of 
the flux quantum, for which the divergences of the energy compensate.

Hence, fractional vortices are quite elusive objects that, in general, cannot be 
observed individually. Yet their integer flux bound states have finite energy, 
and thus are observable
\footnote{
Here, one could see some kind of an analogy with the quark matter that constitutes 
the nuclei: The individual quarks carry a fraction of the electric charge and they 
are linearly confined to form bound states with an integer charge.
}. 
Such integer flux carrying bound states are termed \emph{composite} vortices. 
There are basically two qualitatively different possibilities to form integer flux 
composite objects. The first is to superimpose the singularities of all constituting 
fractional vortices, and the resulting objects are thus \emph{singular} composite 
vortices. The other possibility is to form a bound state for which the individual 
singularities do not overlap. The ensuing objects are thus \emph{coreless} topological 
defects. As detailed in this chapter, these feature additional topological invariants, 
that can discriminate them from singular defects. Because of these additional topological 
properties, these coreless defects are often termed \emph{skyrmions}.

Fractional vortices are not only important as they are the building blocks of more 
complex topological excitations, they are also the cornerstone of the thermodynamical 
properties of multicomponent systems. In single-component superconductors, the 
superconducting phase transition was demonstrated to be driven by the proliferation  
of thermally excited vortex loops \cite{Peskin:78,Dasgupta.Halperin:81}. Likewise, 
in multicomponent superconductors, this is the proliferation of fractional vortices 
that drives the superconducting phase transitions, as demonstrated for London 
superconductors \cite{Smiseth.Smorgrav.ea:05,Smorgrav.Babaev.ea:05,Herland.Babaev.ea:10}, 
or in Ginzburg-Landau \cite{Smorgrav.Smiseth.ea:05a}. In the presence of an external 
field, fractional vortices also play a role in the melting of vortex lattices 
\cite{Smorgrav.Smiseth.ea:05a}. 
Similarly the thermodynamic properties of multicomponent superfluids strongly depend 
on the role of fractional vortices \cite{Dahl.Babaev.ea:08a,Dahl.Babaev.ea:08b,
Dahl.Babaev.ea:08}.

This chapter, about the topological properties of multicomponent superconductors thus 
heavily relies on the concept of fractional vortices. They are indeed crucial in our 
understanding of the responses of the multicomponent systems. 
Below is a plan that details the structure of this chapter, followed by a brief summary 
of the author's contributions about these topological properties.

\subsection*{Plan of the Chapter}

As a starting point, the Section \ref{Sec:Fractional-vortices} addresses the question 
of the flux quantization in multicomponent superconductors. It is shown there, that 
the flux quantization formally allows the existence of fractional vortices. Their 
basic properties are also discussed. In multicomponent systems, the fractional vortices 
are field configurations, where only a single condensate has a phase winding, while the 
others do not. The energy of individual fractional vortices is divergent. However, 
as already emphasized, the energetic divergence of fractional vortices disappears if 
they form bound states.

This implies that, in bulk systems, only composite objects have finite energy. 
The Section \ref{Sec:Addtional-topological} further develops on the topological 
properties of the composite topological defects for multicomponent superconductors. 
In particular, there exist a hidden topological invariant associated with the topology 
of the complex projective space, that characterizes coreless topological defects. 
This invariant, that classifies the maps $\Real^2\to\groupCP{N-1}$, discriminates 
coreless from singular vortices. In the case of a two-component system, the target 
$\groupCP{1}$ space can be identified with the unit two-sphere $\groupS{2}$. 
The topological invariant can thus be interpreted as the Hopf index, and can be used
to characterize knotted vortices in two-component superconductors.

The flux quantization implies that these additional invariant are non-zero, as long as 
not all of the superconducting condensates simultaneously vanish. That is, as long as the 
fractional vortices in the different components do not overlap. The interaction between 
fractional vortices is discussed in Section \ref{Sec:Fractional-vortices:Interaction}. 
Because the interaction between fractional vortices is attractive, the observation of 
coreless topological defects is rather difficult. As explained later on, various mechanisms 
can compensate the attraction between fractional vortices and thus lead the formation 
of coreless defects. These coreless defects that consist in a bound state of fractional 
vortices are often termed \emph{skyrmions}. This terminology originates in the existence 
of a formal relation between the two-component Ginzburg-Landau models and the Skyrme-Faddeev 
model. The relation between these two models is explained in 
Section \ref{Sec:Fractional-vortices:Skyrme-Faddeev}.

The Section \ref{Sec:Skyrmions} presents various situations, origintating in different  
physical mechanisms, that allow the stabilization of coreless defects rather than singular 
vortices.
First, in Section \ref{Sec:Mixtures}, in a model of mixtures of condensates with 
commensurate charges, introduced in \CVcite{Garaud.Babaev:14a}. In this rather exotic 
model, the different superconducting condensates can feature different electric charges 
(\ie different coupling to the gauge field). There, fractional vortices are naturally 
split, and thus form coreless bound states. This model can be applied to describe the 
superconducting state for liquid metallic deuterium, where the electronic Cooper pairs 
coexist with a Bose-Einstein condensate of deuterons.

Next, in Section \ref{Sec:Andreev-Bashkin}, the dissipationless intercomponent drag, 
known as the Andreev-Bashkin effect, is demonstrated to be responsible for the existence 
of skyrmions \CVcite{Garaud.Sellin.ea:14}. Furthermore, the dissipationless drag can 
also stabilize knotted bound states of fractional vortices \CVcite{Rybakov.Garaud.ea:19}. 
These knots, characterized by the Hopf index, are hence termed \emph{hopfions}. 
Interestingly, these hopfions remind Kelvin's earlier idea of knotted vortices 
of luminiferous aether to explain classification of atoms.

Next, the properties of topological defects that occur in superconducting $\sis$ states 
are discussed in Section \ref{Sec:Chiral-Skyrmions}. As discussed in more details in 
Chapter \ref{Chap:TRSB}, these $\sis$ states break a discrete $\groupZ{2}$ symmetry 
associated with the time-reversal symmetry, in addition to the usual $\groupU{1}$ gauge 
symmetry. 
The spontaneous breakdown of a discrete symmetry is associated with formation of domain 
walls. Following \CVcite{Garaud.Babaev:14} these domain walls can be formed by thermal 
quench, and geometrically stabilized against collapse. As demonstrated in 
\CVcite{Garaud.Carlstrom.ea:11} and \CVcite{Garaud.Carlstrom.ea:13}, the complex 
interaction between domain-walls and fractional vortices leads to the existence of new 
skyrmionic states.

\subsection*{Summary of the results that are discussed in this chapter}

\begin{itemize}
\setlength\itemsep{0.025em}

\item In \CVcite{Garaud.Babaev:12} and \CVcite{Garaud.Babaev:15a}, we showed that 
the $p_x\!+ip_y$ superconducting state allows for skyrmionic excitations characterized 
by the homotopy invariants of the $\groupS{2}\!\to\!\groupS{2}$ maps. They can be 
alternatively understood as vortices carrying two quanta of the magnetic flux, that are 
energetically favoured as compared to single-quanta vortices \CVcite{Garaud.Babaev:15a}. 
These two-quanta vortices form hexagonal lattices in an external field 
\CVcite{Garaud.Babaev.ea:16}. Close to $\Hc{2}$ the hexagonal lattices of two-quanta 
vortices dissociate into square lattices of single quantum vortices 
\CVcite{Garaud.Babaev.ea:16}, and this picture persists beyond the mean field 
approximation \CVcite{Krohg.Babaev.ea:21}.

\item Demonstration of an unconventional magnetic response in interface superconductors 
with a strong Rashba spin-orbit coupling \CVcite{Agterberg.Babaev.ea:14}. In the clean 
limit, interface superconductors, such as SrTiO$_3$/LaAlO$_3$, are ideal candidates 
to observe coreless defects characterized by homotopy invariants of 
$\groupS{2} \to \groupS{2}$ maps, in addition to those of $\groupS{1} \to \groupS{1}$ maps. 
Similar skyrmionic states also exist in parity-odd nematic superconductors 
\CVcite{Zyuzin.Garaud.ea:17}.

\item Identification of the topological properties of flux-carrying topological defects 
in mixtures of charged condensates that have different (commensurate) electric 
charges \CVcite{Garaud.Babaev:14a}. Such situation is expected to appear for example 
in liquid metallic deuterium.

\item Prediction of a new phase in $\groupU{1}\!\times\!\groupU{1}$ superconductors 
with interspecies dissipationless drag \CVcite{Garaud.Sellin.ea:14}. The dissipationless 
current interaction renders vortices unstable in favour of skyrmions whose long-range 
interaction substantially modifies magnetization processes. These models of 
superconductivity with disspationless drag support stable knotted vortices 
\CVcite{Rybakov.Garaud.ea:19}. These knots share many properties with the knots in 
luminiferous aether conjectured by Kelvin.

\item Discovery of new kind of stable topological solitons in three-component 
superconductors with spontaneously broken time-reversal symmetry 
\CVcite{Garaud.Carlstrom.ea:11}, and \CVcite{Garaud.Carlstrom.ea:13}. These flux 
carrying topological defects, characterized by $\groupCP{2}$ topological invariants 
are skyrmions. Their observation could signal superconducting states that break the 
time-reversal symmetry, for example in some iron based superconductors, as well as 
in Josephson-coupled bilayers of $s_\pm$ and ordinary $s$-wave superconductor.

\end{itemize}

\section{Flux quantization and fractional vortices}
\label{Sec:Fractional-vortices}

In multicomponent superconductors, the flux quantization relation is modified 
compared to that of single-component superconductors (see the background discussion 
in Section \ref{Sec:Single-component:Quantization}). This modified relation implies 
the existence of vortices that carry arbitrary fractions of the elementary flux quantum 
$\Phi_0$, without violating the flux quantization itself. To illustrate this feature 
of multicomponent systems, let consider here a restriction of the generic free energy 
\Eqref{Eq:General:FreeEnergy}, in the absence of mixed gradient terms. Namely, the 
gradient coupling matrix $\kappa_{ab}$ is the identity $\kappa_{ab}=\delta_{ab}$, 
and the Ginzburg-Landau free energy reads as 
\Equation{Eq:Quantization:FreeEnergy:1}{
\F/\F_0=\bigintsss \frac{1}{2}\big|\Curl\A\big|^2 
+\sum_a\frac{1}{2}\big|\D\psi_a\big|^2 
+ V(\Psi,\Psi^\dagger)	\,.
} 
Here again, $\psi_a=|\psi_a| e^{i\varphi_a}$ are complex fields representing the 
superconducting condensates, labelled by the index $a=1,2,\cdots,N$. For the moment, 
the specific structure of the potential $V(\Psi,\Psi^\dagger)$ is rather unimportant. 
The potential will be specified later when it is necessary. 
Again, besides the potential term, the condensates are indirectly coupled by 
the electromagnetic interaction via the gauge derivative in the kinetic term 
$\D=\Grad+ie\A$. The Amp\`ere-Maxwell equation \Eqref{Eq:General:Maxwell} now 
reads as 
\Equation{Eq:Quantization:Maxwell}{
\Curl\B+\J=0 \,,
}
where the supercurrent is
\Equation{Eq:Quantization:Current}{
\J\equiv e\sum_a|\psi_a|^2\big(\Grad\varphi_a+e\A\big)
=e^2\varrho^2\A+e\sum_a|\psi_a|^2\Grad\varphi_a \,,
~~~\text{with}~~\varrho^2=\sum_a|\psi_a|^2	\,,
}
here $\varrho^2$ is the total superconducting density. Here again, the total 
superconducting current $\J$ can be decomposed in terms of the contributions 
of the partial currents $\J^\oa$ carried by the individual condensate $\psi_a$, as
\Equation{Eq:Quantization:Current:2}{
\J=\sum_a\J^\oa \,,
~~~~~\text{with}~~~~\J^\oa=e|\psi_a|^2\big(\Grad\varphi_a+e\A\big)	\,.
}

\subsection{Separation in charged and neutral modes}

To understand the role of the fundamental excitations (\ie fractional 
vortices), the Ginzburg-Landau free energy \Eqref{Eq:Quantization:FreeEnergy:1}
can be rewritten into \emph{charged} and \emph{neutral} modes, 
by expanding the kinetic term as 
\Align{Eq:Quantization:Kinetic:1}{
\sum_a\big|\D\psi_a\big|^2 &= \sum_a\big(\Grad|\psi_a|\big)^2 
							 +\sum_a|\psi_a|^2\big(\Grad\varphi_a+e\A)^2 \\
							&= \sum_a\big(\Grad|\psi_a|\big)^2 
							 +\sum_a|\psi_a|^2\big(\Grad\varphi_a)^2
							 +\A\cdot\Big(e^2\A\sum_a|\psi_a|^2
							 		+2e\sum_a|\psi_a|^2\Grad\varphi_a \Big)\,.
}
Now, using the definition of the current \Eqref{Eq:Quantization:Current}, allows 
to eliminate the vector potential, and the kinetic term thus reads as:
\Align{Eq:Quantization:Kinetic:2}{
\sum_a\big|\D\psi_a\big|^2 &= \sum_a\big(\Grad|\psi_a|\big)^2 
							 +\sum_a|\psi_a|^2\big(\Grad\varphi_a)^2 \\
						&+\frac{1}{e^2\varrho^2}	 
						\Big(\J-e\sum_a|\psi_a|^2\Grad\varphi_a \Big)\cdot\Big(\J+e		
						\sum_a|\psi_a|^2\Grad\varphi_a \Big)  \\
 &= \sum_a\big(\Grad|\psi_a|\big)^2 
				 +\sum_a|\psi_a|^2\big(\Grad\varphi_a)^2 
				+\frac{\J^2}{e^2\varrho^2}
				-\frac{\big(\sum_a|\psi_a|^2\Grad\varphi_a\big)^2}{\varrho^2} \\
 &= \sum_a\big(\Grad|\psi_a|\big)^2 +\frac{\J^2}{e^2\varrho^2}
			+\frac{1}{\varrho^2}\sum_{a,b}
			|\psi_a|^2|\psi_b|^2\Grad\varphi_a\cdot(\Grad\varphi_a-\Grad\varphi_b)\\
	 &= \sum_a\big(\Grad|\psi_a|\big)^2 +\frac{\J^2}{e^2\varrho^2}
			+\sum_{a,b>a}\frac{|\psi_a|^2|\psi_b|^2}{\varrho^2}
			(\Grad\varphi_a-\Grad\varphi_b)^2	\,.
}
Hence the kinetic energy can be expressed in terms of three contributions:
the density term, the charged mode that involves the current $\J$, and the neutral 
mode which involves only the relative phase $\varphi_{ab}:=\varphi_{b}-\varphi_{a}$
between condensates. 
The free energy now can be written as
\Equation{Eq:Quantization:FreeEnergy:3}{
\F/\F_0=\bigintsss \frac{\B^2}{2} 
+\sum_a\frac{1}{2}\big(\Grad|\psi_a|\big)^2 
+\frac{\J^2}{2e^2\varrho^2}
+\sum_{a,b>a}\frac{|\psi_a|^2|\psi_b|^2}{2\varrho^2}(\Grad\varphi_{ab})^2
+ V(\Psi,\Psi^\dagger)	\,.
}

\subsection{Factional vortices}
\label{Sec:Factional-vortices}

The existence of vortices carrying an arbitrary fraction of the flux quantum follows 
from the evaluation of the flux for a multicomponent superconductors. The Stokes' 
theorem implies that the flux of the magnetic field through a given area ${\cal A}$ 
can be expressed as the line integral over the contour ${\cal C}$ which bounds that 
area $\Phi=\int_{\cal A}\B\!\cdot\!d{\bs S}=\oint_{\cal C} \A\!\cdot\!d{\bs\ell}$. 
Given the definition of the current \Eqref{Eq:Quantization:Current}, the vector 
potential can be expressed in term of the current $\J$ and of the individual phase 
gradients $\Grad\varphi_a$. Hence, the magnetic flux reads as
\Equation{Eq:Quantization:Flux:1}{
   \Phi=\frac{1}{e^2\varrho^2}
    \bigointsss_{\!\!\!\!\!\!{\cal C}} 
    \left(\J-e\sum_{a}|\psi_a|^2\Grad\varphi_a\right)\!\cdot\!d{\bs\ell}\,.
} 
Given a large contour ${\cal C}$, finite energy considerations imply that the total 
current $\J$ vanishes on that contour (Meissner screening implies that the current 
is exponentially suppressed), and that the individual densities $|\psi_a|$ are constant 
to their ground state value. Indeed, the spontaneous breakdown of the $\groupU{1}$ 
symmetry implies that the vector potential $\A$ is massive, and so is the current 
$\J$ (\ie the Meissner effect). The first term in \Eqref{Eq:Quantization:Flux:1} 
thus vanishes, and the flux reads as 
\Equation{Eq:Quantization:Flux:2}{
   \Phi=\frac{-1}{e\varrho^2}
    \bigointsss_{\!\!\!\!\!\!{\cal C}} 
    		\sum_{a}|\psi_a|^2\Grad\varphi_a\!\cdot\!d{\bs\ell}
=\frac{\Phi_0\sum_{a}|\psi_a|^2}{2\pi\varrho^2}
    \bigointsss_{\!\!\!\!\!\!{\cal C}} \Grad\varphi_a\!\cdot\!d{\bs\ell}\,,
}
where $\Phi_0=2\pi/e$ is the flux quantum 
\footnote{
Here, the orientation of the closed integration path is chosen 
so that the flux is positive. 
}.

Each of the condensate has to be single valued, hence the phase of the complex fields 
$\varphi_a$ winds only an integer number of times $n_a$ and thus 
$\oint_{\cal C}\Grad\varphi_a\!\cdot d{\bs\ell}=2\pi n_a$. The individual winding
number $n_a$ of a condensate is independent of the winding of the other condensates. 
It thus makes sense to consider the possibility where only a single condensate, say 
$\psi_a$ has a unit winding: $\oint_{\cal C}\Grad\varphi_a\!\cdot d{\bs\ell}=2\pi$, 
while all the other condensates have zero winding: 
$\oint_{\cal C}\Grad\varphi_b\!\cdot d{\bs\ell}=0$ (for $b\neq a$). 
It results that the configuration for which only one of the condensates has a nonzero 
winding, carries the flux $\Phi_a=\Phi_0|\psi_a|^2/\varrho^2$. Such a configuration 
thus carries only a fraction $|\psi_a|^2/\varrho^2$, of the elementary 
flux quantum $\Phi_0$. 
Conversely, if all components have the same winding number ($n_1=n_2=\cdots\equiv n$), 
then the flux is quantized: $\Phi=n\Phi_0$. 

The configurations, with a winding in only a single condensate, and that hence carry 
only a fraction the elementary flux quantum $\Phi_0$ are called \emph{fractional 
vortices} \cite{Babaev:02}. As earlier mentioned, these objects occupy a central 
place in the statistical properties of multicomponent superfluids \cite{Babaev:05,
Dahl.Babaev.ea:08,Dahl.Babaev.ea:08a,Dahl.Babaev.ea:08b,Herland.Babaev.ea:10} and of
multicomponent superconductors \cite{Babaev.Sudbo.ea:04,Babaev:04a,Smorgrav.Babaev.ea:05,
Smorgrav.Smiseth.ea:05a,Smiseth.Smorgrav.ea:05,Babaev.Ashcroft:07}.

Note however that the existence of vortex excitations carrying only a fraction of the 
flux quantum, does not contradict the traditional arguments for the quantization of 
the magnetic flux. Indeed when considering straight vortex lines, it turns out that 
fractional vortices have an infinite energy per unit length.
This follows from that a single fractional vortex induces a nonzero winding of the 
relative phases in the neutral modes of \Eqref{Eq:Quantization:FreeEnergy:3}. 
Thus straight fractional vortices have logarithmically divergent energy.
Indeed, consider for example the simplest case where only $n_1=1$ and $n_b=0$ 
($b\neq 1$). Then the part of the free energy containing gradients of the relative 
phase is 
\Equation{Eq:Quantization:Neutral}{
\cdots+\sum_{b\neq1}\frac{|\psi_1|^2|\psi_b|^2}{2\varrho^2}
			(\Grad\varphi_{1b})^2 +\cdots \,.
}
At a large distance $r$ from the core, the densities are approximately 
constant and thus the contribution of the neutral sector is approximated by 
\Equation{Eq:Quantization:Logarithmic}{
\sim \int_{r_0}^r r^\prime dr^\prime 
	\left(\frac{1}{r^\prime}\partial_\theta\varphi_{1b}\right)^2
\sim \int_{r_0}^r \frac{dr^\prime}{r^\prime}
\sim\log \frac{r}{r_0}	\,,
}
Where $r$ and $\theta$ are respectively the radius, and the polar angle of cylindrical 
coordinates. Note that if $V(\Psi^\dagger,\Psi)$ features phase-locking terms 
(\ie potential terms involving the relative phase $\varphi_{ab}$, like the Josephson 
coupling), then the logarithmic divergence of the energy becomes linear. This is for 
example discussed in details, in the appendix of \CVcite{Garaud.Carlstrom.ea:13}. 
In any case, the  divergence due to the winding in the relative phase puts a strong 
energy penalty on the existence of fractional vortices. 

\begin{wrapfigure}{R}{0.5\textwidth}
\hbox to \linewidth{ \hss
\includegraphics[width=.975\linewidth]{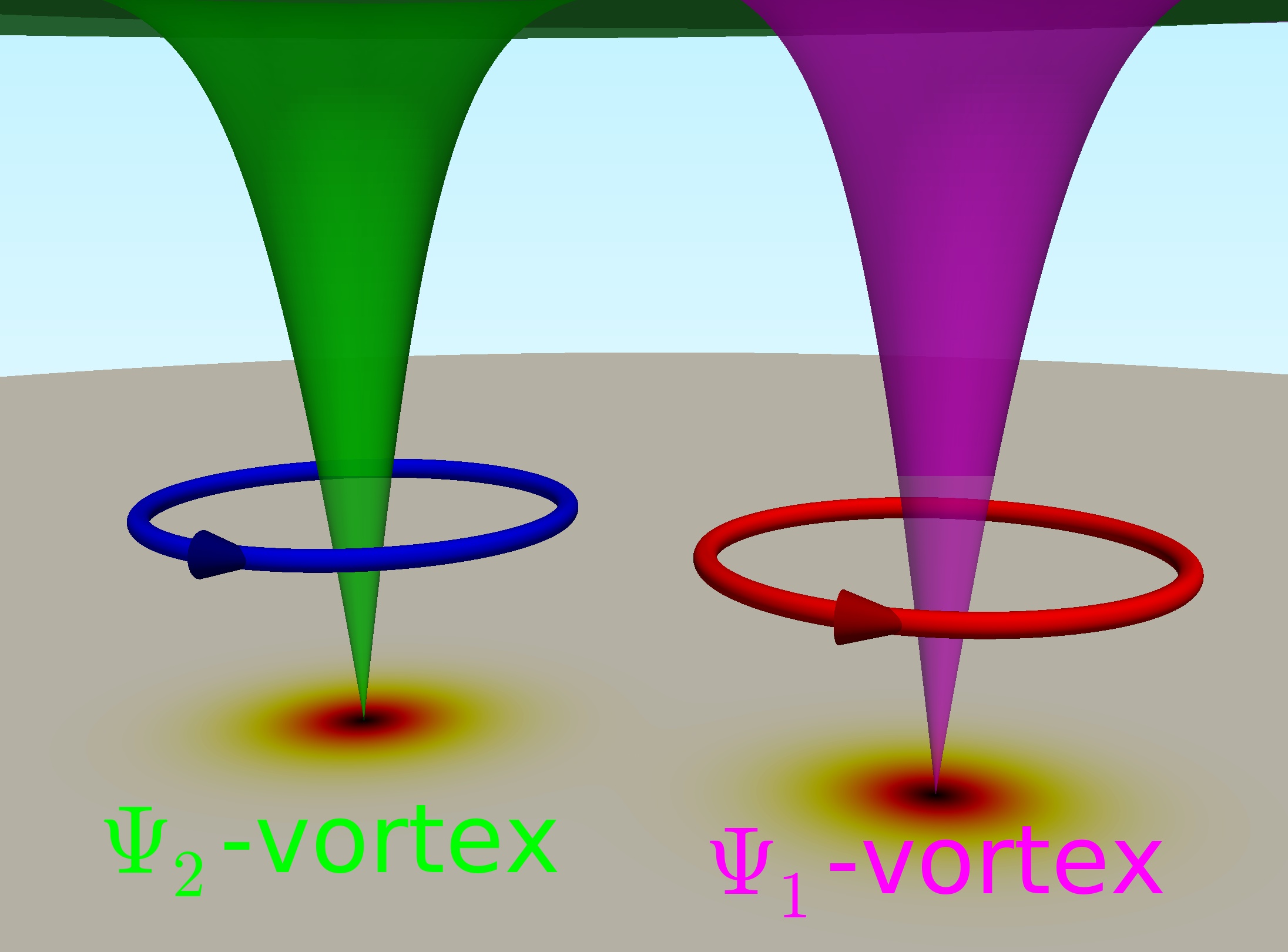}
\hss}
\caption{
Schematic illustration of a composite vortex, in the case of a two-component system.
The cones show the densities $|\psi_a|$ that vanish at the vortex core. The closed 
arrows denote the phase winding, and the circulation of the supercurrents. 
}
\label{Fig:Composite-vorticity}
\end{wrapfigure} 
On the other hand, if \emph{all} condensates have the same winding number, then 
there is no winding of the relative phases. As a results, the long-range contribution 
of the neutral modes vanishes, and the corresponding energy is finite. Thus, the only 
configurations that yield a finite energy per unit length are those where all condensates 
have the same winding, and hence carry an integer flux. As a result, the configurations 
with a fractional flux cannot be excited in bulk superconductors. Note however that they 
can be stabilized near boundaries \cite{Silaev:11,Agterberg.Babaev.ea:14}, in mesoscopic 
samples \cite{Babaev:02,Bluhm.Koshnick.ea:06,Chibotaru.Dao.ea:07,Chibotaru.Dao:10,
Geurts.Milosevic.ea:10} or in samples with geometrically trapped domain walls 
\cite{Garaud.Babaev:14}.
In bulk systems, the condition for the finiteness of the energy, thus allows only 
topological defects that carry integer flux quanta. These objects, made of fractional 
vortices in each of the components, are thus \emph{composite} objects. Such a composite 
vortex is sketched in \Figref{Fig:Composite-vorticity}.

\begin{wrapfigure}{L}{0.45\textwidth}
\hbox to \linewidth{ \hss
\includegraphics[width=.975\linewidth]{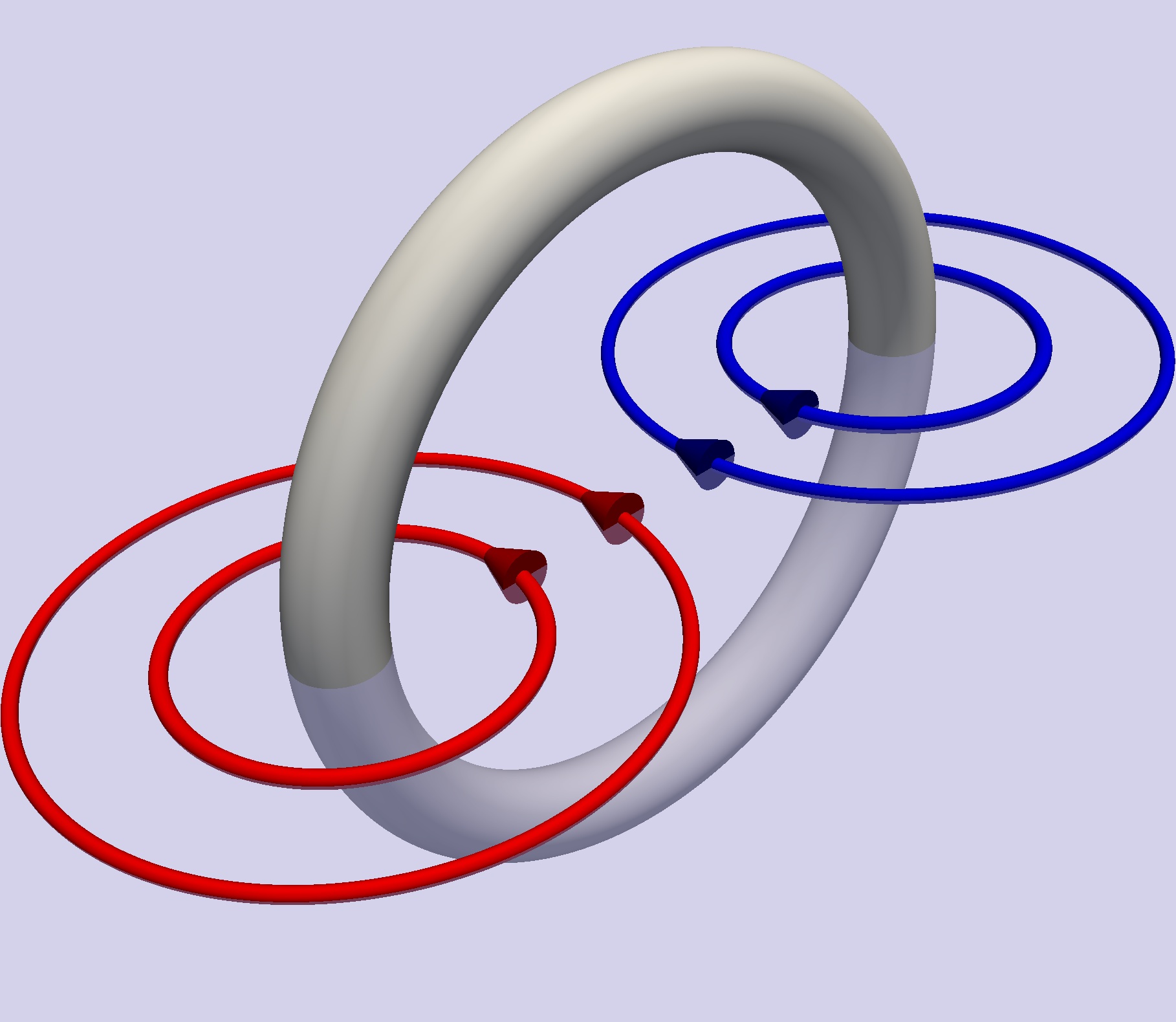}
\hss}
\caption{
Schematic illustration of a vortex loop denoted by the gray tube. The closed 
arrows denote the phase winding, and the circulation of the supercurrents. 
Clearly, a vortex loop has no net winding at large distance.
}
\label{Fig:Vortex-loop}
\end{wrapfigure} 
Remark here that this derivation is essentially two-dimensional. This means that 
this divergence occurs in either in two dimensions or in three-dimensional systems 
with translation invariance along the the third direction. This is reason for the
emphasis on \emph{straight} vortices. In three-dimensional systems, fractional vortices 
oriented along the third direction have a divergent energy per unit length. So they 
will still be energetically penalized. On the other hand, loops of fractional vortices 
always have a finite energy. This can heuristically be understood as follows: 
As sketched in \Figref{Fig:Vortex-loop}, a section of a vortex loop can be seen as
pair of a vortex and an anti-vortex. In a plane such a vortex/anti-vortex pair has 
no net winding at large distance, so it is topologically trivial, and the energy 
is finite. Moreover because it is topologically trivial, there is no topological 
protection, so a vortex and an anti-vortex attract each other until they  
annihilate. Similarly, a vortex loop is topologically trivial, since it has 
no net winding at large distance. A vortex loop tend to collapse because of its 
line tension, quite similarly to a vortex/anti-vortex pair that annihilate each other.

In two-dimensions, vortices, either fractional or composite, are characterized by 
$\groupS{1}\to\groupS{1}$ topological maps. The first circle $\groupS{1}$ denotes the 
closed path faraway from the vortex core (that is homeomorphic to a circle) while the 
second one (the target circle) corresponds to $\groupU{1}$ rotations. Heuristically, 
the $\groupS{1}\to\groupS{1}$ maps have the following meaning: they count how many 
times the target circle is covered while going along the closed path faraway from 
the vortex core (\ie the number of phase windings). Importantly, this number can be 
calculated just by inspecting the closed path faraway for the vortex core. This is 
because the associated density of the topological invariant is a total divergence. 
As discussed later on, this is not the case of the extra invariants.

Depending on how the constituting fractional vortices are located relative 
to each other, there exist two qualitatively different ways to construct the 
composite topological defects. More precisely, as discussed below, the topological 
properties depend on whether their individual singularities overlap, or not.
If the singularities do not overlap, the composite topological defect is coreless 
and can be characterized by an additional, topological invariant. This motivates 
the generic terminology of \emph{skyrmions} \cite{Garaud.Carlstrom.ea:13}.
On the other hand, if all fractional vortices are co-centred, the resulting 
bound state is a singular (multicomponent) vortex.

\subsection{Additional topological properties in multicomponent systems}
\label{Sec:Addtional-topological}

In addition to the winding number, which is the only topological invariant in 
single-component superconductors, multicomponent superconductors can be characterized 
by extra topological properties. As previously emphasized, the $\groupU{1}$ topological 
invariant is associated with the total phase winding at spatial infinity. Depending 
on the nature of the topological defect under consideration, the additional invariant 
is of different nature. If the objects considered are straight, line-like, topological 
defects being the bound state of straight fractional vortices, the additional invariants 
are given as surface integral characterizing skyrmions. If on the other hand, the 
objects consist of closed loops of fractional vortices, the additional invariant 
is given as a volume integral characterizing hopfions. Both skyrmion and hopfion 
numbers are discussed below. Note that while the skyrmion number is well define for 
any number of superconducting condensates, the hopfion number is formally defined 
only for two-component superconductors.

\subsubsection{\texorpdfstring{$\groupCP{N-1}$}{CPN} topological invariant -- Skyrmion number}
\label{Sec:CPN-charge}

The winding number, which is defined as a line integral over a closed path, is 
associated with the maps $\groupS{1}\to\groupS{1}$. It is related to the elements 
of the first homotopy of the circle: $n\in\pi_1(\groupS{1})=\groupZ{}$. In contrast, 
multicomponent superconductors can be characterized by an additional $\groupCP{N-1}$ 
topological index, which is defined as an integral over the plane. Given the 
$N$-component complex vector $\Psi$, the $\groupCP{N-1}$ topological index is 
\CVcite{Garaud.Carlstrom.ea:13}
\Equation{Eq:CPN:Charge}{
   \Q(\Psi)=\int_{\mathbb{R}^2}\frac{i\varepsilon_{ji}}{2\pi|\Psi|^4} \Big[
   |\Psi|^2\partial_i\Psi^\dagger\partial_j\Psi
   +\Psi^\dagger\partial_i\Psi\partial_j\Psi^\dagger\Psi
   \Big]dxdy\,,
}
where $\varepsilon$ is the Levi-Civita symbol.
Provided $\Psi\neq0$ (\ie if singularities do not overlap), the $\groupCP{N-1}$ 
index $\Q(\Psi)$ is an integer number and it is equal to the number of flux quanta: 
$\Q(\Psi)=\int B/\Phi_0=n$ ($\Phi_0$ being the flux quantum and $n$ the number of flux 
quanta) \cite{Garaud.Carlstrom.ea:13}. It results that for a singular vortex, where 
all the superconducting condensates simultaneously vanish (\ie $\Psi=0$), the skyrmion 
number $\Q(\Psi)=0$. On the other hand, if the singularities are non-overlapping 
(\ie $\Psi\neq0$), then $\Q(\Psi)\in \groupZ{}$ and the quantization condition holds. 
Then $\Q(\Psi)$ is a useful quantity that can differentiate between singular vortices 
and skyrmions (which are coreless defects).

It should be emphasized again that unlike the flux-quantization condition
\Eqref{Eq:Quantization:Flux:2}, the integral formula for the topological charge 
$\Q(\Psi)$ above is valid only for field configurations for which $\Psi$ never 
vanishes. The flux is still quantized for ordinary singular vortices, for which 
$\Psi$ vanishes, but it is no longer associated with the topological charge $\Q(\Psi)$, 
rather with the $\groupU{1}$ topological invariant related to the total phase 
winding at spatial infinity (the usual winding number).

Note that the topological number $\Q(\Psi)$ is calculated as an integral over the plane 
$\Real^2$. Hence it formally characterizes either two-dimensional systems ($\Real^2$), 
or three-dimensional systems with translation invariance normal to the plane (\ie 
$\Real^2\times\Real$). In the later case, $\Q(\Psi)$ should be interpreted as a
linear density of topological charge.

\paragraph{Hints of the demonstration:}
For the rigorous derivation of the flux and topological charge 
quantization, see \CVcite{Garaud.Carlstrom.ea:13}. Using, the definition 
\Eqref{Eq:Quantization:Current} of the total supercurrent $\J$, and the relation 
$\Im(\Psi^\dagger\Grad\Psi)=\sum_a|\psi_a|^2\Grad\varphi_a$, the gauge field reads as
\Equation{Eq:Quantization:Current:GF}{
\A=\frac{1}{e^2\varrho^2}\Big(\J-e\sum_a|\psi_a|^2\Grad\varphi_a\Big) 
  =\frac{1}{e^2\varrho^2}\Big(\J-e\Im(\Psi^\dagger\Grad\Psi)    \Big)
\,,
}
where $\varrho^2=\sum_a|\psi_a|^2=\Psi^\dagger\Psi$ is the total superconducting density. 
It follows that the magnetic field can be expressed as
\SubAlign{Eq:Quantization:Current:B}{
B_k&=\frac{1}{e}\varepsilon_{kij}\left\lbrace
\partial_i\left(\frac{\J}{e\varrho^2}\right)
-\partial_i\left(\sum_a\frac{|\psi_a|^2\Grad\varphi_a}{\varrho^2}\right)  
\right\rbrace \label{Eq:Quantization:Current:B:a}\\
&=\frac{1}{e}\varepsilon_{kij}\left\lbrace
\partial_i\left(\frac{\J}{e\varrho^2}\right)
+\frac{i}{\varrho^4}\left[
\varrho^2\nabla_i\Psi^\dagger\nabla_j\Psi
+(\Psi^\dagger\nabla_i\Psi)(\nabla_j\Psi^\dagger\Psi)
\right]  
\right\rbrace \label{Eq:Quantization:Current:B:b}
\,,
}
where $\varepsilon_{ijk}$ is the Levi-Civita symbol. Going from the first to 
the second line of \Eqref{Eq:Quantization:Current:B} is merely done by developing 
the second term and by eliminating the contributions that are symmetric under 
$i\leftrightarrow j$ (since they are contracted with the Levi-Civita symbol which 
is antisymmetric). Obviously the flux of $\B$ is quantized. Indeed, applying the 
Stokes theorem to the equation \Eqref{Eq:Quantization:Current:B:a} yields the relation 
\Eqref{Eq:Quantization:Flux:1}. Now, computing the flux from the second equation 
\Eqref{Eq:Quantization:Current:B:b} gives 
\Equation{Eq:Quantization:Flux:1:new}{
   \Phi=\frac{1}{e^2\varrho^2}
    \bigointsss_{\!\!\!\!\!\!{\cal C}} 
    \J\!\cdot\!d{\bs\ell}
    -\bigintsss    
	\frac{i\varepsilon_{ij}}{e\varrho^4}\Big[ \varrho^2\nabla_i\Psi^\dagger\nabla_j\Psi
	+(\Psi^\dagger\nabla_i\Psi)(\nabla_j\Psi^\dagger\Psi)\Big] dxdy
    \,.
}
Since the Meissner current vanishes asymptotically, this determines the relation between 
the topological charge \Eqref{Eq:CPN:Charge} and the magnetic flux $\int B=\Phi_0\Q(\Psi)$ 
with the flux quantum $\Phi_0=2\pi/e$. Again one can see that $\Q$ is a surface integral, 
since the 2-form integrand is not closed. 

\paragraph{Remark:}
It is important to stress here that the definition of the magnetic field 
\Eqref{Eq:Quantization:Current:B} features two contributions. The first term is the  
contribution from the standard superconducting currents, while the second term appears 
only for multicomponent system. This additional term, which involves relative density 
gradients and relative phase gradients, is responsible for various unusual phenomena 
presented later in Chapter \ref{Chap:TRSB}.

\subsubsection{Hopfions in two-component superconductors}

The above discussion considered the topological properties of straight topological 
defects, where the extra invariant is calculated as a surface integral in the plane 
perpendicular to the vortex line. It is also possible to define an additional 
invariant when vortices form close loops, instead of straight defects.
However, this is possible only for two-component superconductors. 
Topological considerations imply that knotted vortex loops in two-component 
superconductors are characterized by an integer topological index $\I$ (see, 
\eg, discussions given in Refs.~\cite{Faddeev:75,Faddeev.Niemi:97,
Babaev.Faddeev.ea:02,Babaev:02b,Jaykka.Hietarinta.ea:08,Radu.Volkov:08,Babaev:09}). 
As for the previous discussion, this index is well defined when fractional vortices 
in the different components do not intersect ($\Psi\neq0$). Here, the superconducting 
degrees of freedom are cast in a 4-dimensional unit vector ${\bs\zeta}=(
\Re\,\psi_1,\Im\,\psi_1,\Re\,\psi_2,\Im\,\psi_2)/\sqrt{\Psi^\dagger\Psi}$.
Note that for ${\mathbf\zeta}$ to be well defined, there should be no zeros of 
$\Psi$, \ie no overlap between core centres of fractional vortices in both 
components.
The finiteness of energy implies that a superconductor should be in the ground state 
at spatial infinity. It follows that infinity is identified with a single field 
configuration (up to gauge transformations). Hence, the vector field ${\bs\zeta}$ 
is a map from the one-point compactified space to the target 3-sphere 
${\bs\zeta}:\groupS{3}\,[\cong\Real^3\cup\{\infty\}]\to\groupS{3}_\Psi$. 
Maps between 3-spheres fall into disjoint homotopy classes, the elements of the 
third homotopy group $\pi_3(\groupS{3}_\Psi)$, which is isomorphic to the integers: 
$\pi_3(\groupS{3}_\Psi)=\groupZ{}$. Hence, ${\bs\zeta}$ is associated with an 
integer number, the degree of the map ${\bs\zeta}$, which counts how many times the 
target sphere $\groupS{3}_\Psi$ is wrapped while covering the whole $\Real^3$ space. 
Field configurations are thus characterized by the topological index 
\Equation{Eq:Hopf:Degree}{
\I :=\mathrm{deg}{\bs\zeta} = -\frac{1}{12 \pi^2}\int_{\mathbb{R}^3}
	\varepsilon_{ijk}\varepsilon_{abcd}\,\zeta_a
	\partial_i \zeta_b
	\partial_j \zeta_c
	\partial_k \zeta_d
	d{\bf r}\,, 
} 
where $\varepsilon$ is the Levi-Civita symbol. As long as ${\bs\zeta}$ is well defined, 
that is unless $\Psi$ has zeros, the index $\I$ is always an integer. Note that, as 
discussed for example in \cite{Jaykka.Hietarinta.ea:08}, the degree of ${\mathbf\zeta}$, 
$\I$ is equal to the Hopf charge of the combined Hopf map 
$h\circ{\mathbf\zeta}:\groupS{3}\to \groupS{2}$.

\subsection{Interactions between fractional vortices}
\label{Sec:Fractional-vortices:Interaction}

The finiteness of the energy dictates that fractional vortices cannot exist individually, 
and should thus form composite objects carrying an integer flux. These maybe be 
characterized by the additional invariant $\Q(\Psi)$, if $\Psi\neq0$. To determine 
whether singularities overlap or not, and thus whether singular vortices or skyrmions 
are favoured, it is necessary at this point to determine how the fractional vortices 
interact together. 
The Ginzburg-Landau free energy describing multicomponent superconductors is a 
non-linear theory and thus detailed investigation of the properties of topological 
defects typically requires numerical simulations. However, analysing of the properties 
in the London limit provides valuable insight on the behaviour of the topological 
excitations. As demonstrated below, in the London limit, straight fractional vortices 
can be mapped to interacting point charges in two dimensions.

The London limit, assumes that the condensates have a constant density 
$|\psi_a|=\mathrm{const}$ everywhere, except at the small cut-off representing 
the vortex core. Using the Amp\`ere's law \Eqref{Eq:Quantization:Current} to replace 
the current by the magnetic field in Eq.~\Eqref{Eq:Quantization:FreeEnergy:3}, the 
free energy further simplifies
\Equation{Eq:Quantization:FreeEnergy:London}{
\F/\F_0=\bigintsss \frac{1}{2}\left(\B^2 + \frac{1}{e^2\varrho^2} |\Curl\B|^2\right) 
+\sum_{a,b>a}\frac{|\psi_a|^2|\psi_b|^2}{2\varrho^2}(\Grad\varphi_{ab})^2
	\,.
}   	
In principle the potential $V(\Psi,\Psi^\dagger)$ can feature phase-locking terms 
that depend on the relative phase $\varphi_{ab}$ between different condensates. 
In the following discussion, it is assumed for simplicity that there are no such 
terms, and thus that the potential is simply constant and can be dismissed.
The interaction energy of non-overlapping fractional vortices can be approximated, 
in this London limit, by considering separately the \emph{charged} and \emph{neutral} 
modes. First of all, the vector calculus identity 
\Equation{Eq:Quantization:Identity}{
|\Curl\B|^2=\B\cdot\Curl\Curl\B-\Div(\B\times\Curl\B)\,,
}
helps to rewrite the \emph{charged} sector of 
\Eqref{Eq:Quantization:FreeEnergy:London} as
\Equation{Eq:Quantization:FreeEnergy:Charged:1}{ 
   \F_{\mbox{\tiny Charged}}=\bigintsss \frac{\B}{2}
   \left(\B+\frac{1}{e^2\varrho^2}\Curl\Curl\B\right)\,.
}
The London equation for a point-like vortex carrying a flux $\Phi_a$, and  
located at $\x_a$, is
\Equation{Eq:Quantization:London}{
   \lambda^2\Curl\Curl\B+\B=\Phi_a\delta(\x-\x_a)\,,
}
where $\lambda$ is the London penetration length defined as 
$\lambda^{-2}=e^2\varrho^2$. The corresponding solutions of the London equation 
\Eqref{Eq:Quantization:London} are given in terms of $K_0$, the modified Bessel 
of the second kind, as
\Equation{Eq:Quantization:London:sol}{
   \B_a(\x)=\frac{\Phi_a}{2\pi\lambda^2}
   K_0\left(\frac{|\x-\x_a|}{\lambda}\right)\,.
}
In the case of two vortices carrying fluxes $\Phi_a$ and $\Phi_b$, and located at 
$\x_a$ and $\x_b$, the source term in London equation reads as 
$\Phi_a\delta(\x-\x_a)+\Phi_b\delta(\x-\x_b)$, and the magnetic field is the 
superposition of two contributions $\B(\x)=\B_a(\x)+\B_b(\x)$. As a results, 
the corresponding energy of the charged sector
 \Align{Eq:Quantization:FreeEnergy:Charged:2}{
    \F_{\mbox{\tiny Charged}}&=\bigintsss\frac{1}{2}
   (\B_a+\B_b )\big[\Phi_a\delta(\x-\x_a)+\Phi_b\delta(\x-\x_b)\big] \nonumber \\
   &=   \frac{\Phi_a\Phi_b}{2\pi\lambda^2}K_0\left(\frac{|\x_2-\x_1|}{\lambda}\right)
   +E_{va}+E_{vb} \,,
}
where $E_{va}\equiv\int\B_a(\x_a)\Phi_a/2$ stands for the (self-)energy of the 
vortex $a$. It results that, the interaction energy between two vortices in 
components $a$ and $b$ reads as
\Equation{Eq:Quantization:Interaction:Charged}{ 
   E^{(int),\mbox{\tiny Charged}}_{ab}=\frac{2\pi|\psi_a|^2|\psi_b|^2}{\varrho^2}
   K_0\left(\frac{|\x_a-\x_b|}{\lambda}\right)\,.
}
The interaction through the charged sector is thus a screened interaction given by 
the modified Bessel function. This interaction is always positive for any $a$, $b$ 
having the same sign of the vorticity. It then determines, a repulsive interaction 
between any kind of fractional vortices with co-directed winding. That is vortices, 
repel while a vortex and an anti-vortex attract each other. The interaction through 
the \emph{neutral} sector, on the other hand, is attractive (resp. repulsive) for 
fractional vortices in the different (resp. same) components $a$ and $b$. 
The interaction here is logarithmic in the separation of the vortices. Note again 
that if there exist phase-locking potential terms, then the interaction is linear 
in the separation. This was for example discussed in detail, in the appendix of 
\CVcite{Garaud.Carlstrom.ea:13}.
The energy associated with the \emph{neutral} sector of 
\Eqref{Eq:Quantization:FreeEnergy:London} reads 
\Equation{Eq:Quantization:FreeEnergy:Neutral:1}{ 
   \F_{\mbox{\tiny Neutral}}=\sum_{a,b>a}
\frac{|\psi_a|^2|\psi_b|^2}{2\varrho^2}\bigintsss(\Grad\varphi_{ab})^2\,.
}
At sufficiently large distance, a phase winding around some singularity located 
at the point $\x_a$, is well approximated by $\varphi_a=\theta$. Where here $\theta$ 
is the polar angle, and thus 
\Equation{Eq:Quantization:GradPhi}{
\Grad\varphi_a=\frac{\etheta}{|\x-\x_a|}=\bs z\times\Grad\ln|\x-\x_a|\,.
}
The interaction between fractional vortices in different condensates, respectively 
located at $\x_a$ and $\x_b$, is calculated as follows: Expanding the neutral sector
\Eqref{Eq:Quantization:FreeEnergy:Neutral:1}, the interacting part reads as
\Align{Eq:Quantization:Interaction:Neutral:1}{ 
   E^{(int),\mbox{\tiny Neutral}}_{ab}&=-\frac{|\psi_a|^2|\psi_b|^2}{\varrho^2}
\bigintsss\Grad\varphi_a\cdot\Grad\varphi_b \nonumber = 
\frac{|\psi_a|^2|\psi_b|^2}{\varrho^2}\bigintsss\varphi_a\Delta\varphi_b
\nonumber \\
&=\frac{|\psi_a|^2|\psi_b|^2}{\varrho^2}
\bigintsss\ln|\x-\x_a|\delta(|\x-\x_b|)
=2\pi\frac{|\psi_a|^2|\psi_b|^2}{\varrho^2}\ln|\x_b-\x_a|
\,.
}
Similarly, the interaction between two vortices in the same condensate $a$ is 
computed by requiring that the phase is the sum of the individual phases 
$\varphi_a=\varphi_a^\oo+\varphi_a^\ot$, while $\varphi_b=0$. Then the 
interaction reads as
\Equation{Eq:Quantization:Interaction:Neutral:2}{ 
   E^{(int),\mbox{\tiny Neutral}}_{aa}=
-2\pi\frac{|\psi_a|^2\sum_{c\neq a}|\psi_c|^2}{\varrho^2}\ln|\x_a^\ot-\x_a^\oo|
\,. 
}
Thus the interaction, via the neutral sector, between in different condensates is 
logarithmically attractive, while it is repulsive for vortices in the same condensate.

\begin{figure}[!htb]
\hbox to \linewidth{ \hss
\includegraphics[width=.5\linewidth]{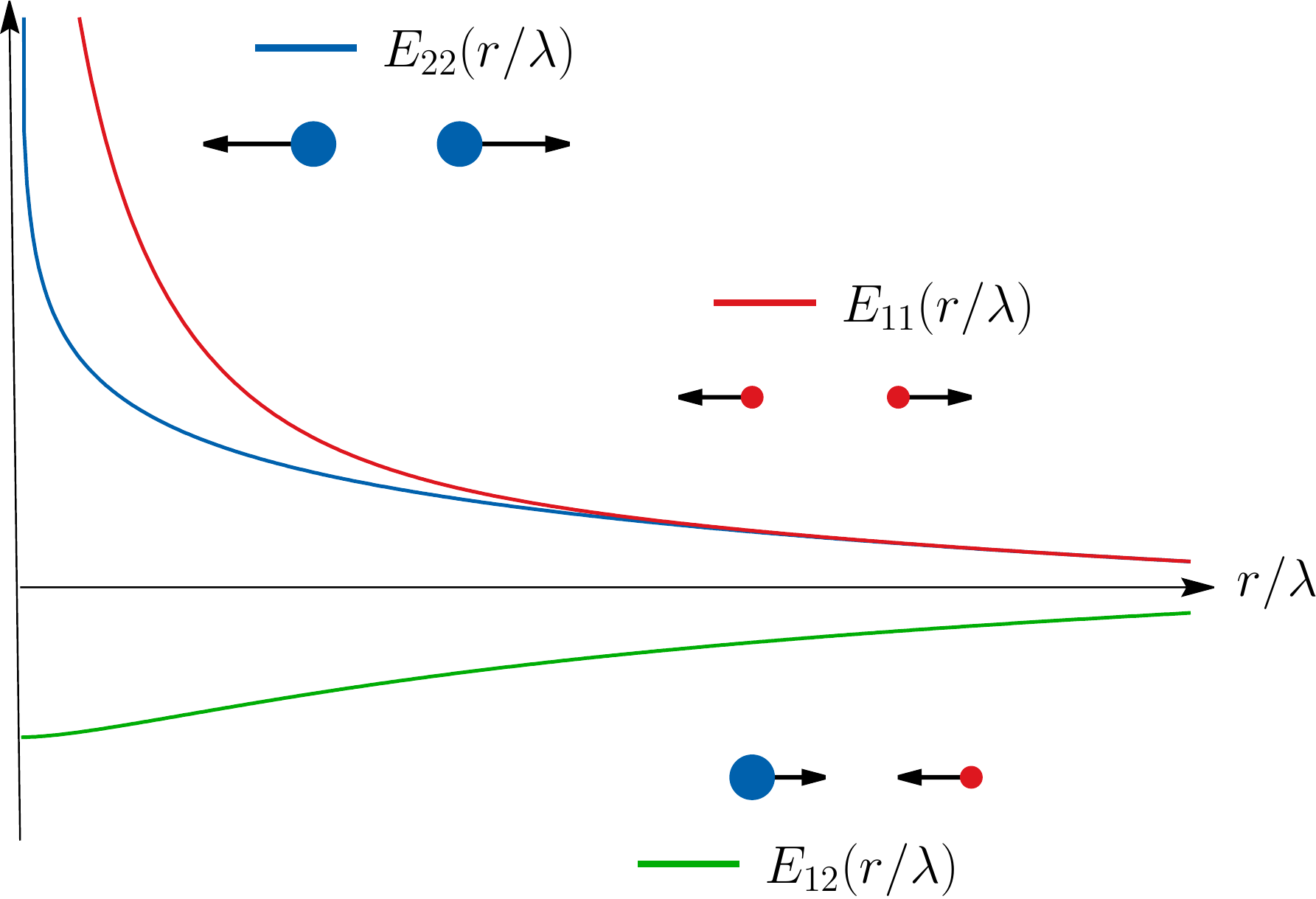}
\hss
\includegraphics[width=.475\linewidth]{Composite-vorticity-cocentred}
\hss}
\caption{
The left panel shows the interaction energies \Eqref{Eq:Quantization:Interaction} 
(with $m=0.2$) between point-like charges associated with vortices in the different 
condensates. The blue (big) dot represents the vortex in $\psi_1$ while the red 
(small) dots represent the vortices in $\psi_2$. Alike charges always repel while 
different charges attract via the long-range logarithmic attraction. 
The right panel displays a schematic illustration of a composite vortex where the 
singularities are co-centred due to the long-range attraction between the fractional 
vortices in different condensates. Note that, in general, there are no reasons 
for the cores of different components to have same size.
}
\label{Fig:London-interaction}
\end{figure}

To summarize, noting the separation $r\equiv|\x_a-\x_b|$ and $R$ the sample's size, 
the interaction energy between fractional vortices in different condensates is 
\Equation{Eq:Quantization:Interaction:Unlike}{
\frac{E^{(int)}_{ab}}{2\pi}=\frac{|\psi_a|^2|\psi_b|^2}{\varrho^2}
    \Big(\ln \frac{r}{R} +K_0\left(\frac{r}{\lambda}\right)\Big)\,.
}
On the other hand, interactions between fractional vortices in the same condensates 
are 
\Equation{Eq:Quantization:Interaction:Like}{
\frac{E^{(int)}_{aa}}{2\pi}= 
-\frac{|\psi_a|^2\sum_{c\neq a}|\psi_c|^2}{\varrho^2} \ln\frac{r}{R} 
+\frac{|\psi_a|^4}{\varrho^2} K_0\left(\frac{r}{\lambda}\right)\,.
}
Equations \Eqref{Eq:Quantization:Interaction:Unlike} and  
\Eqref{Eq:Quantization:Interaction:Like} thus give the different interactions 
between fractional vortices in different condensates. This can be illustrated 
in the case of a two-component superconductor. Choosing the energy scale to be 
$2\pi|\psi_1|^2|\psi_2|^2/\varrho^2$ and defining the parameter $m$ as the ratio 
of densities, $m =|\psi_1|^2/|\psi_2|^2$, the interaction between the fractional 
vortices in the various condensates reads as
\Equation{Eq:Quantization:Interaction}{
E_{11}=\ln\frac{R}{r}+mK_0\left(\frac{r}{\lambda}\right) 			\,,~~
E_{22}=\ln\frac{R}{r}+\frac{1}{m}K_0\left(\frac{r}{\lambda}\right)	\,,~~
E_{12}=-\ln\frac{R}{r}+K_0\left(\frac{r}{\lambda}\right) \,.
}
Thus, the vortex matter in the London limit of a two-component superconductor is 
described by a 2-parameter family $(m,R)$. Figure \ref{Fig:London-interaction} 
shows the profiles of the different interactions \Eqref{Eq:Quantization:Interaction}
between the different kind of fractional vortices.

Note that as $r \to 0$, the modified Bessel functions diverges as $-\ln r$. 
This means that for the interaction between fractional vortices in different 
condensates $E_{12}$, the divergence of the log term is compensated by the 
divergence of the Bessel function. They exactly cancel at $r = 0$, as can be 
seen in\Figref{Fig:London-interaction}. The interaction between the vortices 
in the same condensates is repulsive. 
In multicomponent superconductors, where all condensates have the same number 
of vortices, the vortices in different condensates will attract each other to 
form a bound state of co-centered vortices that minimizes the energy cost of 
the neutral sector \cite{Babaev:02,Smiseth.Smorgrav.ea:05}. As mentioned earlier, 
when there are phase-locking potential terms, then the attraction between fractional 
vortices is no more logarithmic, and it becomes linear, and the confinement becomes 
even stronger.

This means that, at least in the London limit, the tendency for fractional vortices 
(in the simplest model used here), is to simply overlap and thus to form a bound 
state of co-centred vortices, in order to minimize the energy of the neutral sector. 
Subsequently, the charged sector can be minimized by maximizing the separation 
between the co-centered vortices. That is by forming a triangular Abrikosov 
lattice of singular composite vortices. Returning to the discussion about skyrmions 
and vortices, this implies that superconducting condensates tend to simultaneously 
vanish (\ie $\Psi=0$), and thus that the $\groupCP{N-1}$ index $\Q$ \Eqref{Eq:CPN:Charge} 
is zero. 
As a result, forming skyrmions as lowest energy topological excitations is not 
a trivial task, and it typically requires additional ingredients to overcome this 
tendency to form co-centred composite vortices. This is discussed later in the 
Section \ref{Sec:Skyrmions}.

Note that the above remarks about finiteness of the energy and interaction between 
fractional vortices apply for straight vortices in bulk superconducting materials. 
That is, the energetic divergence of the fractional vortices formally occurs only 
in infinite systems. Fractional vortices can yet be stabilized due to finite-size 
effects as for example in mesoscopic samples \cite{Babaev:02,Bluhm.Koshnick.ea:06,
Chibotaru.Dao.ea:07,Geurts.Milosevic.ea:08,Chibotaru.Dao:10,Geurts.Milosevic.ea:10}. 
They can also be thermodynamically stabilized near sample boundaries due to their 
interaction with the Meissner currents \cite{Silaev:11}. This results in a modified 
Bean-Livingston barrier with complex partial vortex penetration as shown in 
\CVcite{Agterberg.Babaev.ea:14}.

Importantly, as mentioned earlier, the loops of vortices carrying a fractional flux 
have only a finite energy. As a result, in three dimensions loops of fractional 
vortices although quite energetic are still formally possible and, although 
dynamically unstable, can for example be thermally excited. Fractional vortex 
loops, actually play a central role in the critical properties of multicomponents 
superconductors \cite{Smiseth.Smorgrav.ea:05,Smorgrav.Smiseth.ea:05a,Herland.Babaev.ea:10}, 
and superfluids \cite{Smorgrav.Babaev.ea:05,Dahl.Babaev.ea:08a,Dahl.Babaev.ea:08b}.

\paragraph{To briefly summarize:}
The elementary topological excitations in multicomponent superconductors are fractional 
vortices. Finite energy considerations dictate that only bound states of fractional 
vortices totalling an integer flux should form in bulk systems, and the intervortex 
interactions promote co-centered vortices. 
However, if the individual singularities do not overlap, such a bound state is 
a coreless defect called a skyrmion, and it is characterized by an additional hidden 
$\groupCP{N-1}$ topological invariant. Scenarii where skyrmions are favoured 
over singular vortices will be discussed in section \ref{Sec:Skyrmions}.
Prior to that, in the next section, it is further discussed that in the particular 
case of two-component systems, the topological properties of the model can also 
be understood using a mapping to a nonlinear $\sigma$-model.

\section{Duality mapping to a Skyrme-Faddeev model, coupled to a massive vector 
\texorpdfstring{$\J$}{J}} 
\label{Sec:Fractional-vortices:Skyrme-Faddeev}

It is interesting to note that generic model of two-component superconductivity 
can be mapped to an easy-plane non-linear $\sigma$-model \cite{Babaev.Faddeev.ea:02,
Babaev:09}. This mapping further motivates the terminology \emph{skyrmion} to 
denote the bound state of well separated fractional vortices. To achieve this 
mapping, it is first of all convenient to consider the free energy expressed in 
terms of charged and neutral modes \Eqref{Eq:Quantization:FreeEnergy:3}. In the 
case of two-components, it reads as
\Equation{Eq:SF:FreeEnergy:1}{
\F/\F_0=\bigintsss \frac{\B^2}{2} 
+\sum_{a=1,2}\frac{1}{2}\big(\Grad|\psi_a|\big)^2 
+\frac{\J^2}{2e^2\varrho^2}
+\frac{|\psi_1|^2|\psi_2|^2}{2\varrho^2}
			(\Grad\varphi_{12})^2
+ V(\Psi,\Psi^\dagger)	\,.
}
Next, the \emph{pseudo-spin unit vector} $\bs n$ taking value on the sphere $\groupS{2}$, 
can be defined as the projection of the superconducting degrees of freedom $\Psi$ 
onto spin-$1/2$ Pauli matrices $\bs\sigma$:
\Equation{Eq:SF:Projection}{
 {\bs n}\equiv (n_x,n_y,n_z)
 =\frac{\Psi^\dagger\bs \sigma\Psi}{\Psi^\dagger\Psi}\,,
 ~~\text{where}~~
 \Psi^\dagger=(\psi_1^*,\psi_2^*)\, .
}
To rewrite the free energy \Eqref{Eq:Quantization:FreeEnergy:3} in terms of the 
pseudo-spin $\bs n$, the total density $\varrho$ and the gauge invariant current $\J$, 
the following identity is useful 
\Equation{Eq:SF:Identity:1}{
\frac{\varrho^2}{4}\partial_kn_a\partial_kn_a+(\Grad\varrho)^2= 
\frac{|\psi_1|^2|\psi_2|^2}{\varrho^2}(\Grad\varphi_{12})^2
+\sum_{a=1,2}(\Grad|\psi_a|)^2 \,.
}
Moreover, noting the relation 
\Equation{Eq:SF:Identity:2}{4
\varepsilon_{ijk}\partial_i\left(
\sum_{a=1,2}\frac{|\psi_a|^2}{\varrho^2}\partial_j\varphi_a\right) = 
\varepsilon_{ijk}\varepsilon_{abc}n_a\partial_in_b\partial_jn_c ,
}
together with the definition of the current \Eqref{Eq:Quantization:Current}, 
the magnetic field can be expressed as
\Equation{Eq:SF:Identity:3}{
B_k=\frac{1}{e}\varepsilon_{ijk}\left(\partial_i\left( \frac{J_j}{e\varrho^2} \right)
-\frac{1}{4}\varepsilon_{abc}n_a\partial_in_b\partial_jn_c \right)\,.
}
The free energy \Eqref{Eq:SF:FreeEnergy:1} can thus be written as
\Equation{Eq:SF:FreeEnergy:2}{
\frac{\F}{\F_0}\!=\!\bigintsss \frac{1}{2}(\Grad\varrho)^2
	+\frac{\varrho^2}{8}\partial_k n_a\partial_k n_a
  	+\frac{\J^2}{2e^2\varrho^2}+V(\varrho,{\bs n})	
+\frac{1}{2e^2}\left[\varepsilon_{ijk}\left(
\partial_i \left(\frac{J_j}{e\varrho^2}\right)
	-\frac{1}{4}\varepsilon_{abc}n_a\partial_i n_b\partial_j n_c \right)\right]^2
 \,.
}
In all generality, potential term $V$ depends on the total density $\varrho$ 
and on the pseudo-spin ${\bs n}$. The potential, which introduces anisotropies 
on the target 2-sphere, explicitly reads as
\Equation{Eq:SF:Potential}{
V(\varrho,{\bs n})=\frac{\varrho^2}{2}\big(a_1+a_2n_x\big)
+\frac{\varrho^4}{4}\big(b_1+2b_2n_x+b_3n_x^2+b_4n_z^2\big)\,.
}
Depending on the values of the coefficients $a_i$ and $b_i$, the potential is 
either an easy-plane or an easy-axis. The explicit dependence of $a_i$ and $b_i$, 
on the original potential coefficients $\alpha_{ab}$ and $\beta_{abcd}$ is not 
very important at that point. The following relations may however be useful
to identify the various coefficients
\Align{Eq:SF:Relations}{
\varrho^2&=|\psi_1|^2+|\psi_2|^2 \,, ~~
\varrho^4=	|\psi_1|^4+|\psi_2|^4+2|\psi_1|^2|\psi_2|^2	\,, ~~
\varrho^4n_z^2=|\psi_1|^4+|\psi_2|^4-2|\psi_1|^2|\psi_2|^2\,, \nonumber \\
\varrho^4(1-n_z^2)&=\varrho^4(n_x^2+n_y^2)=4|\psi_1|^2|\psi_2|^2 \,, ~~
\varrho^4(1+n_z^2)=2\big(|\psi_1|^4+|\psi_2|^4\big) \,, \nonumber \\
\varrho^2n_x&=|\psi_1||\psi_2|\cos\varphi_{12} \,, ~~
\varrho^4(n_x^2-n_y^2)=|\psi_1|^2|\psi_2|^2\cos2\varphi_{12} \,.
}

\paragraph{Topological properties:}
By rewriting the theory using the dual variables, it becomes possible to provide an 
alternative understanding of the $\groupCP{1}$ topological charge \Eqref{Eq:CPN:Charge}. 
The pseudo-spin $\bs n$ \Eqref{Eq:SF:Projection} is a map from the one-point 
compactification of the plane ($\Real^2\cup\{\infty\}\simeq \groupS{2} $) onto the 
two-sphere target space spanned by $\bs n$. That is ${\bs n}:\groupS{2}\to\groupS{2}$, 
which is classified by the homotopy class $\pi_2(\groupS{2})\in\groupZ{}$, thus defining 
the topological invariant  
\Equation{Eq:SF:Charge}{
   \Q({\bs n})=\frac{1}{4\pi} \int_{\Real^2}
   {\bs n}\cdot\partial_x {\bs n}\times \partial_y {\bs n}\,\,
  d x d y \,.
}
As before, since $\bs n$ is ill-defined when $\Psi=0$, ordinary (composite) singular 
vortices, have $\Q({\bs n})=0$. Core-split vortices, on the other hand, have an integer 
topological charge $\Q({\bs n})\in\groupZ{}$ which coincides with the number of carried 
flux quanta. In a way, $\Q({\bs n})$ counts the number of times the pseudo-spin texture 
of ${\bs n}$ wraps the target two-sphere. 
It is worth emphasizing that the topological charge \Eqref{Eq:SF:Charge} is 
an integer, when integrated over the infinite plane $\Real^2$, or at 
least an large enough domain $\Omega\subset\Real^2$. 

\begin{figure}[!htb]
\hbox to \linewidth{ \hss
\includegraphics[width=.4\linewidth]{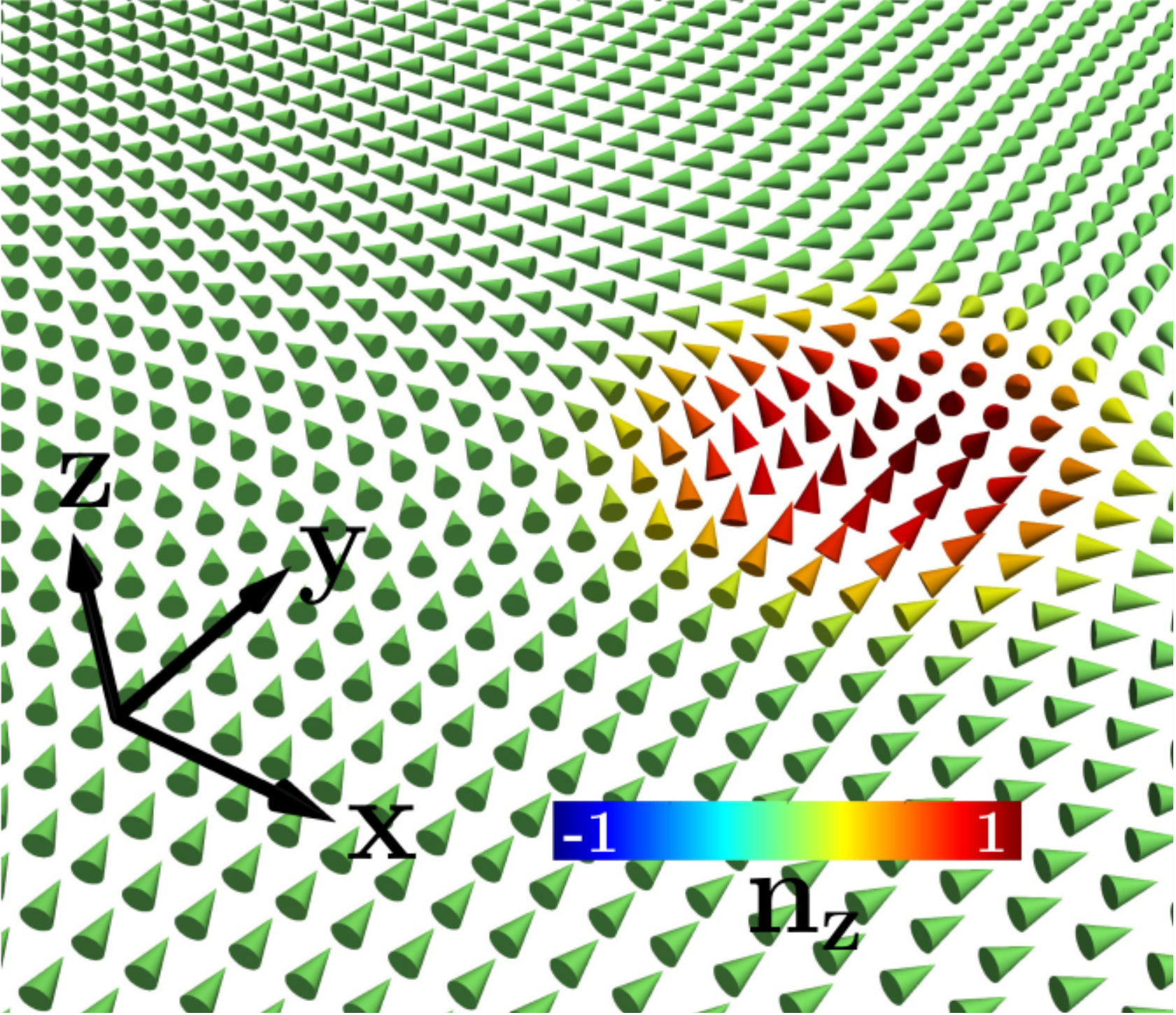}
\hss
\includegraphics[width=.4\linewidth]{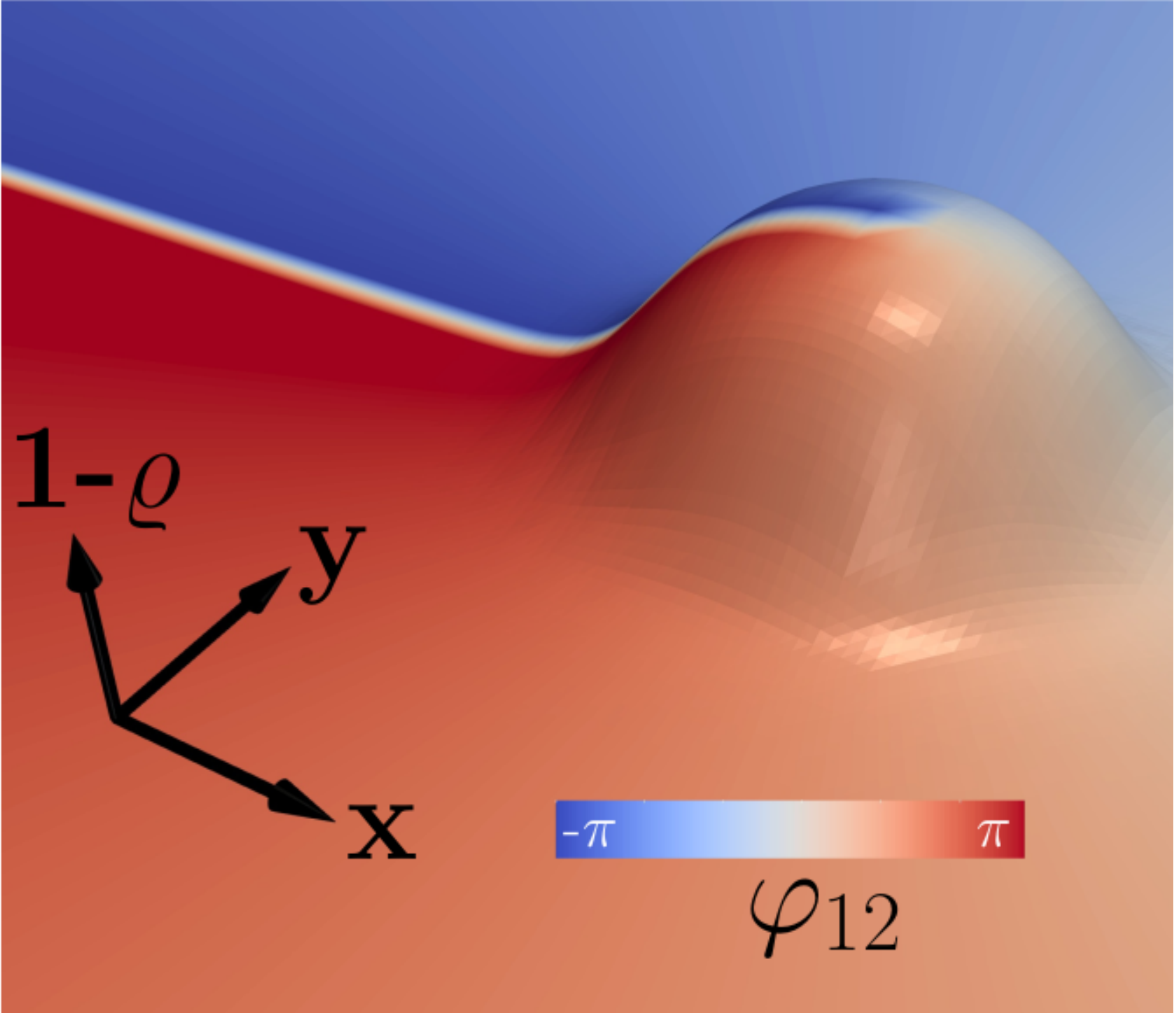}
\hss}\vspace{0.2cm}
\hbox to \linewidth{ \hss
\includegraphics[width=.4\linewidth]{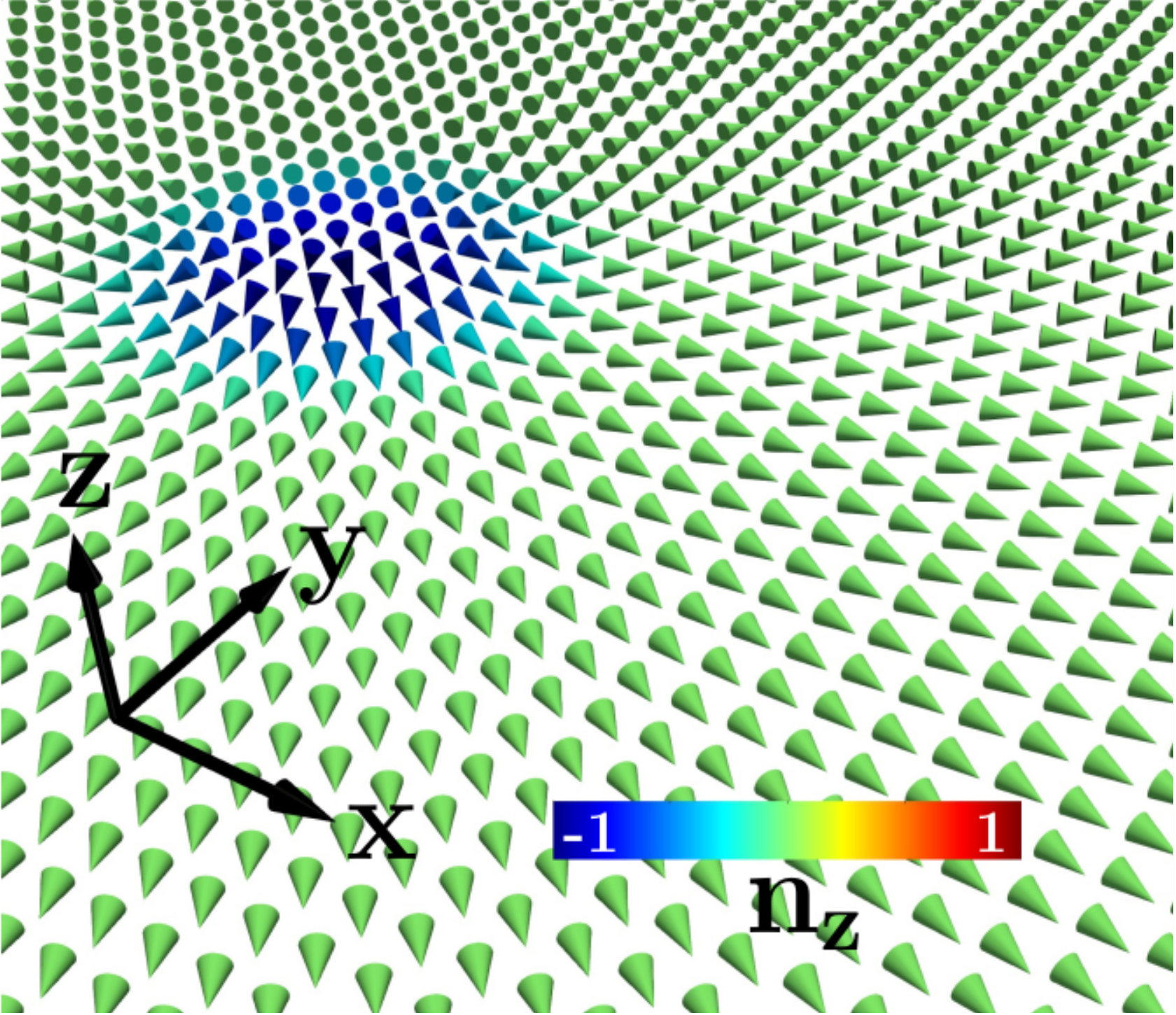}
\hss
\includegraphics[width=.4\linewidth]{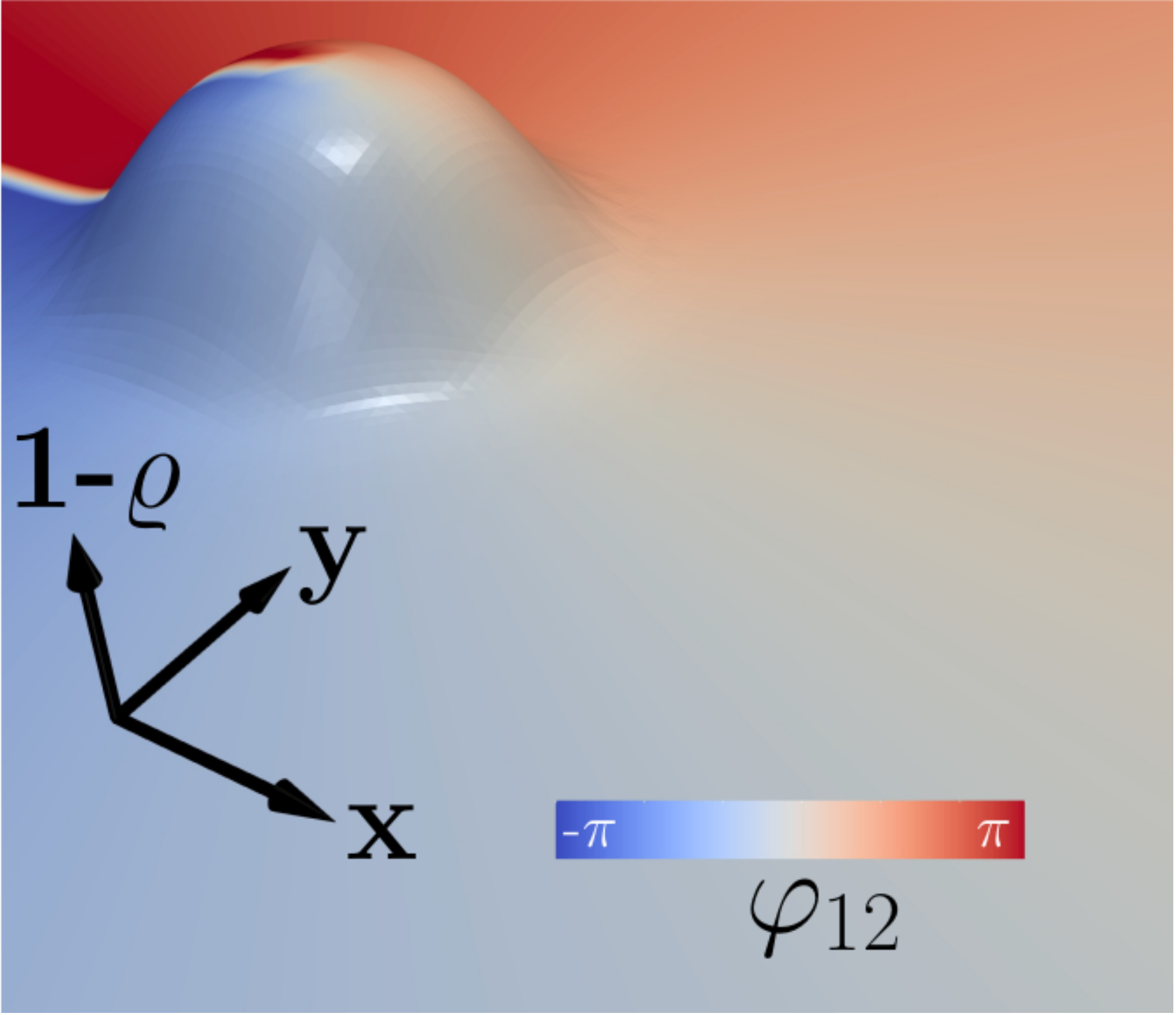}
\hss}\vspace{0.2cm}
\hbox to \linewidth{ \hss
\includegraphics[width=.4\linewidth]{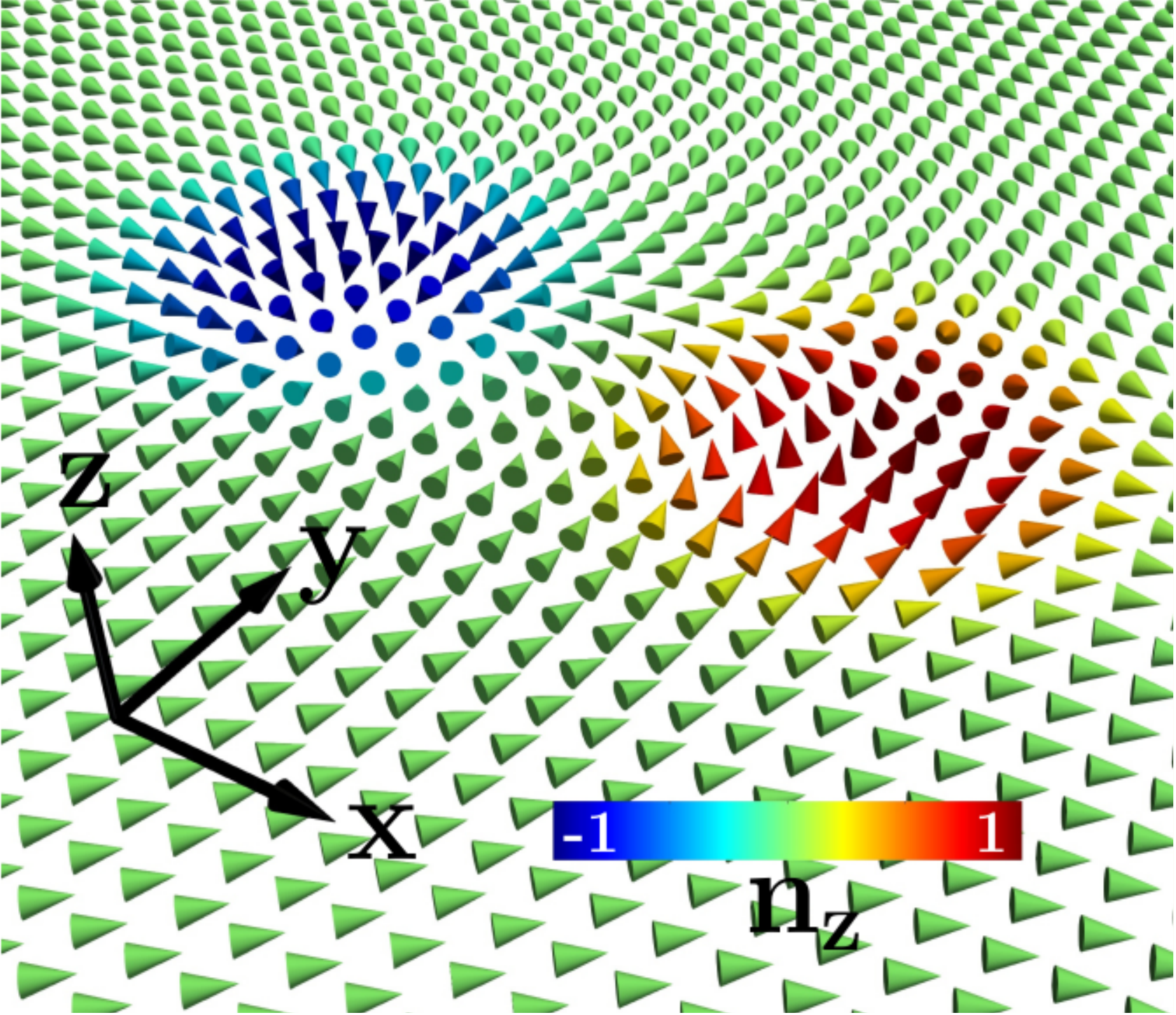}
\hss
\includegraphics[width=.4\linewidth]{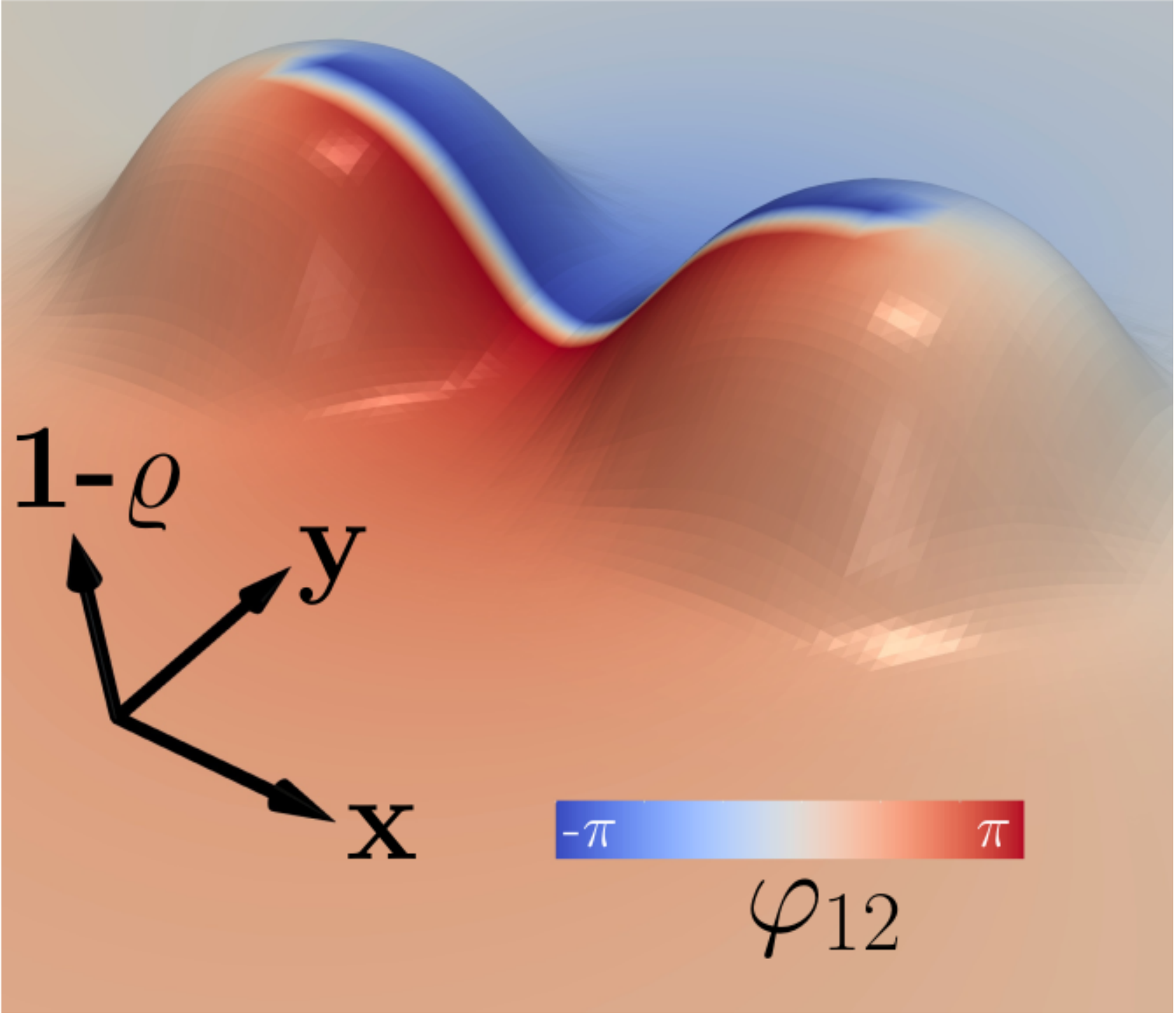}
\hss}
 	\begin{picture}(0,0)%
		\put(10,360){\rotatebox{90}{\Huge meron}}
		\put( 4,185){\rotatebox{90}{\Huge anti-meron}}
		\put( 4, 40){\rotatebox{90}{\Huge skyrmion}}
	\end{picture}

\caption{
Illustration of (anti-)merons and skyrmion solutions. The left panels show the 
pseudo-spin $\bs n$ calculated from the Ginzburg-Landau quantities, using the relation 
\Eqref{Eq:SF:Projection:2}. The arrows show $\bs n$ and the color scale encodes the 
value of $n_z$. The right panels display the corresponding Ginzburg-Landau quantities. 
Namely the elevation of the surface stems for the density depletion $1-\varrho$, while 
the colouring indicates the value of the relative phase $\varphi_{12}$.
}
\label{Fig:Schematic:Skyrmion}
\end{figure}
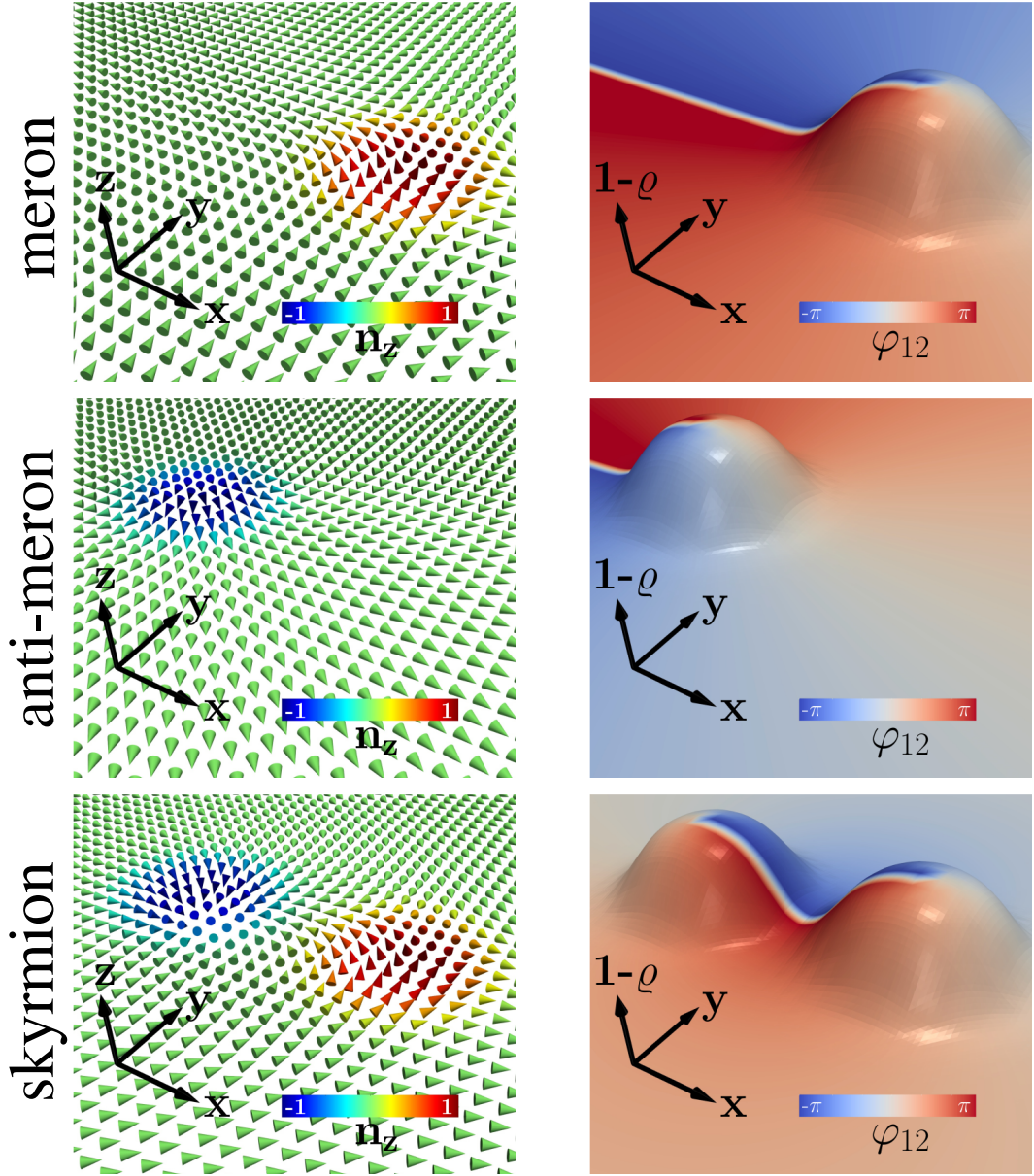

\vspace{2cm}

\paragraph{Pseudo-spin textures. Skyrmions and Merons:}
The core-split vortices thus have a skyrmionic character, with a topological 
charge $\Q({\bs n})\in\groupZ{}$. In terms of the degrees of freedom of the 
two-component Ginzburg-Landau theory \Eqref{Eq:SF:FreeEnergy:1}, the pseudo-spin 
reads as 
\Equation{Eq:SF:Projection:2}{
 {\bs n}= \left( \frac{2|\psi_1||\psi_2|}{|\psi_1|^2+|\psi_2|^2}\cos\varphi_{12} ,
				 \frac{2|\psi_1||\psi_2|}{|\psi_1|^2+|\psi_2|^2}\sin\varphi_{12} ,
				 \frac{|\psi_1|^2-|\psi_2|^2}{|\psi_1|^2+|\psi_2|^2}
				 \right)\, .
}
From this, it is obvious that the core of a vortex in $\psi_1$ maps to ${\bs n}=(0,0,-1)$, 
while a core in $\psi_2$ gives ${\bs n}=(0,0,1)$. The vortex cores thus map to the 
poles of the target sphere. 

Moreover, the pseudo-spin associated to a single fractional vortex covers only half 
of the target sphere. Such configurations are sometimes called merons. More precisely, 
a fractional vortex in $\psi_2$ covers only the north hemisphere is a meron, while a 
vortex $\psi_1$ covers only the south hemisphere is an anti-meron. This interpretation 
of the fractional vortices as (anti-)merons is displayed in 
\Figref{Fig:Schematic:Skyrmion}. Since the pseudo-spin do not wrap entirely the target 
two-sphere, the associated topological charge \Eqref{Eq:SF:Charge} is trivial: 
$\Q({\bs n})=0$. Both the pseudo-spin and the Ginzburg-Landau quantities show that 
the relative phase have a net winding, and thus have divergent energy as discussed 
earlier.

The last line of figure~\ref{Fig:Schematic:Skyrmion} displays a skyrmion configuration 
corresponding to core-split vortices. The skyrmion, which is a bound state of 
non-overlapping fractional vortices, can be seen as the bound state of a meron 
and an anti-meron. Unlike merons, a skyrmion do not have a net winding of the relative 
phase $\varphi_{12}$. Thus skyrmions, have finite energy. Moreover, the pseudo-spin 
of a skyrmion do wrap entirely the target two-sphere. It follows that the associated 
topological charge \Eqref{Eq:SF:Charge} is integer, here $\Q({\bs n})=1$.

Note that the configurations that are displayed here, are only illustrations of the 
topological properties of (anti-)merons and skyrmion configurations. They are 
configurations with the appropriate topological properties, but they are \emph{not} 
solutions (\ie minima) of the Ginzburg-Landau theory \Eqref{Eq:SF:FreeEnergy:1}. 
Indeed, as discussed earlier, the fractional vortices in the different components 
(and thus the merons and anti-merons) typically attract each other. Because of the 
attraction, merons and anti-merons can annihilate, thus converting a skyrmion into 
a singular vortex where the individual singularities superimpose. Yet, as discussed 
in the next section, there can exist various mechanism to stabilize skyrmions.


\section{Existence of skyrmions and exotic vortex states} 
\label{Sec:Skyrmions}

To summarize briefly, and repeat the topological properties presented in the above 
sections, the elementary topological excitations in multicomponent superconductors are 
fractional vortices. That is, field configurations with independent phase windings
in either of the components, which carry only a fraction of the elementary flux 
quantum $\Phi_0$. Because they have a winding of the relative phase between the 
different components, fractional vortices have divergent energy per unit length. 
Thus finite energy requirements dictate that composite vortices (i.e. bound states 
of fractional vortices in each component) are the only configurations with finite 
energy per unit length. As a result these composite vortices, carrying an integer 
flux quantum, are the only configurations that can form in bulk multicomponent 
superconductors. 
Depending on the relative positions of the individual singularities, the resulting 
composite vortex is either termed \emph{singular}, if the singularities are 
superimposed, or \emph{coreless}, if they do not. Composite coreless vortices are 
characterized by the additional topological invariant associated with the complex 
projective space $\groupCP{N-1}$ \Eqref{Eq:CPN:Charge}. The existence of the additional 
invariant motivates the terminology \emph{skyrmion} to denote a coreless vortex.
The propensity to form either singular or coreless vortices depends on the 
interaction energy between the fractional vortices. The London limit calculations, 
in the most elementary case, state that fractional vortices are logarithmically 
bound and that the interaction energy is minimized when they superimpose. 
This raises the natural question of whether the skyrmions (the coreless vortices) 
can form at all, and be favoured compared to (Abrikosov) singular vortices? In other 
word is there a mechanism that provides a short-range repulsion between the 
fractional vortices, so that they form a bound state with a finite separation?
Note that composite singular defects still feature interesting new properties. 
These are discussed later in Chapter~\ref{Chap:Semi-Meissner}.

So, in order to form skyrmions, a stabilizing mechanism is needed to counteract the 
tendency of the singularities to superimpose. It turns out that there exist various 
ways to circumvent the confinement of fractional vortices. For example this can 
originate in non-linear effects beyond the London approximation, or by additional 
interacting terms in the free energy. 
It follows that, despite the naive arguments, skyrmions actually exist in 
many superconducting systems. These mechanisms that enforce the core-splitting, 
stabilize skyrmions against collapsing into a singular Abrikosov vortex.
One could alternatively say that because of the extra terms, the Abrikosov vortex 
is unstable and decay into a coreless defect: a skyrmion. 
Various possibility are detailed below.

\subsection{Skyrmions in mixtures of commensurately charged condensates} 
\label{Sec:Mixtures}

Typically, in condensed matter systems the multiple superconducting condensates are 
coupled to the gauge field via the same gauge coupling which is the charge of the 
Cooper pairs (twice the charge of their constituent fermion). Still, it is formally 
possible to consider the case of a mixture of condensates with different electric 
charges. Such a situation is expected to appear for example in the superconducting 
state of liquid metallic deuterium, where the deuteron is a charge-$1$ boson which 
can form a Bose-Einstein condensate, that coexists with the Cooper pairs of electrons 
and/or protons \cite{Oliva.Ashcroft:84,Oliva.Ashcroft:84a,Bedaque.Buchoff.ea:11}. 
The Bose-Einstein condensate of deuterons carries only once the charge of its 
constituent boson, while Cooper pairs carry twice the charge of their constituent 
fermion. There, the electronic Cooper pairs carry a charge $-2e$ while the Bose-Einstein 
condensate of deuteron carries an electric charge $+e$.
Interestingly, the flux quantization and finite energy considerations similar to 
that discussed in the Section \ref{Sec:Fractional-vortices}, dictate that the 
electric charges should be commensurate, for the vortex solutions to have finite 
energy \CVcite{Garaud.Babaev:14a}. This implies that the composite vortices carrying 
an integer flux quantum are bound states of a different number of fractional vortices 
in both components. Moreover, the resulting bound state is typically a core-less
defect, and these skyrmions can be characterized by a pseudo-spin texture, and by 
a hidden $\groupCP{1}$ invariant similar to those discussed above. This was analyzed 
in details in \CVcite{Garaud.Babaev:14a}.

To account for the possibility to have a mixture of two superconducting condensates 
carrying different charges, the gauge derivative in the kinetic term is chosen to 
have different gauge couplings for the two individual condensates: 
$\D\psi_a=(\Grad+ie_a\A)\psi_a$ (here $a=1,2$). The individual gauge couplings are 
conveniently parametrized as $e_a=eg_a$ where $g_a$ are integer numbers 
$g_a\in\groupZ{}$, and $e$ fixes the scale of the electric charges. Applying this 
parametrization to liquid metallic deuterium, the couplings are $(g_1,g_2)=(1,2)$ 
and $\psi_1$ stands for the deuteron condensate while $\psi_2$ (carrying twice 
the electric charge of $\psi_1$) denotes the electronic Cooper pairs. When considering 
the possibility that the condensates can have different charges, via the redefinition 
of the gauge derivative, the superconducting current \Eqref{Eq:Quantization:Current:2} 
reads as $\J=e\sum_{a}g_a|\psi_a|^2(\Grad\varphi_a+eg_a\A)$. Since 
the supercurrent $\J$ is screened and that it decays exponentially, the magnetic 
flux \Eqref{Eq:Quantization:Flux:2} reads as 
\Equation{Eq:Quantization:Flux:RV}{
   \Phi=\frac{-1}{e\sum_bg_b^2|\psi_b|^2}
    \bigointsss_{\!\!\!\!\!\!{\cal C}} 
    		\sum_{a}g_a|\psi_a|^2\Grad\varphi_a\!\cdot\!d{\bs\ell}
=\frac{\Phi_0\sum_{a}g_a|\psi_a|^2}{2\pi\sum_bg_b^2|\psi_b|^2}
    \bigointsss_{\!\!\!\!\!\!{\cal C}} \Grad\varphi_a\!\cdot\!d{\bs\ell}\,,
}
where $\Phi_0=2\pi/e$ is the flux quantum and the closed integration path is 
chosen for the flux to be positive. The condensates $\psi_a$ being complex fields, 
they wind an integer number of time, and $(n_1,n_2)$ denotes the field configurations 
with the winding $n_a$ of the condensate $\psi_a$. A fractional vortex in the 
condensate $a$ carries a fraction $\Phi_a/\Phi_0=g_a|\psi_a|^2/\sum_bg_b^2|\psi_b|^2$ 
of the magnetic flux. As long as $g_1\neq g_2$, the condition for the quantization 
of the flux, is that each condensate should wind $n_a = g_a$ times. This follows from 
\Equation{Eq:Quantization:Flux:RV:2}{
\sum_an_a\frac{\Phi_a}{\Phi_0}=
	\frac{\sum_an_ag_a|\psi_a|^2}{\sum_bg_b^2|\psi_b|^2}
	=1\,,~~~\text{iff}~~n_a = g_a \,.
}
A separation into charged and neutral modes, similar to that derived above  
\Eqref{Eq:Quantization:FreeEnergy:3}, dictates that only the configurations that 
carry an integer flux ($n_a=g_a$) have finite energy \cite{Garaud.Babaev:14a}.
This can be seen by requiring the absence of logarithmic divergence, and thus that 
the neutral mode has no winding 
\Equation{Eq:Quantization:Flux:RV:FiniteEnergy}{
0  =\bigointsss_{\!\!\!\!\!\!{\cal C}} 
	\Grad(g_1\varphi_2-g_2\varphi_1)\!\cdot\!d{\bs\ell}
   = (g_1n_2-g_2n_1)  \,.
}
For a configuration carrying a single flux quantum (and since $g_a$ and $n_a$ are 
integers), this implies that $n_a=g_a$. Thus the finite energy configurations winds 
$g_1$ times in $\psi_1$ and $g_2$ times in $\psi_2$. Note that this condition 
implies that the total flux is integer \Eqref{Eq:Quantization:Flux:RV:2}.
On the other hand, when $g_1n_2-g_2n_1\neq0$, the screening is incomplete and the 
associated energy grows with the system size. Such vortex configurations are thus 
thermodynamically unstable in bulk systems.

\begin{figure}[!htb]
\hbox to \linewidth{ \hss
\includegraphics[width=.495\linewidth]{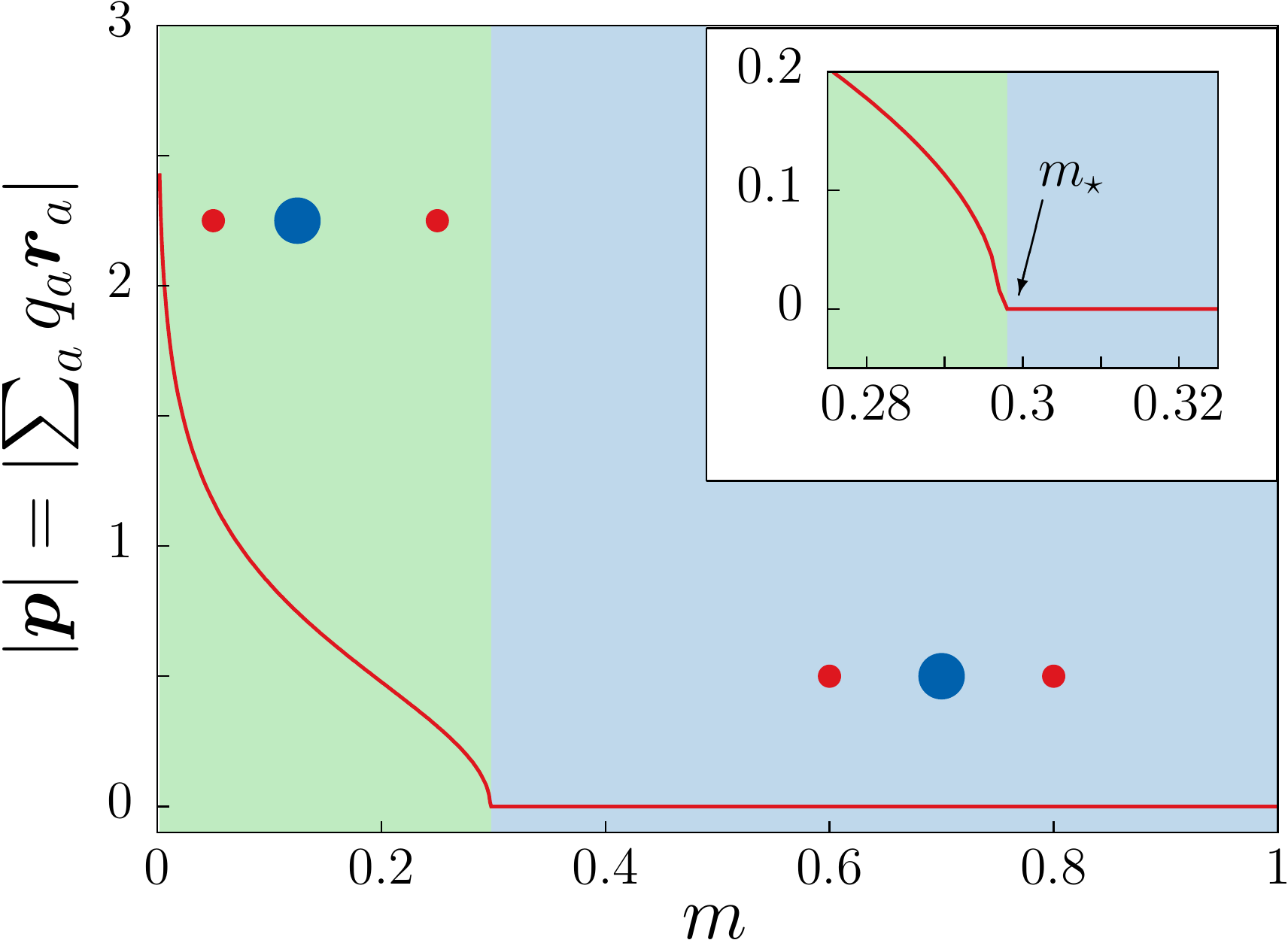}
\hss
\includegraphics[width=.475\linewidth]{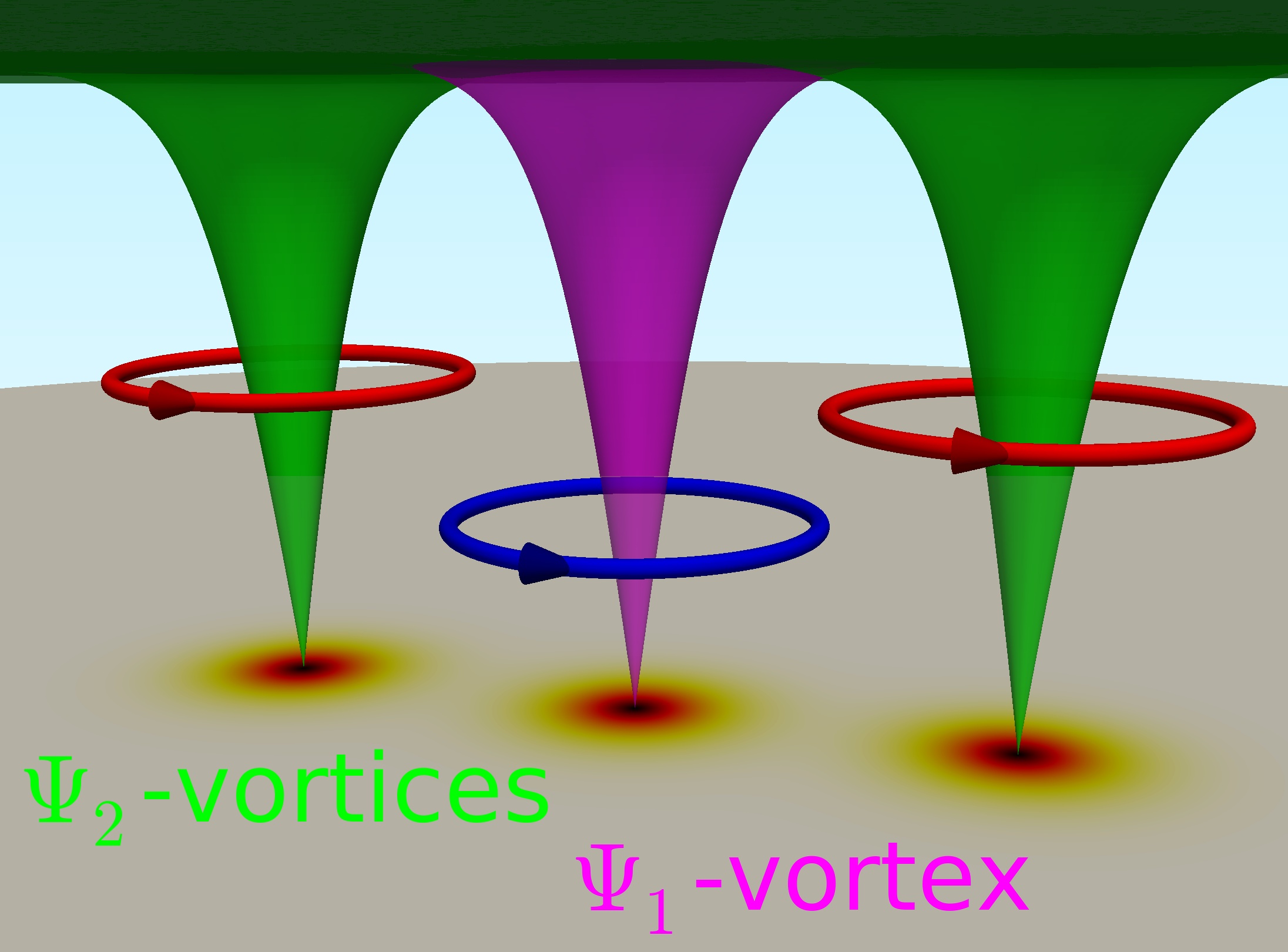}
\hss}
\caption{
The left panel shows the structure of a single skyrmion (thus corresponding to the 
charges $g_1=1$ and $g_2=2$), obtained by minimizing the corresponding interaction 
energy \Eqref{Eq:Quantization:Interaction:RV}. It shows the parametric dependence 
of the dipole moment defined in point charges mapping with charges $q_2=g_1$ and 
$q_1=-g_2$. The composite vortex structure undergoes a phase transition from a 
dipolar phase below $m_\star$ to a quadrupolar phase.
The right panel displays a schematic illustration of the a composite vortex 
with a single fractional vortex in the condensate $\psi_1$ bound with two 
vortices in $\psi_2$.
}
\label{Fig:London-interaction:RV}
\end{figure}

\vspace{1cm}

The condensates with different charges favour bound states of fractional vortices 
with a finite separation, unlike the inter-vortex interaction derived in Section 
\ref{Sec:Fractional-vortices:Interaction} \CVcite{Garaud.Babaev:14a}. Indeed, 
similar calculation shows that the intervortex interactions are characterized by 
the ratio of the charges in the different condensates $s=\frac{g_1}{g_2}$.
The interaction between fractional vortices reads as 
\Equation{Eq:Quantization:Interaction:RV}{
E_{11}=\frac{1}{s}\ln\frac{R}{r}
		+\frac{m}{s}K_0\left(\frac{r}{\lambda}\right) 		    	\,,~~
E_{22}=s\ln\frac{R}{r}+\frac{s}{m}K_0\left(\frac{r}{\lambda}\right)	\,,~~
E_{12}=-\ln\frac{R}{r}+K_0\left(\frac{r}{\lambda}\right) \,.
}
It follows that the vortex matter in the London limit of a two-component 
superconductor with incommensurate charges is described by a 3-parameter family 
$(m,s,R)$. The individual interactions are thus similar to that illustrated in 
\Figref{Fig:London-interaction}, and that vortices in different condensates 
attract each other to form a bound state of co-centered vortices while vortices 
in the same condensate always repel each other. However, since both condensates 
have a different number of fractional vortices, the system has to compromise 
between the fact that vortices in the different condensates tend to overlap, 
while vortices in similar condensates repel each other. As illustrated in 
\Figref{Fig:London-interaction:RV}, the integer flux carrying defect is thus 
a molecule-like bound state of split fractional vortices.

\begin{figure}[!htb]
\hbox to \linewidth{ \hss
\includegraphics[width=.75\linewidth]{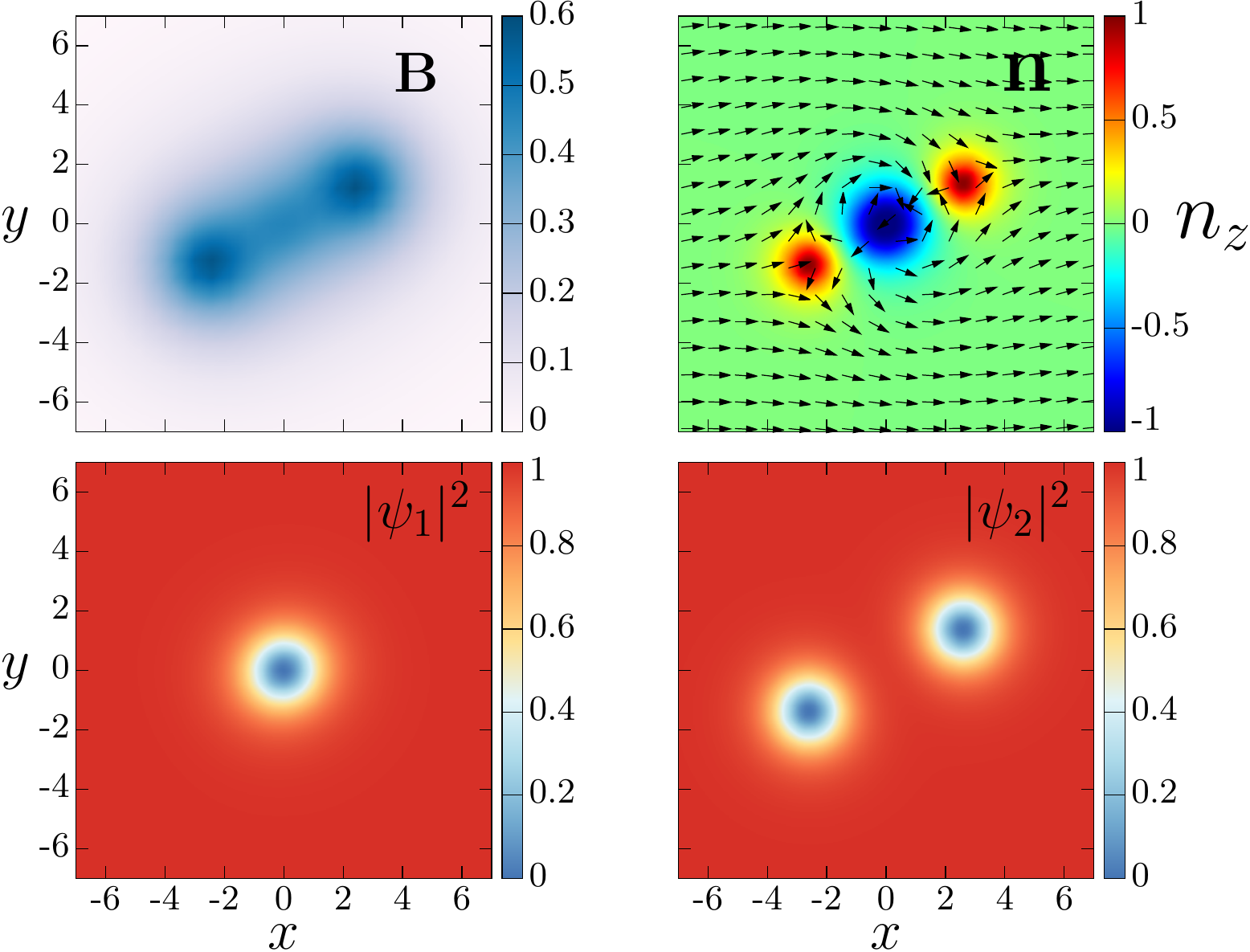}
\hss}
\caption{
Molecule-like topological excitations carrying a single flux quantum, for the different 
couplings $(g_1,g_2)=(1,2)$. 
The parameters of the Ginzburg-Landau energy \Eqref{Eq:Quantization:RV:Potential}
are $\alpha_{aa}=-1$, $\beta_{aa}=1$ , and $e=0.2$. 
The displayed quantities on the top row are the magnetic field $\B$ and the pseudo-spin 
$\bs n$. The bottom row shows the densities of the two components $|\psi_a|^2$. 
}
\label{Fig:Vortex:Rational}
\end{figure}

\paragraph{Skyrmions beyond the London limit.}
The analysis in the London limit predicts the existence of coreless topological defects 
with different number of vortices in the different condensates. The analysis beyond the 
London approximation requires to specify the potential for the Ginzburg-Landau energy
\Eqref{Eq:Quantization:FreeEnergy:3}, as for example
\Equation{Eq:Quantization:RV:Potential}{
V(\Psi,\Psi^\dagger) = \sum_{a=1}^2
\left(\alpha_{aa}|\psi_a|^2+\frac{\beta_{aa}}{2}|\psi_a|^4 \right) \,.
}
The coreless vortex solutions are obtained by minimizing the Ginzburg-Landau energy with 
a suitable initial guess (see the numerical methods in the Appendix~\ref{App:numerics}). 
These solutions do exist beyond the London approximation, as can be seen in Figure 
\ref{Fig:Vortex:Rational}. 
This is the simplest excitations carrying a unit flux quantum, for different couplings 
$(g_1,g_2)=(1,2)$.

Many other solutions for various couplings $(g_1,g_2)$, and the possible relevance to 
describe superconducting state for liquid metallic deuterium (LMD), were considered in 
\CVcite{Garaud.Babaev:14a}. This model of two-condensates with arbitrary charges was 
generalized, together with the spectrum of the topological excitation in 
\cite{Chatterjee.Gudnason.ea:20}.

\paragraph{Topological properties.} The two-component model with different charges, 
feature similar properties than the usual two-component model. However, because of 
the different gauge couplings, the superconducting degrees of freedom should be 
collected as 
\Equation{Eq:SF:Rational}{
\Psi^\dagger=\left(\psi_1^{*g_2},\psi_2^{*g_1}\right) \,.
}
Provided this modification, the charge $\Q(\Psi)$ \Eqref{Eq:CPN:Charge}, projection to 
the pseudo-spin $\bs n$ \Eqref{Eq:SF:Projection}, and the associated charge $\Q(\bs n)$ 
\Eqref{Eq:SF:Charge} are still valid.

\subsection{Skyrmions and hopfions stabilized by the Andreev-Bashkin effect} 
\label{Sec:Andreev-Bashkin}

The dissipationless inter-component drag, known as the Andreev-Bashkin effect
\cite{Andreev.Bashkin:75}, is a generic interaction in multicomponent superconductors 
and superfluids, but it is often neglected. Namely, in superfluid mixtures with two 
components, the current of a given component ${\bs j}_{1,2}$ generically depends on 
both superfluid velocities ${\bs v}_{1,2}$. This is due to the inter-component 
interactions occurring between particles, and the relations between currents and 
velocities are:
\Equation{Eq:Andreev-Bashkin:Current}{
{\bs j}_1 = \rho_{11}{\bs v}_1+\rho_{12}{\bs v}_2\,,~~~\text{and}~~~
{\bs j}_2 = \rho_{22}{\bs v}_2+\rho_{21}{\bs v}_1\,.
}
The off-diagonal coefficients $\rho_{12}$ and $\rho_{21}$ determine the fraction of the 
density of one of the superfluid components carried by the superfluid velocity of the 
other. They thus describe the \emph{inter-component drag}. The inter-component coefficients 
$\rho_{ab}$ can be large (compared to the intra-component coefficients $\rho_{aa}$), 
for example in strongly correlated systems \cite{Kuklov.Svistunov:03,
Soyler.Capogrosso-Sansone.ea:09,Sellin.Babaev:18}, Fermi-liquids mixtures \cite{Sjoberg:76,
Chamel:08}, or in spin-triplet superconductors and superfluids \cite{Leggett:75}.
More general drag interactions, the \emph{dissipationless vector drag}, where recently 
argued to be occur in certain kind of optical lattices \cite{Syrwid.Blomquist.ea:21}.

\vspace{1cm}

The dissipation-less drag \Eqref{Eq:Andreev-Bashkin:Current}, in a two-component 
model of superconductivity, can be accounted for by adding current-current interaction
\Equation{Eq:Andreev-Bashkin:FreeEnergy}{
	\cdots+\sum_{a,b=1,2}\frac{\mu_{ab}}{2} {\bs J}_a\cdot{\bs J}_b + \cdots \,,
}
to the generic Ginzburg-Landau free energy density. This was discussed in details 
for example in \cite{Garaud.Sellin.ea:14,Rybakov.Garaud.ea:19}. There, 
$\J_a=\Im(\psi_a^*\D\psi_a)$, stands for the individual currents in the absence of 
drag. The off-diagonal coefficients $\mu_{12}=\mu_{21}$ of the current coupling 
matrix $\hat{\mu}$ describe the intercomponent drag. The total current, which is 
the sum of the individual supercurrents, have a similar structure to that described 
in \Eqref{Eq:Andreev-Bashkin:Current}. For now, assuming the absence of mixed-gradient  
terms in the free energy \Eqref{Eq:General:FreeEnergy} (coefficients $\kappa_{ab}=0$ 
if $a\neq b$). When adding the current-current interactions 
\Eqref{Eq:Andreev-Bashkin:FreeEnergy}, the total current reads as: 
\Equation{Eq:Andreev-Bashkin:CurrentGL}{
\J/e=\sum_{a}w_a|\psi_a|^2(\Grad\varphi_a+e\A) \,,
~~\text{where}~~~w_a=\kappa_{aa}+\sum_{b}\mu_{ab}	|\psi_b|^2\,.
}
The role of $w_a$ is to re-weight the individual currents. More precisely, the 
charged and neutral modes in \Eqref{Eq:Quantization:FreeEnergy:3} are renormalized 
differently due to the inter-component drag.

\begin{figure}[!htb]
\hbox to \linewidth{ \hss
\includegraphics[width=.5\linewidth]{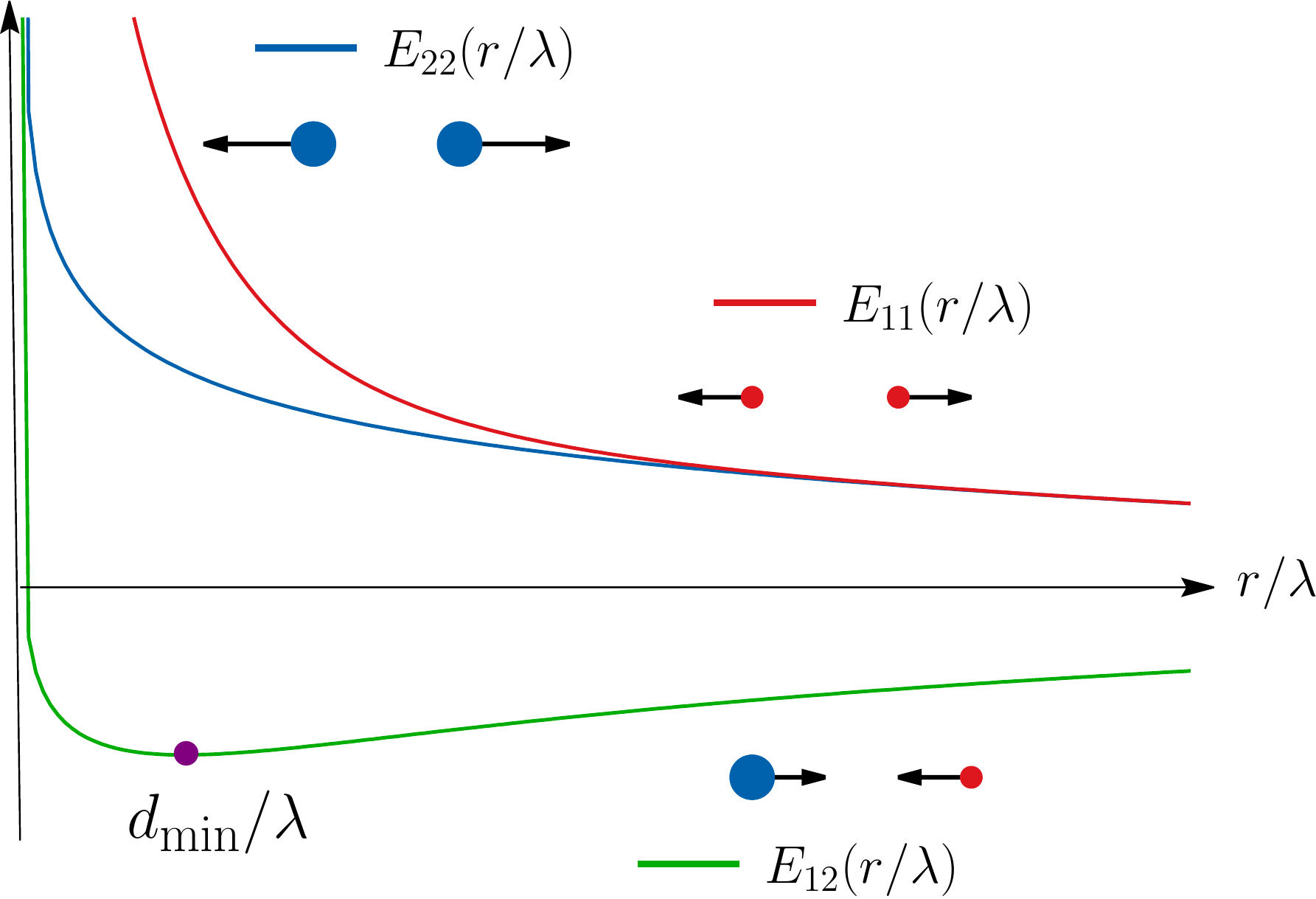}
\hss
\includegraphics[width=.475\linewidth]{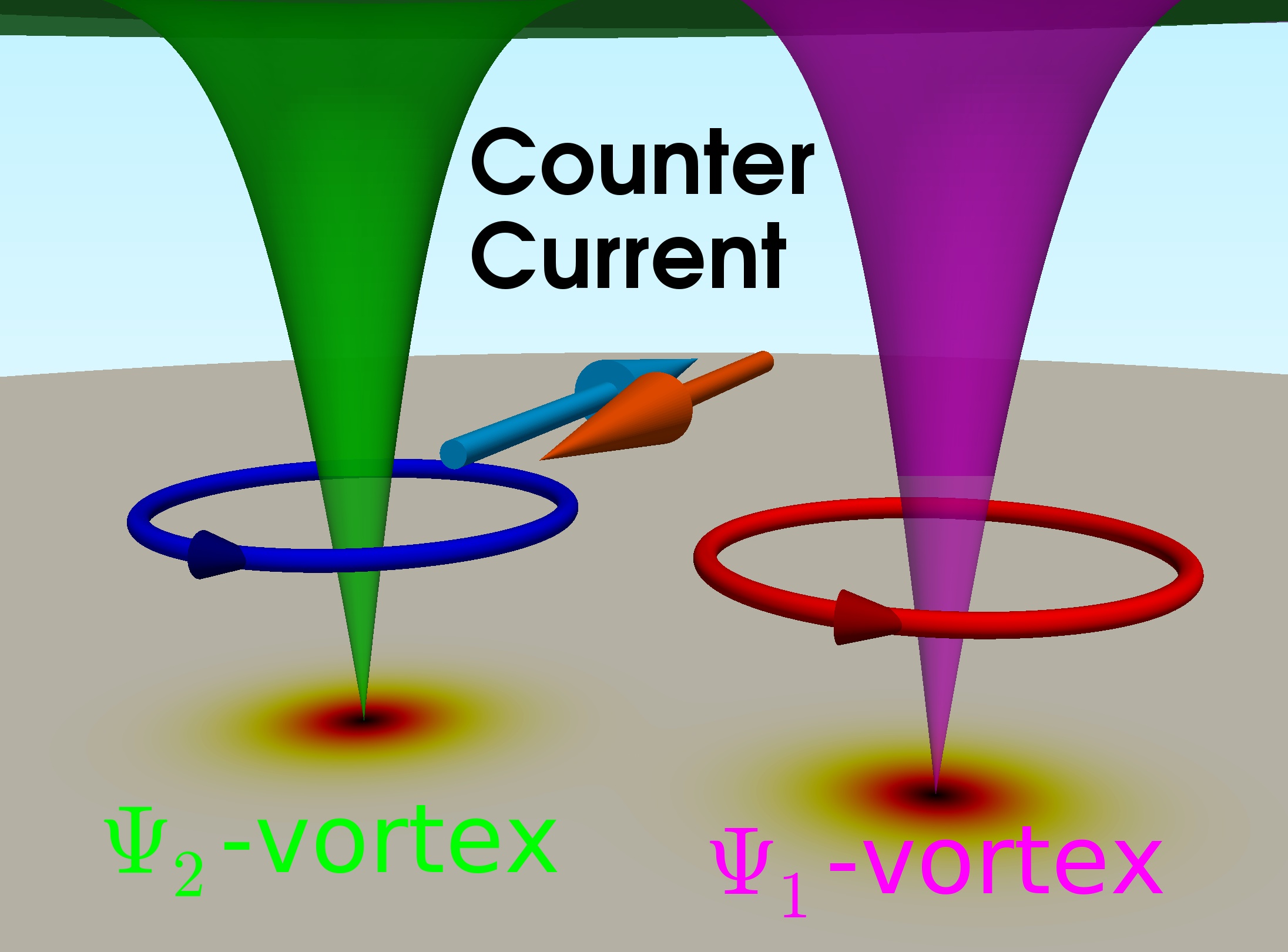}
\hss}
\caption{
The left panel shows the interaction energies \Eqref{Eq:Quantization:Interaction:AB} 
(with $m=0.2$, $w=1.8$) between the point-like charges associated with vortices in 
different condensates. The blue (big) dot represents the vortex in $\psi_1$ while 
the red (small) dots represent the vortices in $\psi_2$. Alike charges always repel, 
while different charges interact with a long-range logarithmic attraction, and a 
short-range repulsion.  
The right panel displays a schematic illustration of the Andreev-Bashkin effect in 
a composite vortex
}
\label{Fig:London-interaction:AB}
\end{figure}

An analysis of the inter-vortex interaction, similar to that derived in Section 
\ref{Sec:Fractional-vortices:Interaction}, shows that the fractional vortices should 
form a bound state with a finite separation \CVcite{Garaud.Sellin.ea:14}. Namely, 
in the London limit, the intervortex interactions \Eqref{Eq:Quantization:Interaction} 
are characterized by two additional parameters $w>1$ and $m>0$, that depend on the 
drag. In the case where $\mu_{11}=\mu_{22}=\mu_{12}\equiv\mu$, the interaction between 
the fractional vortices in the various condensates reads as \cite{Garaud.Sellin.ea:14}:
\Equation{Eq:Quantization:Interaction:AB}{
E_{11}=\ln\frac{R}{r}+mwK_0\left(\frac{r}{\lambda}\right) 			\,,~~
E_{22}=\ln\frac{R}{r}+\frac{w}{m}K_0\left(\frac{r}{\lambda}\right)	\,,~~
E_{12}=-\ln\frac{R}{r}+wK_0\left(\frac{r}{\lambda}\right) \,.
}
Here, $w=1+\mu\varrho^2$, $m=|\psi_1|^2/|\psi_2|^2$ and $\lambda^{-1}=e\sqrt{w\varrho^2}$. 
$R$ stands for the system size and $\varrho^2=\sum_a|\psi_a|^2$ is the total density. 
Thus the vortex matter in the London limit of a two-component superconductor with 
dissipationless drag is described by a 3-parameter family $(m,w,R)$. The figure 
\ref{Fig:London-interaction:AB} shows the profiles of the different interactions 
\Eqref{Eq:Quantization:Interaction:AB} between the different kind of fractional vortices.
Because of the additional parameter $w$, the repulsive and attractive contributions 
in $E_{12}$ do not compensate at $r=0$ any longer. Instead, they cancel at a finite 
separation, so that even in the London limit, the interactions tend to form skyrmions.

\begin{figure}[!htb]
\hbox to \linewidth{ \hss
\includegraphics[width=.95\linewidth]{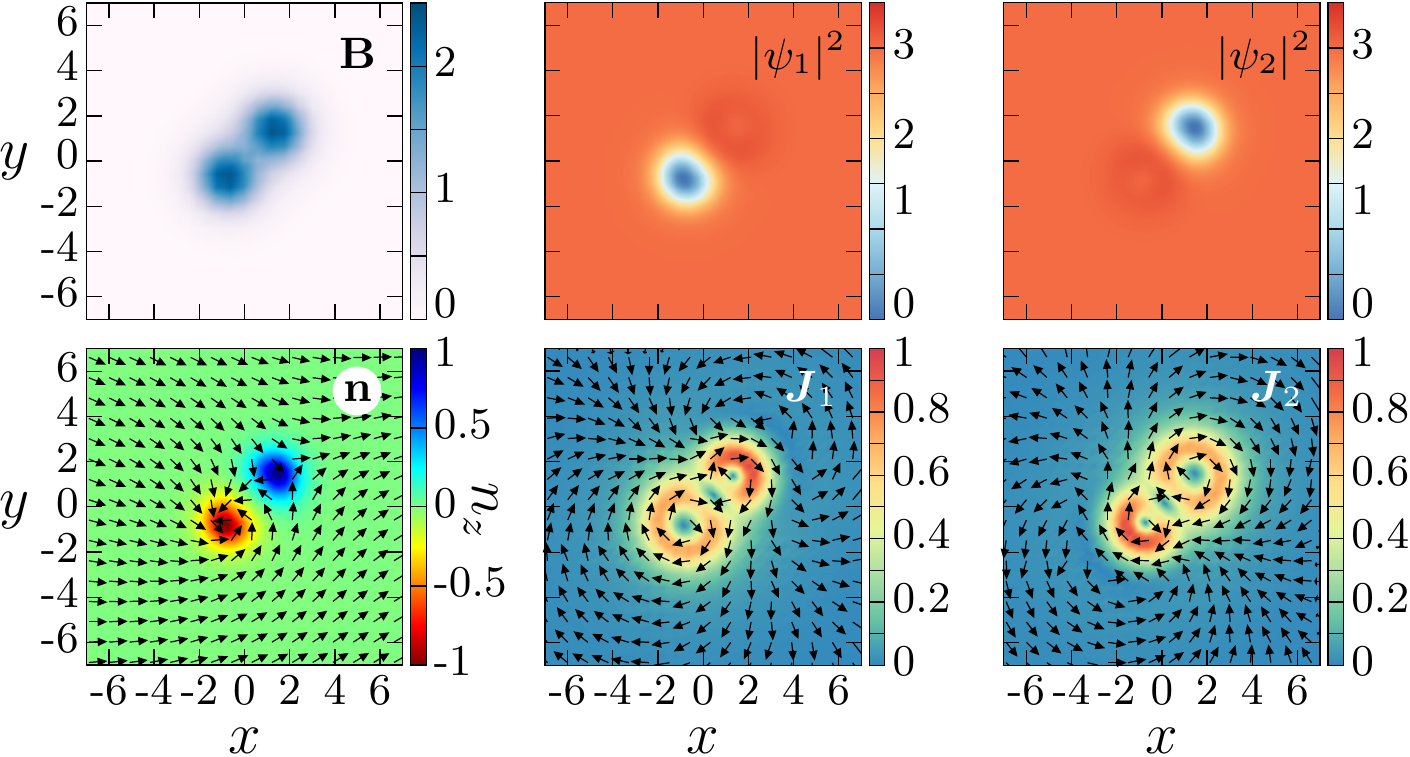}
\hss}
\caption{
A skyrmion solution of two-component superconductor with dissipationless drag interaction
\Eqref{Eq:Andreev-Bashkin:FreeEnergy}, with the couplings $\mu_{11}=\mu_{22}=\mu_{12}=2$. 
The parameters of the simple potential interaction \Eqref{Eq:Quantization:RV:Potential} 
are $(\alpha_{aa},\beta_{aa})=(-3,1)$, and the gauge coupling is $e = 0.2$.
The displayed quantities on the top left panel is the magnetic field. The other panels 
on the top row, show the individual densities of the superconducting condensates 
$|\psi_1|^2$ and $|\psi_2|^2$. This is clearly a coreless defect, since both singularities 
do not overlap. 
The bottom left panel shows the associated pseudo-spin texture $\bs n$ 
\Eqref{Eq:SF:Projection}, while the other panels on the bottom row, are the individual 
currents $\J_a$.
}
\label{Fig:Vortex:Andreev-Bashkin}
\end{figure}

\paragraph{Skyrmions beyond the London limit:}
Physically the drag term $\mu_{12}\J_1\cdot\J_2$, promotes counter-flow compared to 
co-flowing currents. More precisely, the Andreev-Bashkin term gives a penalty on 
co-directed currents $\J_1$ and $\J_2$ (when $\mu_{12}>0$). Note that, as illustrated 
in \Figref{Fig:London-interaction:AB}, if the individual singularities do not overlap, 
the region in-between features counter-directed currents. It follows that the drag term 
(when $\mu_{12}>0$) reduces the energy of the configuration by maximizing the region 
where individual currents are counter-directed. Hence the Andreev-Bashkin term promotes 
the region of counter-directed currents and thus favours core-splitting.

The above calculations in the London limit thus predict that fractional vortices 
form a bound state where the individual singularities do not overlap. As illustrated 
in the Figure~\ref{Fig:Vortex:Andreev-Bashkin}, such solutions also exist beyond the 
London approximation (see more examples in \CVcite{Garaud.Sellin.ea:14}). The panels 
showing the individual densities clearly demonstrate that the numerically obtained 
solution feature non-overlapping vortices in both components. 
Moreover, the drag effect appears clearly from the panels showing the individual currents. 
Indeed, the \Figref{Fig:Vortex:Andreev-Bashkin} shows that the vortex in $\psi_1$ induces 
a non-zero circulating current in $\psi_2$. This can be seen by defining the 
supercurrent associated with a given condensate 
\Equation{Eq:Andreev-Bashkin:CurrentGL:2}{
     \J_a/e  =\left(1+\mu|\psi_a|^2\right)\Im(\psi_a^*\D\psi_a)	
	    +\mu|\psi_a|^2 \Im(\psi_b^*\D\psi_b)	\,, 
	    ~~~\text{with}~~b\neq a  \,.
}

As demonstrated in \CVcite{Garaud.Sellin.ea:14}, the two-component model with 
dissipationless drag interaction \Eqref{Eq:Andreev-Bashkin:FreeEnergy}, can be mapped 
to a Skyrme-Faddeev model similarly to the discussion in Section
\ref{Sec:Fractional-vortices:Skyrme-Faddeev}:  
\Equation{Eq:SF:FreeEnergy:AB}{
\frac{\F}{\F_0}\!=\!\bigintsss \frac{1}{2}(\Grad\varrho)^2
	+\frac{\varrho^2}{8}\partial_k n_a\partial_k n_a
  	+\frac{\J^2}{2we^2\varrho^2}+V(\varrho,{\bs n})	
+\frac{1}{2e^2}\left[\varepsilon_{ijk}\left(
\partial_i \left(\frac{J_j}{we\varrho^2}\right)
	-\frac{1}{4}\varepsilon_{abc}n_a\partial_i n_b\partial_j n_c \right)\right]^2
 .
}
The difference with the mapping \Eqref{Eq:SF:FreeEnergy:2} is the role of 
$w=1+\mu\varrho^2$ that renormalizes the terms involving the total current $\J$. 
The coreless vortices are thus characterized, as compared to singular vortices, by the 
additional topological charge $\Q(\Psi)$ \Eqref{Eq:CPN:Charge} or $\Q({\bs n})$ 
\Eqref{Eq:SF:Charge} that is quantized with $\Q=n$ (with $n$ being the number of carried 
flux quanta). Numerically calculated topological charge is found to be integer (with 
typical error of order $10^{-4}-10^{-5}$).

\vspace{1cm}

The pseudo-spin texture $\bs n$ \Eqref{Eq:SF:Projection} of the corresponding 
skyrmion tube is also displayed on the bottom left panel in the Figure 
\ref{Fig:Vortex:Andreev-Bashkin}. The skyrmion here consists of one meron and 
an anti-meron, standing for the fractional vortices in the individual components, 
similar to those illustrated in \Figref{Fig:Schematic:Skyrmion}. The mapping of 
the fractional vortices to point particles suggests the existence of long-range 
inter-skyrmions dipolar forces. Indeed, this mapping interprets the fractional 
vortices as point particles with opposite charges. 
A bound state of such charges thus features a dipole moment as well. The detailed 
analysis of the two-component $\groupU{1}\!\times\!\groupU{1}$ superconductors with 
interspecies dissipationless drag, in Ref.~\CVcite{Garaud.Sellin.ea:14}, confirmed 
this by various numerical simulations, in the full Ginzburg-Landau regime.

The magnetic field of these skyrmions, as can be seen in \Figref{Fig:Vortex:Andreev-Bashkin}, 
is different from that of singular vortex. Moreover, the interaction between vortices 
is long-ranged dipolar \cite{Garaud.Sellin.ea:14}. 
This results in an unconventional magnetic response in low fields which features 
vortex lattices lacking the conventional hexagonal structure. Importantly, the 
magnetization process, as well shows unconventional properties with square lattice 
growing inward from boundaries of the sample \CVcite{Garaud.Sellin.ea:14}. Such 
unusual properties can in principle be identified easily for example in scanning 
SQUID measurements.

\begin{figure}[!htb]
  \hbox to \linewidth{ \hss
\includegraphics[width=.975\linewidth]{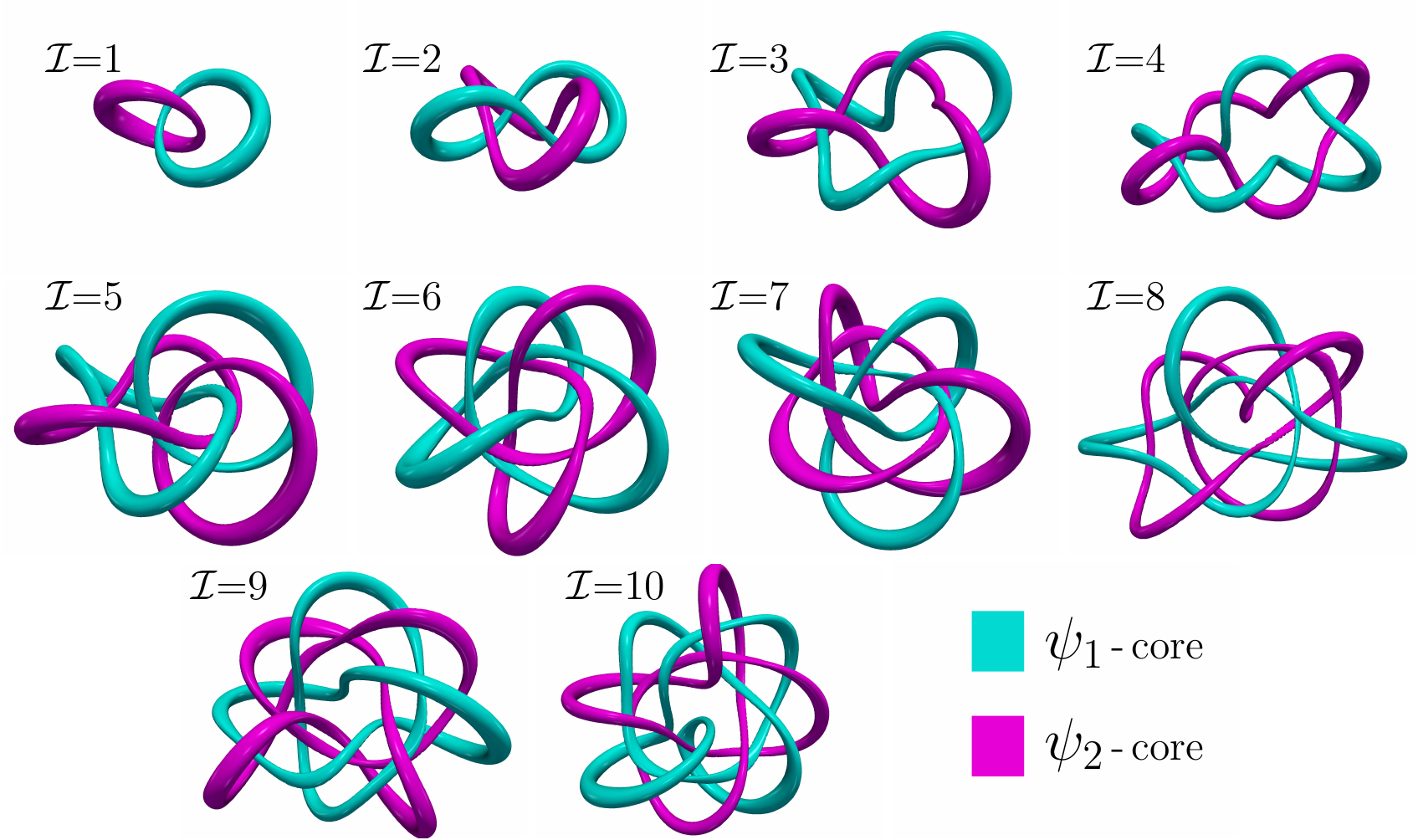}
\hss}
\caption{
Solutions for stable knots with topological charges $\I$=1-10. The various panels 
display the core structure where the cyan and magenta tubes are surfaces enclosing 
the cores of the fractional vortices in each component. When increasing the 
index $\I$, the topological structure of the vorticity becomes increasingly more 
complex\protect\footnotemark. 
}
\label{Fig:Knots}
\end{figure}

\paragraph{Hopfions beyond the London limit:}
The intuition suggests that twisted loops of such skyrmions might have much more 
complicated interaction, and possibly would not completely decay, as the loops of 
conventional vortices do. Such a twisted skyrmion loop consists of braided 
closed loops of fractional vortices. The numerical minimization, starting from various 
initial states of knotted and linked vortex loops, demonstrate that knotted vortices 
can indeed be stabilized because of the Andreev-Bashkin terms 
\CVcite{Rybakov.Garaud.ea:19}.
These simulations are, in a way, related to the relaxation processes of vortex 
tangles formed due to fluctuations. Indeed, in the absence of an external field, 
closed loops form dynamically, for example due to quenches or to thermal fluctuations. 
Thus the superconducting state here can support an infinite number of (meta-)stable 
solutions corresponding to topologically different ways to tie vortex knots. The 
Figure \ref{Fig:Knots} shows $10$ stable knotted vortex loops, termed hopfions, 
with the smallest values of the topological index~$\I$ \Eqref{Eq:Hopf:Degree}.

Note that unlike the previously discussed skyrmions, the hopfions here are characterized 
\emph{only} by the hidden topological invariant $\I$ \Eqref{Eq:Hopf:Degree}. Indeed 
the skyrmions, which are straight topological defects, are characterized by the global 
$\groupU{1}$ charge (the total winding) and by the additional topological charge 
$\Q(\Psi)$ \Eqref{Eq:CPN:Charge} or $\Q({\bf n})$ \Eqref{Eq:SF:Charge}. Knots on the 
other hands are trivial regarding the $\groupU{1}$ charge (the total winding is zero), 
while the degree $\I$ \Eqref{Eq:Hopf:Degree} is an integer.

\footnotetext{
Animations showing the structure of these knotted vortices, and their formation can 
be found in the Supplemental Material of \CVcite{Rybakov.Garaud.ea:19}, or at the 
following \href{http://www.theophys.kth.se/~garaud/knots.html}{webpage}.
}

The knotted solutions displayed in \Figref{Fig:Knots} are represented by their core 
structure, and labelled by the corresponding value of invariant $\I$ 
\Eqref{Eq:Hopf:Degree}. The cyan and magenta tubes enclose the singularities in the 
different condensates, and thus indicate the core structure. The knotted solutions are 
also characterized by a complicated structure of knotted lines of the magnetic field $\B$. 
These are not displayed here, but can be found from \CVcite{Rybakov.Garaud.ea:19}.
The solutions with $\I=1-4$ consist of two fractional vortex loops, linked together and 
twisted around each other a varying number of times. The solution with $\I=5$ is a bound 
state of two pairs of linked fractional vortex loops, and $\I=6$ consists of two linked 
trefoil knots. In all these configurations, there is a symmetry of exchanging the cores 
of the different condensates. Interestingly, for higher values of the topological index 
$\I=7-10$, the vorticity of the different components are inequivalent, and this symmetry 
is no longer available. For example, the $\I=7$ knot consists in a trefoil knot in one 
component, linked to two twisted fractional vortex loops of the other component.
These solutions thus form some kind of isomers.

In contrast to the London limit, the vortex splitting to support skyrmions in the 
Ginzburg-Landau regime requires a critical strength of the dissipationless drag.
Similarly, the coefficients of the Andreev-Bashkin term have to be sufficiently 
strong for the knotted solutions to be stable. Namely, the Andreev-Bashkin couplings 
$\hat{\mu}$ should be substantially larger than the usual gradient couplings 
$\kappa_{aa}$. It is not obvious why such terms should dominate. However, 
this can be realised in various scenarios, and especially in the vicinity of some 
critical points \CVcite{Rybakov.Garaud.ea:19}. 
For example near the phase transition to paired phases caused by strong correlations 
\cite{Kuklov.Svistunov:03,Sellin.Babaev:18}. There, the ratio of the stiffnesses of 
counter-flows and co-flows vanishes. This implies that, close enough to the critical 
point, the Andreev-Bashkin coupling can be arbitrarily strong \cite{Sellin.Babaev:18}.
Strong Andreev-Bashkin couplings can also occur close to the phase transition to 
Fulde-Ferrel-Larkin-Ovchinnikov (FFLO) state \cite{Fulde.Ferrell:64,
Larkin.Ovchinnikov:65}, where the gradient couplings $\kappa_{aa}$ change signs
(see, e.g., \cite{Buzdin.Kachkachi:97}), and the Andreev-Bashkin interaction 
remains non-zero. It results that close to a FFLO phase transition, even systems 
with relatively weak Andreev-Bashkin interactions $\mu_{ab}$, satisfy the 
requirements of the disparities of the coefficients $\hat{\kappa}$ and $\hat{\mu}$.

\subsection{Chiral skyrmions -- Vortex splitting on domain-walls} 
\label{Sec:Chiral-Skyrmions} 

Here, we present another mechanism that allows for the stabilization of coreless 
defects. The discussion here is heuristic, and these aspects will be addressed 
more quantitatively in the Chapter \ref{Chap:TRSB}. Namely, the properties of the 
domain-walls will be reconsidered in the Section \ref{Sec:TRSB:Domain-walls}, 
while the properties of the skyrmions will be addressed in the Section 
\ref{Sec:TRSB:Chiral-Skyrmions}.

Domain-walls are the topological defects associated with the spontaneous breakdown 
of a discrete $\groupZ{2}$ symmetry (see e.g. \cite{Manton.Sutcliffe,Vachaspati,
Vilenkin.Shellard,Rajaraman,Shnir:18}). The domain-walls are field configurations 
that interpolate between the (discrete) degenerate vacua of the theory. As discussed 
here, in the framework of multicomponent superconductors, the domain-walls typically 
interact non-trivially with the other topological defects such as vortices. This leads 
to a new kind of topological defects that can share properties of both domain-walls 
and vortices. In particular, the interaction with domain-walls provides a mechanism
to split vortex cores, and thus a possibility to stabilize coreless defects as the 
bound state of fractional vortices interacting with domain-walls 
\cite{Garaud.Carlstrom.ea:11,Garaud.Carlstrom.ea:13,Garaud.Babaev:14}.

\begin{figure}[!htb]
\hbox to \linewidth{ \hss
\includegraphics[width=.5\linewidth]{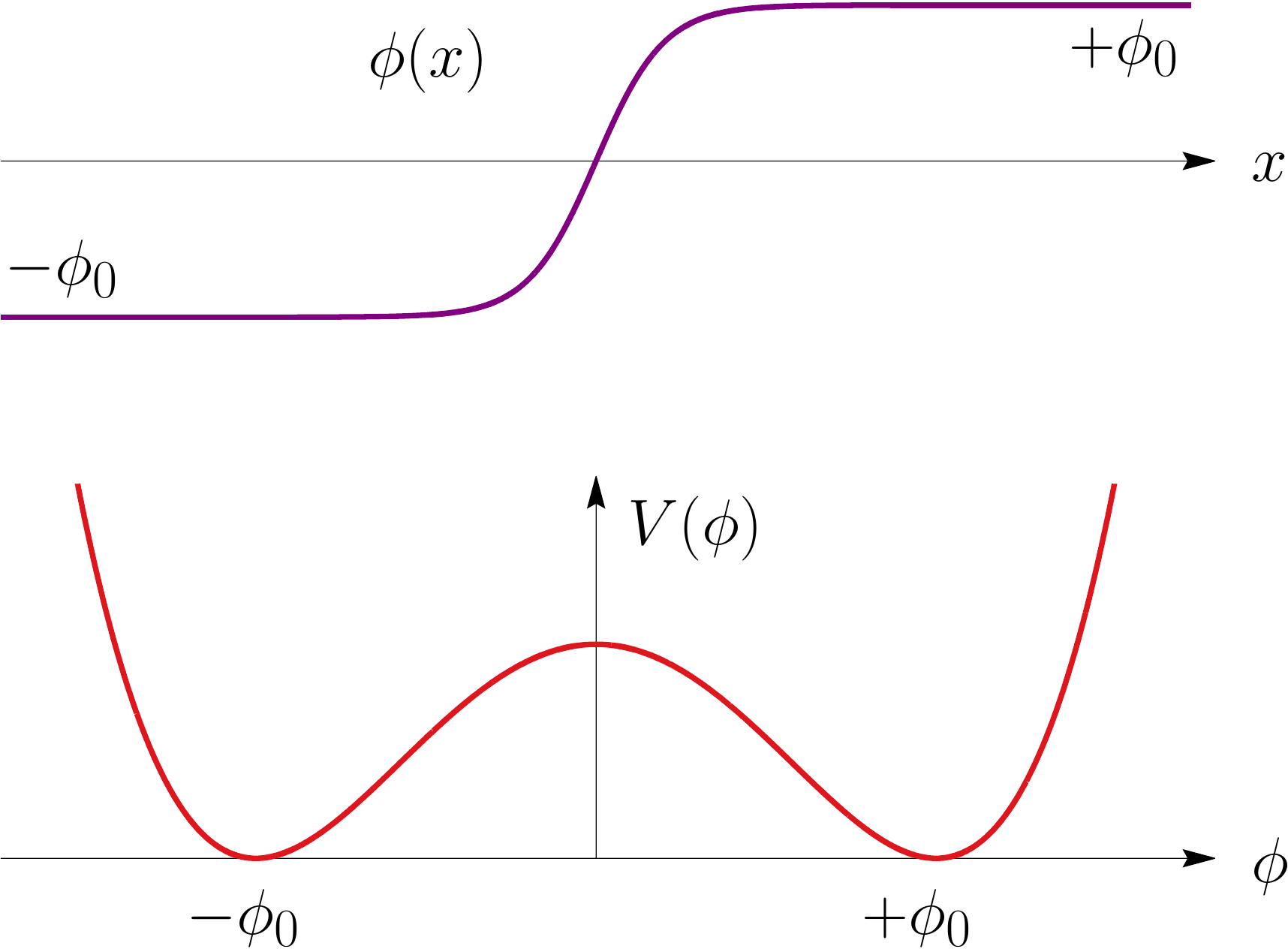}
\hss
\includegraphics[width=.4125\linewidth]{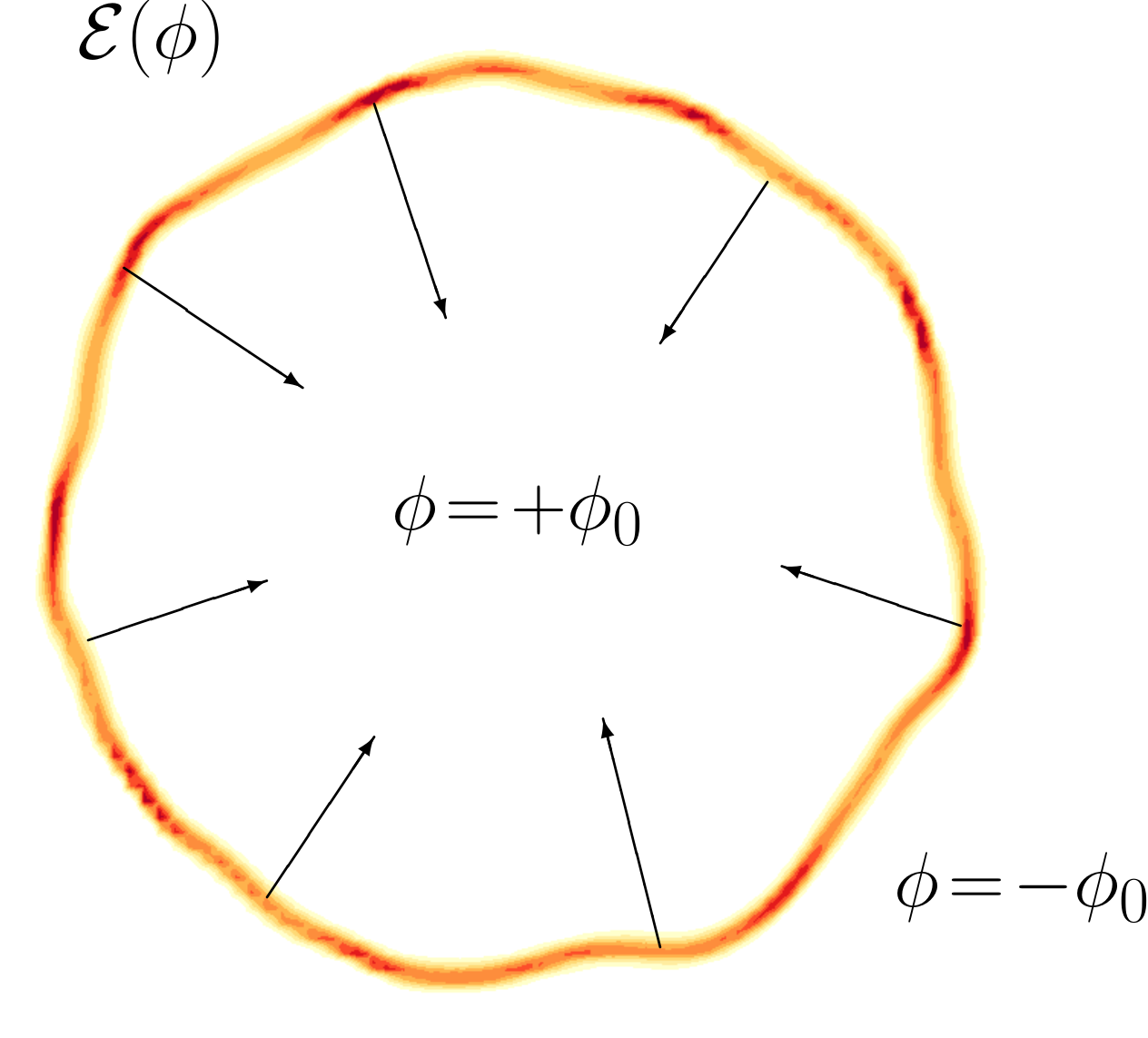}
\hss}
\caption{
The left panel shows the potential energy (bottom), with a discrete degeneracy of 
the ground state: $\pm\phi_0$; and a typical one-dimensional domain-wall solution 
that interpolates between the two inequivalent ground states (top). 
The right panel displays a closed domain-wall, where the inner region is in one 
ground-state while the exterior falls into the other ground state. The closed 
domain-wall tends to collapse because of its line tension (the energy per 
unit-length $\cal E$).
}
\label{Fig:DW:1}
\end{figure}

\vspace{1cm}

The domain-walls, also referred to as \emph{kinks}, thus appear in models where the 
ground state has a discrete degeneracy. The $\phi^4$-model is a textbook example 
of a theory that features domain-wall solutions (for detailed textbook discussion 
see \eg \cite{Manton.Sutcliffe,Vachaspati,Vilenkin.Shellard,Rajaraman,Shnir:18}). 
As a reminder, this is a theory of a real scalar field with a quartic self-interacting 
potential $V(\phi)\propto (\phi^2-\phi_0^2)^2$. As illustrated in \Figref{Fig:DW:1}, 
the ground state of this potential has a discrete degeneracy: $\pm\phi_0$. 
In one spatial dimension, the domain-walls feature topological protection and thus 
cannot transform into either of the ground-states. Indeed, the configurations that 
interpolate between $-\phi_0$ and $+\phi_0$ fall into disjoint homotopy classes, 
and thus no finite energy transformation can transform it to a constant ground state, 
see \Figref{Fig:DW:1}. 
In two dimensions, on the other hand, the Hobart-Derrick theorem implies that 
there are no static finite-energy solutions \cite{Hobart:63,Derrick:64}. Indeed, 
either the field asymptotically fall into degenerate ground states and in that 
case the domain-wall is infinitely long, or the domain-wall is closed with a finite 
length, but topological trivial. 
Thus closed domain-walls in two dimensions, being topologically trivial, do not 
enjoy any kind of topological protection. The energy of a closed domain-wall is 
finite and proportional to its length. Hence, as illustrated on the right panel of 
the Figure \ref{Fig:DW:1}, since a closed domain-wall has no topological protection, 
it dynamically collapses due to its line tension (similarly to a vacuum bubble that 
shrinks because of its surface tension).

There exist various possibilities for a superconducting state to spontaneously break 
a discrete $\groupZ{2}$ symmetry. One of them, is to break the time-reversal symmetry. 
The breakdown of the time-reversal symmetry typically occurs due to the competition 
between different phase-locking terms, between the different condensates. Aspects of 
time-reversal symmetry breaking in multicomponent superconductors, and especially 
the so-called $\sis$ state, are developed in details later in the Chapter \ref{Chap:TRSB}.
The details of the spontaneous breakdown of the time-reversal symmetry, may differ 
depending on the underlying properties of the model under consideration. For example, 
depending on whether the system is described by a two- or three-component Ginzburg-Landau 
theory, the terms responsible for the spontaneous breakdown of the time-reversal 
symmetry will be different (see the discussion in the Chapter \ref{Chap:TRSB}).
Considering again the scalar multiplet $\Psi$, the superconducting ground state is 
the field configuration $\Psi_0$ defined as $\Psi_0:=\argmin V(\Psi,\Psi^\dagger)$.
Heuristically, the time-reversal symmetry of the ground-state can be understood
as the invariance $\Psi_0$ under complex conjugation (up to global $\groupU{1}$ 
transformations). Conversely, a superconducting state that spontaneously breaks the 
time-reversal symmetry satisfies the condition
\Equation{Eq:BTRS:Condition}{
\Psi_0^*\Exp{i\chi}\neq\Psi_0~~\forall\chi\,.
}

\vspace{1cm}

Depending on the number of components, the difference between time-reversal symmetric 
(TRS) and broken time-reversal symmetry (BTRS) states, can be illustrated with the 
following simple examples (for normalized states $\Psi_0^\dagger\Psi_0=1$). 
For example, in the case of two components:
\Align*{
\Psi_0&=(1,- 1)/\sqrt{2}&:~~~\,\Psi_0^*&=(1,- 1)/\sqrt{2}=   \Psi_0	&
	&\Longrightarrow  \text{TRS\phantom{B}}		\\
\Psi_0^\pm&=(1,\pm i)/\sqrt{2}&:~\Psi_0^{\pm*}&=(1,\mp i)/\sqrt{2}=\Psi_0^\mp
	\neq\Psi_0^\pm &	&\Longrightarrow \text{BTRS}			\,.	\\ 
}
Similarly, the case of three-components can be illustrated by:
\Align*{
\Psi_0&=(1,1,- 1)/\sqrt{3}&:~~~\,\Psi_0^*&=(1,1,- 1)/\sqrt{3}=   \Psi_0	&
	&\Longrightarrow  \text{TRS\phantom{B}}		\\
\Psi_0^\pm&=(1,\Exp{\pm i\pi/3},\Exp{\mp i\pi/3})/\sqrt{3}&:~\Psi_0^{\pm*}&=
	(1,\Exp{\mp i\pi/3},\Exp{\pm i\pi/3})/\sqrt{3}=\Psi_0^\mp
	\neq\Psi_0^\pm & &\Longrightarrow \text{BTRS}	\,.
}
The superconducting states that break the time-reversal symmetry are thus not 
invariant under complex conjugation. More precisely, the time-reversal operations 
(the complex conjugation) sends one ground state onto the other. Symmetry-wise, 
in addition  to the spontaneous breakdown of the $\groupU{1}$ gauge symmetry, 
the time-reversal symmetry breaking states also break the discrete $\groupZ{2}$ 
symmetry associated with the complex conjugation. Using the above notations, 
$\Psi_0^\pm$ are thus two disconnected ground-states related to each other by the 
time-reversal operations $\Psi_0^{\pm*}=\Psi_0^\mp$. Hence the theory supports 
domain-walls as those illustrated in the Figure \ref{Fig:DW:1}.

As discussed in more details in the Chapter \ref{Chap:TRSB}, the frustrated competition 
between phase-locking terms in superconductors with two, three (or more) components 
yields ground states where the relative phases $\varphi_{ab}=\varphi_{b}-\varphi_{a}$ 
are neither $0$ nor $\pi$. Since it is not invariant under complex conjugation, such 
a ground-state spontaneously breaks the time-reversal symmetry \cite{Ng.Nagaosa:09,
Stanev.Tesanovic:10}. Indeed, for a two-component superconductor, the inter-component 
Josephson interaction $\propto |\psi_a||\psi_b|\cos\varphi_{ab}$ either locks or 
anti-locks the phases, so that the ground-state relative phase is respectively $0$ 
or $\pi$. With more than two components, each inter-component Josephson coupling 
favours (anti-)locking of the two corresponding phases. However, these Josephson terms 
can collectively compete so that optimal phases are neither locked nor anti-locked. 
Let consider here a simple potential for a three-component model with competing 
phase-locking terms
\Equation{Eq:ChiralSK:potential}{
 V=\sum_{a=1}^3\Bigg\{-|\psi_a|^2+\frac{1}{2}|\psi_a|^4 +
 \sum_{b>a}^3|\psi_a||\psi_b|\cos\varphi_{ab}\Bigg\}\,.
}
The invariance under complex conjugation implies that, the potential energy does not 
change if the sign of all relative phases is changed, $\varphi_{ab}\to-\varphi_{ab}$. 
It follows that if any of the relative phases $\varphi_{ab}$ is not an integer 
multiple of $\pi$, then the ground state has an additional discrete $\groupZ{2}$ 
degeneracy. This is the case of the potential \Eqref{Eq:ChiralSK:potential} that 
admits two possible ground states: $\Psi_0^+$ with 
$\varphi_{12}=2\pi/3,\;\varphi_{13}=-2\pi/3$ or $\Psi_0^-$ with
$\varphi_{12}=-2\pi/3,\;\varphi_{13}=2\pi/3$. This potential thus allows for 
domain-walls that interpolate between $\Psi_0^+$ and $\Psi_0^-$.

\begin{figure}[!htb]
\hbox to \linewidth{ \hss
\includegraphics[width=.5\linewidth]{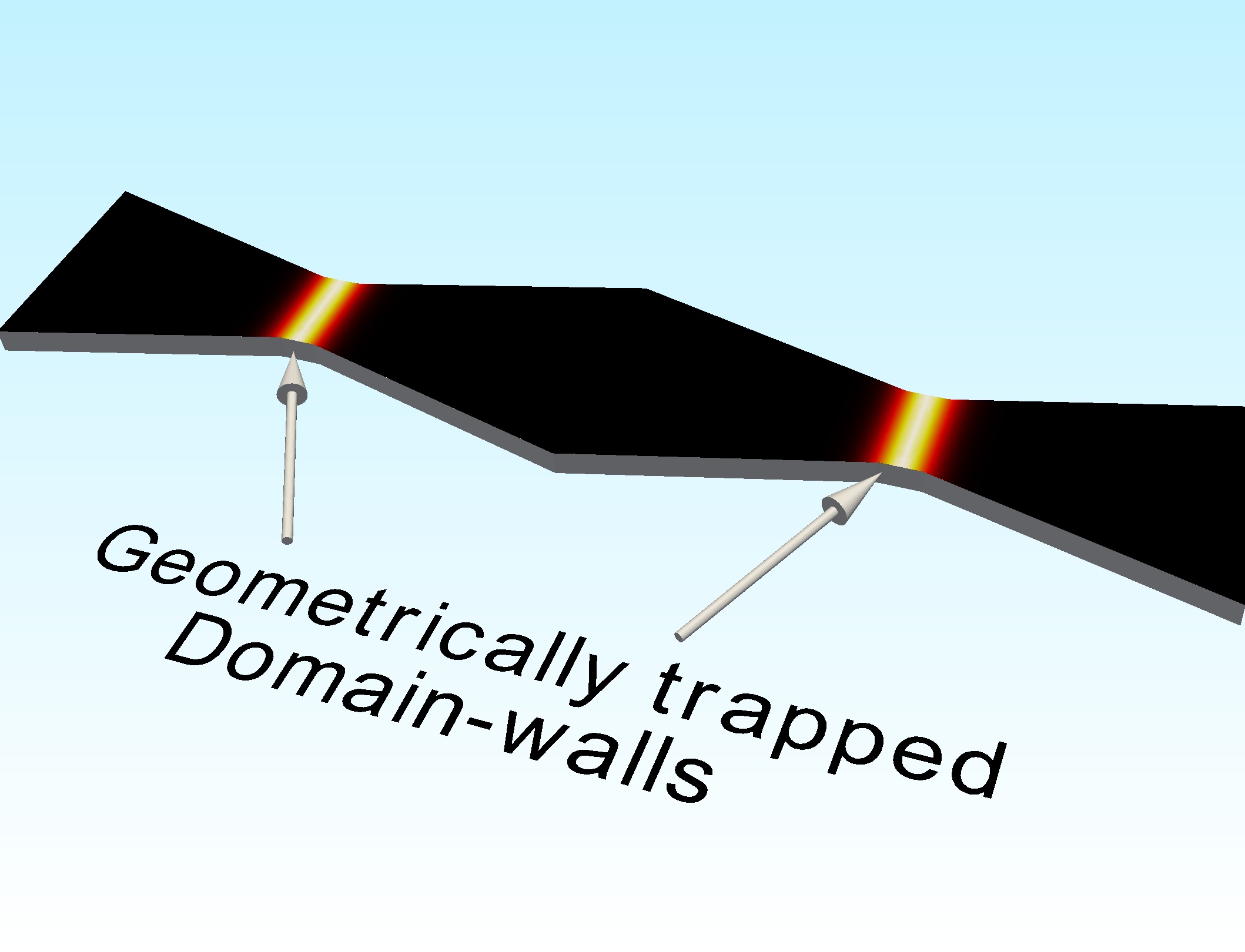}
\hss}
\caption{
This sketches the principle of the geometric stabilization of domain-walls in 
non-convex domains \CVcite{Garaud.Babaev:14}. The domain-walls, for example formed 
via the Kibble-Zurek mechanism in a rapid quench, are geometrically trapped. Indeed, 
for a domain-wall to escape it should increases its length, which is energetically 
costly.
}
\label{Fig:DW:2}
\end{figure}

The domain-walls can typically form during cooling processes via the Kibble-Zurek 
mechanism \cite{Kibble:76,Zurek:85}. As mentioned previously, closed domain-walls 
in two spatial dimensions are unstable and collapse for dynamical reasons. Similarly 
domain-walls in finite domains decay for the same reasons. However, as demonstrated 
in \CVcite{Garaud.Babaev:14} they can be stabilized by geometric barrier in non-convex 
domains, or by the existence of pinning centres. Such a mechanism for the geometric 
stabilization of domain walls, illustrated in \Figref{Fig:DW:2}, may help for their 
observability. If during a quench a domain-wall ending on the short (non-convex) 
parts of the domain, is created via the Kibble-Zurek mechanism, it can relax to a 
(meta)stable configuration. Indeed, for both ends of the domain-wall to join and 
then collapse to zero size, the domain-wall would have to first increase its length. 
Unlike the domain-walls in chiral $p$-wave superconductors, that carry a uniform 
magnetic field orbital momentum of Cooper pairs (see e.g. \cite{Ferguson.Goldbart:11,
Raghu.Kapitulnik.ea:10,Vadimov.Silaev:13}), domain-walls between $\sis$ states carry 
a magnetic field only locally \CVcite{Garaud.Babaev:14}. Namely it features opposite 
local field at its ends, with a total net flux through the sample that is zero.

\vspace{1cm}along the walls 

The fractional vortices are confined together, due to the energy cost associated with the 
neutral sector. Note that since the phase-locking terms in \Eqref{Eq:ChiralSK:potential} 
provide a potential energy for the relative phases, the attraction here is linear and 
not logarithmic as derived in \Eqref{Eq:Quantization:Interaction:Unlike}.
A domain-wall interpolating between two time-reversal symmetry broken states is, 
by definition the region where the phase-locking is the least optimal. As a result, 
it allows to accommodate a more favourable relative phase by splitting the integer 
flux singular vortices into fractional flux vortices. Moreover since the superfluid 
density is suppressed on a domain-wall, because of the field gradients, it can pin 
vortices. Thus while in the bulk, the fractional vortices are confined because of 
the strong linear attractive interaction, they repel each other when bound to a 
domain-wall (see detailed discussion in \cite{Garaud.Carlstrom.ea:11,
Garaud.Carlstrom.ea:13,Garaud.Babaev:14}, and later in Section 
\ref{Sec:TRSB:Chiral-Skyrmions}).
This implies that the magnetization processes can be strongly altered if a system 
has some pre-existing domain-walls that are geometrically trapped 
\CVcite{Garaud.Babaev:14}. Indeed, since some density components are depleted at 
the domain wall, the vortex entry for the corresponding component costs much less 
energy there, than from other parts of the boundaries. It follows that the first 
vortex entry can occur at lower fields than $\Hc{1}$, since they have to overcome 
a smaller Bean-Livingston barrier on the domain-wall than in the bulk.

The interaction between fractional vortices confined on the domain-walls between 
different time-reversal symmetry broken $\sis$ states is repulsive. On the other 
hand, closed (bare) domain-walls, as sketched in \Figref{Fig:DW:1}, collapse due 
to their line tension. This thus hints to a scenario where both tendency would 
compromise, leading to an object with a closed domain-wall of finite size, stabilized 
by the repulsion between confined fractional vortices. As demonstrated in 
\CVcite{Garaud.Carlstrom.ea:11} and \CVcite{Garaud.Carlstrom.ea:13}, such composite 
objects can indeed be formed either as metastable states, or as thermodynamically stable 
excitations, in external fields. Moreover, as these flux carrying topological defects 
consist in non-overlapping fractional vortices, they carry a quantized $\groupCP{2}$ 
topological invariant \Eqref{Eq:CPN:Charge}. These composite objects, that are 
bound states of domain-walls and fractional vortices are called chiral $\groupCP{2}$ 
skyrmions. The prefix \emph{chiral} refers to the fact that the fractional vortices 
order differently along the wall, in the different time-reversal symmetry broken states 
\CVcite{Garaud.Carlstrom.ea:13}.

The chiral $\groupCP{2}$ skyrmions can be formed via various scenario, as for example 
during field-cooled quenches in external fields \cite{Garaud.Babaev:14}. Since the have 
very specific magnetic signatures, they may be observed (and easily discriminated from 
singular vortices) for example in high-resolution scanning SQUID, Hall, or magnetic 
force microscopy measurements \cite{Garaud.Carlstrom.ea:11,Garaud.Carlstrom.ea:13}.

\subsection{Skyrmions stabilized by condensate repulsion}
\label{Sec:Bi-quadratic}

It was previously emphasized that the existence of skyrmions requires a stabilizing 
mechanism to counteract the predisposition of the singularities to superimpose. 
Some of the mechanisms presented above can split the vortex cores, even in the London 
limit. This is for example the case of condensates with commensurate charges in 
Sec.~\ref{Sec:Mixtures}, or with dissipationless drag introduced in 
Sec.~\ref{Sec:Andreev-Bashkin}. The core-splitting mechanism is still effective 
beyond the London limit. Roughly speaking, these mechanisms for core splitting rely 
on modifications on the kinetic terms. On the other hand, the mechanism that splits 
the fractional vortices confined on a domain-wall, introduced in 
Sec.~\ref{Sec:Chiral-Skyrmions}, has no counterpart in the London limit. The stabilization 
of the chiral $\groupCP{2}$ skyrmions is thus purely a nonlinear effect. This section, 
presents another mechanism which is also purely nonlinear, and that originates only 
in the potential terms; namely the bi-quadratic inter-component coupling. 
This interaction which impedes the coexistence of the condensates is relevant for 
a broad variety of models.

Using the generic structure of the potential term \Eqref{Eq:General:FreeEnergy}, 
the bi-quadratic interactions are the terms which are quadratic in two of the individual 
densities. Namely they read as
\Equation{Eq:Bi-quadratic}{
  \cdots+\sum_{a,b>a} \beta_{ab}|\psi_a|^2|\psi_b|^2+\cdots\,,
}
where the coefficients correspond to $\beta_{ab}:=\beta_{abab}$ in  
\Eqref{Eq:General:FreeEnergy}. Depending on the sign of the bi-quadratic coupling 
$\beta_{ab}$, this interaction either promotes ($\beta_{ab}<0$) or impedes 
($\beta_{ab}>0$) the coexistence of both condensates densities $|\psi_a|^2$ and 
$|\psi_b|^2$. For example, if the bi-quadratic interaction is repulsive ($\beta_{ab}>0$) 
and strong, it can promote a phase separation where one of the condensate vanishes 
in the ground state (for example $|\psi_a|\neq0$ and $|\psi_b|=0$).
These terms not only modify the ground state properties, but also alter the 
interactions between the elementary topological defects. Indeed, the repulsive
bi-quadratic interaction (when $\beta_{ab}>0$) typically promotes core splitting 
of the composite vortices. That is, it enforces the situation where the singularities 
of the ensuing fractional vortices do not overlap.
This kind of inter-component interactions obviously cannot be accounted by the 
the London limit, where densities are constant.

\begin{figure}[!htb]
\hbox to \linewidth{ \hss
\includegraphics[width=.95\linewidth]{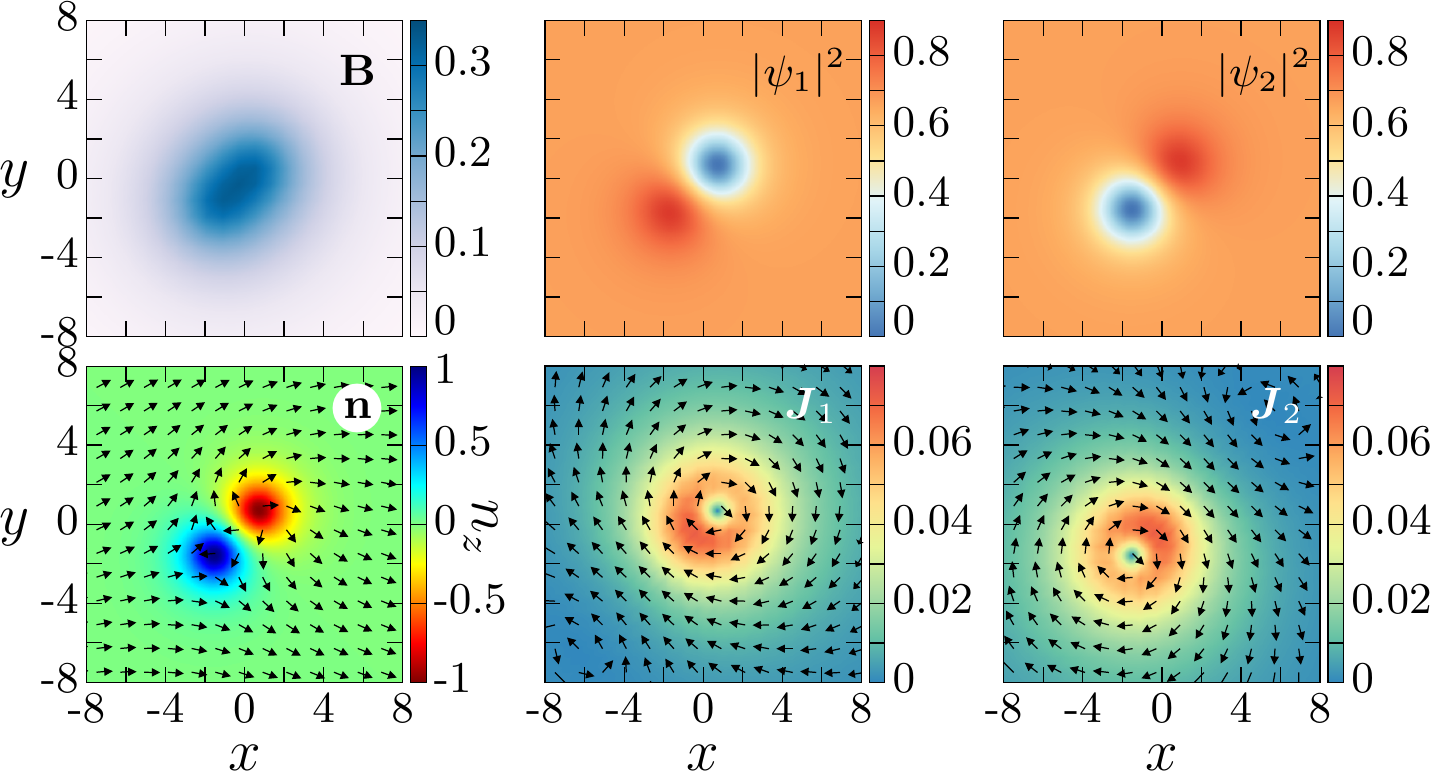}
\hss}
\caption{
A skyrmion solution of a two-component superconductor, stabilized by the repulsive 
bi-quadratic interaction \Eqref{Eq:Bi-quadratic}. The parameters of the later detailed 
potential \Eqref{Eq:Type-1.5:AnalyticEx:Potential} are $(\alpha_{aa},\beta_{aa})=(-1,1)$, 
$\beta_{12}>0$ and the gauge coupling is $e = 0.25$.
The displayed quantity on the top left panel is the magnetic field. The other panels 
on the top row, show the individual densities of the superconducting condensates 
$|\psi_1|^2$ and $|\psi_2|^2$. 
The bottom left panel shows the associated pseudo-spin texture $\bs n$ 
\Eqref{Eq:SF:Projection}, while the other panels on the bottom row, are the individual 
currents $\J_a$.
}
\label{Fig:Vortex:Biquadratic}
\end{figure}

As illustrated in the Figure \ref{Fig:Vortex:Biquadratic}, the bi-quadratic interaction 
indeed can favour core splitting. It is clear from the panels showing the individual 
densities that the solution feature non-overlapping vortices in both components, thus 
implying that the vortex is coreless. It follows that again it is characterized by 
the additional invariant $\Q(\Psi)=1$ \Eqref{Eq:CPN:Charge}.
The skyrmion here consists of one meron and an anti-meron of the pseudo-spin texture 
$\bs n$ \Eqref{Eq:SF:Projection}. Again these stand for the fractional vortices in the 
individual components, similarly to \Figref{Fig:Schematic:Skyrmion}.

These skymrions have a magnetic field that is not axially symmetric, and thus may be 
detectable by local magnetic field measurement, such as scanning SQUID.
Furthermore, similarly to the skyrmions stabilized by the Andreev-Bashkin term in 
Sec. \ref{Sec:Andreev-Bashkin}, the relative phase exhibits a dipolar mode that is 
long-ranged. These skyrmions thus interact non trivially together.

As already mentioned, the bi-quadratic interaction \Eqref{Eq:Bi-quadratic} in the 
phenomenological Ginzburg-Landau models, may occur in a broad variety of contexts 
describing different microscopic physics. Below, is a brief overview of the different 
microscopic systems which yield a bi-quadratic interaction that stabilizes skyrmions.

\paragraph{Skyrmions in interface superconductors.}
Superconductors with strong Rashba spin-orbit coupling can exhibit skyrmions similar 
to that displayed in \Figref{Fig:Vortex:Biquadratic}. More precisely, in a weak-coupling 
theory for a clean superconductor with isotropic pairing interactions ($s$-wave pairing), 
an in-plane Zeeman field, and a strong Rashba spin-orbit coupling, the resulting 
Ginzburg-Landau theory features bi-quadratic interaction \Eqref{Eq:Bi-quadratic}. 
As discussed in details in \CVcite{Agterberg.Babaev.ea:14}, this applies for example 
to interface superconductors, such as SrTiO$_3$/LaAlO$_3$.

\paragraph{Skyrmions in immiscible mixtures of two condensates.}
The bi-quadratic interactions can also be responsible for the spontaneous breakdown 
of a discrete $\groupZ{2}$ symmetry, different than the competition between phase-locking 
terms discussed in Sec. \ref{Sec:Chiral-Skyrmions}. For example, if a global $\groupSU{2}$ 
symmetry is explicitly broken by bi-quadratic interactions \Eqref{Eq:Bi-quadratic}. 
Namely, for a potential term 
$V=\Lambda(\Psi^\dagger\Psi-\Psi_0^2)^2+\delta|\psi_1|^2|\psi_2|^2$. 
There the global $\groupSU{2}$ symmetry of the potential is explicitly broken down to 
$\groupU{1}\!\times\!\groupU{1}\!\times\!\groupZ{2}$. If the symmetry breaking parameter 
$\delta>0$, then the condensates cannot coexist and the ground state is either $(\Psi_0,0)$ 
and $(0,\Psi_0)$. This thus describes an immiscible mixture of two condensates. 
As a result the theory also supports domain-walls which can combine to the vorticity 
and result into skyrmions \CVcite{Garaud.Babaev:14b}. This is easily understood that 
when one of the component is suppressed (\eg at a vortex core), then it is beneficial 
to condense the other component there. They can form giant skyrmions carrying several 
flux quanta even in a type-2 regime. As discussed \CVcite{Garaud.Babaev:14b} and 
\CVcite{Garaud.Babaev:15}, this strongly affect the magnetization properties. 

\paragraph{Skyrmions in nematic superconductors}
The Ginzburg-Landau theory that describes some odd-parity nematic superconductors 
feature skyrmions similar to that displayed \Figref{Fig:Vortex:Biquadratic}. 
As argued in \CVcite{Zyuzin.Garaud.ea:17}, Cu$_x$Bi$_2$Se$_3$ is a candidate material 
for the existence of such skyrmions.
Note however that the appropriate Ginzburg-Landau theory to describing such nematic 
superconductors, also features additional kinetic terms that are anisotropic and mix 
the different components. These extra terms read as \CVcite{Zyuzin.Garaud.ea:17}
\Equation{Eq:Nematic:mixed}{
	 \kappa_1\Re\left[ (D_x\psi_1)^*D_x\psi_2 - (D_y\psi_1)^*D_y\psi_2 \right]	
    +\kappa_2\Im\left[ (D_x\psi_1)^*D_y\psi_2 + (D_y\psi_1)^*D_x\psi_2 \right]	\,.
}
Such term definitely alter the structure of the skyrmions, and their interactions. 
However, this is again the bi-quadratic interaction that is the principal ingredient 
of the vortex splitting.

\paragraph{Skyrmions in chiral $p$-wave superconductors}
Chiral $p$-wave superconductors also feature bi-quadratic interaction that promotes 
core splitting. Like for the nematic superconductors, they also feature additional 
kinetic terms of the form \Eqref{Eq:Nematic:mixed}. 
As demonstrated in \CVcite{Garaud.Babaev:12}, the Ginzburg-Landau models for $\pip$ 
superconductors, do support skyrmion excitation. Depending on the regions of the 
parameter space they can be energetically favoured as compared to singular vortices. In 
particular, it was demonstrated that charge-2 skyrmions can be interpreted as two-quanta 
vortices, and that they are always preferred over isolated single quanta vortices 
\CVcite{Garaud.Babaev:15a}. Lattices of these two-quanta vortices form spontaneously in 
an external magnetic field. In high applied field, the lattice of two-quanta vortices 
dissociates into a lattice of single quantum vortices \CVcite{Garaud.Babaev.ea:16}, and 
this picture persists beyond the mean field approximation \CVcite{Krohg.Babaev.ea:21}.


\graphicspath{{Plots/04-Type-1.5/}}
\chapter{Type-1.5 superconductivity}	
\label{Chap:Semi-Meissner}

Superconductors with multicomponent order parameters, not only allow for a rich 
zoology of topological defects that have no counterpart in single-component systems
(fractional vortices, skyrmions, hopfions, etc), but they also allow for richer 
kind of interactions between them. 
As introduced in the previous chapter, the elementary topological excitations in 
multicomponent superconductors are vortices carrying a fraction of the flux quantum. 
These combine to form composite defects (either singular or coreless) so that, 
in the bulk, the only finite-energy excitations carry an integer flux. Regardless 
of their core structure, the interactions between the topological defects are ruled, 
to a large extent, by the characteristic length-scales of the theory (or equivalently 
the mass spectrum). 
Single-component superconducting condensates are characterized by a the coherence 
length $\xi$ associated with the density variations (Anderson-Higgs mode). 
Multicomponent order parameters, on the other hand, typically feature several 
length-scales. While the associated scalar modes are typically attractive, the 
charged modes, associated with penetration depth $\lambda$ of the gauge field, 
mediate repulsion between flux carrying defects. See the discussion about 
single-component superconductors in the Appendix \ref{App:Single-Component}.

\begin{figure}[!htb]
\hbox to \linewidth{ \hss
\includegraphics[width=.675\linewidth]{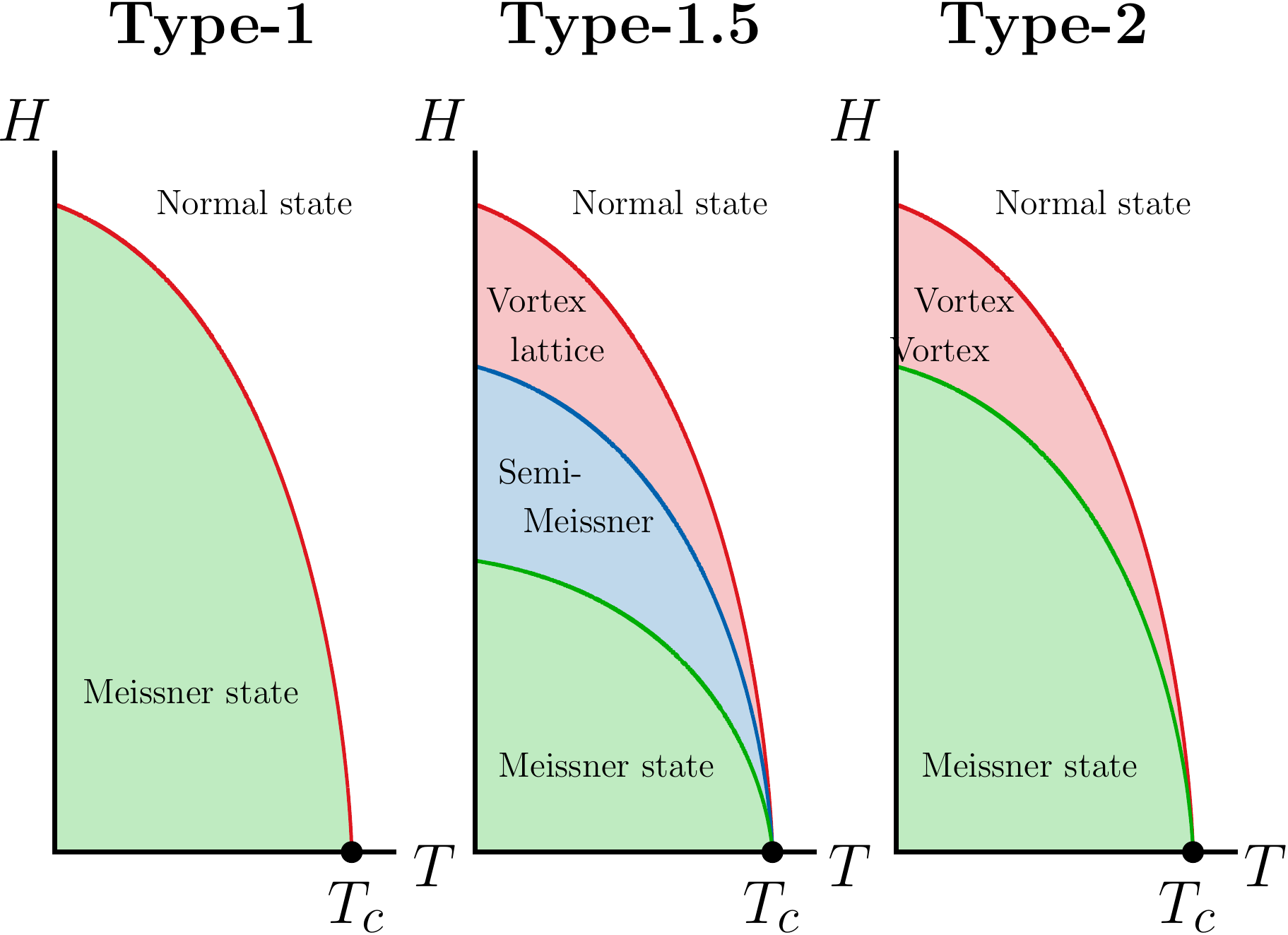}
\hss}
\hbox to \linewidth{ \hss
\includegraphics[width=.65\linewidth]{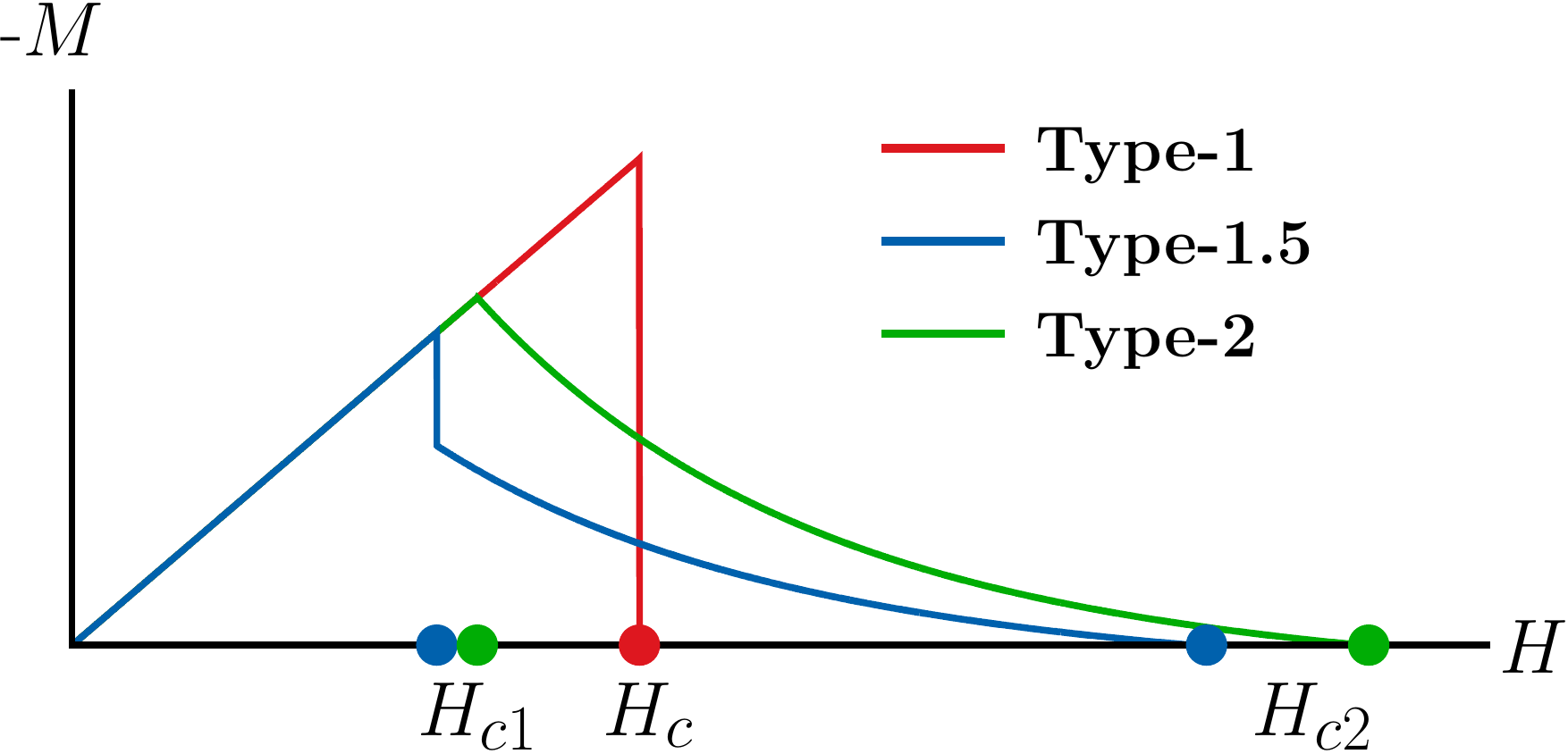}
\hss}
\caption{
Phase diagrams of superconductors in type-1, type-2 and type-1.5 regimes.
} \label{Fig:Types-diagrams}
\end{figure}

The textbook classification divides superconductors into two classes, depending on 
their behaviour in an external field. This classification is quantified by the 
dimensionless Ginzburg-Landau parameter $\kappa$ defined as the ratio of both 
fundamental length-scales $\kappa=\lambda/\xi$. When $\sqrt{2}\lambda<\xi$ (type-1), 
superconductors expel low magnetic field (the Meissner state), while macroscopic 
normal domains are formed when large fields are applied \cite{Ginzburg.Landau:50,
Gennes}. On the other hand type-2 superconductors, for which $\xi<\sqrt{2}\lambda$, 
feature thermodynamically stable vortex excitations \cite{Abrikosov:57}. More precisely, 
the magnetic field is expelled below some critical value $\Hc{1}$. Above this value, 
and until the destruction of superconductivity at the second critical field $\Hc{2}$, 
type-2 superconductors form lattices or liquids of vortices carrying a flux quantum. 
These behaviour are summarized in \Figref{Fig:Types-diagrams} and in tables 
\ref{Table:type-1.5:a} and \ref{Table:type-1.5:b}. 
The energy cost of a boundary between normal and superconducting states is positive 
in the type-1 regime. The absence of thermodynamically stable vortices, and the 
formation of macroscopic normal domains, thus follows from the minimization of 
the interface energy. It results that intervortex forces are purely attractive 
(thus vortices collapse to a giant vortex). On the other hand, type-2 superconductors 
support thermodynamically stable single-quanta vortices, as the boundary energy 
between normal and superconducting states is negative. The interaction between 
vortices is purely repulsive, and they form (triangular) Abrikosov lattices 
\footnote{
As stated in the introduction, these lattices can be seen as crystal realisation 
of the vortex-matter. This in a sense resonates with Kelvin's ides.
}
\cite{Abrikosov:57}. In the Ginzburg-Landau theory, at the critical value 
$\kappa=1/\sqrt{2}$ (called the Bogomol'nyi point), vortices do not interact 
\cite{Kramer:73,Bogomolnyi:76}. There, the current-current repulsion exactly 
compensates the core-core attraction at all distances.

\vspace{1.5cm}

Unlike in single-component superconductors, it is not possible to construct a single 
dimensionless parameter for multicomponent superconductors featuring several coherence 
lengths $\xi_a$. Hence the usual type-1/type-2 dichotomy is insufficient to capture 
the whole physics, and to classify multicomponent superconductors. Indeed, since the 
coherence lengths $\xi_a$ associated with the superconducting condensates are typically 
different, the penetration depth $\lambda$ can formally be an \emph{intermediate} 
length-scale: $\xi_1\!<\!\cdots\!<\!\sqrt{2}\lambda\!<\!\cdots\!<\!\xi_N$. For such a 
length-scale hierarchy, the modes associated with the length-scales larger than 
$\sqrt{2}\lambda$  provide long-range attraction as in type-1 superconductors. 
On the other hand, the modes with length-scales shorter than $\sqrt{2}\lambda$ enable 
short-range repulsion, as in the type-2.

Consider, for example, a two-component superconductor satisfying this length-scale 
hierarchy: $\xi_1<\sqrt{2}\lambda<\xi_2$. There, the mode associated with the 
largest length-scale $\xi_2$ should provide a long-range intervortex attraction
(loosely speaking due to the ``outer cores" overlap). On the other hand, the 
current-current and electromagnetic interaction associated with $\lambda$ provides 
a short-range repulsive interaction. The competition between these behaviours 
opens the possibility of a non-monotonic intervortex interacting potential that 
is long-range attractive (as in type-1 regime) and short-range repulsive (as in 
type-2 regime). This compromise between the type-1 and type-2 behaviours motivated 
the \emph{type-1.5} terminology of such states \cite{Moshchalkov.Menghini.ea:09}. 
The hierarchy where $\lambda$ is an intermediate length-scale is a necessary, 
although not sufficient condition, to realize non-monotonic intervortex interactions. 
Yet, if realized, the non-monotonic forces result in a preferred intervortex distance
such that two vortices form a bound state, and that many vortices coalesce to 
form vortex aggregate (clusters), coexisting with macroscopic domains of Meissner 
(vortexless) state: the \emph{semi-Meissner} state \cite{Babaev.Speight:05}.

\begin{table*}
\begin{tabular}{|p{0.18\linewidth}||p{0.25\linewidth}|p{0.25\linewidth}|p{0.25\linewidth}|}
\hline
  & {\bf Characteristic lengths-scales} 
  & {\bf Intervortex interaction} 
  & {\bf Energy of $N$-quantum vortex }

\\ \hline \hline
 {\bf  Single-component type-1 } 
 & Penetration length $\lambda$ and  coherence length $\xi$, with
 	$\lambda/\xi< 1/\sqrt{2}$
 & Attractive 
 & $\frac{E(N)}{N}<\frac{E(N-1)}{N-1}$. \newline{}
 Vortices collapse onto an $N$-quantum single mega-vortex.

\\ \hline
 {\bf  Single-component type-2 } 
 & Penetration length $\lambda$ and  coherence length $\xi$, with
 	$\lambda/\xi> 1/\sqrt{2}$ 
 & Repulsive 
 & $\frac{E(N)}{N} >\frac{E(N-1)}{N-1}$. \newline{} 
	$N$-quantum vortex decays into $N$ infinitely separated 
	single-quantum vortices

\\ \hline\hline
 {\bf  Multicomponent type-1.5 } 
 & Multiple length scales $\xi_a$, 
  and the penetration length $\lambda$. 
  Non-monotonic vortex interaction occurs 
  when  $\xi_1\leq\cdots<\sqrt{2}\lambda\leq\cdots\leq\xi_N$ 
 & Non-monotonic: long-range attractive and short-range repulsive 
 & $N$-quantum vortices decay into vortex clusters. Isolated single-quantum 
 	vortices attract to form a cluster

\\ \hline
\end{tabular}
\caption{Characteristics of bulk clean superconductors in the type-1, type-2 and type-1.5 
		regimes. Here the most common units are used in which the value of the 
		Ginzburg-Landau parameter $\kappa$ which separates type-1 and type-2 regimes  
		in a single-component theory is $\kappa_c=1/\sqrt{2}$.
}
\label{Table:type-1.5:a}
\end{table*}

\begin{table*}
\begin{tabular}{|p{0.18\linewidth}||p{0.275\linewidth}|p{0.25\linewidth}|p{0.25\linewidth}|}
\hline
  & {\bf Superconducting/normal state interface energy} 
  & {\bf Magnetic field required to form a vortex} 
  & {\bf Phases in an external magnetic field }
  
\\ \hline \hline
 {\bf  Single-component type-1 } 
 & Positive 
 & Larger than the thermodynamical critical magnetic field 
 & (i) Meissner state at low fields; \newline
 (ii) Macroscopically large normal domains at 
 elevated fields. First order phase transition 
 Meissner$\to$Normal

\\ \hline
 {\bf  Single-component type-2 } 
 & Negative 
 & Smaller than the thermodynamical critical magnetic field 
 & (i) Meissner state at low fields, \newline
   (ii) Vortex lattices/liquids at larger fields.  \newline
   Second order phase transitions: Meissner$\to$Vortex and Vortex$\to$Normal 
   (at the level of mean-field theory).

\\ \hline \hline
 {\bf  multicomponent type-1.5 } 
 & Negative~SC/N~interface energy inside vortex clusters 
 but positive energy of the cluster's boundary 
 & Either: (i) smaller than the thermodynamical critical magnetic field or 
 \newline
 (ii) larger than critical magnetic field for single vortex but smaller than 
 critical magnetic field for a vortex cluster of a certain critical size
 & (i) Meissner state at low fields,\newline
  (ii) Macroscopic phase separation into vortex clusters coexisting with Meissner 
       domains at intermediate fields \newline
  (iii) Vortex lattices/liquids at larger fields.  
  Vortices form via a first order phase transition. The transition from vortex states 
  to normal state is second order.

\\ \hline
\end{tabular}
\caption{Continued characteristics of the type-1, type-2 and type-1.5 regimes. 
	The corresponding magnetization curves in these regimes are displayed on 
	\Figref{Fig:Types-diagrams}.
}
\label{Table:type-1.5:b}
\end{table*}

\vspace{1cm}

The magnetic response of single-component superconductors can be classified by considering 
interface energy between the superconducting and the normal state (see details in 
Appendix~\ref{App:Single-Component}). In the type-1.5 regime of multicomponent systems, 
such an argument cannot be straightforwardly applied. Indeed, the energy per vortex 
greatly depends on whether or not the vortex is located inside a cluster. Indeed, 
inside a cluster (where the vortices are placed in a minimum of the interaction potential), 
the energy per flux quantum is smaller than that for an isolated vortex. Hence, the 
formation of a single isolated vortex might be energetically unfavourable, while the 
formation of vortex clusters can be favourable. 
Moreover, besides the energy of the vortices inside the clusters, there appears 
additional characteristics associated with the energy of the boundary of the cluster
itself. In other words systems with inhomogeneous vortex states are characterized by 
several different interfaces, some of which having positive and some with negative 
free energy. 
Overall, this is the non-monotonic intervortex interaction which defines the essential 
properties of the type-1.5 regime, but this is not a state-defining one. Indeed, the 
attraction between vortices can arise under certain circumstances in single-component 
materials as well. However, in the case of the type-1.5 regime, the long-range attraction 
is a consequence of multiple coherence lengths and comes with several new physical effects 
discussed below.

The previous chapter introduced the generic framework of multicomponent systems, 
and of their topological properties. There was a particular focus on the various 
interactions that can result in coreless topological defect. Here the central point 
is about singular defects where the core of constituting individual vortices overlap.

\vspace{-.5cm}

\subsection*{Plan of the Chapter}

As stated in the introduction of this chapter, the essential ingredient for the 
type-1.5 regime in multicomponent superconductors, is to have a hierarchy such that 
the penetration depth is an intermediate length-scale. Thus, as a starting point, 
the Section \ref{Sec:Type-1.5:Length-scales} will present the general framework for 
the analysis of the length-scales. This follows from the analysis of the eigenspectrum 
of the (linear) perturbation operator around the ground state. The general framework 
is then supplemented with a particular example that can be addressed analytically.

The length-scales, defined from the eigenspectrum of the perturbation operator,  
determine the asymptotic behaviour of vortex matter. In particular, as detailed in 
the Section \ref{Sec:Type-1.5:Intervortex-forces}, this controls the long-range 
interaction between vortices \cite{Speight:97}. If the penetration depth is an 
intermediate length-scale, then the intervortex forces are long-range attractive and 
short-range repulsive. This strongly suggests that vortices should form bound states 
with a preferred separation.

When the length-scale hierarchy allows for non-monotonic intervortex forces, 
then vortices can aggregate together thus forming \emph{vortex clusters} surrounded 
by vortex-less regions of Meissner state. Few examples of such clusters are exemplified  
in Section \ref{Sec:Type-1.5:Clusters}.

Next, the Section \ref{Sec:Type-1.5:Formation} examines the possible mechanism that 
should lead to the formation of vortex clusters; and the various models where 
this was observed. The possible experimental signatures of the vortex clusters, 
and their relevance are presented there as well.

\subsection*{Summary of the results that are relevant for this chapter}

\begin{itemize}

\item Finding of a new kind of multibody intervortex forces in multiband superconductors 
\CVcite{Carlstrom.Garaud.ea:11}. The non-monotonic intervortex interactions lead to 
the formation of vortex clusters surrounded by macroscopic Meissner domains (vortexless 
state). The structure formation can be highly impacted by non-pairwise intervortex 
interactions, originating in the nonlinear superposition of vortices. The non-monotonic 
intervortex forces also result in cluster formation in three-band superconductors 
\CVcite{Carlstrom.Garaud.ea:11a}. See reviews in \CVcite{Babaev.Carlstrom.ea:12}.
Non-monotonic intervortex forces can also occur in superconducting systems with competing 
order parameters. That is when the intercomponent interactions prohibit the coexistence of 
two condensates in the ground state \CVcite{Garaud.Babaev:14b} and 
\CVcite{Garaud.Babaev:15}. 

\item Explanation of the vortex coalescence in an putative two-band model for \SRO 
superconducting material \CVcite{Garaud.Agterberg.ea:12}. We argued that the observed 
vortex coalescence in \SRO can be explained by non-monotonic interactions originating 
in the multiband nature of \SRO. Vortex coalescence in \SRO received experimental 
support from $\mu$SR measurements in \href{http://dx.doi.org/10.1103/PhysRevB.89.094504}
{Phys. Rev. B 89, 094504 (2014)} \cite{Ray.Gibbs.ea:14}.

\item Prediction of an unconventional magnetic response in interface superconductors 
with a strong Rashba spin-orbit coupling \CVcite{Agterberg.Babaev.ea:14}. We demonstrate 
microscopically that in the clean limit interface superconductors, such as 
SrTiO$_3$/LaAlO$_3$, can exhibit formation of vortex clusters.

\item In a series of works on the microscopic properties of dirty two-band superconductors, 
we demonstrated that they feature regions of the parameter space, where the hierarchy of 
length-scales allows in principle the formation of vortex clusters due to the vicinity 
of an hidden second order phase transition within the superconducting state 
\CVcite{Silaev.Garaud.ea:17}. This should similarly occur in clean three-band systems 
\CVcite{Garaud.Silaev.ea:17}. Numerical simulations show that it indeed occurs, and that 
this results in peculiar signals that can be discriminated from other scenarios via 
global measurements of the response of muon-spin-rotation experiments 
\CVcite{Garaud.Corticelli.ea:18a}.

\end{itemize}

\section{Length-scales in multicomponent systems}
\label{Sec:Type-1.5:Length-scales}

The possibility that the interaction between vortices in multicomponent systems 
can be non-monotonic relies on the non-trivial hierarchy of the length-scales of 
the theory. It is also based on the asymptotic interaction between vortices already 
introduced in the previous section \ref{Sec:Fractional-vortices:Interaction}. 
Before reviewing the intervortex interactions in the context of the semi-Meissner 
state, we review below the general framework, and the analysis of the length-scales
of multicomponent Ginzburg-Landau theory.

\vspace{1.5cm}

To illustrate the underlying mechanism for the semi-Meissner state, let consider here 
a restriction of the generic free energy \Eqref{Eq:General:FreeEnergy}, for $N$ 
superconducting condensates coupled via various intercomponent interactions
\SubAlign{Eq:Type-1.5:FreeEnergy:1}{
\F/\F_0&=\bigintsss \frac{1}{2}\big|\Curl\A\big|^2 
+\sum_{a=1}^N\frac{1}{2}\big|\D\psi_a\big|^2 + V(\Psi,\Psi^\dagger)	
\,,  \label{Eq:Type-1.5:FreeEnergy:1:a} \\
\text{where}~V(\Psi,\Psi^\dagger)&= \sum_{a=1}^N
\left(\alpha_{aa}|\psi_a|^2+\frac{1}{2}\beta_{aa}|\psi_a|^4 \right)
+\sum_{a=1}^N\,\sum_{b>a}^N\alpha_{ab}\big(\psi_a^*\psi_b+\psi_b^*\psi_a\big)	
 \\
&+\sum_{a=1}^N\,\sum_{b>a}^N\beta_{ab}|\psi_a|^2|\psi_b|^2
+\sum_{a=1}^N\,\sum_{b>a}^N\frac{\gamma_{ab}}{2}
		\big(\psi_a^{*2}\psi_b^2+\psi_b^{*2}\psi_a^2\big)	
\,.\label{Eq:Type-1.5:FreeEnergy:1:b}
}
For simplicity again, we consider the absence of mixed gradient terms (namely 
$\kappa_{ab}=\delta_{ab}$). The role of mixed gradient terms
\footnote{
Mixed gradient terms are kinetic terms that mix different component and/or different 
directions: $(D_i\psi_a)^*D_j\psi_b+cc$. 
}, in the context of the semi-Meissner state, was discussed in 
details in \cite{Carlstrom.Babaev.ea:11}. Also the possibility to eliminate mixed 
gradient terms in two-component systems was discussed in \cite{Garaud.Corticelli.ea:18a,
Garaud.Silaev.ea:17}. The role of mixed gradients, and the possibility for anisotropies 
was also considered, see \cite{Winyard.Silaev.ea:19a} and \cite{Speight.Winyard.ea:19}
for extensions to chiral $p$-wave superconductivity.
In principle other interaction terms are allowed on symmetry grounds, but the potential 
\Eqref{Eq:Type-1.5:FreeEnergy:1:b} is general enough for the current discussion.
The ground state values of the fields  $|\psi_a|$ and $\varphi_{ab}$ of free energy 
\Eqref{Eq:Type-1.5:FreeEnergy:1} are found by minimizing its potential energy: 
\Equation{Eq:Type-1.5:ground-state}{
\Psi_0=\underset{\Psi\in\Complex^N}{\mathrm{argmin}}~V(\Psi,\Psi^\dagger)  \,.
}
The ground state is determined by the system of equations given by the variations of 
the potential, with respect to the physical degrees of freedom $|\psi_a|$ and $\varphi_a$.

\subsection{Length-scales}

The length-scales that characterize the superconducting degrees of freedom are called 
the coherence lengths, while that associated with the gauge field is the London 
penetration depth. These length-scales are the exponents that characterise how the 
ground state is recovered from an infinitesimal perturbation.
The analysis of the length-scales, or equivalently of the mass spectrum, of the theory 
is achieved by investigating the linear response to infinitesimally small perturbation
\Equation{Eq:Type-1.5:Expansion}{
\psi_a = (u_a + \eps f_a)
\exp\left\{i\left(\bvarphi_a+\eps\frac{\phi_a}{u_a}\right)\right\}	
\,,~~~\text{and}~~\A=\eps{\bs a} \,.
}
Here, $u_a$ and $\bvarphi_a$  are respectively the ground state densities and phases 
introduced in the previous section. $f_a\equiv f_a(\x)$ are the density amplitudes, 
while $\phi_a\equiv \phi_a(\x)$ are the normalized phase amplitudes. The amplitudes
${\bs a}\equiv{\bs a}(\x)$ characterize the fluctuations of the gauge field around 
the ground state.

Inserting the expansion \Eqref{Eq:Type-1.5:Expansion} inside the free energy 
\Eqref{Eq:Type-1.5:FreeEnergy:1}, and collecting order by order in $\eps$ determines 
the fluctuation operator. More precisely, the zeroth order in $\eps$ defines the 
energy of the ground state, and the first order in $\eps$ determines by the ground state. 
The quadratic term in $\eps$ defines the perturbation operator whose eigenvalues 
determine the mass spectrum of the theory.
The fluctuations are characterized by a system of Klein-Gordon equations for the $2N$ 
condensate fluctuations ($N$ densities plus $N$ phases), plus one Proca equation for 
the gauge field. In the gauge where $\Div{\bs a}=0$, the fluctuation operator reads as
\Equation{Eq:Type-1.5:KG}{
\frac{1}{2}\Upsilon^T\left(-\Grad^2 + {\cal M}^2\right)\Upsilon
\,,~~~\text{where}~~~
\Upsilon=(f_1,\cdots, f_N,\phi_1,\cdots,\phi_N,{\bs a})^T\,.
}
Here ${\cal M}^2$ is the (squared) mass matrix that can be read from 
\SubAlign{Eq:Type-1.5:KGmass}{
&\Upsilon^T{\cal M}^2\Upsilon= e^2\sum_{a=1}^Nu_a^2{\bs a}^2 
+\sum_{a=1}^N 2(\alpha_{aa}+3\beta_{aa}u_a^2)f_a^2 
+\sum_{a=1}^N\sum_{b>a}^N 2\alpha_{ab}f_af_b\cos\bvarphi_{ab } \\
&+\sum_{a=1}^N\sum_{b>a}^N\frac{2\alpha_{ab}}{u_au_b}\Big\{
(u_af_b+f_bu_a)\big(\phi_bu_a-\phi_au_b\big)\sin\bvarphi_{ab}
-\frac{1}{2}\big(\phi_bu_a-\phi_au_b\big)^2\cos\bvarphi_{ab}\Big\}
	\\
&+\sum_{a=1}^N\sum_{b>a}^N2\big(\beta_{ab}+\gamma_{ab}\cos2\bvarphi_{ab}\big)
\big(u_a^2f_b^2 + 4u_au_bf_af_b  + u_b^2f_a^2\big) \\
&-\sum_{a=1}^N\sum_{b>a}^N4\gamma_{ab}
\Big\{\big(u_b\phi_a-u_a\phi_b\big)^2\cos2\bvarphi_{ab}
+2(u_af_b+f_bu_a)\big(\phi_bu_a-\phi_au_b\big)\sin2\bvarphi_{ab}
\Big\}
\,.
}
The eigenspectrum of the matrix ${\cal M}^2$ determines the squared masses of the 
excitations and the corresponding normal modes. The inverse of each mass $m_a$ defines a 
characteristic length-scale $\ell_a:=1/m_a$ of the theory. Note that the eigenspectrum of 
${\cal M}^2$ always contains a zero mode, which is associated to the Goldstone boson that 
gives the mass to the gauge field.

Remark that there are alternative possibilities to investigate the length-scale 
spectrum of the theory. For example, by inserting a perturbative expansion in 
terms of the \emph{gauge invariant} physical fields inside the free energy 
\Eqref{Eq:Quantization:FreeEnergy:3}, expressed in terms of charged and neutral 
modes. Such an approach is explained in details in the Section~\ref{Sec:Background:GS} 
of the Appendix~\ref{App:Single-Component}, in the case of single-component 
Ginzburg-Landau theory.

\paragraph{Penetration depth:}
Clearly, the gauge field fluctuations always decouple from the fluctuations associated 
with the superconducting state. The (squared) mass of the gauge field is given by the 
total density as $m_\A=e^2\sum_{a=1}^Nu_a^2$. The associated length-scale, the penetration 
depth is thus $\lambda=1/e\sqrt{\sum_{a=1}^Nu_a^2}$. It is important to note that, given 
a ground state, the penetration depth can be adjusted to any value by tuning the gauge 
coupling $e$.

\paragraph{Length-scale hierarchy:}
It is important to stress that the modes associated to the superconducting fluctuations 
are in general all coupled together. This means that, the matrix ${\cal M}^2$ is 
in general a dense, symmetric, square matrix. Unless ${\cal M}^2$ is a multiple of 
the identity, the mass spectrum cannot be fully degenerate. This implies that 
there is a hierarchy of the eigenmasses: 
\Equation{Eq:Type-1.5:Hierarchy:Mass}{
m_0=0 \leq m_I \leq m_{II} \leq \cdots \leq m_{2N-1} \,,
}
and at least one the inequality is a strict inequality. The first mass $m_0=0$ is the 
Goldstone zero mode associated with the global symmetry mentioned above. It follows 
that there is a hierarchy of the physical length-scales 
\Equation{Eq:Type-1.5:Hierarchy:Length}{
 \ell_I \geq \ell_{II} \geq \cdots \geq \ell_{2N-1} \,, 
}
where at least one the inequality is a strict inequality. 
Note that in certain situations, because of the underlying symmetry, the mass matrix 
can become block-diagonal. This is for example the case which is discussed below. 
The underlying symmetry is thus associated with a zero mode, but this does not 
change the fact that, in general, the mass spectrum is non-degenerate.

Now, taking the penetration depth into account, there are only three possible hierarchies 
of the length-scales : 
\begin{itemize}
\item all the coherence lengths are larger than $\lambda$ (which is a type-1 behaviour)

\item all the coherence lengths are smaller than $\lambda$ (which is a type-2 behaviour)

\item the penetration depth $\lambda$ is an intermediate length-scale 
(the so-called ``type-1.5" regime \cite{Moshchalkov.Menghini.ea:09})

\end{itemize}

Since the penetration depth can be adjusted to any value by tuning the gauge 
coupling $e$, the hierarchy \Eqref{Eq:Type-1.5:Hierarchy:Length} implies that for 
\emph{any} set of parameters of the theory, it is possible to find a value of 
$e$ such that $\lambda$ is an intermediate length-scale
\Equation{Eq:Type-1.5:Hierarchy}{
 \ell_I \geq \cdots \geq \lambda \geq \cdots \geq \ell_{2N-1} \,.
}
Of course, this does not imply that such choice of parameter is realized in the 
nature. The physical realizability of the length-scale hierarchy 
\Eqref{Eq:Type-1.5:Hierarchy} is discussed later. The discussion of the hierarchy of 
length-scales can further be extend to anisotropic models \cite{Winyard.Silaev.ea:19a}.
As discussed below, since the intervortex interactions are related to the long-range 
asymptotics, such a hierarchy of the length-scales suggests long-range attractive, 
short-range repulsive intervortex forces.

Remark that the eigenmasses $m_a$, or equivalently the length-scales $\ell_a$, can be 
used to determine the coherence lengths $\xi_a=\sqrt{2}/m_a=\sqrt{2}\ell_a$. The factor 
$\sqrt{2}$ factor in the definition of coherence length is a matter of convention. 
This convention is that where the non-interacting regime (the Bogomol'nyi regime 
\cite{Bogomolnyi:76}) is $\kappa=1/\sqrt{2}$ for single-component superconductors 
\cite{Tinkham}, see Appendix \ref{App:Single-Component}. The length-scale hierarchy 
\Eqref{Eq:Type-1.5:Hierarchy} thus becomes:
\Equation{Eq:Type-1.5:Hierarchy:b}{
 \xi_I \geq \cdots \geq \sqrt{2}\lambda \geq \cdots \geq \xi_{2N-1} \,.
}

\paragraph{Coherence lengths:}
The expression of the mass matrix \Eqref{Eq:Type-1.5:KGmass} is very generic, and 
not all aspects are relevant for the present discussion. In particular, depending 
on the properties of the ground state, the perturbation operator can greatly simplify. 
There exist two-qualitatively different ground states: Ground states with 
\emph{trivial} phase-locking, \ie $\bvarphi_{ab}=0,\pi$, and ground state with 
\emph{non-trivial} relative phases $\bvarphi_{ab}\neq0,\pi$. The later case features 
new properties, and will be discussed in more details later in the Chapter 
\ref{Chap:TRSB} about the superconducting states that break the time-reversal 
symmetry. So, let focus here on the case of \emph{trivial} phase-locking, where the 
ground state relative phases are $\bvarphi_{ab}=0,\pi$. There, since $\sin\bvarphi_{ab}=0$, 
the normalized phase amplitudes $\phi_a$ decouple from the density amplitudes $f_a$.

The mass of the density amplitudes $f_a$ are thus given by the eigenvalues of 
${\cal M}_{ff}^2$ defined from
\SubAlign{Eq:Type-1.5:KGmass:ff}{
&\Upsilon_f^T{\cal M}_{ff}^2\Upsilon_f= \sum_{a=1}^N 2(\alpha_{aa}+3\beta_{aa}u_a^2)f_a^2 
+\sum_{a=1}^N\sum_{b>a}^N 2\alpha_{ab}f_af_b\cos\bvarphi_{ab} \\
&+\sum_{a=1}^N\sum_{b>a}^N2\big(\beta_{ab}+\gamma_{ab}\cos2\bvarphi_{ab}\big)
\big(u_a^2f_b^2 + 4u_au_bf_af_b  + u_b^2f_a^2\big)\,.
}
Hence, as long as $\alpha_{ab}\neq0$, or $\beta_{ab}\neq0$, or $\gamma_{ab}\neq0$, 
the density modes are in general mixed. It follows that the characteristic length-scales 
of the density fields are associated with the linear combinations of the 
fields, see \eg \cite{Babaev.Carlstrom.ea:10,Carlstrom.Babaev.ea:11,Silaev.Babaev:11}. 
Physically this implies that disturbing one of the density fields necessarily perturbs 
the others. This also implies that in a vortex, the long-range asymptotics of all density 
fields is governed by the same exponent, corresponding to a mixed mode with the smallest 
mass.

The masses of the normalized phase amplitudes $\phi_a$, on the other hand are given by 
the eigenvalues of ${\cal M}_{\phi\phi}^2$ defined from 
\SubAlign{Eq:Type-1.5:KGmass:Legget}{
\Upsilon_\phi^T{\cal M}_{\phi\phi}^2\Upsilon_\phi
&=-\sum_{a=1}^N\sum_{b>a}^N\Big(\frac{\alpha_{ab}}{u_au_b}\cos\bvarphi_{ab}
+4\gamma_{ab}\cos2\bvarphi_{ab}\Big)\big(u_b\phi_a-u_a\phi_b\big)^2 \\
&=-\sum_{a=1}^N\sum_{b>a}^N\Big(\alpha_{ab}u_au_b\cos\bvarphi_{ab}
+4\gamma_{ab}u_a^2u_b^2\cos2\bvarphi_{ab}\Big)\hat{\phi}_{ab}^2 
\,.
}
Here, the (non-normalized) relative phase amplitudes 
$\hat{\phi}_{ab}:=\frac{\phi_b}{u_b}-\frac{\phi_a}{u_a}$ have been introduced, as they 
directly relate to the fluctuations of the relative phases. This is the mass of the 
Leggett's mode \cite{Leggett:66}, and the associated length sets the scale at which 
a perturbed phase difference recovers its ground state values. For discussion of these 
collective excitations in two-band superconductors see, \eg, \cite{Sharapov.Gusynin.ea:02}. 
Observation of the Legget mode in MgB$_2$ was reported in \cite{Blumberg.Mialitsin.ea:07}.

Note that for the single-component Ginzburg-Landau model, the coherence length is 
occasionally indirectly assessed. For example through the overall size of the vortex 
core or from the slope of the order parameter near the center of the vortex core. 
These estimates give consistent results, only in some special cases. For example, 
even in the simplest single-component $s$-wave superconductors, away from $T_c$ all 
these definitions give inconsistent results \cite{Gygi.Schlueter:91}. 
In the multicomponent systems discussed here, the physics of the length-scales is 
more complicated. Thus they should not be a priori expected to be easily assessable 
from quantities such as the slope of the order parameter near the vortex center.

\subsection{A simple illustrative example} 

In general, neither the ground state, nor the mass spectrum can be addressed 
analytically. As a simple illustrative example, we consider here the case of 
a two-component Ginzburg-Landau model \Eqref{Eq:Type-1.5:FreeEnergy:1} 
with $N=2$ and where $\alpha_{12}=\gamma_{12}=0$. The potential thus reads as 
\Equation{Eq:Type-1.5:AnalyticEx:Potential}{
V(\Psi,\Psi^\dagger)= \sum_{a=1}^2
\left(\alpha_{aa}|\psi_a|^2+\frac{1}{2}\beta_{aa}|\psi_a|^4 \right)
+\beta_{12}|\psi_1|^2|\psi_2|^2 \,.
}
In this example, which was investigated in great details in \CVcite{Garaud.Babaev:15}, 
both the ground state and the mass spectrum can be found analytically. The condensates 
in \Eqref{Eq:Type-1.5:AnalyticEx:Potential} are directly coupled together by only via 
a bi-quadratic (density-density) interaction potential term when $\beta_{12}\neq0$. 
Since there is no coupling of the relative phase $\varphi_{12}$, the theory features 
a $\groupUU$ symmetry of the potential for generic values of the coupling constants
\footnote{
The symmetry of the theory is enlarged to $\groupU{1}\times\groupU{1}\times\groupZ{2}$, 
for special values of the parameters $\alpha_{11}=\alpha_{22}$, and 
$\beta_{11}=\beta_{22}$. 
}.
For positive values of $\beta_{12}$,  the biquadratic interaction is repulsive, \ie 
the condensates tend to suppress each other.

\begin{figure}[!htb]
\hbox to \linewidth{ \hss
\includegraphics[width=.75\linewidth]{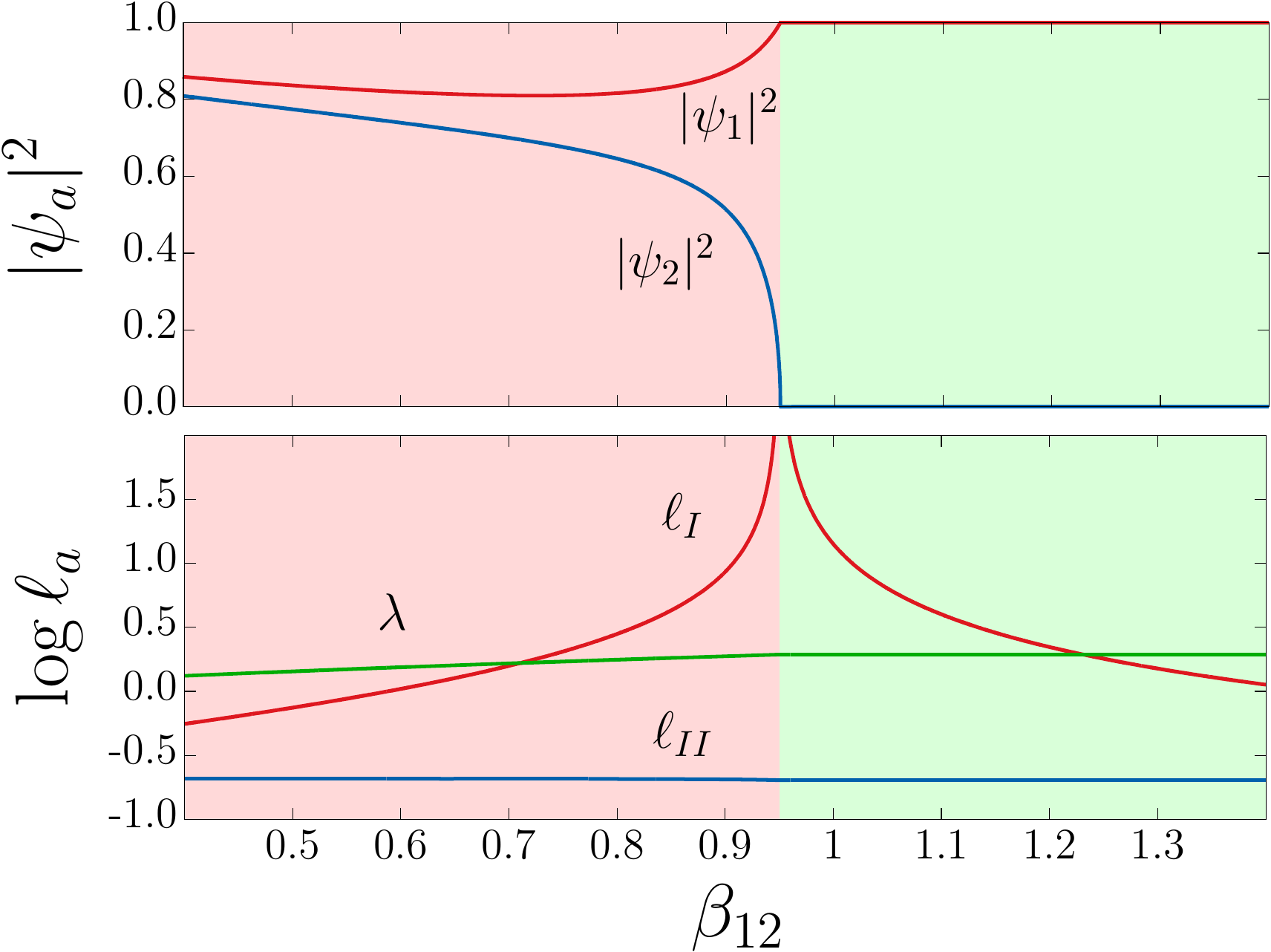}
\hss}
\caption{
The ground state and length-scales of the two-component model 
\Eqref{Eq:Type-1.5:AnalyticEx:Potential} as functions of the bi-quadratic density 
coupling $\beta_{12}$, for the parameters $(a_{11},\beta_{11})=(-1,1)$, and 
$(a_{22},\beta_{22})=(-0.95,1)$. 
The top panel shows the ground state densities $|\psi_a|^2$, while the bottom panel shows 
the relevant length-scales. 
Depending on the value of $\beta_{12}$, the ground state is either in a miscible 
phase (red background) or in an immiscible phase (green background).
There is clearly a length-scale hierarchy $\ell_{I}>\ell_{II}$, 
and the penetration depth $\lambda$ can be either the largest length-scale, or intermediate 
in the vicinity of the symmetry changing transition. Here $e=1$, but since gauge coupling 
scales the value of $\lambda$, it is clear that there always exist regimes that satisfy 
the length-scale hierarchy $\ell_{I}>\lambda>\ell_{II}$.
}
\label{Fig:Length-scales-2CGL}
\end{figure}

The ground state \Eqref{Eq:Type-1.5:ground-state}, for the particular potential 
\Eqref{Eq:Type-1.5:AnalyticEx:Potential}, is defined by the ground state densities 
$u_a$ that satisfy: 
\Equation{Eq:Type-1.5:AnalyticEx:Extrema}{
2\left(\alpha_{11}+\beta_{11}u_1^2+\beta_{12} u_2^2\right)u_1 =0\,~~~\text{and}~~~
2\left(\alpha_{22}+\beta_{22}u_2^2+\beta_{12} u_1^2\right)u_2 =0\,.
}
Apart from the normal state ($u_1=u_2=0$), there are two qualitatively different 
solutions of \Eqref{Eq:Type-1.5:AnalyticEx:Extrema} 
\footnote{
Note that for the extrema to be a minimum, the eigenvalues of the Hessian matrix 
must be positive. 
}
the \emph{miscible}-phase for which both condensates have a non-zero ground state density 
($u_1,u_2\neq0$), and the \emph{immiscible}-phase for which only one condensate has a 
non-zero ground state density: either $u_1\neq0$ and $u_2=0$ or $u_1=0$ and $u_2\neq0$. 
Assuming that $\alpha_{aa}<0$ and $\beta_{aa}>0$, the qualitatively 
different stable phases determined by \Eqref{Eq:Type-1.5:AnalyticEx:Extrema} 
\SubAlign{Eq:Type-1.5:AnalyticEx:Phases}{
\textbf{miscible-phase:~}& (u_1^2,u_2^2)=
\left(
\frac{\alpha_{22}\beta_{12}-\alpha_{11}\beta_{22}}{\beta_{11}\beta_{22}-\beta_{12}^2}
\,,
\frac{\alpha_{11}\beta_{12}\alpha_{22}\beta_{11}}{\beta_{11}\beta_{22}-\beta_{12}^2}	
\right)	 
\label{Eq:Type-1.5:AnalyticEx:Phases:a} \\
&\text{if~}	
\beta_{11}\beta_{22}>\beta_{12}^2\,,~	\alpha_{22}\beta_{12}-\alpha_{11}\beta_{22}>0 
\text{~and~}\alpha_{11}\beta_{12}-\alpha_{22}\beta_{11}>0\,. \nonumber\\
\textbf{immiscible-phase:~}& (u_1^2,u_2^2)=
\left(\frac{-\alpha_{11}}{\beta_{11}}\,,0\right)
~~\text{or}~~
\left(0,\frac{-\alpha_{22}}{\beta_{22}}\right)
\label{Eq:Type-1.5:AnalyticEx:Phases:b}	\\
&\text{if~}\alpha_{22}\beta_{11}-\alpha_{11}\beta_{12}>0
~~\text{or}~~\alpha_{11}\beta_{22}-\alpha_{22}\beta_{12}>0
\,. \nonumber
}
The ground state in the miscible phase spontaneously breaks the $\groupUU$ symmetry. 
In the immiscible phase, only one of the $\groupU{1}$ is spontaneously broken while 
the other, associated with the suppressed condensate, remains unbroken.
The different ground states phases are illustrated in \Figref{Fig:Length-scales-2CGL}.
The top panel of \Figref{Fig:Length-scales-2CGL} shows the ground state densities as 
functions of $\beta_{12}$ that parametrizes the bi-quadratic density interaction.
Depending on its value, the ground state is either in a miscible phase or in an 
immiscible phase, and there is a parametric transition between both phases.

The density and relative phase sectors of the perturbation operator decouple. 
Moreover, in agreement with the $\groupUU$ symmetry of the theory, the perturbations 
of the relative phase \Eqref{Eq:Type-1.5:KGmass:Legget}, the Leggett mode, are massless 
${\cal M}_{\phi\phi}^2=0$. The remaining non-trivial perturbations are defined by the 
mass matrix
\Equation{Eq:Type-1.5:AnalyticEx:MassMatrix}{
{\cal M}_{ff}^2= 2\left(\begin{array}{c c}
\alpha_{11}+3\beta_{11}u_1^2+\beta_{12} u_2^2	& 2\beta_{12} u_1u_2 \\
2\beta_{12} u_1u_2		&\alpha_{22}+3\beta_{22}u_2^2+\beta_{12} u_1^2
\end{array}\right)\,,
}
whose eigenvalues are $m^2_I$ and $m^2_{II}$ are non-degenerate. As a result, as 
illustrated on the bottom panel of \Figref{Fig:Length-scales-2CGL}, the corresponding 
length-scales $\ell_I$ and $\ell_{II}$ are also non-degenerate. At the transition 
between the miscible, and the immiscible phase, the smallest mass $m^2_I$ becomes 
massless, and thus $\ell_I$ diverges here. 
On the other hand, the penetration depth $\lambda=1/e\sqrt{u_1^2+u_2^2}$ is always 
finite, even at the transition between the two phases. This implies in particular 
that the length-scale hierarchy \Eqref{Eq:Type-1.5:Hierarchy} can always be realized 
in a close enough vicinity of the transition.

It is important to emphasize again that $\ell_{I,II}$ corresponds to hybridized modes 
and cannot be attributed to a given condensate separately. That is, $m^2_{I,II}$ are 
the decay rates of a linear combination of $\psi_1$ and $\psi_2$.

\subsection{Long-range intervortex forces}
\label{Sec:Type-1.5:Intervortex-forces}

\begin{figure}[!htb]
\hbox to \linewidth{ \hss
\includegraphics[width=.75\linewidth]{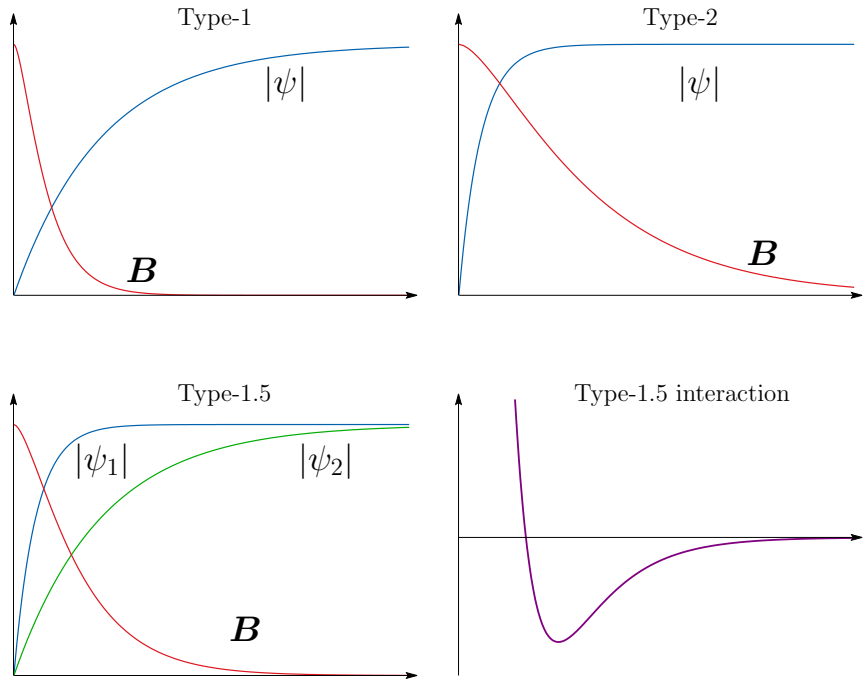}
\hss}
\caption{
The panels on the top row illustrate the vortex profiles in single-component 
type-1 and type-2 regimes. The bottom left panel shows a vortex profile when the 
penetration depth is an intermediate length-scale. This allows, in principle, for 
non-monotonic intervortex forces as illustrated in the bottom right panel.
}\label{Fig:Profiles}
\end{figure}

\begin{wrapfigure}{R}{0.4\textwidth}
\hbox to \linewidth{ \hss
\includegraphics[width=.975\linewidth]{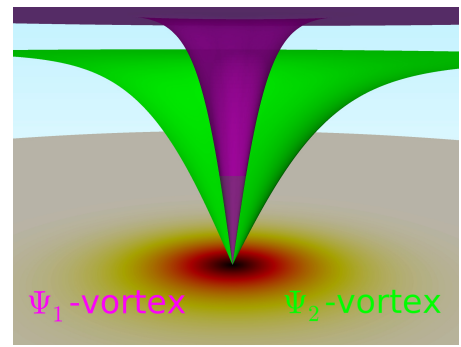}
\hss}
\caption{
A composite vortex with two co-centred fractional vortices, in the case of a 
two-component system. Here the singularities of both condensates overlap.
}\label{Fig:Cocentred}
\end{wrapfigure}

The analysis of the perturbation operator \Eqref{Eq:Type-1.5:KG}, 
\Eqref{Eq:Type-1.5:KGmass} not only determines the mass spectrum, the length-scales, 
and the hybridization of the various modes, but it also determines the asymptotic 
intervortex forces.
The typical vortex profiles are illustrated in \Figref{Fig:Profiles}.
The asymptotic behaviour of vortices is determined by the same linearized 
theory. Assuming that the singularities in the different components overlap, as sketched 
in \Figref{Fig:Cocentred}, the linearized theory yields the following long-range 
intervortex interaction 
\cite{Speight:97,Babaev.Speight:05,Babaev.Carlstrom.ea:10,Carlstrom.Babaev.ea:11}:
\Equation{Eq:Type-1.5:Asymptotics}{
E_\mathrm{int}(r)=C_\lambda K_0(r/\lambda) - \sum_{a=I,II,\cdots}C_aK_0(r/\ell_a)  \,,
}
where $K_0$ is the modified Bessel function of the second kind, and $r$ is the distance 
that separates two vortices. The coefficients $C_\lambda$ and $C_a$ depend on the 
eigenstates of the perturbation operator (the normal modes) and on the nonlinearities
\footnote{
Note that in the case of zero modes, the corresponding coefficient $C$ is zero, 
so it do not mediate any interaction.
}.
The first term describes the repulsion driven by the magnetic and current-current 
interactions. The second term, associated with the scalar fields, is attractive.
Thus, at very large distance $r$, the intervortex interaction $E_\mathrm{int}(r)$ is 
dominated by whichever term corresponds to the largest length-scale. Or equivalently, 
by the smallest of the masses. The length-scales $\lambda$ and $\ell_a$ thus
determine whether the vortices attract or repel at long range. The overall asymptotic 
intervortex interaction \Eqref{Eq:Type-1.5:Asymptotics}, when the length-scale hierarchy 
allows for non-monotonic forces, is represented in the bottom right panel of 
\Figref{Fig:Profiles}.

When the penetration depth $\lambda$ is an intermediate length-scale, \ie for a hierarchy 
of the length-scale \Eqref{Eq:Type-1.5:Hierarchy}, the vortices feature a long-range 
tail of the scalar fields. This results in long-range attractive intervortex forces 
(dominated by the core-core interactions). On the other hand, at the intermediate 
scales specified by the penetration depth of the magnetic field, the interactions 
are dominated by the current-current interactions which are repulsive. It follows 
that the long-range intervortex interacting potential \Eqref{Eq:Type-1.5:Asymptotics} 
predicted by the linear theory can be long-range attractive and short-range repulsive.  
This results in non-monotonic intervortex forces, see the calculations in different 
models \cite{Babaev.Speight:05,Babaev.Carlstrom.ea:10,Carlstrom.Babaev.ea:11,
Silaev.Babaev:11,Carlstrom.Garaud.ea:11,Carlstrom.Garaud.ea:11a,Garaud.Agterberg.ea:12,
Speight.Winyard:21}. These forces can promote the formation of a bound state of vortices. 
In such a bound state, the distance separating the vortices does not directly follow 
from the linearized theory, but it is determined by full nonlinear theory.

Importantly, in multicomponent superconductors, there can be second order phase 
transitions within the superconducting state. That is, phase transitions for which 
the total density, and thus the penetration depth $\lambda$ are finite. This is for 
example the case of the transition between miscible and immiscible phases displayed 
in \Figref{Fig:Length-scales-2CGL}, see \eg \cite{Garaud.Babaev:15}. Transitions 
to the time-reversal symmetry breaking $s+is$ state, discussed in more details in the 
Chapter~\ref{Chap:TRSB} also occur within the superconducting state, see \eg 
\cite{Silaev.Garaud.ea:17,Garaud.Silaev.ea:17,Garaud.Corticelli.ea:18a}. Since 
second order phase transitions are associated with a divergent length-scale, and that
$\lambda$ remains finite, the penetration depth is always an intermediate length-scale 
close enough to the transition. This implies that in the vicinity such transition 
the length-scale hierarchy \Eqref{Eq:Type-1.5:Hierarchy} is always satisfied.

The analysis of the long-range intervortex forces thus opens the possibility for 
non-monotonic interaction between composite vortices, that are long-range attractive 
and short-range repulsive (see \eg \cite{Babaev.Speight:05,Babaev.Carlstrom.ea:10,
Carlstrom.Babaev.ea:11}). Thus, the long-range properties can be easily determined by 
the analysis of the length-scales of the model. However, the length-scale analysis 
provides only a necessary, yet not sufficient condition for the existence of 
non-monotonic forces beyond the linear approximation \cite{Babaev.Carlstrom.ea:10,
Carlstrom.Babaev.ea:11}. It is thus necessary to investigate the intervortex forces 
at the nonlinear level. This was investigated in details at the level of the 
Ginzburg-Landau theory, and it was demonstrated that non-monotonic forces exist for 
various kind of potential and also survives the existence of mixed-gradient term 
\cite{Babaev.Carlstrom.ea:10,Carlstrom.Babaev.ea:11}. It was further demonstrated that 
this behaviour can also appear in fully microscopic models \cite{Silaev.Babaev:11}.

In practice, the demonstration that non-monotonic forces exist at the nonlinear level 
can be done in the following way. In a descretized system, start with an ansatz that 
describes to separated vortices (see \eg Appendix~\ref{App:numerics:IG}, for discussion 
of such an ansatz). Then relax all degrees of freedom except those corresponding to 
the core location. After convergence, the resulting configuration is that of two 
vortices at a certain distance. The energy of such a configuration is thus the energy 
of two isolated plus the interaction energy between the vortices. Repeating the procedure 
for various vortex separations provides the interaction energy between the vortices. 
This procedure was used to determine that two-component superconductors indeed 
can support non-monotonic forces at the non-linear level \cite{Babaev.Carlstrom.ea:10,
Carlstrom.Babaev.ea:11,Lin.Hu:11b}. The same procedure was also used to investigate the 
interactions between three and four vortices \cite{Babaev.Carlstrom.ea:12,Edstrom:13}, 
and that the nonlinear superposition can result in unusual additional effects.
Note also that the criteria that determine the conditions for non-monotonic intervortex 
interaction can also be supplemented with the analysis of the surface energy between 
superconducting and normal states \cite{Chaves.Komendova.ea:11}.

Large systems where vortices are approximated as point particles with the effective 
non-monotonic interaction revealed an extremely rich physics of vortex structure. 
Indeed, the multi-scale nature of the interaction allows for vortex arrangements 
that are much more complicated than the Abrikosov lattices that occur in purely 
repulsive systems. These include vortex clusters and stripe phases 
\cite{OlsonReichhardt.Reichhardt.ea:10,Dao.Chibotaru.ea:11,Drocco.OlsonReichhardt.ea:13}, 
stripe and gossamer phases \cite{Sellin.Babaev:13}, honeycomb and square lattices 
\cite{Meng.Varney.ea:14}, multi-stripe phases, polymer phases, void phases
\cite{Meng.Varney.ea:17}. This richness of the large scale organizations have been 
confirmed in simulations of type-1/type-2 bilayers \cite{Komendova.Milosevic.ea:13}.

\section{Vortex clusters} 
\label{Sec:Type-1.5:Clusters}

The discrepancy of the length-scales thus opens the possibility for long-range attractive 
and short-range repulsive intervortex forces. It was thus demonstrated that the 
non-monotonic forces do survive at the nonlinear level \cite{Babaev.Carlstrom.ea:10,
Carlstrom.Babaev.ea:11}. The non-monotonic forces indicate a preferred separation between 
two vortices. Hence they should form a bound state. When there are more than two-flux 
quanta, this means that they should form aggregates, or clusters of various shape. 
This expectation is supported by the analysis of point particle vortices with multi-scale 
interaction \cite{OlsonReichhardt.Reichhardt.ea:10,Dao.Chibotaru.ea:11,
Drocco.OlsonReichhardt.ea:13,Sellin.Babaev:13,Meng.Varney.ea:14,Meng.Varney.ea:17}.

\begin{figure}[!htb]
\hbox to \linewidth{ \hss
\includegraphics[width=.95\linewidth]{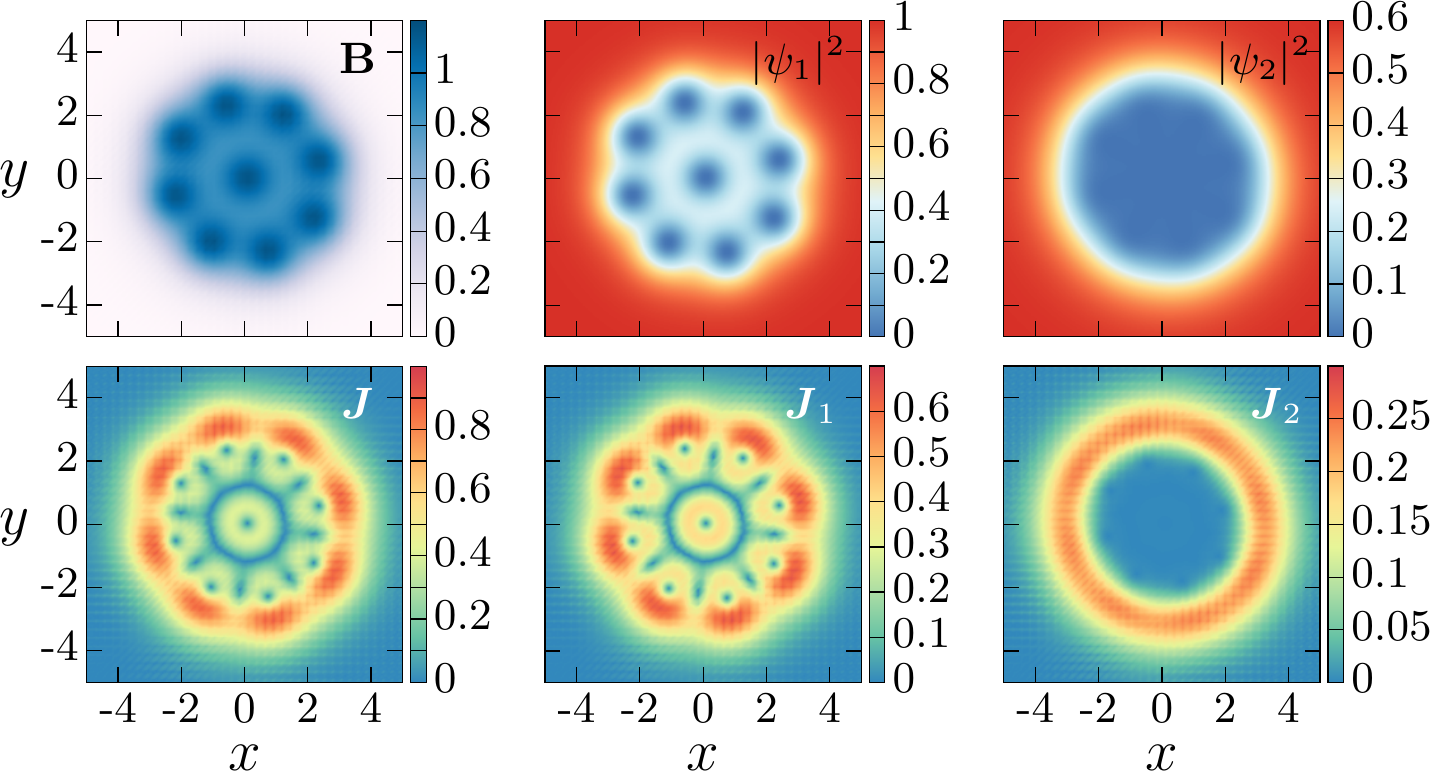}
\hss}
\caption{
A vortex cluster solution of a two-component superconductor, carrying 9 flux quanta. 
The parameters of the simple potential interaction \Eqref{Eq:Type-1.5:FreeEnergy:1:b} 
are $(\alpha_{11},\beta_{11})=(-1,1)$, $(\alpha_{22},\beta_{22})=(-0.6,1)$, and the gauge 
coupling is $e = 1.5$. Here $\alpha_{12}=\beta_{12}=\gamma_{12}=0$, thus the theory as a 
$\groupUU$ symmetry.
The displayed quantities on the top left panel is the magnetic field. The other panels 
on the top row, show the individual densities of the superconducting condensates 
$|\psi_1|^2$ and $|\psi_2|^2$. 
%
The bottom left panel shows the total current, while the other panels on the bottom row, 
are the individual currents $\J_a$. The color map here shows the magnitude of the 
supercurrents $|\J_a|$.
%
This is a close view of the cluster, but the actual numerical grid is larger. 
}
\label{Fig:Vortex-cluster-2CGL}
\end{figure}

These strong indications of the existence of clusters should however be investigated 
in details, at the nonlinear level of the original Ginzburg-Landau theory. The numerical 
minimization of multi-vortex states, in the regime of non-monotonic interactions, shows 
that two-component Ginzburg-Landau models indeed allow the formation vortex clusters 
\CVcite{Carlstrom.Garaud.ea:11}, as illustrated for example in 
\Figref{Fig:Vortex-cluster-2CGL} 
\footnote{
The displayed numerically obtained solutions are typically a close view of the relevant 
quantities. In general, except when considering mesoscopic domains, the actual numerical 
grid is chosen to be much larger. This rules out any kind of complicated stabilizing 
boundary effects.
}.
The discretization method, the choice of the numerical algorithm, and the general 
procedure are presented in details in the Appendix \ref{App:numerics}. Yet this can 
be outlined as follows: The (nonlinear) minimization algorithm typically converges 
easily to the solution, provided a suitable initial guess. In short, an initial 
field configuration which winds $n$ times in each of the condensate will converge 
to a configuration that carries $n$ flux quanta (see detailed discussion in Section 
\ref{App:numerics:IG}). Provided the parameter set allows for non-monotonic forces 
this typically results in a cluster carrying $n$ flux quanta, as that displayed in 
\Figref{Fig:Vortex-cluster-2CGL}.

As can be seen from the density panels of \Figref{Fig:Vortex-cluster-2CGL}, the first 
component clearly forms nine vortices that repel each other like in type-2 superconductors. 
The second component, on the other hand, is almost fully depleted inside the vortex 
cluster (like in the type-1). Finally, there is a clear overlap of the magnetic field of 
the different vortices. Here, the interesting feature is that the cluster clearly 
comprises between the tendencies of the different condensates. Indeed, the first 
component tends to form a regular vortex lattice, while the second condensate favours 
a circular normal domain with a supercurrent predominantly located on the boundary.
This competition between forming a circular cluster, and a triangular lattice leads 
to a variety of clusters which are neither an hexagonal lattices nor a fully circular 
boundary. For more details of the various possible cluster, see 
\CVcite{Carlstrom.Garaud.ea:11}.

\begin{figure}[!htb]
\hbox to \linewidth{ \hss
\includegraphics[width=.95\linewidth]{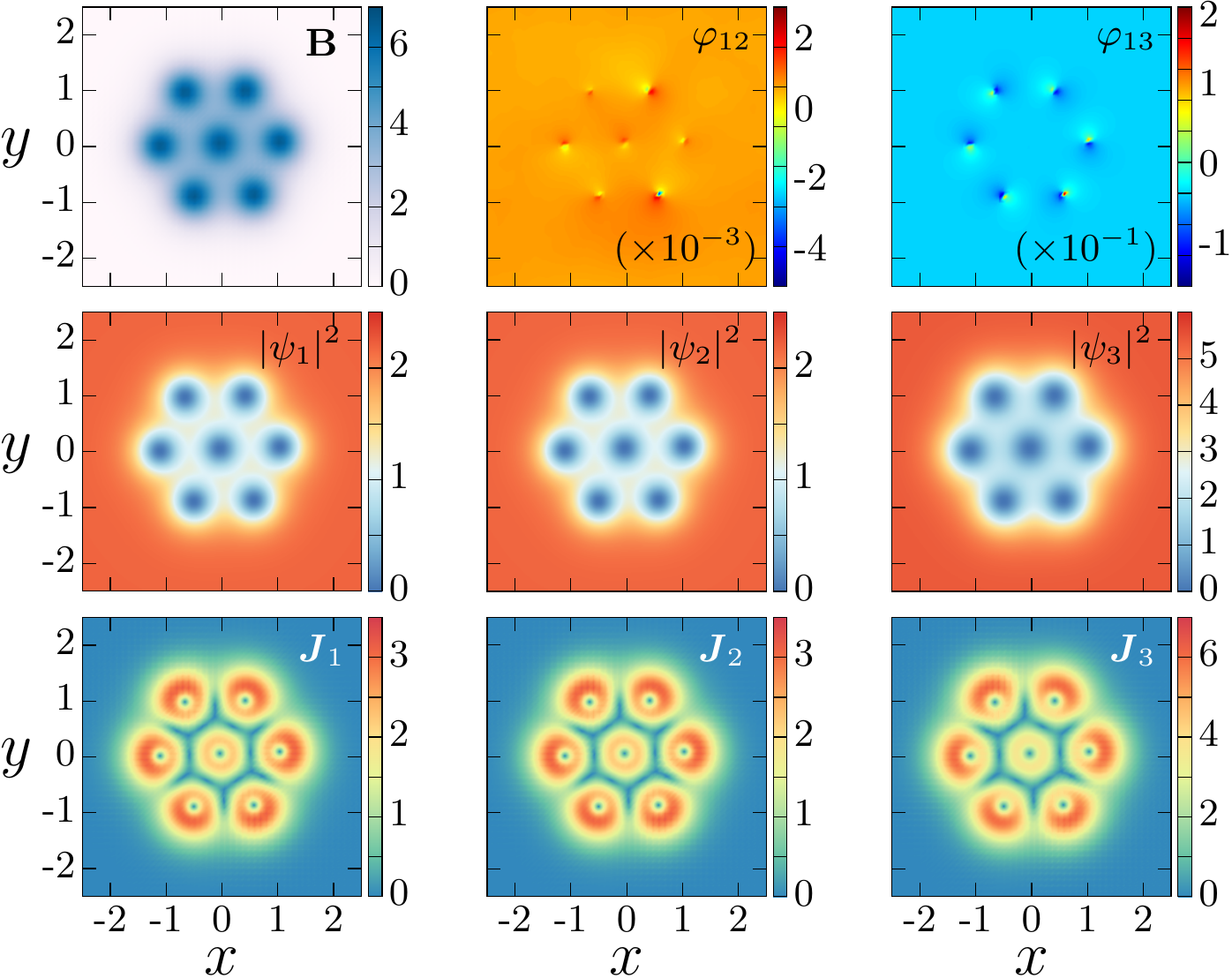}
\hss}
\caption{
A vortex cluster solution of three-component superconductor, with the parameters of the 
potential term \Eqref{Eq:Type-1.5:FreeEnergy:1:b} are 
$(\alpha_{11},\beta_{11})=(\alpha_{22},\beta_{22})=(-3,3)$, 
$(\alpha_{33},\beta_{33})=(2,0.5)$, $\alpha_{12}=2.25$ $\alpha_{13}=\alpha_{23}=-3.7$
and the gauge coupling is $e = 1.35$. The other parameters are $\beta_{ab}=\gamma_{ab}=0$.
The displayed quantities on the top row are the magnetic field and two relative phases 
$\varphi_{12}$ and $\varphi_{13}$. The middle row shows the densities of the three 
components $|\psi_a|^2$, from which it is clear that the cores in the different components 
have different sizes. The panels on the bottom row displays the associated super-currents 
$\J_a$. 
Interestingly, as can be seen from the relative phase panels, the cores in the different 
components do not fully overlap at the exterior of the cluster.
}
\label{Fig:Vortex-cluster-3CGL}
\end{figure}

The existence of vortex aggregates due to non-monotonic intervortex forces easily 
promotes to more than two components. The figure~\ref{Fig:Vortex-cluster-3CGL} shows 
an example of a vortex cluster in the case of a three-components model. The case of 
vortex clusters occurring in three-components models was demonstrated in details in 
\CVcite{Carlstrom.Garaud.ea:11a}. 
Unlike the cluster illustrated in \Figref{Fig:Vortex-cluster-2CGL} all components of 
the cluster in \Figref{Fig:Vortex-cluster-3CGL} form vortices with different core sizes.
While the cluster in \Figref{Fig:Vortex-cluster-2CGL} had to comprise between type-1-like 
and type-2-like behaviours, the cluster in \Figref{Fig:Vortex-cluster-3CGL} easily fits 
with a triangular lattice.

\paragraph{Remark about the hybridization:} 
In single-component Ginzburg-Landau model, the coherence length can easily be guessed 
from the size of the vortex cores. This is not true in general for in multicomponent 
models. Indeed, as already discussed in the previous section, the length-scale analysis 
shows that the various modes are in general hybridized, and thus that it is not possible 
to attribute a single length-scale to a given condensate. This is especially the case when 
there are phase-locking terms. Thus inspection of the vortex core sizes can easily the 
discrepancy of the length-scales. The cluster illustrated in \Figref{Fig:Vortex-cluster-3CGL} 
is a good example of this: The cores in the different condensates seems to have almost 
the same sizes. Yet there is a strong cluster structure.

\section{Formation of vortex clusters. Realization of semi-Meissner states}
\label{Sec:Type-1.5:Formation}

The vortex clusters presented above are found numerically by the minimization of the 
Ginzburg-Landau free energy, given an initial guess with the appropriate winding, 
see details in the Appendix~\ref{App:numerics}.
They thus represent vortex clusters surrounded by voids of the Meissner state. Hence 
the terminology \emph{semi-Meissner state}. The properties of the semi-Meissner state 
can also be investigated in an external applied field $\He$, by minimizing the Gibbs
free energy, 
\Equation{Eq:Type-1.5:Gibbs}{
G=F-\int\B\cdot\He\,,
}
instead of the Helmholtz free energy $F$ \Eqref{Eq:Type-1.5:FreeEnergy:1}. 
Thus simulations in increasing values of the external field $\He$, reproduce the 
magnetization curves sketched in \Figref{Fig:Types-diagrams}. 
Complementary to the investigation of the magnetization properties, are simulations 
that mimic field cooled experiments. Indeed for realistic models, the various parameters 
are temperature dependent. For example, the parameters of the quadratic terms depend 
on the temperature $T$ as $\alpha_{aa}\equiv\alpha_{aa}(T)=\alpha_{aa}^{(0)}(T/\Tc{,a}-1)$. 
Here $\alpha_{aa}^{(0)}$ is the value of the coefficient at zero temperature, while 
$\Tc{,a}$ is the characteristic temperature where the condensate $\psi_a$ becomes active.  
Here ``active", means: The temperature below which $\psi_a$ would condense, if decoupled 
from the other condensates.

A magnetization process thus accounts to a vertical line in the $H-T$ phase diagrams 
sketched in \Figref{Fig:Types-diagrams}. A field cooled experiment on the other hand 
accounts to an horizontal line in the $H-T$ phase diagrams. In practice, in numerical 
simulations, a magnetization process is realized by minimizing the Gibbs energy 
\Eqref{Eq:Type-1.5:Gibbs}, at a given value of the applied field $\He$. After convergence, 
then the value of $\He$ is increased/decreased, and the procedure is repeated for the 
new value of the external field. Likewise, a field cooled experiment is simulated by 
by minimizing the Gibbs energy in an external field $\He$, at a given value of the 
parameters $\alpha_{aa}(T)$. Then, after convergence, the temperature modified, and 
the minimization is restarted for the new parameters  $\alpha_{aa}(T+\delta T)$. 
Surely, simulating such processes is much more time consuming than the construction 
of (isolated) topological defects.

It should be expected that the magnetic response near the second critical field $\Hc{2}$, 
or in the vicinity of the critical temperature, reduces to single-component. See for 
example the discussions \cite{Silaev.Babaev:11,Babaev.Silaev:12,Babaev.Silaev:12a} or 
the reviews \cite{Babaev.Carlstrom.ea:12,Babaev.Carlstrom.ea:17,Babaev.Carlstrom.ea:17a}. 
This can also be seen from the high field behaviour of the magnetization curve in 
\Figref{Fig:Types-diagrams}. Indeed, in high field, there is no way to tell the difference 
between the magnetization curve of type-1.5 and usual type-2. Heuristically this follows 
from the fact that the intervortex distance in a vortex lattice become small. In that case, 
the vortex attractionis too weak, and the vortices are not ``free'' to form aggregates.

\begin{figure}[!htb]
\hbox to \linewidth{ \hss
\includegraphics[width=.95\linewidth]{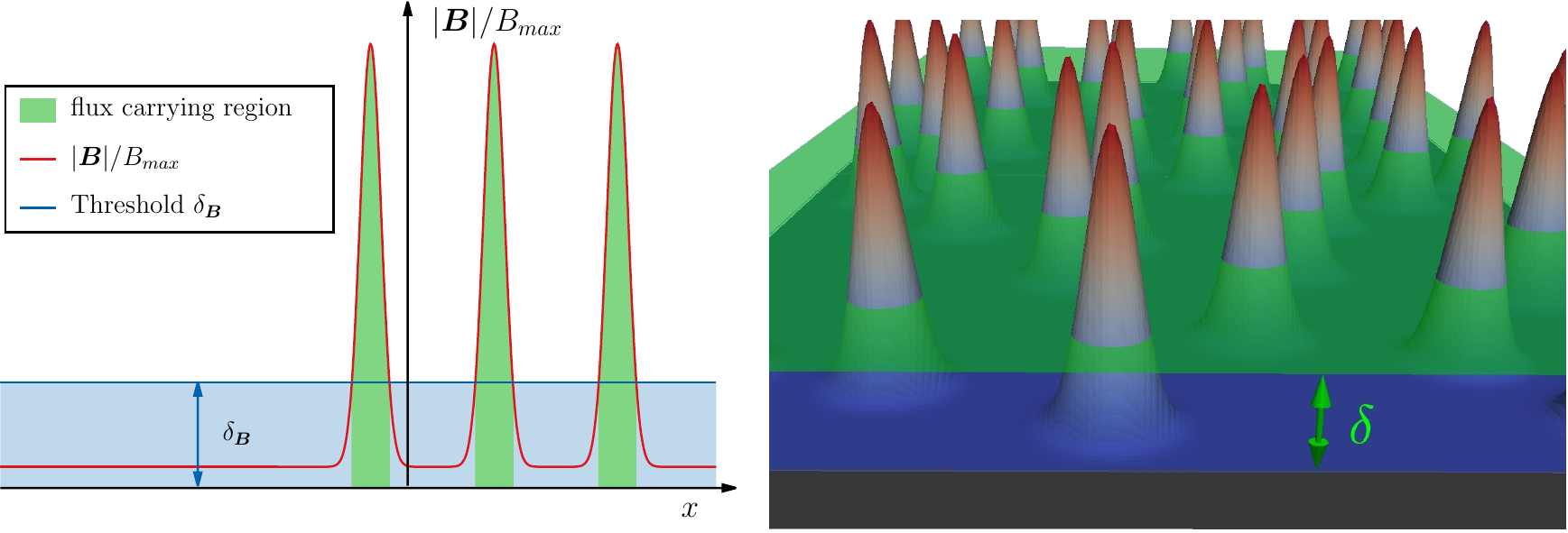}
\hss}
\caption{
Principle to determine the fraction of sample containing flux. For a given 
distribution of the magnetic field, the regions for which the magnetic 
field is below a threshold value $\delta_\B B_{max}$ are said to be in the 
Meissner state. Conversely, regions for which $|\B|>\delta_\B B_{max}$ are 
the \emph{flux carrying regions}. 
}
\label{Fig:Measures}
\end{figure}

However, at lower temperature the non-monotonic forces can set in and lead to the 
formation of vortex clusters. This is again sketched in \Figref{Fig:Types-diagrams}.
Thus, in a field cooled experiment the Abrikosov lattice can collapse into clusters 
at low temperatures. This was confirmed for example in simulations for clean interface 
superconductors \CVcite{Agterberg.Babaev.ea:14}. Measurement of the flux carrying area 
thus show that clusters form. Indeed, the flux carrying area, illustrated in 
\Figref{Fig:Measures}, is defined as 
\Equation{Eq:Flux-area}{
\text{Flux carrying area}=\int \Theta\left( \B/B_\mathrm{max}-\delta_\B \right) \,.
}
Here $\Theta$ is the Heaviside step function, $\delta_\B$ some tolerance, and 
$B_\mathrm{max}$ is the maximal value of the magnetic field.
The local internal flux density of the flux carrying regions is the value 
of the magnetic field, averaged over the \emph{flux carrying regions}. This 
can be formally defined as: 
\Equation{Eq:Flux-density}{
\text{Internal flux density}= 
\frac{\int \Theta\big(|\B|/B_{max}-\delta_\B\big)|\B|}
{\int \Theta\big(|\B|/B_{max}-\delta_\B\big)}\,.
}
The internal flux density should show a strong peak where the attractive intervortex 
forces are strongest and where thus the clusters are the most compact. These effects for 
the clusterization was discussed in \CVcite{Garaud.Corticelli.ea:18a}, in the context 
of dirty two-band superconductors. 

\subsection*{Models with semi-Meissner states}

In all the discussions above, all parameters of the Ginzburg-Landau theory are given 
as free parameters. This is of course not the case in real life, where the various 
parameters cannot in general be chosen independently. Thus, taking into account the 
underlying microscopic properties of a given model of superconductivity constraints 
the relation between the various parameters. In particular, the various parameters 
are temperature dependent. It follows that the conditions for the realization of 
the \emph{type-1.5 regime}, may be satisfied only in a given interval of the 
temperature. Moreover, it is quite expectable that if vortex bound states are formed, 
their typical size should also be temperature dependent.

Various works investigated the possibility to realize semi-Meissner states from 
realistic microscopic theory of multi-band superconductors, see e.g. 
\cite{Silaev.Babaev:11,Silaev.Babaev:12,Frank.Lemm:16}. 
Also models for clean interface superconductors, such as SrTiO$_3$/LaAlO$_3$ predict 
the unconventional magnetic response due to non-monotonic intervortex forces
\CVcite{Agterberg.Babaev.ea:14}.

Also dirty two-band superconductors can feature a second order phase transition to 
the $\sis$ state that breaks the time-reversal symmetry see e.g. 
\CVcite{Silaev.Garaud.ea:17} and \CVcite{Garaud.Corticelli.ea:18a}. 
In the vicinity of such a transition, the necessary length-scale hierarchy should be 
realized, thus opening the possibility for cluster formation. This was discussed in 
more details \CVcite{Carlstrom.Garaud.ea:11a} and \CVcite{Garaud.Corticelli.ea:18a}.
See also the related discussion in the next chapter.

\subsection*{Applicability and experimental relevance}

In the regime where the penetration depth is an intermediate length-scale, vortices 
can thus have a long-range attractive and short-range repulsive interaction. It 
follows that vortices can aggregate into clusters surrounded by Meissner domains. 
Experimental works using magnetic decoration, scanning SQUID and scanning Hall 
probes measurements reported inhomogeneous vortex distribution on clean 
superconducting material MgB$_2$ \cite{Moshchalkov.Menghini.ea:09,Nishio.Dao.ea:10,
Dao.Chibotaru.ea:11}. Being a two-band material, the very inhomogeneous distribution 
of vortices was attributed to non-monotonic forces originating from the competing type-1 
and type-2 behaviours. This regime was termed ``type-1.5 superconductivity" 
\cite{Moshchalkov.Menghini.ea:09}. When cycling field, the observation that the 
vortex clusters form in different parts of the sample, allows to rule out the 
alternative explanation that the observed aggregation is due to pinning.

Several experimental works reported local vortex coalescence in layered-perovskite 
superconductor Sr$_2$RuO$_4$ \cite{Dumont.Mota:02,Bjornsson.Maeno.ea:05,
Dolocan.Veauvy.ea:05,Hasselbach.Dolocan.ea:07}. The semi-Meissner state was 
proposed to be responsible for the aggregates observed in Sr$_2$RuO$_4$. This 
proposal followed from both theoretical \CVcite{Garaud.Agterberg.ea:12}, and 
experimental scanning SQUID \cite{Hicks.Kirtley.ea:10}, and muon-spin rotation 
measurements \cite{Ray.Gibbs.ea:14}. There, it was observed that the vortex clusters 
contract upon temperature decrease, well below $T_c$. This behaviour hints for 
attractive inter-vortex forces, rather than pinning, to be responsible for the cluster 
formation. Note that earlier experiment also reported attractive inter-vortex
forces, which were attributed to domain walls trapping vortices \cite{Dumont.Mota:02}. 
However signatures domain-walls were not observed in surface probes 
\cite{Hicks.Kirtley.ea:10}.

The scenario of vortex coalescence attributed to the semi-Meissner regime 
received experimental support as well in other superconducting material such as 
LaPt$_3$Si \cite{Kawasaki.Watanabe.ea:13,Fujisawa.Yamaguchi.ea:15}. Further 
theoretical predictions argued that the type-1.5 regime could be realized as well for 
certain interface superconductors SrTiO$_3$/LaAlO$_3$ \CVcite{Agterberg.Babaev.ea:14}, 
and should generically be present near transitions from $s$ to $\sis$ pairing states
in iron-based superconductors \CVcite{Carlstrom.Garaud.ea:11a} and 
\CVcite{Carlstrom.Garaud.ea:11a}; or in the context of dirty two-band materials  
\CVcite{Garaud.Corticelli.ea:18a}. 

For other works on this and related subjects see \eg \cite{Wang:10,Dao.Chibotaru.ea:11,
Li.Nishio.ea:11,Gutierrez.Raes.ea:12,Varney.Sellin.ea:13,Drocco.OlsonReichhardt.ea:13,
Edstrom:13,Meng.Varney.ea:14,Garaud.Babaev:15,Forgacs.Lukacs:16,Meng.Varney.ea:17}.

\paragraph{Controversy about the existence of type-1.5 regime.}
Note that the possibility to realize type-1.5 regime was heavily debated on 
theoretical grounds \cite{Geyer.Fernandes.ea:10,Kogan.Schmalian:11,Babaev.Silaev:12a,
Kogan.Schmalian:12}. In my view the question is sorted and there is no doubt that 
this phenomenon do exist. This follows the accumulation of theoretical and numerical 
works demonstrating the existence of this regime.
Moreover one of the earlier works claiming of inexistence of the type-1.5 regime 
\cite{Kogan.Schmalian:11,Kogan.Schmalian:11} bases its conclusions on an erroneous 
set of equations; for the detailed demonstration, see \cite{Babaev.Carlstrom.ea:17a}. 
Using the equations obtained by this incorrect derivation, several further papers 
propagated misconceptions about the possibility of having a type-1.5 regime. 
Having in mind that the objections, against the existence of type-1.5 regime, originate 
into incorrect equations, there is no doubt about the theoretical relevance of such 
a physics.
While there is no reason to doubt about the theoretical possibility of the type-1.5 
regime, there is no guarantee that this is realized in nature. Note however, that 
as mentioned above, several experiments are consistent with that physics.


\doPrint{ \newpage \thispagestyle{empty}\ \newpage }{ }
\graphicspath{{Plots/05-TRSB-sis/}}
\chapter{Superconducting states that Break the Time-Reversal Symmetry}
\label{Chap:TRSB}

Theories that describe the physics of superconductors (or superfluids) are invariant under 
complex conjugation. This invariance is usually referred to as the time-reversal symmetry 
(TRS). As already emphasized, the properties of multicomponent superconducting states 
can be qualitatively different from their simplest single-band $s$-wave counterparts. 
In multicomponent systems, the time-reversal symmetry can be spontaneously broken by 
the ground state. That is, the ground state is \emph{not} invariant (up to global phase 
rotations) under complex conjugation.
Such states can appear if the relative phase, between the superconducting gap functions in 
the different bands, differ from $0$ or $\pi$ \cite{Balatsky:00,Lee.Zhang.ea:09,
Platt.Thomale.ea:12,Stanev.Tesanovic:10,Fernandes.Millis:13,Agterberg.Barzykin.ea:99,
Ng.Nagaosa:09,Lin.Hu:12,Carlstrom.Garaud.ea:11a,Bobkov.Bobkova:11,Maiti.Chubukov:13,
Maiti.Sigrist.ea:15}. 
It results that, in addition to the breakdown of the usual $\groupU{1}$ gauge symmetry, 
such superconducting states feature a discrete $\groupZ{2}$ degeneracy, associated with 
the spontaneous breakdown of the time-reversal symmetry (BTRS).

Spontaneously broken time-reversal symmetry (BTRS) states attracted much interest in the 
context of unconventional spin-triplet superconducting models, especially the $p_x+ip_y$  
state, which have been intensively studied in relation with layered perovskite
superconductor \SRO. Another time-reversal symmetry breaking state, which attracted more 
recently a lot of attention, is the $\sis$ superconducting state. Indeed, it received 
a strong theoretical support in relation with some iron-based superconductors, and in 
particular with hole-doped Ba$_{1-x}$K$_x$Fe$_2$As$_2$ \cite{Hirschfeld.Korshunov.ea:11,
Maiti.Korshunov.ea:11,Maiti.Korshunov.ea:11a,Maiti.Korshunov.ea:12,Maiti.Chubukov:13}.
The $\sis$ state is of particular interest, as it is the simplest time-reversal symmetry 
breaking extension of the most abundant $s$-wave state. It is a complex admixture of 
distinct superconducting states, with the same symmetry, that compete through phase-locking 
terms. In pnictides, it is believed to originate in the competition between different 
pairing channels \cite{Maiti.Chubukov:13}, but could as well be engineered on interfaces 
of superconducting bilayers \cite{Ng.Nagaosa:09}.

The spontaneous breakdown of the time-reversal symmetry has various interesting physical 
consequences. Some of which, like the existence of domain-walls, where earlier discussed 
in Chapter \ref{Chap:Topological-defects}. Iron-based superconductors 
\cite{Kamihara.Watanabe.ea:08} are among the most promising materials for the observation 
of the time-reversal symmetry breaking $\sis$ states that originate in the multiband 
character of superconductivity and several competing pairing channels. 
Indeed, the experimental data show that in the hole-doped 122 compounds 
Ba$_{1-x}$K$_x$Fe$_2$As$_2$, the symmetry of superconducting state can change depending 
on the doping level $x$. A typical band structure of Ba$_{1-x}$K$_{x}$Fe$_2$As$_2$ 
consists of two hole pockets at the $\Gamma$ point and two electron pockets at $(0,\pi)$ 
and $(\pi,0)$. At moderate doping level $x\sim 0.4$ various measurements, including 
ARPES \cite{Ding.Richard.ea:08,Khasanov.Evtushinsky.ea:09,Nakayama.Sato.ea:11}, 
thermal conductivity \cite{Luo.Tanatar.ea:09} and neutron scattering experiments 
\cite{Christianson.Goremychkin.ea:08}, are consistent with the hypothesis of an $s_\pm$ 
state where the superconducting gap changes sign between electron and hole pockets.
On the other hand, the symmetry of the superconducting state at strong doping 
$x\rightarrow 1$ is not so clear, regarding the question whether the $d$ channel dominates 
or if the gap retains $s_\pm$-symmetry changing sign between the inner hole bands at the 
$\Gamma$ point \cite{Maiti.Korshunov.ea:11,Maiti.Korshunov.ea:11a}. Indeed, there are 
evidences that $d$-wave pairing channel dominates \cite{Reid.Juneau-Fecteau.ea:12,
Reid.Tanatar.ea:12,Tafti.Juneau-Fecteau.ea:13,Tafti.Clancy.ea:14} while other ARPES 
data were interpreted in favour of an $s$-wave symmetry \cite{Okazaki.Ota.ea:12,
Watanabe.Yamashita.ea:14}.
In both situations this implies the possible existence of an intermediate complex state 
that compromises between the behaviours at moderate and high doping. Depending on whether 
$d$ or $s$ channel dominates at strong doping such a complex state is named $\sis$ or 
$\sid$. Both of these superconducting states break the time-reversal symmetry.

The $\sis$ state is isotropic and preserves crystal symmetries \cite{Maiti.Chubukov:13}. 
On the other hand, the $\sid$ state breaks the $C_4$ symmetry, while it remains 
invariant under combination of time-reversal symmetry operation and $C_4$ rotations. 
Being anisotropic, it is thus qualitatively different from $\sis$ state. Note that the 
$\sid$ superconducting state is also qualitatively different from the (time-reversal 
preserving) $s+d$ states that attracted interest in the context of high-temperature 
cuprate superconductors (see e.g. \cite{Joynt:90,Li.Koltenbah.ea:93,Berlinsky.Fetter.ea:95}). 
It also contrasts with $d+id$ state, which can appear in the presence of an external 
magnetic field, and that violates both parity and time-reversal symmetries 
\cite{Balatsky:00, Laughlin:98} . 
While it is an interesting scenario, possibly relevant for pnictides, the properties of 
the $\sid$ state will not be further considered here. The focus will be put on the 
analysis of $\sis$ superconducting state. This state is of particular theoretical 
interest, being the simplest extension of the most abundant $s$-wave state, that breaks 
the time-reversal symmetry. Also, it is expected to arise from various microscopic physics 
\cite{Stanev.Tesanovic:10,Platt.Thomale.ea:12,Suzuki.Usui.ea:11,Chubukov.Efremov.ea:08,
Maiti.Chubukov:13,Ahn.Eremin.ea:14}. The $\sis$ state could as well be fabricated on 
demand on the interfaces of superconducting bilayers \cite{Ng.Nagaosa:09}. 

The experimental observation of the $\sis$ or $\sid$ time-reversal symmetry breaking 
states is challenging. Indeed, this requires probing the relative phases between the 
components of the order parameter in different bands, which is a challenging task.
For example the $\sis$ state does not break the point group symmetries and 
is therefore not associated with an intrinsic angular momentum of the Cooper pairs. 
Consequently it cannot produce a local magnetic field and thus is {\it a priori} 
invisible for conventional methods like muon spin relaxation and polar Kerr 
effect measurements that were for example used to probe time-reversal breaking 
$\pip$ superconducting state in e.g. Sr$_2$RuO$_4$ compound \cite{Mackenzie.Maeno:03}. 
Several proposals were voiced, each with various limitations, for indirect observation 
of BTRS signatures in pnictides. These, for example, include the investigation of the 
spectrum of the collective modes which includes massless \cite{Lin.Hu:12} and mixed 
phase-density \cite{Carlstrom.Garaud.ea:11a,Stanev:12,Maiti.Chubukov:13,
Marciani.Fanfarillo.ea:13} excitations.
Also, it was proposed to consider the properties of exotic topological excitations 
such as skyrmions and domain walls \cite{Garaud.Carlstrom.ea:11,Garaud.Carlstrom.ea:13,
Garaud.Babaev:14}, unconventional mechanism of vortex viscosity \cite{Silaev.Babaev:13}, 
formation of vortex clusters \cite{Carlstrom.Garaud.ea:11a}, exotic reentrant and 
precursor phases induced by fluctuations \cite{Bojesen.Babaev.ea:13,Bojesen.Babaev.ea:14,
Carlstrom.Babaev:15,Hinojosa.Fernandes.ea:14}.
Moreover, spontaneous currents were predicted to exist near non-magnetic impurities in 
anisotropic superconducting $\sid$ states \cite{Lee.Zhang.ea:09,Maiti.Sigrist.ea:15} 
or in samples subjected to strain \cite{Maiti.Sigrist.ea:15}. The latter proposal 
actually involves symmetry change of $\sis$ states and relies on the presence of 
disorder, which can typically have uncontrollable distribution.
It was also pointed out that the time-reversal symmetry breaking $\sis$ 
state features an unconventional contribution to the thermoelectric effect 
\cite{Silaev.Garaud.ea:15}. Related to this an experimental set-up, based on a local 
heating was recently proposed \cite{Garaud.Silaev.ea:16}. The key idea being that 
local heating induces local variations of the relative phases, which further yield 
an electromagnetic response that is directly observable.

\subsection*{Plan of the Chapter}

The existence of the $\sis$ state can originate from various mechanisms, including the 
competition between different pairing channels, or impurity scattering. These aspects 
will not be addressed here. Instead we will discuss how this $\sis$ state appears in 
phenomenological Ginzburg-Landau models. Discussions of the microscopic origin of the 
$\sis$ state, and its relation to Ginzburg-Landau models was for example discussed in 
\CVcite{Silaev.Garaud.ea:15} and \CVcite{Garaud.Silaev.ea:17}. See also 
\CVcite{Silaev.Garaud.ea:17} and \CVcite{Garaud.Corticelli.ea:18a} for discussions of 
the $\sis$ state that originates in impurity scattering.

As a starting point, the Section \ref{Sec:TRSB:Frustration} presents the mechanism 
of \emph{phase frustration}, that is responsible for the spontaneous breakdown of 
the time-reversal symmetry, in three-component superconductors. Namely, this is the 
competition between different phase-locking terms that can result in the $\sis$ state.
Since the time-reversal symmetry is a discrete operation, its spontaneous breakdown 
implies that the ground state features a discrete degeneracy. The properties of such 
ground state will be analyzed in Section \ref{Sec:TRSB:GS}, while the corresponding 
length-scales are derived in Section \ref{Sec:TRSB:LS}. The case of a two-component 
$\sis$ state will also be addressed in this section.

Next, the properties of topological defects that can occur in superconducting states 
which break the time-reversal symmetry are addressed in the Section 
\ref{Sec:TRSB:Topological-defects}. There is some overlap with the discussions 
of Section \ref{Sec:Chiral-Skyrmions} of the Chapter \ref{Chap:Topological-defects}.
The fact that the ground state breaks a discrete symmetry implies that the theory 
allow for domain-wall excitations. These domain-walls between different time-reversal 
symmetry broken states are constructed in Section \ref{Sec:TRSB:Domain-walls}.
The domain-walls interact non-trivially with the vortex matter. As detailled in Section 
\ref{Sec:TRSB:Chiral-Skyrmions}, closed domain-walls can form bound states with vortices. 
As discussed in the Chapter \ref{Chap:Topological-defects}, since the resulting composite 
defects are coreless, they feature extra topological properties: they are the chiral 
$\groupCP{2}$ skyrmions.

Finally, in the thermoelectric properties of the $\sis$ state will be discussed in 
Section \ref{Sec:TRSB:Thermoelectric}. The new thermoelectric properties of the 
$\sis$ state are consequences of the modified current, and magnetic relations in 
multicomponent superconductors.

\subsection*{Summary of the results that are relevant for this chapter}
\begin{itemize}
\setlength\itemsep{0.025em}

\item In \CVcite{Garaud.Silaev.ea:17} we have demonstrated that the mean field theories 
for the $s\!+\!is$ superconducting states, that break the time-reversal symmetry, are 
quantitatively consistent with microscopic multi-band models. We have further demonstrated 
that the $s\!+\!is$ state can also appear in two-band systems due to impurity scattering 
\CVcite{Silaev.Garaud.ea:17}. Within the quasiclassical approximation we show that the 
$s\!+\!is$ state forms as an intermediate phase at the impurity-driven crossover between 
$s_\pm$ and $s_{++}$ states. We further established in \CVcite{Garaud.Silaev.ea:17} 
and \CVcite{Garaud.Corticelli.ea:18a} that the $s\!+\!is$ domain is surrounded by a 
line of second-order phase transition, which implies the existence of a soft mode with 
a divergent length-scale. The other coherence lengths remain finite at this transition 
and thus there is an infinite disparity of coherence lengths, which may lead to unusual 
vortex physics with non-monotonic forces \CVcite{Garaud.Corticelli.ea:18a} and 
\CVcite{Garaud.Corticelli.ea:18}.

\item In \CVcite{Silaev.Garaud.ea:15} and \CVcite{Garaud.Silaev.ea:16} we have 
demonstrated that the existence of time-reversal symmetry broken states measurably 
impacts the thermoelectric response of superconductors. In \CVcite{Silaev.Garaud.ea:15} 
we predicted that superconductors which break the time-reversal symmetry feature a giant 
thermoelectric effect of principally different nature than that in single-component 
superconductors. It originates in thermally induced intercomponent counterflows, in 
contrast to the counterflows of normal and superconducting currents in the classical 
Ginzburg mechanism.
We have further demonstrated in \CVcite{Garaud.Silaev.ea:16} that these unconventional 
thermoelectric properties can be used to induce experimentally observable magnetic and 
electric fields by local heating of candidate materials. The induced fields are sensitive 
to the presence of domain-walls and crystalline anisotropy, while nonstationary heating 
process produces an electric field and a charge imbalance in the different bands
\CVcite{Garaud.Silaev.ea:16}.

\item Prediction of the experimental signatures of domain-wall structures that form 
during quenches, via the Kibble-Zurek mechanism, in superconductors with broken 
time-reversal symmetry originating in $\sis$ gap structure \CVcite{Garaud.Babaev:14}. 
As it is a discrete symmetry, the spontaneous breakdown of the time-reversal symmetry 
in the $\sis$ state, dictates that it possess domain wall excitations. We also discuss 
the influence of geometrically stabilized domain-walls on the magnetization processes.

\item Discovery of a new kind of stable topological solitons in three-component 
superconductors that spontaneously breaks the time-reversal symmetry 
\CVcite{Garaud.Carlstrom.ea:11} and \CVcite{Garaud.Carlstrom.ea:13}. These flux 
carrying topological defects, are characterized by a hidden topological charge, 
associated with the topology of the complex projective space $\groupCP{N-1}$. These  
$\groupCP{2}$ skyrmions can spontaneously form in field cooled experiment
\CVcite{Garaud.Babaev:14}, when the cooling process goes through the phase transition 
to the time-reversal symmetry broken state.

\item Findings of a new kind of collective mode in three-band superconductors with 
broken time-reversal symmetry \CVcite{Carlstrom.Garaud.ea:11a}. This collective mode is 
associated with mixed phase-density collective excitations. Thus it is different from 
the Leggett's mode.

\end{itemize}

\section{Phase frustration in three-band superconductors}
\label{Sec:TRSB:Frustration}

The time-reversal symmetry breaking $\sis$ states occurs due to the competition between 
different pairing channels. At the level of the mean field theory, this manifests itself 
through the existence of competing phase-locking terms. When different phase-locking 
terms cannot be simultaneously satisfied, the system may comprise via non-trivial 
relative phases between the condensates \cite{Ng.Nagaosa:09,Stanev.Tesanovic:10,
Tanaka.Yanagisawa:10,Hu.Wang:12}. It was indeed  observed that the inclusion of 
a third superconducting condensate leads to qualitatively different physics compared 
to two-component systems \cite{Ng.Nagaosa:09,Stanev.Tesanovic:10,Tanaka.Yanagisawa:10,
Hu.Wang:12}. As detailed below, the states with non-trivial relative phases 
(\ie when at least one is neither $0$ nor $\pi$) break the time-reversal symmetry.

To illustrate how phase frustration leads to the spontaneous breakdown of the 
time-reversal symmetry, we consider here a restriction of the generic free energy 
\Eqref{Eq:General:FreeEnergy}, for three superconducting condensates coupled via 
bilinear Josephson interaction: 
\SubAlign{Eq:TRSB:3CGL:FreeEnergy:1}{
\F/\F_0&=\bigintsss \frac{1}{2}\big|\Curl\A\big|^2 
+\sum_{a=1}^3\frac{1}{2}\big|\D\psi_a\big|^2 + V(\Psi,\Psi^\dagger)	
\,,  \label{Eq:TRSB:3CGL:FreeEnergy:1:a} \\
\text{where}~V(\Psi,\Psi^\dagger)&= \sum_{a=1}^3
\left(\alpha_a|\psi_a|^2+\frac{1}{2}\beta_a|\psi_a|^4 \right)
+\sum_{a=1}^3\,\sum_{b>a}^3\eta_{ab}(\psi_a^*\psi_b+\psi_b^*\psi_a)	
\,,\label{Eq:TRSB:3CGL:FreeEnergy:1:b}
}
Here again, $\psi_a=|\psi_a| e^{i\varphi_a}$ are complex fields representing the three 
superconducting condensates labelled by the indices $a,b=1,2,3$. For simplicity, mixed 
gradient terms are not considered (namely $\kappa_{ab}=\delta_{ab}$). 
Besides the electromagnetic coupling, the different condensates are directly coupled via 
the (bilinear) Josephson interaction: 
\Equation{Eq:TRSB:3CGL:Josephson}{
\eta_{ab}(\psi_a^*\psi_b+\psi_b^*\psi_a) = 
2\eta_{ab}|\psi_a||\psi_a|\cos(\varphi_{ab})\,,
}
where $\varphi_{ab}=\varphi_b-\varphi_a$ denotes the relative phase between two 
condensates.
Such a multicomponent Ginzburg-Landau free energy can, in certain cases, be derived 
microscopically at temperatures close to $T_c$ (for a review see \cite{Gurevich:07}, see 
also \CVcite{Garaud.Silaev.ea:17}). Note that the existence of three superconducting 
bands is not by any means a sufficient condition for a system to have Ginzburg-Landau 
expansion like that displayed in \Eqref{Eq:TRSB:3CGL:FreeEnergy:1}. 
However many of the question of phase-frustration which is considered here, do not 
require to be in the high-temperature region where the Ginzburg-Landau expansion 
\Eqref{Eq:TRSB:3CGL:FreeEnergy:1} can be formally justified. 
In the following, the minimal Ginzburg-Landau model \Eqref{Eq:TRSB:3CGL:FreeEnergy:1} 
is used as a convenient framework to discuss qualitatively the physics of phase 
frustration, and how it leads to time-reversal symmetry breaking. 
Discussions on microscopic physics aspects can be found, for example, in 
\CVcite{Garaud.Corticelli.ea:18}, \CVcite{Garaud.Silaev.ea:17}, and
\CVcite{Silaev.Garaud.ea:17}.
Note that three-component models feature in principle additional terms allowed by 
symmetry, \eg bi-quadratic terms in density (see for example 
\cite{Garaud.Carlstrom.ea:13}). However these terms play a little role in the phase 
frustration that leads to the breakdown of the time-reversal symmetry. Hence we focus 
here on the simplest model.

The quartic coefficients $\beta_a$ are positive so that the free energy 
\Eqref{Eq:TRSB:3CGL:FreeEnergy:1} is bounded from below. On the other hand, the quadratic 
coefficients $\alpha_a$ change signs at some characteristic temperatures which are 
generally different for all components. Below this characteristic temperature 
$\alpha_a<0$ and the band is said to be active, while above it, $\alpha_a>0$ and the 
band is passive. Nevertheless, passive bands can feature nonzero superfluid density 
because of the interband Josephson tunnelling terms with the coupling $\eta_{ab}$. 
For the superconducting state to exist, the determinant of the second order couplings 
should be negative \footnote{
Using the notations of \Eqref{Eq:General:FreeEnergy}, the criterion for the existence 
of the superconducting state is $\det \hat{\alpha}<0$ where $\hat{\alpha}$ denotes the 
(symmetric) matrix of the coefficient $\alpha_{ab}$ of the bilinear terms 
$\alpha_{ab}\psi_a^*\psi_b$.
}.
Hence, for the potential \Eqref{Eq:TRSB:3CGL:FreeEnergy:1:b}, the superconducting states 
exists if
\Equation{Eq:TRSB:3CGL:GS:Criterion}{
\alpha_1\alpha_2\alpha_3+2\eta_{12}\eta_{13}\eta_{23} 
-\alpha_1\eta_{23}^2 -\alpha_2\eta_{13}^2 -\alpha_3\eta_{12}^2 <0 \,.
}

\begin{wrapfigure}{R}{0.5\textwidth}
\hbox to \linewidth{ \hss
\includegraphics[width=.975\linewidth]{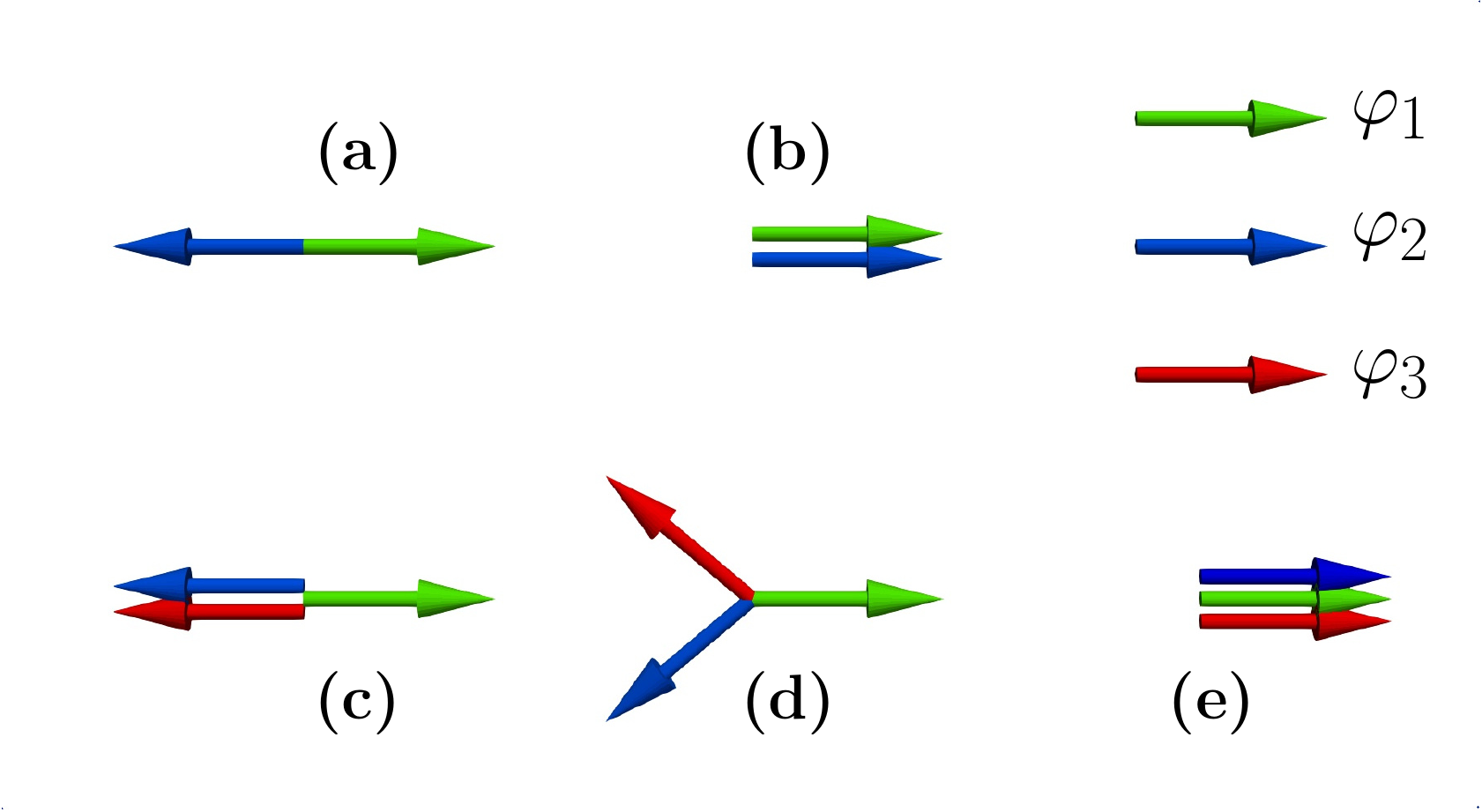}
\hss}
\caption{
Schematic illustration of phase locking patterns. 
Panels (a) and (b) show the two possible phase locking for two components; 
either zero or $\pi$.
Panels (c), (d) and (e) display the possible, qualitatively different, phase locking 
in the case of three components. Panels (c) and (e), the relative phases are said to 
be trivial (\ie either zero or $\pi$). The case of the panel (d) is a three frustrated 
component superconductor, with non-trivial phase locking.
}
\label{Fig:Schematic:Phase-locking}
\end{wrapfigure}

When the parameters satisfy the condition \Eqref{Eq:TRSB:3CGL:GS:Criterion}, systems 
with more than two condensates can exhibit \emph{frustration} of the competing 
Josephson coupling terms. When $\eta_{ab}<0$, a given Josephson interaction term 
\Eqref{Eq:TRSB:3CGL:Josephson} is minimal for zero relative phase $\varphi_{ab}$, 
and this coupling is a ``phase-locking". On the other hand, when $\eta_{ab}>0$, the 
Josephson interaction is minimal for the relative phase $\varphi_{ab}=\pi$, and this 
coupling is a ``phase-antilocking"
\footnote{
In two-component systems, there exists a ``parametric'' symmetry with respect to the 
sign change of the Josephson coupling $\eta_{ab}\to -\eta_{ab}$. Indeed, such change 
can be compensated by an overall change of the relative phase 
$\varphi_{ab}\to\varphi_{ab}\pm\pi$, so that the system recovers the same interaction.
However, in systems with more than two condensates there is generally no such symmetry. 
}.
The behaviour of the individual phase-locking terms between two condensates is illustrated 
in \Figref{Fig:Schematic:Phase-locking}. There, the panel (a) shows a relative phase 
$\varphi_{12}=\pi$, and thus corresponds to $\eta_{12}>0$. Panel (b) on the other hand 
shows the zero relative phase that minimizes the Josephson interaction for a negative 
Josephson coupling $\eta_{12}<0$.

Now, in the case of three condensates, depending on the signs of the couplings $\eta_{ab}$, 
the individual Josephson terms cannot always be simultaneously minimal. Indeed, the 
relative phases are not all independent; for example 
\Equation{Eq:TRSB:3CGL:GS:Phases}{
\varphi_{23}:=\varphi_{3}-\varphi_{2}=\varphi_{3}-\varphi_{1}+\varphi_{1}-\varphi_{2}
:=\varphi_{13}-\varphi_{12}.
}
Such a situation, 
where the phase-locking terms compete with each other, and where the system comprises 
between the individual tendencies, is called \emph{frustrated}. Note that, as detailed 
below, frustration in three-component systems like \Eqref{Eq:TRSB:3CGL:FreeEnergy:1} is 
a \emph{necessary} (yet not sufficient) condition for the spontaneous breakdown of the 
time-reversal symmetry.
The condition for frustration can be determined in terms of the sign of the 
Josephson couplings $\eta_{ab}$. Indeed there are four principal situations, 
summarized in the \Tabref{Tab:Frustration}. 
Note that frustration also occurs in the general case of arbitrary number of components. 
There are more possibilities for such systems to be frustrated, which allows for even 
richer physics than in three-components. Frustration in four-component systems was 
discussed in details in \cite{Weston.Babaev:13}.

\begin{table}[!ht]
\centering
\begin{tabular}{ c||c|c } 
\hline
Case & Sign of $\eta_{12},\eta_{13},\eta_{23}$ & Ground state phases \\ 
\hline
1& $- - -$ & $\varphi_1=\varphi_2=\varphi_3$ \\
2& $- - +$ & Frustrated  \\ 
3& $- + +$ & $\varphi_1=\varphi_2=\varphi_3+\pi$ \\
4& $+ + +$ & Frustrated
\\ \hline
\end{tabular}
\caption{
Illustration of the qualitatively different  representative situation in the case of 
three-component superconducting condensates coupled via bilinear Josephson interaction. 
Depending of the sign of the $\eta_{ab}$, there are four principal situations. 
The ground state phases are given when the signatures do not lead to frustration.
}
\label{Tab:Frustration}
\end{table}

\paragraph{Example of phase frustration and broken time-reversal symmetry:}
The simplest illustration of phase frustration is the case where the coefficients of the  
three components are completely symmetric, and all Josephson couplings are repulsive. 
For example, with the individual couplings $\alpha_a=-1$, $\beta_a=1$ for $a=1,2,3$, 
and the Josephson couplings $\eta_{12}=\eta_{13}=\eta_{23}=-1$. Each individual 
phase-locking term favours relative phases that equal $\pi$. This is however 
impossible, since, \eg
\Equation*{
\text{if}~~~\varphi_{12}=\varphi_{13}=\pi\,,~~\text{then}~~~
\varphi_{23}\equiv\varphi_{13}-\varphi_{12}=0\neq \pi
~~~~~(\text{with}~~\varphi_{ab}=\varphi_{b}-\varphi_{a})	\,.
}
Hence, since the individual phase-locking terms cannot be simultaneously satisfied, 
the system is said to be \emph{frustrated}. Instead, the ground state has all the 
relative phases equal to $\pm2\pi/3\mod 2\pi$. More precisely, it can be shown that 
\SubAlign{Eq:Frustration:GS}{
\text{either}~~~
\varphi_{12}&=+2\pi/3\,, & \varphi_{13}&=-2\pi/3\,,& \varphi_{23}&=+2\pi/3\mod 2\pi\,,
\label{Eq:Frustration:GS:a}
\\ \text{or}~~~
\varphi_{12}&=-2\pi/3\,, & \varphi_{13}&=+2\pi/3\,,& \varphi_{23}&=-2\pi/3\mod 2\pi\,.
\label{Eq:Frustration:GS:b}
}

\begin{wrapfigure}{R}{0.5\textwidth}
\hbox to \linewidth{ \hss
\includegraphics[width=.975\linewidth]{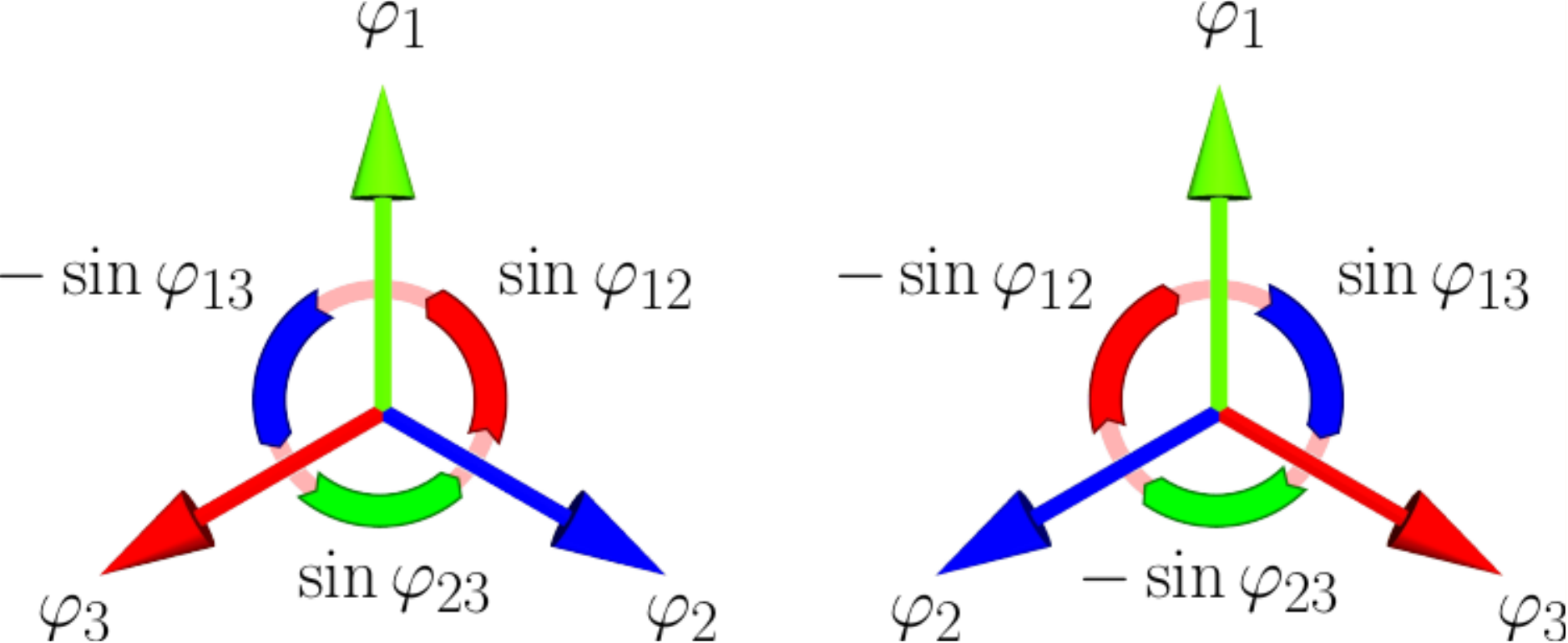}
\hss}
\caption{
Schematic illustration of the two inequivalent ground states \Eqref{Eq:Frustration:GS}. 
Since both configurations are mirror of each others, these states are also referred to 
as \emph{chiral states}.
The circulating arrows denote circulating interband tunneling currents.
}
\label{Fig:Schematic:Chiral-states}
\end{wrapfigure}
Here, not only the system is frustrated, but also it features a discrete degeneracy. 
Indeed, since the relative phases are gauge invariant, both configurations 
\Eqref{Eq:Frustration:GS:a} and \Eqref{Eq:Frustration:GS:b} cannot be transformed into 
each other by a global phase rotation. Both these configurations are displayed in 
\Figref{Fig:Schematic:Chiral-states}. They are related to each other by the complex 
conjugation, associated with the time-reversal transformation $\mathcal{T}(\psi_a)=\psi_a^*$. 
Since $\mathcal{T}(\varphi_{ab})=-\varphi_{ab}$, the two ground states are indeed related 
to each other by the time-reversal transformation. Hence the ground state is said to 
spontaneously break the time-reversal symmetry.

The ground state here, thus features a discrete $\groupUZ$ degeneracy. In a phase 
transition the ground state randomly picks either of the chiralities, and this may 
lead to the formation of domain-walls via the Kibble-Zurek mechanism \cite{Kibble:76,
Zurek:85}. This was further discussed in \CVcite{Garaud.Babaev:14} and in 
Sec.~\ref{Sec:Chiral-Skyrmions}. See also the continued discussion in 
Sec.~\ref{Sec:TRSB:Domain-walls}.
Under certain conditions, the spontaneous breakdown of the time-reversal symmetry also 
allows for composite topological excitations which are bound states of closed domain 
walls and vortices \cite{Garaud.Carlstrom.ea:11,Garaud.Carlstrom.ea:13}, see 
Sec.~\ref{Sec:Chiral-Skyrmions} and the continued discussion in 
Sec.~\ref{Sec:TRSB:Chiral-Skyrmions}.

\subsection{Ground state of a three-component superconductor}
\label{Sec:TRSB:GS}

As discussed above, frustration occurs depending on the signatures of the different 
Josephson couplings, as summarized in Table~\ref{Tab:Frustration}. Frustration is a 
necessary, yet not sufficient, condition for the spontaneous breakdown of the time-reversal 
symmetry. The breakdown occurs depending on the relative values of the Josephson couplings.
The dependence of the ground state, with respect to a given coupling illustrates this. 
The ground state values of the fields  $|\psi_a|$ and $\varphi_{ab}$ of the free energy 
\Eqref{Eq:TRSB:3CGL:FreeEnergy:1} are found by minimizing its potential energy: 
\Equation{Eq:TRSB:3CGL:Ground-state}{
\Psi_0=\underset{\Psi\in\Complex^N}{\mathrm{argmin}}~V(\Psi,\Psi^\dagger)  \,,
~~\text{with}~~
 V=\sum_{a=1}^3\Bigg\{\alpha_a|\psi_a|^2+\frac{1}{2}\beta_a|\psi_a|^4 +
 \sum_{b>a}^3\eta_{ab}|\psi_a||\psi_b|\cos\varphi_{ab}\Bigg\}\,.
}
The ground state is determined by the system of equations given by the variations of 
the potential, with respect to the physical degrees of freedom $|\psi_a|$ and $\varphi_a$.
The variation with respect to the densities read explicitly
\SubAlign{Eq:TRSB:3CGL:Ground-state:Eq:Densities}{
\frac{\delta V}{\delta|\psi_1|}:=
2\alpha_1|\psi_1|+2\beta_1|\psi_1|^3 + \eta_{12}|\psi_2|\cos\varphi_{12}
									 + \eta_{13}|\psi_3|\cos\varphi_{13}  &=0 \,,\\
\frac{\delta V}{\delta|\psi_2|}:=
2\alpha_2|\psi_2|+2\beta_2|\psi_2|^3 + \eta_{12}|\psi_1|\cos\varphi_{12}
									 + \eta_{23}|\psi_3|\cos\varphi_{23}  &=0 \,,\\
\frac{\delta V}{\delta|\psi_3|}:=
2\alpha_3|\psi_3|+2\beta_3|\psi_3|^3 + \eta_{13}|\psi_1|\cos\varphi_{13}
									 + \eta_{23}|\psi_2|\cos\varphi_{23}  &=0 \,.
}
Variations of with respect to the phases, on the other hand, read as
\SubAlign{Eq:TRSB:3CGL:Ground-state:Eq:Phases}{
\frac{\delta V}{\delta\varphi_1}:=
 + \eta_{12}|\psi_1||\psi_2|\sin\varphi_{12} 
 + \eta_{13}|\psi_1||\psi_3|\sin\varphi_{13}  &=0  \,, 
 	\label{Eq:TRSB:3CGL:Ground-state:Eq:Phases:a} \\
\frac{\delta V}{\delta\varphi_2}:=
 - \eta_{12}|\psi_1||\psi_2|\sin\varphi_{12} 
 + \eta_{23}|\psi_2||\psi_3|\sin\varphi_{23}  &=0  \,,\\
\frac{\delta V}{\delta\varphi_3}:=
 - \eta_{13}|\psi_1||\psi_3|\sin\varphi_{13}  
 - \eta_{23}|\psi_2||\psi_3|\sin\varphi_{23}  &=0  \,.
}
The potential is invariant under global rotation of all phases. Thus it can be 
convenient to fix the gauge by imposing the value of one of the phases, for example 
$\varphi_1=0$. In that case, the equation $\delta V/\delta\varphi_1=0$ 
Eq.~\Eqref{Eq:TRSB:3CGL:Ground-state:Eq:Phases:a}, becomes trivial as it is a linear 
combination of the other two.
Finding the ground state thus boils down to solving the nonlinear system of five equations. 
The three density equations \Eqref{Eq:TRSB:3CGL:Ground-state:Eq:Densities}, plus the two 
nontrivial equations of \Eqref{Eq:TRSB:3CGL:Ground-state:Eq:Phases}. This cannot, in 
general, be solved analytically and thus requires numerical methods such as Nonlinear 
Conjugate Gradient or Newton-Raphson algorithm.

\paragraph{Alternative to the gauge fixing:}
Remark that another possibility, to determine the ground state is to consider, together 
with the density equations \Eqref{Eq:TRSB:3CGL:Ground-state:Eq:Densities}, the variations 
with respect to the relative phases (which are gauge invariant quantities). This however 
have to be addressed carefully. Indeed, it is important to stress again that the three 
relative phases $\varphi_{ab}$ are \emph{not} independent. It is thus necessary to first 
express one in terms of the other two, \eg $\varphi_{23}=\varphi_{13}-\varphi_{12}$, see 
Eq.~\Eqref{Eq:TRSB:3CGL:GS:Phases}. 
The variations of the potential with respect to the remaining relative phases 
thus yield the equation
\SubAlign{Eq:TRSB:3CGL:Ground-state:Eq:Relative-Phases}{
\frac{\delta V}{\delta\varphi_{12}}:=
 - \eta_{12}|\psi_1||\psi_2|\sin\varphi_{12} 
 + \eta_{23}|\psi_2||\psi_3|\sin(\varphi_{13}-\varphi_{12})  &=0  \,,\\
\frac{\delta V}{\delta\varphi_{13}}:=
 - \eta_{13}|\psi_1||\psi_3|\sin\varphi_{13} 
 - \eta_{23}|\psi_2||\psi_3|\sin(\varphi_{13}-\varphi_{12})  &=0  \,.
}
Note that this alternative is peculiar to systems with more than two components. Indeed, 
when there are only two components, fixing the gauge and working with the relative phase
are completely equivalent.

\paragraph{Practical implementation:} While expressing the complex degrees of freedom 
in terms of moduli and phases $\psi_a=|\psi_a|\Exp{i\varphi_a}$, is physically intuitive, 
it is not convenient for numerical computations. Indeed, the moduli are strictly positive 
quantities, $|\psi_a|\in\Real^+$. This makes the functional non-convex, and thus unsuitable 
for some algorithms, such as the Nonlinear Conjugate Gradient (see details in the Appendix 
\ref{App:numerics:NLCG}). Indeed, the Nonlinear Conjugate Gradient algorithm works when 
the functional is approximately quadratic near the minimum.  This is the case when the 
function is twice differentiable at the minimum and the second derivative is non-singular 
there. This is obviously not always true when considering the moduli. The alternative to 
this issue is to parametrize the superconducting degrees of freedom, in terms or the real 
and imaginary parts of the complex fields $\psi_a=X_a+iY_a$, instead of the moduli and 
phases $\psi_a=|\psi_a|\exp\{i\varphi_a\}$.
When working with real and imaginary parts of the complex fields, gauge fixing of \eg 
$\varphi_1=0$ is achieved by setting $Y_1=0$, since $Y_a\equiv|\psi_a|\sin\varphi_a$.
However, it can also be convenient not to fix the gauge at all, and to consider only 
the gauge invariant quantities.

\begin{figure}[!htb]
\hbox to \linewidth{ \hss
\includegraphics[width=.75\linewidth]{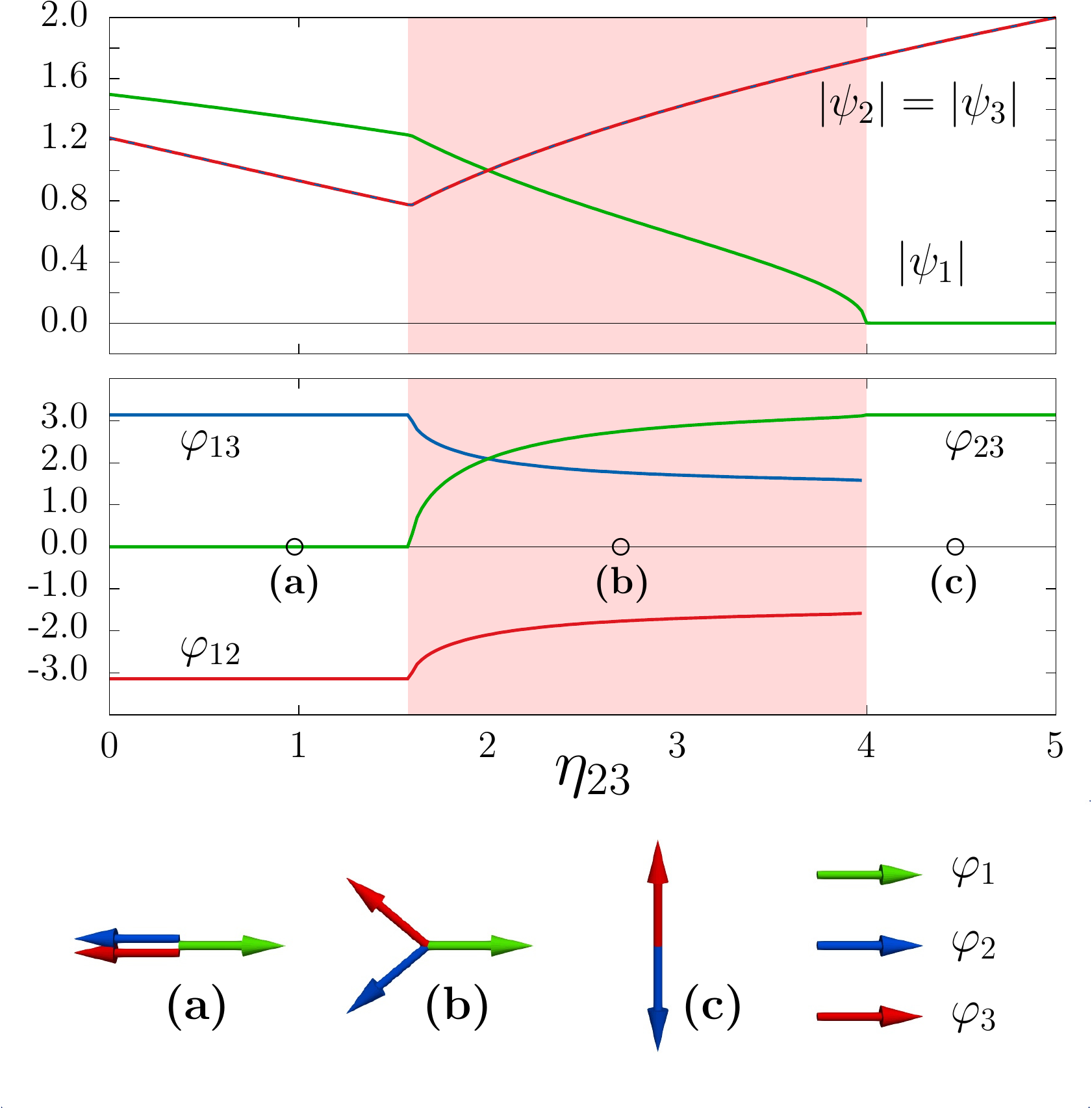}
\hss}
\caption{
Ground state phases of the three-components superconductor 
\Eqref{Eq:TRSB:3CGL:FreeEnergy:1}, as function of $\eta_{23}$.  
The other parameters of the Ginzburg-Landau potential energy are $\alpha_a=1$, $\beta_a=1$,
and $\eta_{12}=\eta_{13}=2$.
The top graph shows the ground state densities $|\psi_a|$, while the bottom graph shows 
the relative phases $\varphi_{ab}$.
All individual couplings are not simultaneously satisfied, so the system is frustrated. 
However, for small coupling $\eta_{23}$ (regime (a)), the relative phases are trivial 
(\ie either 0 or $\pi$). 
For intermediate values of $\eta_{23}$ (regime (b)), all ground state relative phases 
are non-trivial. Hence the ground state features a discrete symmetry: 
$\groupU{1}\times\groupZ{2}$ rather than $\groupU{1}$. Indeed, the energy is invariant 
under $\varphi_{ab}\to-\varphi_{ab}$, but the configuration cannot be continuously 
transformed into each other.
For large Josephson coupling $\eta_{23}$ (regime (c)) the third condensate vanishes, 
and the only remaining relative phase is $\varphi_{12}=\pi$.
The red background denotes the region of broken time-reversal symmetry.
As discussed later on, the both transitions (a)\,$\leftrightarrow$\,(b) and 
(b)\,$\leftrightarrow$\,(c) are of the second order.
Similar diagram showing the ground state of three-component superconductor can be found 
in \CVcite{Carlstrom.Garaud.ea:11a}. 
}
\label{Fig:Diagram:Phase-locking-3CGL}
\end{figure}

Consider, as an example, the situation where all Josephson couplings are positive. 
As can be seen in \Tabref{Tab:Frustration}, this situation is frustrated. The dependence 
of the ground state, with respect to a single Josephson coupling (here $\eta_{23}$) are 
displayed in \Figref{Fig:Diagram:Phase-locking-3CGL}. It shows three qualitatively distinct 
phases. At small values of $\eta_{23}$ (regime (a)), the relative phases are all trivial: 
$[\varphi_{12},\varphi_{13},\varphi_{23}] = [\pi,\pi,0]$. In the opposite limit, for large 
values of $\eta_{23}$ (regime (c)), the one condensate (here $\psi_1$) is fully depleted. 
It follows that only one of the relative phase, here $\varphi_{23}$, is relevant. The 
two remaining phases are $[\varphi_{2},\varphi_{3}] = [-\pi/2,\pi/2]$.
The most interesting phase here is the phase (b). Indeed, as stated above, the relative 
phases here are neither 0 nor $\pi$. As stated above, this implies that the ground state 
features a discrete degeneracy, since it is \emph{not} invariant under the time-reversal 
transformation $\mathcal{T}(\varphi_{ab})=-\varphi_{ab}$. The ground state here thus 
spontaneously breaks the time-reversal symmetry.

\vspace{0.5cm}

Symmetry-wise, in the phases (a) and (c), the ground state features the usual $\groupU{1}$ 
degeneracy associated with the global rotation of all phases. In the phase (b), on the 
other hand, the ground state has an extra discrete degeneracy $\groupZ{2}$ so that the 
overall degeneracy is $\groupUZ$.
Clearly, both phases (a) and (c) are  symmetric under the time-reversal operations
$\mathcal{T}(\varphi_{ab})=-\varphi_{ab}$, while the phase (b) is not.
Moreover, as discussed below, the both transitions (a)\,$\leftrightarrow$\,(b) and 
(b)\,$\leftrightarrow$\,(c) are of the second order. These are thus associated with 
a divergent length-scale.

\subsection{Length-scales of a three-band superconductor}
\label{Sec:TRSB:LS}

The spontaneous breakdown of the time-reversal symmetry, as presented here at the 
phenomenological level of the Ginzburg-Landau theory, is a property of the ground state. 
This however also qualitatively affects the excitations, and in particular in the 
vicinity of the transitions of the time-reversal symmetry broken states.
The detailed analysis of the length-scales and of the associated modes of the perturbation
operator indeed reveals new, interesting properties.
To have a grasp on this, let's follow the procedure of the analysis of the perturbation 
operator described previously in Sec.~\ref{Sec:Type-1.5:Length-scales}. It might also be 
useful to compare with the analysis of the mass spectrum for single-component 
Ginzburg-Landau model in the section Sec:~\ref{Sec:Background:GS} of the 
Appendix~\ref{App:Single-Component}. The analysis of the perturbation operator 
determines the mass spectrum (and thus the length-scales), and the corresponding modes. 
For a detailed discussion of the perturbative spectrum of three-band superconductor, 
see \cite{Carlstrom.Garaud.ea:11a,Carlstrom.Garaud.ea:11aErratum}.

Considering the perturbative expansion in terms of the infinitesimal parameter $\eps$
\Equation{Eq:TRSB:3CGL:Expansion}{
\psi_a = (u_a + \eps f_a)
\exp\left\{i\left(\bvarphi_a+\eps\frac{\phi_a}{u_a}\right)\right\}	\,.
}
Here, $u_a$ and $\bvarphi_a$  are respectively the ground state densities and phases, 
introduced in the previous section. $f_a\equiv f_a(\x)$ are the density amplitudes, 
while $\phi_a\equiv \phi_a(\x)$ are the normalized phase amplitudes. 
For simplicity here, we do not consider the excitations of the gauge field, and focus 
only on the properties of the superconducting degrees of freedom. The fluctuations 
are thus characterized by a system of Klein-Gordon equations for the six condensate 
fluctuations (three densities, plus three phases). The Klein-Gordon system reads as
\Equation{Eq:TRSB:3CGL:KG}{
\frac{1}{2}\Upsilon^T\left(-\Grad^2 + {\cal M}^2\right)\Upsilon
\,,~~~\text{where}~~~
\Upsilon=(f_1,f_2,f_3,\phi_1,\phi_2,\phi_3)^T\,.
}
Here ${\cal M}^2$ is the squared mass matrix that is straightforwardly obtained by 
retaining the quadratic order of the infinitesimal parameter $\eps$, after introducing 
the expansion \Eqref{Eq:TRSB:3CGL:Expansion} into the free energy 
\Eqref{Eq:TRSB:3CGL:FreeEnergy:1}. The squared mass matrix can thus be read from 
\SubAlign{Eq:TRSB:3CGL:KGmass}{
&\Upsilon^T{\cal M}^2\Upsilon=\sum_{a=1}^3 2(\alpha_a+3\beta_au_a^2)f_a^2 
+\sum_{a=1}^3\sum_{b>a}^3 2\eta_{ab}f_af_b\cos\bvarphi_{ab } \\
&+\sum_{a=1}^3\sum_{b>a}^3\frac{2\eta_{ab}}{u_au_b}\Big\{
(u_af_b+f_bu_a)\big(\phi_bu_a-\phi_au_b\big)\sin\bvarphi_{ab}
-\frac{1}{2}\big(\phi_bu_a-\phi_au_b\big)^2\cos\bvarphi_{ab}\Big\}
\,.
} 
The eigenspectrum of the matrix ${\cal M}^2$ determines the squared masses of the 
excitations, the associated length-scales and the corresponding normal modes.

Before quantitatively investigating the mass spectrum, important qualitative properties 
can be determined by carefully examining the structure of the mass matrix. As emphasized 
in the previous section, the current model features essentially different states, 
depending on whether the ground state relative phases are trivial or not.

\paragraph{In the case of a trivial phase-locking,}\ie $\bvarphi_{ab}=0,\pi$, 
the mass matrix \Eqref{Eq:TRSB:3CGL:KGmass} simplifies, and the density amplitudes $f_a$ 
decouple from the normalized phase amplitudes $\phi_a$. Indeed, since $\sin\bvarphi_{ab}=0$ 
in that case, the mass matrix becomes block-diagonal (namely ${\cal M}_{f\phi}$).
The mass of the density amplitudes $f_a$ are thus given by the eigenvalues of 
${\cal M}_{ff}^2$ defined from
\Equation{Eq:TRSB:3CGL:KGmass:ff}{
\Upsilon_f^T{\cal M}_{ff}^2\Upsilon_f=\sum_{a=1}^3 2(\alpha_a+3\beta_au_a^2)f_a^2 
+\sum_{a=1}^3\sum_{b>a}^3 2\eta_{ab}\cos\bvarphi_{ab } f_af_b
\,.
}
Hence, as long as $\eta_{ab}\neq0$, the density modes are in general mixed. It follows 
that, as previously discussed in Sec.~\ref{Sec:Type-1.5:Length-scales}, the characteristic 
length-scales of the density fields are associated with linear combinations of the 
fields, see \eg \cite{Babaev.Carlstrom.ea:10,Carlstrom.Babaev.ea:11,Silaev.Babaev:11}. 
This means physically that disturbing one of the density fields necessarily perturbs 
the others. This also implies that in a vortex, the long-range asymptotics of all density 
fields is governed by the same exponent, corresponding to a mixed mode with the lowest 
mass.

The masses of the normalized phase amplitudes $\phi_a$, on the other hand are given by 
the eigenvalues of ${\cal M}_{\phi\phi}^2$ defined from 
\Equation{Eq:TRSB:3CGL:KGmass:Legget}{
\Upsilon_\phi^T{\cal M}_{\phi\phi}^2\Upsilon_\phi=
\sum_{a=1}^3\sum_{b>a}^3\frac{-\eta_{ab}\cos\bvarphi_{ab}}{u_au_b}
\big(\phi_bu_a-\phi_au_b\big)^2
\equiv-\sum_{a=1}^3\sum_{b>a}^3\eta_{ab}u_au_b\cos\bvarphi_{ab}
\hat{\phi}_{ab}^2
\,.
}
Here again, the (non-normalized) relative phase amplitudes 
$\hat{\phi}_{ab}:=\frac{\phi_b}{u_b}-\frac{\phi_a}{u_a}$ have been introduced, as 
they directly relate to the fluctuations of the relative phase. This defines again the 
mass of the Leggett's mode \cite{Leggett:66}.

\paragraph{On the other hand, for a non-trivial phase-locking,}\ie $\bvarphi_{ab}\neq0,\pi$, 
the density fluctuation are always coupled to the phase fluctuations. Indeed, since there 
$\sin\bvarphi_{ab}\neq0$, the decoupling discuss in the previous paragraph is not possible: 
\Equation{Eq:TRSB:3CGL:KGmass:Coupling}{
\Upsilon^T{\cal M}^2_{f\phi}\Upsilon=
\sum_{a=1}^3\sum_{b>a}^3\frac{2\eta_{ab}\sin\bvarphi_{ab}}{u_au_b}
(u_af_b+f_bu_a)\big(\phi_bu_a-\phi_au_b\big)\,\neq 0\,.
}
It follows that when the time-reversal symmetry is broken, there are no ``phase-only" 
Leggett's modes. Instead there appears a new kind of collective excitations where the 
phase difference modes are hybridized with the density (Higgs) modes  
\cite{Carlstrom.Garaud.ea:11a,Maiti.Chubukov:13,Stanev:12,Marciani.Fanfarillo.ea:13}.
These hybridized normal modes have a complex structure that mixes all density amplitudes, 
to all the phase amplitudes. This is not further discussed here, and the detailed analysis 
of the hybridized modes can be found for example in \CVcite{Carlstrom.Garaud.ea:11a}.
As discussed below, the hybridized modes can be associated with large characteristic 
length-scales even in the case of strong Josephson coupling. 
Note finally that, in principle, there could be accidental decouplings, but possibility 
this will not be discussed here. Such a possibility was discussed in a broader context of 
four-component models \cite{Weston.Babaev:13}.

\begin{figure}[!htb]
\hbox to \linewidth{ \hss
\includegraphics[width=.8\linewidth]{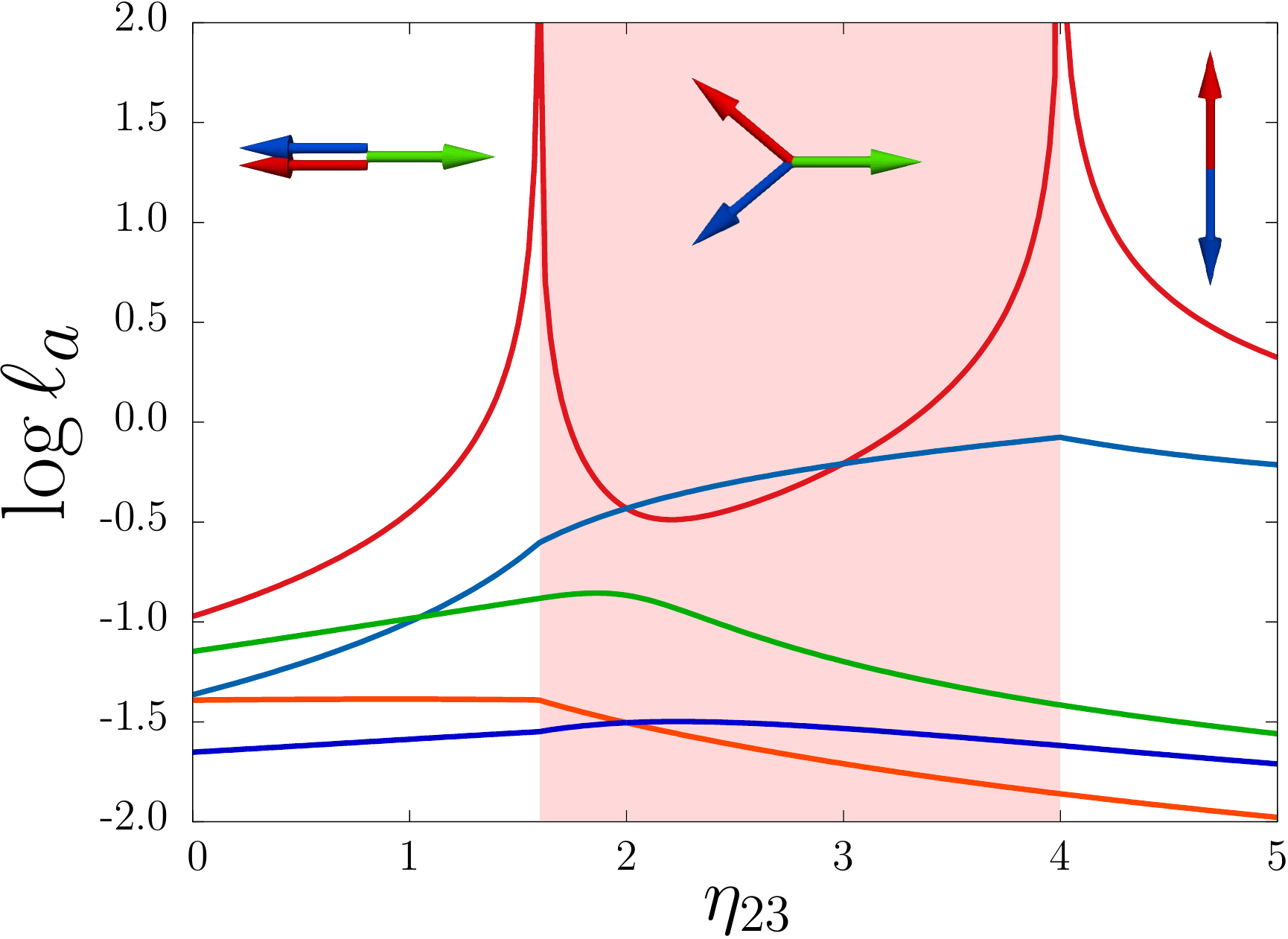}
\hss}
\caption{
Length-scales of the three-components superconductor \Eqref{Eq:TRSB:3CGL:FreeEnergy:1}, 
as functions of the Josephson coupling $\eta_{23}$. The other parameters of the 
Ginzburg-Landau potential energy are $\alpha_a=1$, $\beta_a=1$, and $\eta_{12}=\eta_{13}=2$. 
This parameter set corresponds to the ground states displayed in 
\Figref{Fig:Diagram:Phase-locking-3CGL}. Here, the length-scale of the $\groupU{1}$ 
zero mode is not shown as it corresponds to an unphysical degree of freedom, 
so there are only 5 length-scales associated with the superconducting degrees of freedom. 
The penetration depth, which is not displayed here, is always finite.
Here again one can clearly identify three different regimes. In the central regime, which 
is denoted by the red background, the time-reversal symmetry is broken, while the two 
other regimes are time-reversal symmetric.
The transition from the time-reversal symmetric states to the time-reversal symmetry 
broken states is a symmetry change. It is thus expected to be accompanied by a divergent 
length-scale.
Similar diagram showing the length-scales of three-component superconductor can be found 
in \CVcite{Carlstrom.Garaud.ea:11a}. There are also more details like the various eigenmodes.
}
\label{Fig:Diagram:Lengthscales-3CGL}
\end{figure}

The examination of the structure of the mass matrix thus predicts important 
qualitative properties, especially when the time-reversal symmetry is broken. Now, 
for a more quantitative discussion, the eigenvalue spectrum of ${\cal M}^2$ has to 
be determined given the ground state $u_a$, $\bvarphi_{ab}$. This is easily handle 
numerically with standard linear algebra tools.
The figure~\ref{Fig:Diagram:Lengthscales-3CGL} shows the various length-scales 
associated with the ground state displayed in \Figref{Fig:Diagram:Phase-locking-3CGL}, 
here again as functions the Josephson coupling $\eta_{23}$. The eigenspectrum of 
\Eqref{Eq:TRSB:3CGL:KG}, with the (squared) mass matrix \Eqref{Eq:TRSB:3CGL:KGmass}, 
is the set of $6$ squared masses ${\cal M}_a^2$, whose corresponding lengths 
$\ell_a=1/{\cal M}_a$ are the physical length scales of the model. 
Note again that there is a spontaneously broken $\groupU{1}$ symmetry associated with 
the simultaneous rotation of all phases. This mode has a zero mass, and it can easily 
be decoupled 
\footnote{
In practice it is convenient not to decouple this mode. Indeed, first of all this makes 
the system much simpler to write. Moreover the zero mode provides an estimation of the 
numerical resolution of masses.
}. 
Thus there are only 5 physical lengths that are associated with the superconducting 
condensates. The additional length-scale given by the London penetration depth, is not 
discussed here.

Investigating the ground state data in \Figref{Fig:Diagram:Phase-locking-3CGL}, and 
the associated length-scales in \Figref{Fig:Diagram:Lengthscales-3CGL}, shows again 
that there are three qualitatively distinct regimes, depending on the Josephson coupling 
$\eta_{23}$.
At $\eta_{23}=1.6$ there is a transition between the regime where with the usual 
$\groupU{1}$ degeneracy, and the regime where the ground state has an overall $\groupUZ$
degeneracy. A similar transition between $\groupUZ$ and $\groupU{1}$ states, occurs at 
$\eta_{23}=4$. 
As explained above, in the $\groupU{1}$ regimes, the density modes are mixed and there 
is no mixing between the density modes and the phase modes. It follows that the 
perturbations of the phases and of the densities recover independently of each other.
The fluctuations of the phase modes are the three-component generalization of the 
standard Leggett's modes. In the $\groupUZ$ regime all the modes are mixed, thus any
perturbation of the densities induces a perturbation of the relative phases, and vice versa. 
At the two transition points, $\eta_{23}=1.6$  and $\eta_{23}=4$, there is a divergent 
length-scale. The examination of the corresponding eigenvector shows that in the 
$\groupU{1}$ regimes this a mode is a phase-only. Thus a Leggett mode becomes massless 
at the transitions to the time-reversal symmetry broken states. This was realized in 
the London model in \cite{Lin.Hu:12}, and in general Ginzburg-Landau model 
\cite{Carlstrom.Garaud.ea:11a,Carlstrom.Garaud.ea:11aErratum}. It follows that here, 
the decay of the corresponding perturbation is not exponential, but it is governed by 
a power law.

The perturbation operator thus features a divergent length-scale in the vicinity of both 
transitions to the $\sis$, time-reversal symmetry breaking state. This transition is thus 
of the second order. Indeed, the theory of the mean-field second-order phase transitions 
states that the mass of one of the modes should go to zero, while other length-scales 
remain finite. 
Unlike the phase transition to the normal state, at which the gauge field becomes 
massless, the penetration depth is always finite at these transitions 
\footnote{
The penetration depth here is not displayed. Yet it is clear from the data presented in 
\Figref{Fig:Diagram:Phase-locking-3CGL}, that the total density do not vanish. Since it 
is proportional to mass of the gauge field, this implies that the penetration depth is 
always finite. 
}.

In the vicinity of these transition points, the largest length-scale is anomalously 
large due, to the frustration between Josephson couplings. Since the penetration depth 
remains finite, this implies that there exists always a region, close enough to the phase 
transition, where $\lambda$ is an intermediate length scale. It follows, as discussed 
in the Chapter~\ref{Chap:Semi-Meissner}, that superconductors that are in the vicinity 
of a time-reversal symmetry breaking transition are potentially type-1.5, with long-range 
attractive and short-range repulsive intervortex forces.

\subsection*{Time-reversal symmetry breaking in two-components superconductor}

In this section, we extensively discussed how the $\sis$ time-reversal symmetry breaking 
states occur in three-component superconductors, due to the competition between the bilinear 
Josephson couplings. 
Such a three-component Ginzburg-Landau theory is relevant to describe a three-band 
superconducting state, with an intra-band dominated pairing (see \eg 
\CVcite{Garaud.Silaev.ea:17}). However, the $\sis$ state can also be realized in 
two-component Ginzburg--Landau models. For example, microscopic three-band model, 
with an interband dominated repulsive pairing were suggested to be relevant for 
some iron-based superconductors \cite{Maiti.Korshunov.ea:12,Maiti.Chubukov:13,
Marciani.Fanfarillo.ea:13}. In this case, only two fields can nucleate and the 
relevant model is a two-component Ginzburg--Landau theory, see \eg 
\CVcite{Garaud.Silaev.ea:17}. Similar model can also be realized in dirty two-band 
superconductor, due to impurity scattering \cite{Gurevich:03,Gurevich:07,
Stanev.Koshelev:14}, see also \CVcite{Silaev.Garaud.ea:17}.

The most generic potential free energy \Eqref{Eq:General:FreeEnergy} of a two-component 
Ginzburg--Landau model reads as:
\SubAlign{Eq:TRSB:2CGL:FreeEnergy:Potential}{
V(\Psi,\Psi^\dagger) &= \sum_{a=1}^2
\left(\alpha_{aa}|\psi_a|^2+\frac{\beta_{aa}}{2}|\psi_a|^4 \right)
\label{Eq:TRSB:2CGL:FreeEnergy:Potential:a} \\
&+2(\alpha_{12}+\gamma_{11}|\psi_1|^2+\gamma_{22}|\psi_2|^2)|\psi_1||\psi_2|\cos\varphi_{12}
\label{Eq:TRSB:2CGL:FreeEnergy:Potential:b} \\
&+(\beta_{12}+\gamma_{12}\cos2\varphi_{12})|\psi_1|^2|\psi_2|^2
\label{Eq:TRSB:2CGL:FreeEnergy:Potential:c} \,.
}
This potential can describe an $\sis$ state, and the coefficients of the Ginzburg-Landau 
functional can be calculated from a given set of input microscopic parameters, from the 
relevant microscopic model. See for example \CVcite{Garaud.Silaev.ea:17} for the case of 
a interband dominated pairing for clean superconductors, or \CVcite{Silaev.Garaud.ea:17} 
and \CVcite{Garaud.Corticelli.ea:18a} for dirty two-band superconductors.
Note that in the $\sis$ case, the superconducting condensates are also coupled via mixed 
gradients terms $(\D\psi_1^*\D\psi_2+c.c.)$, which do not impact the ground state. 
However, such terms play a role in the length scales and the associated normal modes. 
Details of the length scales and normal modes of such an $\sis$ superconducting state
was investigated in \CVcite{Garaud.Corticelli.ea:18a}.

Here again, the potential \Eqref{Eq:TRSB:2CGL:FreeEnergy:Potential} cannot always be 
minimized analytically. Yet, qualitative information can be understood by considering 
the different phase locking term. Indeed, the first phase-locking term 
\Eqref{Eq:TRSB:2CGL:FreeEnergy:Potential:b} promotes the relative phase $\varphi_{12}$ 
to be either $0$ or $\pi$, depending the sign of the density-dependent effective coupling 
$(\alpha_{12}+\gamma_{11}|\psi_1|^2+\gamma_{22}|\psi_2|^2)$. The second phase-locking 
term, coupled via $\gamma_{12}$ in \Eqref{Eq:TRSB:2CGL:FreeEnergy:Potential:b} favors 
either $\varphi_{12}=\pm\pi/2$ when $\gamma_{12}>0$, or $\varphi_{12}=k\pi$ (with 
$k\in\Relative$) when $\gamma_{12}<0$.
As for the three-components discussed above, the different phase-locking terms can 
compete with each other, and thus may lead to frustration and to a discrete degeneracy 
of the ground state.

The relative phase is determined by the equation $\delta V/\delta\varphi_{12}=0$:
\Equation*{  
 \big(\alpha_{12}+\gamma_{11}|\psi_1|^2+\gamma_{22}|\psi_2|^2\big)
 |\psi_1||\psi_2|\sin\varphi_{12} 
 +\gamma_{12}|\psi_1|^2|\psi_2|^2\sin2\varphi_{12}=0  \,.
}
This has different solutions in the different  states:
\Equation*{ 
 s_{\pm}:~\varphi_{12}=\pi\,,~~~~ s_{++}:~\varphi_{12}=0
 \,, ~~~~
 s+is:~\varphi_{12}=
 \pm\arccos\left(-\frac{\alpha_{12}+\gamma_{11}|\psi_1|^2+\gamma_{22}|\psi_2|^2}
 {2\gamma_{12} |\psi_1||\psi_2|}\right) \,.
}
The ground state values of the densities are determined by the other 
equations $\frac{\delta V}{\delta|\psi_a|}=0$. The whole nonlinear system 
has to be solve numerically.

\begin{figure}[!htb]
\hbox to \linewidth{ \hss
\includegraphics[width=.65\linewidth]{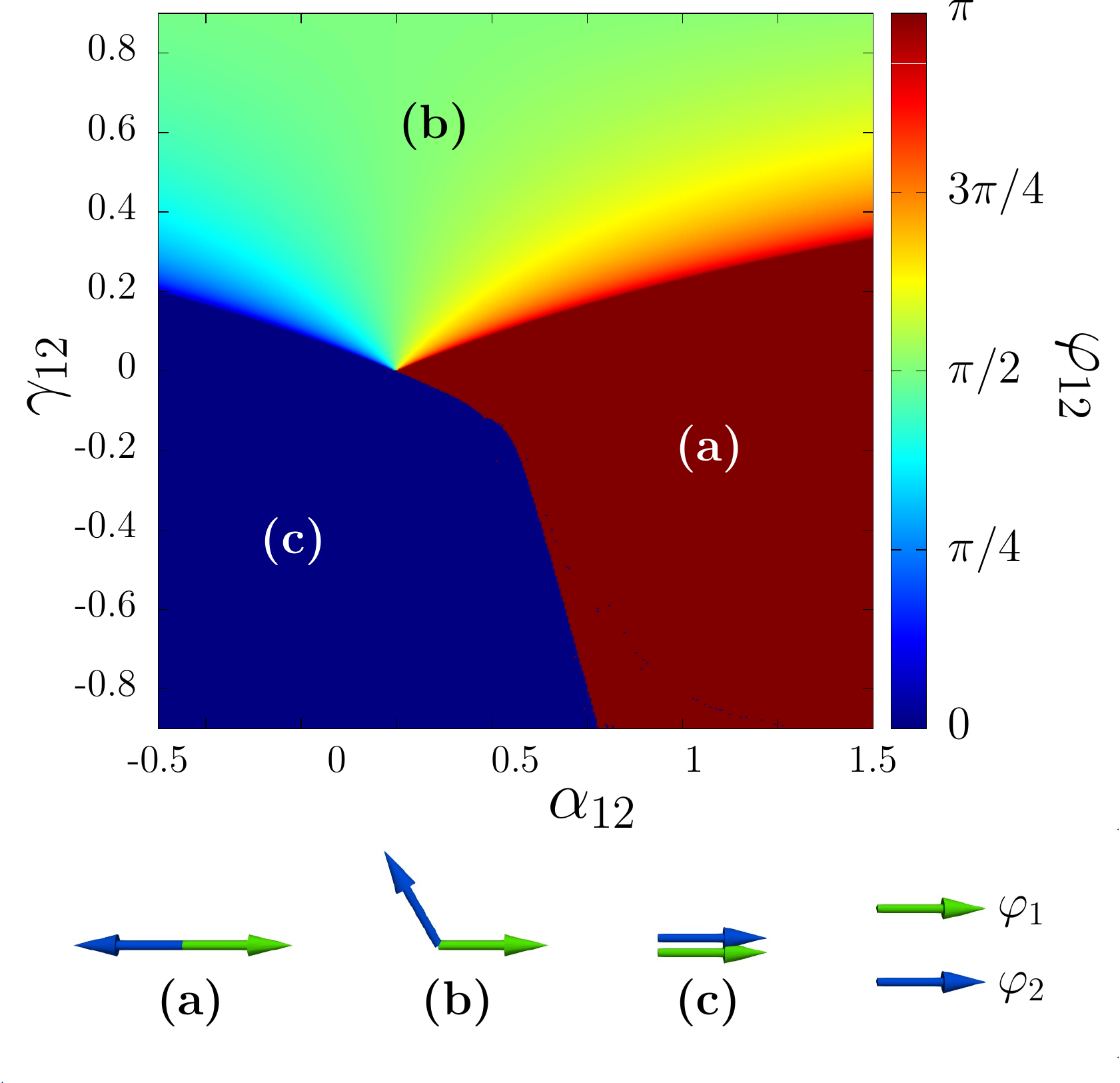}
\hss}
\caption{
%
Ground state phases of the two-components superconductor, as function of 
linear ($\alpha_{12}$) and bilinear ($\gamma_{12}$) Josephson couplings.  
The other parameters of the Ginzburg-Landau potential energy are $\alpha_{aa}=-1$, 
$\beta_{aa}=1$, and $\beta_{12}=\gamma_{11}=\gamma_{22}=0$.
The top panel shows the ground state relative phase $\varphi_{12}$.
Both phase-locking cannot always be simultaneously satisfied, so the system can be 
frustrated. There are two phases for which the relative phase is trivial. 
In the regime (a) $\varphi_{12}=\pi$, while in the regime (c) $\varphi_{12}=0$
In the intermediate regime (b), the ground state relative phase is non-trivial, and the 
ground state features the discrete symmetry $\groupU{1}\times\groupZ{2}$ rather than 
$\groupU{1}$. 
Similar phase diagram is found in \cite{Garaud.Corticelli.ea:18,Garaud.Corticelli.ea:18a} 
where the parameters of the Ginzburg-Landau theory are expressed in terms of the parameters 
of the underlying microscopic theory.
}
\label{Fig:Diagram:Phase-locking-2CGL}
\end{figure}

For example, let's consider the situation where the weighted Josephson couplings 
$\gamma_{aa}=0$, and where the biquadratic density term vanishes ($\beta_{12}=0$). 
The dependence of the ground state, with respect to the bilinear ($\alpha_{12}$) and 
biquadratic ($\gamma_{12}$) Josephson coupling is displayed in 
\Figref{Fig:Diagram:Phase-locking-2CGL}. As for the three-component model, the diagram 
shows three different phases: Two are time-reversal symmetric, and the time-reversal 
symmetry is broken in the third phase.
The time-reversal symmetric states are the $s_\pm$ state (the red region with 
$\varphi_{12}=\pi$) and the $s_{++}$ state (the blue regions with $\varphi_{12}=0$).

The principal message here is that the time-reversal symmetry breaking $\sis$ state,  
that occurs in three-component models due to phase frustration, also have two-component 
counterpart. Here again, this is the frustrated competition between different phase-locking 
terms, that can result in ground states with non-trivial ground state relative phase. 
The principal difference is that in the case of three components, the competition between 
bilinear Josephson couplings is sufficient to result in frustration. On the other hand, 
the existence of higher order Josephson terms is necessary to obtain the frustration.

The transition between both time-reversal symmetric states can occur either via a 
direct crossover (directly from (a) to (c)), or via the intermediate complex $\sis$ 
state, that breaks the time-reversal symmetry (the region (b) of 
\Figref{Fig:Diagram:Phase-locking-2CGL}, where $\varphi_{12}\neq0,\pi$).
Discussions of the phase diagram of the two-component $\sis$ state with the relevant 
length scales, in terms of the parameters of the microscopic theory can be found in 
\CVcite{Silaev.Garaud.ea:17}, \CVcite{Garaud.Corticelli.ea:18a}.

\paragraph{Crossover region.} 
The direct crossover from the $\spm$ state (a) to the $\spp$ state (c) do not break 
the time-reversal symmetry. Moreover, since it is not associated with a symmetry change, 
there are no divergent length-scale here. Yet vortices feature interesting new properties 
in the vicinity of that crossover, and there is a transition in the structure of vortex 
cores \CVcite{Garaud.Silaev.ea:17a}. More precisely, in addition to the common singularity 
of both condensates, the vortices can acquire a circular nodal line around the singular 
point. This nodal line in one of the superconducting condensates results in a peculiar 
``moat"-like profile of the associated condensate. In other words, these new solutions 
realize the $\spm$ state ($\varphi_{12}=\pi$) near the vortex core, while the phase 
locking in the bulk is the $\spp$ state ($\varphi_{12}=0$). See the detailed discussion 
\CVcite{Garaud.Silaev.ea:17a}. As further discussed in \CVcite{Garaud.Corticelli.ea:18}, 
this implies that in an external field, there can be global transitions of the overall 
relative phase.

\paragraph{Other superconducting states that break the time-reversal symmetry.} 
As discussed earlier, the $\sis$ state is the simplest time-reversal symmetry breaking 
extension of the most abundant $s$-wave state. There exist different superconducting 
states, that also break the time-reversal symmetry, as for example the $\sid$, $\did$, 
or $\pip$ states. All these models are described by potential similar to 
\Eqref{Eq:TRSB:2CGL:FreeEnergy:Potential}, and also feature domain wall excitation.
However, as they may break different point group symmetries, they have essentially 
different kinetic terms. For example, see the different kinetic term in the two-component 
$\sis$ state features mixed gradients $(\D\psi_1^*\D\psi_2+c.c.)$ while in the $\sid$ 
state, these are $(D_x\psi_1^*D_x\psi_2-D_y\psi_1^*D_y\psi_2+c.c.)$. As discussed later 
on, this can lead to different responses between these states \CVcite{Garaud.Silaev.ea:16}.
Similarly, the chiral $\pip$ state also have a different structure of the kinetic term.

\section{Topological defects in the time-reversal symmetry breaking states}
\label{Sec:TRSB:Topological-defects}

The topological properties of the superconducting states that break the time-reversal 
symmetry where already partially addressed in Chapter \ref{Chap:Topological-defects}.  
Indeed, in Section \ref{Sec:Chiral-Skyrmions} features a heuristic description of 
domain-walls, and how they can pin vortices. Below, we continue with a more quantitative 
description of the properties of the additional topological defects in time-reversal 
symmetry breaking states. In particular, in relation with the underlying three-component 
models.

\subsection{Domain-walls}
\label{Sec:TRSB:Domain-walls}

As explained in Section \ref{Sec:Chiral-Skyrmions}, domain-walls are the topological 
defects that are naturally associated with the spontaneous breakdown of a discrete 
$\groupZ{2}$ symmetry. Hence it is natural to expect that domain-walls should form in 
superconducting states that break the time-reversal symmetry. Following the analysis 
in \CVcite{Garaud.Babaev:14} these domain walls can be formed by thermal quench, 
and be geometrically stabilized against collapse.

\begin{figure}[!htb]
\hbox to \linewidth{ \hss
\includegraphics[width=.75\linewidth]{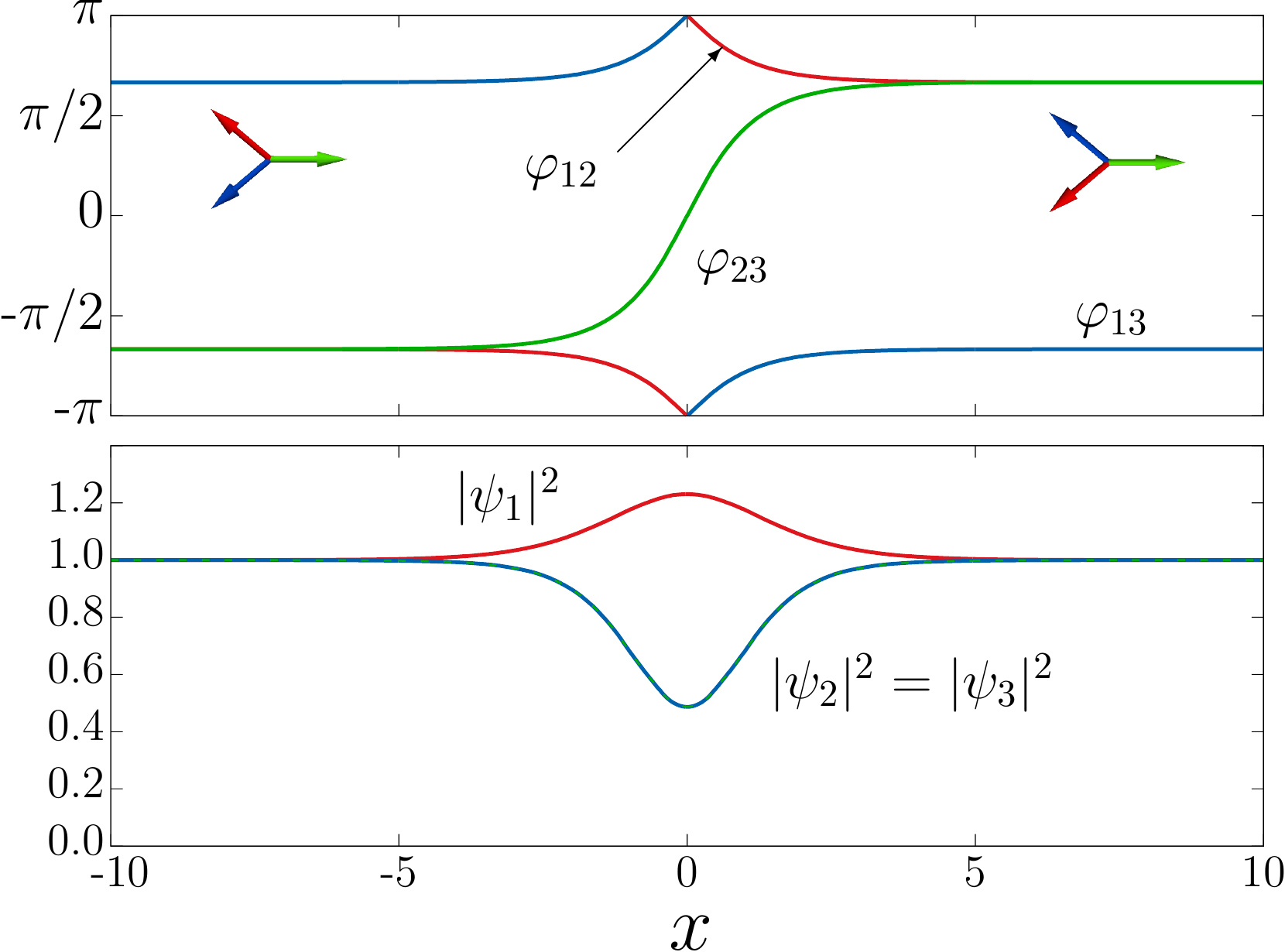}
\hss}
\caption{
A domain-wall solution of a three-component superconductor that breaks the 
time-reversal symmetry. The parameters of the Ginzburg-Landau free energy 
\Eqref{Eq:TRSB:3CGL:FreeEnergy:1} are $\alpha_a=0$, $\beta_a=1$, and 
$\eta_{12}=\eta_{13}=\eta_{23}=1$ and $e=0.1$. The top panel displays the relative 
phases, and the bottom panel shows the densities. 
The domains-wall interpolates between two inequivalent phase-locking. At $x=0$, 
the relative phases are $\varphi_{12}=\varphi_{13}=\pi$ and $\varphi_{23}=0$, thus 
the Josephson couplings term $\cos\varphi_{ab}$ are the most unfavourable there.
}
\label{Fig:Domain-wall:3CGL}
\end{figure}

The figure \ref{Fig:Domain-wall:3CGL} shows a domain-wall solution of a three-component 
Ginzburg-Landau model \Eqref{Eq:TRSB:3CGL:FreeEnergy:1}, when the time-reversal symmetry 
is broken. Here, the domain-wall interpolates between the two inequivalent ground states, 
with different chiralities. These solutions are found numerically by minimizing the free 
energy \Eqref{Eq:TRSB:3CGL:FreeEnergy:1} with an appropriate initial guess. More precisely, 
the domain-walls interpolating between the distinct ground states fall into disjoint 
homotopy classes, and thus no finite energy transformation can transform it to a 
constant ground state (see \eg the textbooks \cite{Manton.Sutcliffe,Vachaspati,
Vilenkin.Shellard,Rajaraman,Shnir:18}). It follows that such an initial state is very 
robust to the minimization of the energy. As discussed in more details in the Section 
\ref{App:numerics:IG} of Appendix \ref{App:numerics}, the knowledge of the topological 
properties is also very useful for the numerical construction of topological defects.

It is clear from the top panel  of \Figref{Fig:Domain-wall:3CGL} that the domain 
interpolates between the two ground states that are complex conjugate of each other.
At the domain-wall, here at $x=0$, the relative phase are such that the Josephson 
couplings terms $|\psi_a||\psi_b|\cos\varphi_{ab}$ there are energetically the most 
unfavourable. As a result, the total density there is reduced, in order to reduce the 
energy cost.

\paragraph{Magnetic signature of the domain-walls:} 
Unlike the domain-walls that appear for example in chiral $p$-wave superconductors, 
the $\sis$ domain-walls are not associated, \emph{in principle}, with a non-zero magnetic 
field. Indeed, the domain-walls in chiral $p$-wave superconductors, carry a uniform 
magnetic field due to the orbital momentum of the Cooper pairs (see e.g. 
\cite{Ferguson.Goldbart:11,Raghu.Kapitulnik.ea:10,Vadimov.Silaev:13}, see also the 
discussion in \cite{Garaud.Babaev.ea:16} and detailed analysis in \cite{Bouhon.Sigrist:10,
Bouhon.Sigrist:14}). 
The domain walls between the two $\sis$ states, on the other hand, are kinks in the 
relative phases. Inspection of the separation in charged and neutral modes 
\Eqref{Eq:Quantization:FreeEnergy:3} makes it natural to expect that the associated 
gradients in the relative phases do not couple to the charged modes. It is however possible, 
as demonstrated in \CVcite{Garaud.Babaev:14}, that domain-walls between $\sis$ states 
carry a magnetic field only locally, with no net flux through the sample. 
Indeed, in multicomponent superconductors the relation between the magnetic field and 
the total current is more complicated than the usual London's magnetostatics relation. 
More precisely, the magnetic field can be expressed as \CVcite{Garaud.Carlstrom.ea:13}, 
\CVcite{Garaud.Babaev:14}:
\Equation{Eq:TRSB:MagneticField}{
B_k =\varepsilon_{kij}\left\lbrace 
\nabla_i\left(\frac{J_j}{e^2\varrho^2}\right) 
+\frac{i}{e\varrho^4}\left(\varrho^2\nabla_i\Psi^\dagger\nabla_j\Psi
+(\Psi^\dagger\nabla_i\Psi)(\nabla_j\Psi^\dagger\Psi) \right) \right\rbrace \,.
}
Here $\J$ is the total Meissner current, $\rho:=(\Psi^\dagger\Psi)^{1/2}$ is the total 
density, and $\Psi^\dagger=(\psi_1^*,\psi_2^*,\psi_3^*)$ is scalar multiplet which 
carries all the superconducting degrees of freedom. For details of the derivation, 
see the related discussion in Section~\ref{Sec:CPN-charge}, and for generalization to 
anisotropic models see \cite{Silaev.Winyard.ea:18}.
The first term in \Eqref{Eq:TRSB:MagneticField} is the standard contribution to the 
magnetic field from the superconducting currents, while the second term is the additional 
contribution due to the inter-component interactions. Importantly, this second term 
depends only on the relative phases and relative densities of the condensates. 
This can be seen more explicitly in the discussion about thermoelectric effects in
Sec.~\ref{Sec:TRSB:Thermoelectric}, for example in equation \Eqref{Eq:TRSB:MagneticField:2}.
As emphasized in \CVcite{Garaud.Babaev:14}, there can be situations where these 
additional contributions are only \emph{partially} screened by the standard London 
contribution, thus resulting in a non-zero signature of the magnetic field.

Note that the relative phase gradients alone do not induce magnetic fields since they 
do not lead to charge transfer in real space. However, a magnetic field do appear if, 
in addition there are relative density gradients that are not collinear to the relative 
phase gradients. Such a local magnetic field is expected to be stronger on domain-walls, 
since the relative phase gradients are stronger there.

In \CVcite{Garaud.Babaev:14}, it was demonstrated that such uncompensated contribution 
can occur if a domain-wall is attached to a boundary with an important curvature, such 
as "bumps" in a non-convex geometry, or pinning centres. The magnetic signatures of 
domain-walls, and the possibility to discriminate $\sis$ domain-walls with $\sid$ 
domain walls, was discussed in \cite{Benfenati.Barkman.ea:20}. As further 
discussed in Sec.~\ref{Sec:TRSB:Thermoelectric}, the possibility to observe this additional 
contribution via thermoelectric properties was also discussed in \CVcite{Silaev.Garaud.ea:15}
and \CVcite{Garaud.Silaev.ea:16} .

\paragraph{Formation of the domain-walls.}
The spontaneous breakdown of the time-reversal symmetry dictates that the $\sis$ state  
possess domain wall excitations. It is well known that going through a phase transition
allows uncorrelated regions to fall into different ground states \cite{Kibble:76,Zurek:85}. 
This is the Kibble-Zurek mechanism for the formation of topological defects. 
For a review of the Kibble-Zurek mechanism in conventional superconductors, see 
\cite{Rivers:01}. Here, while a superconductor goes through the transition to the 
time-reversal symmetry broken state, domain walls are created as different regions fall 
into either of the $\groupZ{2}$ states \cite{Garaud.Babaev:14}. Similarly, domain-walls 
can form in the context of chiral $p$-wave superconductors \cite{Vadimov.Silaev:13}. 
Moreover, the Kibble-Zurek mechanism relates the number of produced topological defects, 
with "speed" of the transition. Heuristically, a rapid transition produces more topological 
defects. 

However, in finite systems, domain-walls can be dynamically eliminated by continuously 
be moved out of the domain. Nevertheless, as demonstrated in \CVcite{Garaud.Babaev:14}, 
they can be stabilized by pinning centres. They can also be stabilized in non-convex 
geometries, as discussed earlier in the Section~\ref{Sec:Chiral-Skyrmions}. The geometric 
stabilization, illustrated in \Figref{Fig:DW:2}, may help for the observability of 
domain-walls formed during a phase transition to the time-reversal symmetry broken 
states.

Furthermore, as discussed shortly, the domain-walls interact non-trivially with vortices. 
In a nutshell, to accommodate the unfavourable relative phase at the domain-wall, they 
tend to confine the vorticity. Moreover the points where the domain walls are attached 
to the boundaries, are easy entry points for vortices to enter the system. 
It follows that the magnetization process, when the zero field configuration features 
a stabilized domain-wall, is different than when domain-walls are initially absent
\CVcite{Garaud.Babaev:14}. This can be observed via the fact the first (fractional) 
vortex entry occurs at much lower fields than the bulk $\Hc{1}$. It can also be seen that 
the vortex matter distributes differently. Hence repeating measurements of the magnetization 
process after rapid cooling (or other kind of quench) could easily identify the presence 
of domain-walls, and consequently signal that the time-reversal symmetry is broken 
\CVcite{Garaud.Babaev:14}.

\subsection{Chiral \texorpdfstring{$\groupCP{2}$}{CP2} skyrmions}
\label{Sec:TRSB:Chiral-Skyrmions}

As mentioned above, the domain-walls interact non-trivially with vortices. 
It follows that, as demonstrated in \CVcite{Garaud.Carlstrom.ea:11} and 
\CVcite{Garaud.Carlstrom.ea:13}, domain-walls can combine with vortices to form
new kind of topological defects called \emph{chiral $\groupCP{2}$ skyrmions}. 
The existence of such states, associated with new non-trivial topological properties, 
was already partially discussed in section~\ref{Sec:Chiral-Skyrmions}. The underlying 
mechanisms of the interaction between domain-walls and vortices are discussed here 
in more details.

As previously explained, and as can be seen from the \Figref{Fig:Domain-wall:3CGL}, 
the relative phases at the domain-wall provide the energetically most unfavourable  
Josephson couplings terms $|\psi_a||\psi_b|\cos\varphi_{ab}$. Moreover the domain-wall 
feature an additional energy cost, associated with a gradient in the relative phase 
$\varphi_{ab}$ in \Eqref{Eq:Quantization:FreeEnergy:3}. It follows that the total density 
is reduced to accommodate the extra energy cost.
As a results, if a vortex is close to a domain-wall, the depletion of the densities 
acts attractively to bind the vortex to the domain-wall.
Moreover, if an integer composite vortex is located on a domain wall, the Josephson 
terms tend to split it into fractional vortices, thus allowing more favourable relative 
phase in between the split fractional vortices. In the absence of domain-walls, fractional 
vortices are linearly confined by the Josephson interaction terms, as discussed in the 
section \ref{Sec:Fractional-vortices:Interaction} or in \CVcite{Garaud.Carlstrom.ea:13}. 
On the other hand, domain-walls tend to confine vortices and to split them into fractional 
vortices that repel each other. This was discussed in details in \CVcite{Garaud.Babaev:14}.

Because of its line tension, as sketched earlier in \Figref{Fig:DW:1}, a closed 
domain-wall collapses to zero size. In contrast, fractional vortices confined on a 
domain-wall repel each other. This opens the possibility for a composite solution 
consisting in a closed domain-wall `decorated' with vortices to be (meta)stable.
Indeed, for large enough penetration depth, the repulsion between the fractional
vortices confined on the domain wall can become strong enough to overcome the 
domain-wall's tension. It thus results in a composite topological defect made of 
$n$ fractional vortices in each condensate $|\psi_a|$, distributed along a closed 
domain-wall. Such a configuration, that carries $n$ flux quanta, is stabilized by the 
competing forces, see \CVcite{Garaud.Carlstrom.ea:11}and \CVcite{Garaud.Carlstrom.ea:13}.

\begin{figure}[!htb]
\hbox to \linewidth{ \hss
\includegraphics[width=.95\linewidth]{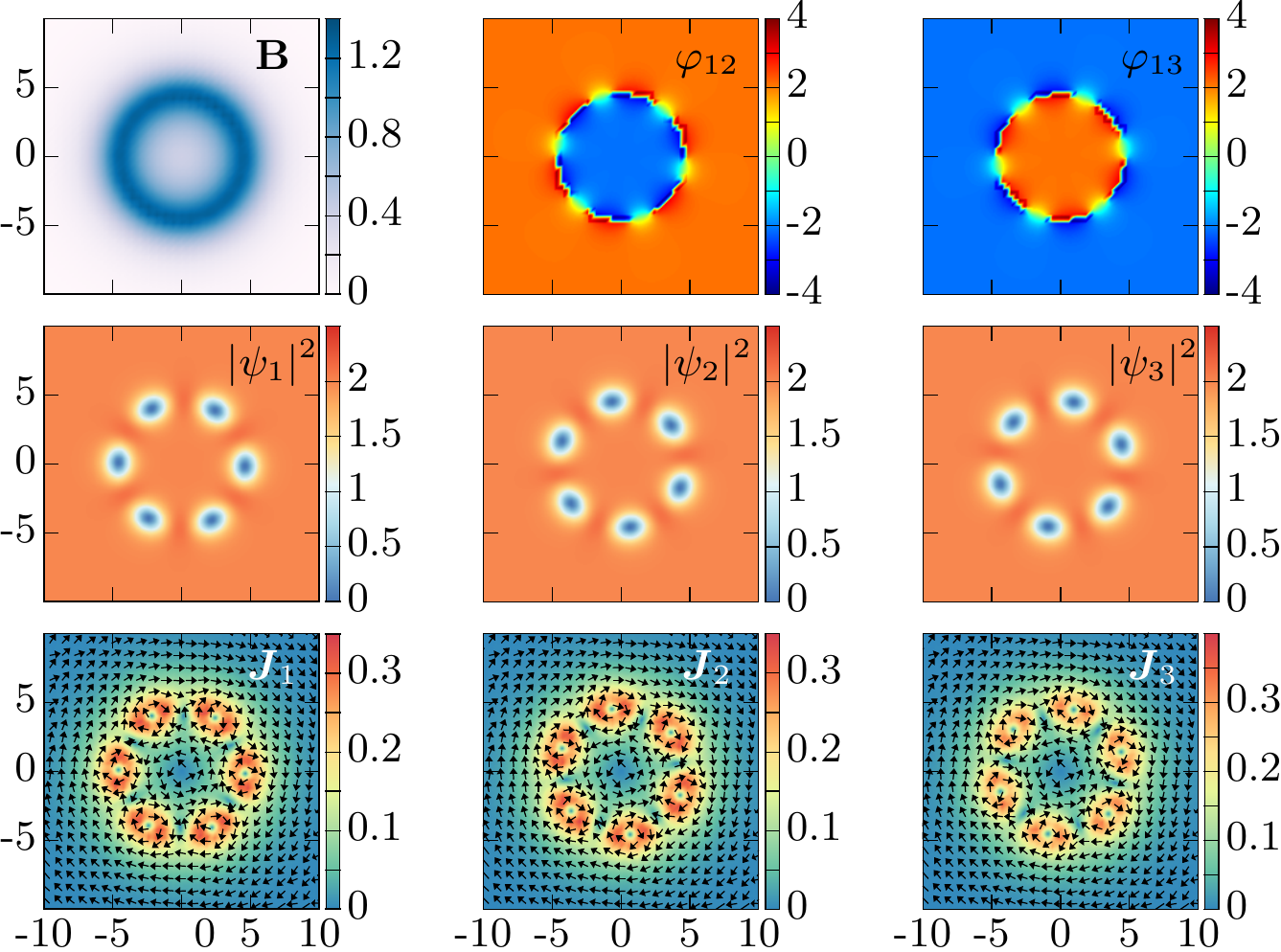}
\hss}
\caption{
A Skyrmion solution of a three-component superconductor that breaks the time-reversal 
symmetry. The parameters of the Ginzburg-Landau energy \Eqref{Eq:TRSB:3CGL:FreeEnergy:1}, 
are $\alpha_{aa}=-1$, $\beta{aa}=1$ $\alpha_{ab}=1$ (with $a\neq b$), and $e=0.25$. 
The solution here carries six flux quanta and thus consists in 18 fractional vortices 
(six for each component). 
The displayed quantities on the top row are the magnetic field and two relative phases 
$\varphi_{12}$ and $\varphi_{13}$. The middle row shows the densities of the three 
components $|\psi_a|^2$, while the bottom row displays the associated super-currents. 
The flux quantization implies that these chiral skyrmions carry a topological charge 
\Eqref{Eq:CPN:Charge} $\Q(\Psi)=6$.
}
\label{Fig:Domain-wall:Skyrmion}
\end{figure}

The figure \ref{Fig:Domain-wall:Skyrmion} shows the details of a composite solution
consisting in $6$ fractional vortices in each of the condensates $\psi_a$, distributed 
along a closed domain-wall. As explained earlier, this numerical solution is obtained 
by minimizing the energy, from an initial configuration that winds $6$ times in each 
of the condensates (see details in Appendix \ref{App:numerics} and Section 
\ref{App:numerics:IG}). 
As can be seen from the three panels showing the densities $|\psi_a|^2$, the cores of 
the different fractional vortices do not overlap, and thus $\Psi(\x)\neq0$ everywhere. 
It follows, as explained in details in Section \ref{Sec:CPN-charge}, that these 
solutions are associated with a non-vanishing $\groupCP{2}$ topological invariant 
$\Q$ \Eqref{Eq:CPN:Charge}. The flux quantization implies that $\Q=6$.
Hence the solution displayed in \Figref{Fig:Domain-wall:Skyrmion}, is called a chiral 
$\groupCP{2}$ skyrmions 
\footnote{
The adjective \emph{chiral} follows from that, far from the vortex cores, one of the 
ground state phase locking is realized (here $\varphi_{12}=-\varphi_{13}=2\pi/3$), 
while the other ground state relative phase is realized inside the topological defect.
It follows that the solution transforms non-trivially under the time-reversal operations. 
For detailed discussion, see \cite{Garaud.Carlstrom.ea:13}.
}
of the three-component model \Eqref{Eq:TRSB:3CGL:FreeEnergy:1}.

Depending on the details of the model, a chiral $\groupCP{2}$ skyrmions carrying $\Q$ 
flux quanta, might be energetically favoured compared to a set of $\Q$ single quanta 
vortices \cite{Garaud.Carlstrom.ea:13}. In such a case, the skyrmions are expected to 
spontaneously form in an external field. By contrast, if the skymions are more energetic 
than vortices they exist as robust metastable solution. Indeed unpinning the vortices 
from the domain-walls is energetically costly and they are quite stable to perturbations 
\cite{Garaud.Carlstrom.ea:11,Garaud.Carlstrom.ea:13}. Metastable skyrmions can form in 
field-cooled experiments via the Kibble-Zurek mechanism, when the transition to 
time-reversal symmetry broken states $\Tc{,\groupZ{2}}$ occurs below the superconducting 
critical temperature $\Tc{}$ \CVcite{Garaud.Babaev:14}. Moreover, in a finite sample a 
skyrmion is surrounded by regular vortices, the later press the skyrmion, thus having a 
stabilizing effect against its decay.

Because of the splitting of the fractional vortices, the magnetic flux is distributed 
along the domain wall. This can be seen in the top left panel of the figure 
\ref{Fig:Domain-wall:Skyrmion}. The choice of the parameters of the Ginzburg-Landau theory 
here are very symmetric, and thus flux is evenly distributed along the domain wall. 
When the parameters are different, then the chiral skyrmions can feature very exotic
signatures of the magnetic field. This is discuss in details in 
\CVcite{Garaud.Carlstrom.ea:11} and \CVcite{Garaud.Carlstrom.ea:13}. Since they have 
very distinct signatures of the magnetic field, they can be observed by scanning SQUID 
or Hall or magnetic force microscopy experiments.

\paragraph{Skyrmions can also exist in other time-reversal symmetry breaking states.} 
As mentioned earlier, there exist superconducting states with different symmetries, 
that also break the time-reversal symmetry. As they also allow for domain-wall excitations, 
it is now rather natural to expect that they can allow for chiral skyrmions as well.
As already discussed at the end of the Sec.~\ref{Sec:Bi-quadratic}, $\pip$ superconductors 
indeed support stable skyrmionic excitations \CVcite{Garaud.Babaev:12}. Some of these 
can be interpreted as vortices carrying two flux quanta \CVcite{Garaud.Babaev:15a}, 
and form lattices see \CVcite{Garaud.Babaev.ea:16} and \CVcite{Krohg.Babaev.ea:21}.
Skyrmions were also shown to have an important role in the magnetization process of 
mesoscopic samples of $\pip$ superconductors \cite{Zhang.Becerra.ea:16,
FernandezBecerra.Sardella.ea:16,Becerra.Milosevic:17}. For similar a discussion of 
skyrmions in the $\sid$ state, see \cite{Zhang.Zhang.ea:20}.

\section{Thermoelectric properties of superconductors that break the time-reversal symmetry}
\label{Sec:TRSB:Thermoelectric}

\begin{wrapfigure}{R}{0.5\textwidth}
\hbox to \linewidth{ \hss
\includegraphics[width=.975\linewidth]{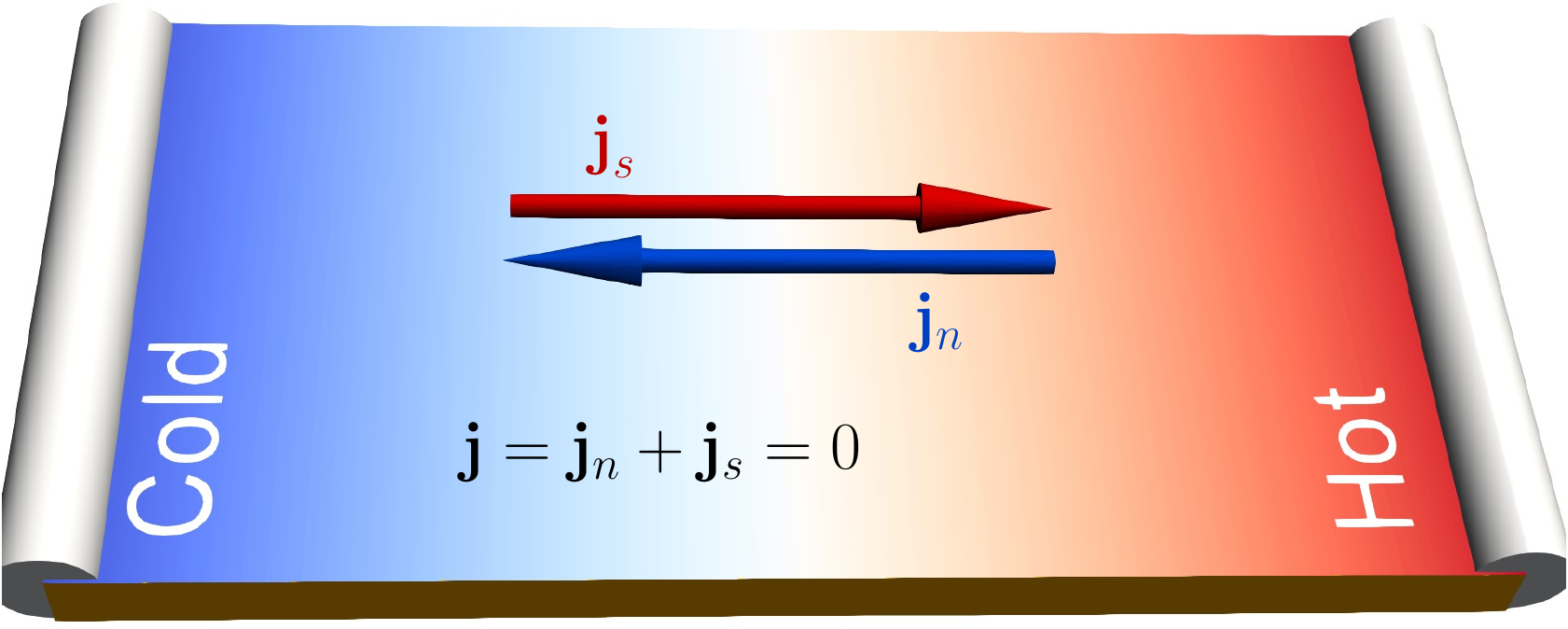}
\hss}
\caption{
Schematic illustration of the Ginzburg mechanism for the thermoelectric effect: 
A temperature gradient results in a charge transfer by thermal quasiparticles, that 
is compensated by the counterflow of the superconducting current.
}
\label{Fig:Schematic:Ginzburg-mechanism}
\end{wrapfigure}

As discussed earlier, the magnetostatic properties of multi-component superconductors 
feature an additional contribution due to the intercomponent interactions. These extra 
contributions can be excited by applying thermal gradients to a multi-component 
superconductor, as demonstrated in \CVcite{Silaev.Garaud.ea:15} and
\CVcite{Garaud.Silaev.ea:16}.

The thermoelectric effects in superconductors, were discussed by Ginzburg in the 
mid 1940's \cite{Ginzburg:44,Ginzburg.Zharkov:78,Ginzburg:04,Ginzburg:98}.
These originate in the charge transfer by thermal quasiparticles 
\cite{VanHarlingen.Heidel.ea:80}, which is compensated by the counterflow of the  
superconducting current. Namely, a temperature gradient applied to a superconductor 
induces an electric current $\J_n = b_n \Grad T$, where $b_n$ is the thermoelectric 
coefficient. This current is carried by quasiparticles that exist at finite temperatures.
In superconductors, the total current also features the contribution from the 
superconducting electrons $\J_s$. It follows that, in contrast to a normal metal, 
the total current a superconductor vanishes in order to obey the Meissner effect, 
the thermoelectric current is cancelled by the superconducting current: $\J=\J_s+\J_n=0$. 
As a result, the counterflow of the superconducting current is $\J_s = -b_n \Grad T$. 
This is sketched in \Figref{Fig:Schematic:Ginzburg-mechanism}.
In the recent years, there was a revival of the interest about this thermoelectric 
effect, see \eg \cite{Loefwander.Fogelstroem:04,Chandrasekhar:09,Ozaeta.Virtanen.ea:14,
Machon.Eschrig.ea:13,Giazotto.Heikkila.ea:15,Shelly.Matrozova.ea:16}.

Remark that as it is determined by the dissipative normal current, a thermally 
induced supercurrent is irreversible, since $\J_s$ changes its sign under the time-reversal 
transformation while $b_n$ and $\Grad T$ remain invariant.
As discussed below, since multicomponent superconductors feature additional contributions 
due to the inter-component interaction, the thermoelectric effect can be substantially 
altered. In particular, in superconductors that break the time-reversal symmetry
\CVcite{Silaev.Garaud.ea:15}.

\subsection{Current relations in multicomponent superconductors}

It was emphasized, in Sec.~\ref{Sec:CPN-charge} and in Sec.~\ref{Sec:TRSB:Domain-walls}, 
that the magnetic field gets an additional contribution to the Meissner currents 
(see \eg Eq.\Eqref{Eq:TRSB:MagneticField}). This extra contribution, which is due to 
the inter-component interactions, depends on the gradients of relative phases and 
relative densities. 
Similarly, the superconducting current features an extra contribution due to relative 
phase gradients. This can be seen by rewriting the current in terms of the total phase 
and of the relative phases. For an arbitrary number of components $N$, the individual 
phases can be written as 
\Equation{Eq:Currents:Phases}{
\varphi_a = \varphi_\Sigma + \frac{1}{N}\sum_{b\neq a}\varphi_{ab}	
\,,~~\text{where}~~~\varphi_\Sigma=\frac{1}{N}\sum_{a=1}^N\varphi_{a} 
\,,~~\text{and}~~~\varphi_{ab}:=\varphi_{b}-\varphi_{a}\,.	
}
The total current $\J=e\sum_a|\psi_a|^2(\Grad\varphi_a+e\A)$ thus reads as
\SubAlign{Eq:Currents:Decomposed:1}{
\J/e&=\varrho^2\big(\Grad\varphi_\Sigma+e\A\big) 
	-\frac{1}{N}\sum_a\sum_{b\neq a}|\psi_a|^2\Grad\varphi_{ab} \\
	&=\varrho^2\big(\Grad\varphi_\Sigma+e\A\big) 
	+\frac{1}{N}\sum_a\sum_{b> a}\big(|\psi_b|^2-|\psi_a|^2\big)\Grad\varphi_{ab} \,,
} 
where again $\varrho^2=\sum_a|\psi_a|^2$. Finally, introducing the notation 
${\bf Q}_\Sigma:=\Grad\varphi_\Sigma-e\A$, the current reads as
\Equation{Eq:Currents:Decomposed:2}{
\J=e\varrho^2{\bf Q}_\Sigma
	+\frac{e}{N}\sum_a\sum_{b> a}\big(|\psi_b|^2-|\psi_a|^2\big)\Grad\varphi_{ab} \,.
} 
The first term here is a usual Meissner current while the second part describes the 
charge transfer by the counter-currents of the different superconducting condensates.

\subsection{Thermoelectric relations in multicomponent superconductors}

The key idea behind the multicomponent thermoelectric effect is that, a since the 
relative phases are generically temperature-dependent $\varphi_{ab}=\varphi_{ab}(T)$, 
a temperature bias generates a relative phase gradient between the different components.
Assuming that temperature gradients are small, so that the order parameter is determined 
by the local temperature, the phase relation \Eqref{Eq:Currents:Phases} becomes
\Equation{Eq:Currents:Phases:2}{
\varphi_a = \varphi_\Sigma + \Gamma_a(T)\Grad T
\,,~~\text{where}~~~\Gamma_a(T)=\frac{1}{N}\sum_{b\neq a}\frac{d\varphi_{ab}(T)}{dT}\,,
}
and again $\varphi_\Sigma=\frac{1}{N}\sum_{a=1}^N\varphi_{a}$. The coefficients  
$\Gamma_a (T)$ are called the \emph{thermophase coefficients}. It follows that
the current \Eqref{Eq:Currents:Decomposed:2} can be written as
\Equation{Eq:Currents:Decomposed:3}{
\J=e\varrho^2\big({\bf Q}_\Sigma +\Gamma(T)\Grad T\big)
\,,~~\text{where}~~~\Gamma(T)=\sum_a
\frac{|\psi_a(T)|^2}{N\varrho(T)^2}\Gamma_a(T) \,.
}

\begin{wrapfigure}{R}{0.5\textwidth}
\hbox to \linewidth{ \hss
\includegraphics[width=.975\linewidth]{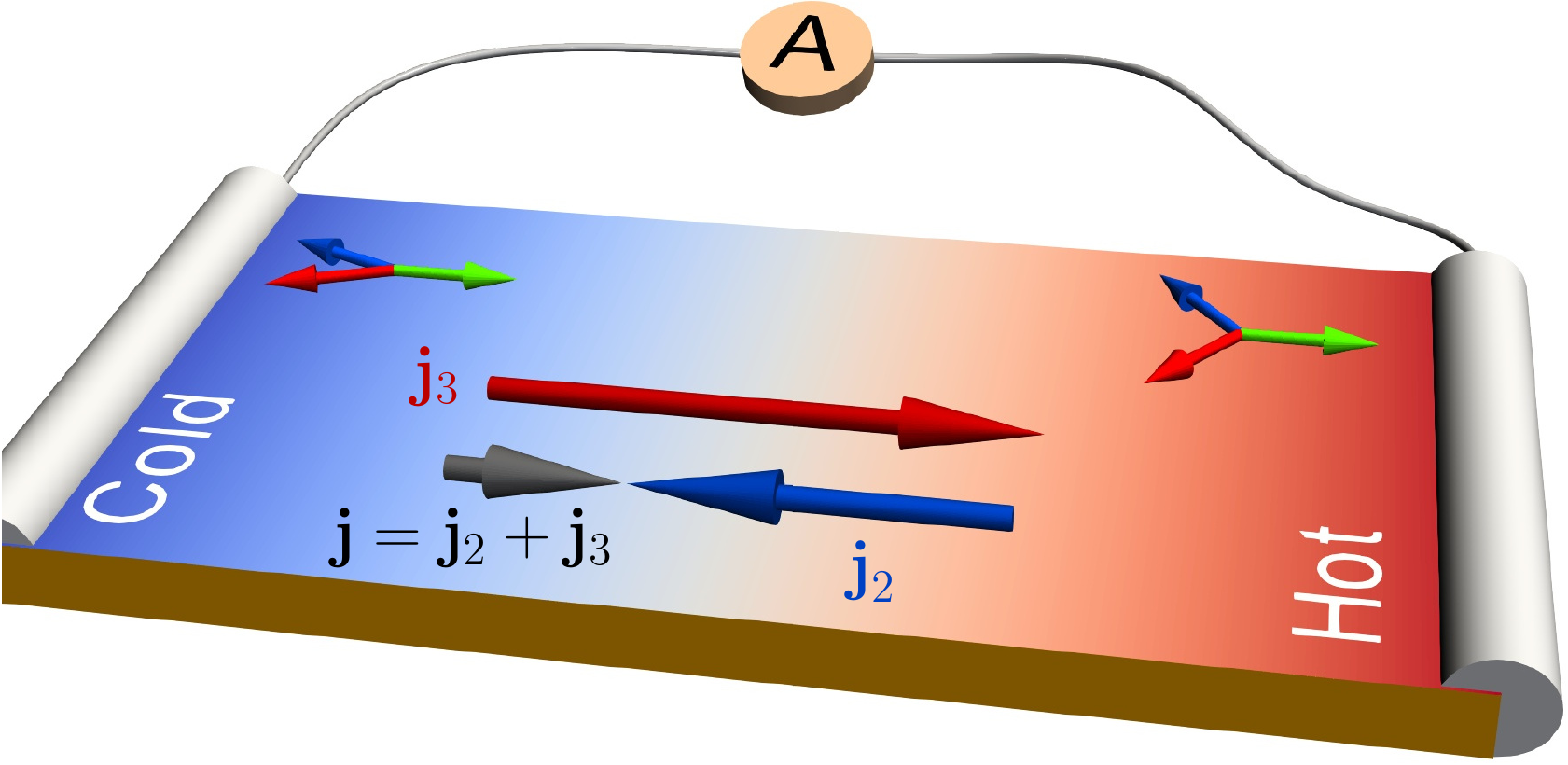}
\hss}
\caption{
A superconducting sample subject to thermal gradients. The thermophase effect 
appears because of the temperature-dependent intercomponent relative phase 
$\varphi_{ab}=\varphi_{ab}(T)$. In the case of a superconducting state that breaks 
the time-reversal symmetry, the total thermally induced current $\J$ will have 
opposite directions for different TRSB states.
}
\label{Fig:Schematic:Thermophase}
\end{wrapfigure}
Note that the thermophase coefficients are odd under the time-reversal transformation: 
${\cal T}(\Gamma_a)=-\Gamma_a$. This implies that, in a superconductor that breaks the 
time-reversal symmetry, the thermophase coefficients are opposite for the different 
$\sis$ states. Thus the thermally induced superconducting currents are sensitive to 
the time-reversal transformation. Hence, given a temperature bias, the currents flow 
in opposite directions for the different time-reversal symmetry broken states 
\CVcite{Silaev.Garaud.ea:15}. This is sketched in \Figref{Fig:Schematic:Thermophase}.

As demonstrated in \CVcite{Silaev.Garaud.ea:15}, the thermophase coefficients can be 
large in the vicinity of the time-reversal symmetry breaking transition $T_{\groupZ{2}}$. 
Moreover, the discussed thermoelectric effect generically dominates at low temperatures 
in the $\sis$ state. More precisely the new contribution is important in the vicinity 
of the time-reversal symmetry breaking phase transition $T_{\groupZ{2}}$, which can occur 
at much lower temperature than $T_c$. There, the usual contribution to the Ginzburg 
mechanism due to the thermal quasiparticles is typically extinct.

\subsection{Magnetic and electric fields induced due to thermal gradients}

Similarly to the total current \Eqref{Eq:Currents:Decomposed:2}, the magnetic field 
$\B$ features additional contributions as already demonstrated in 
\Eqref{Eq:TRSB:MagneticField} (see also the corresponding discussion 
in Sec.~\ref{Sec:CPN-charge}). Using the notations of \Eqref {Eq:Currents:Decomposed:1}
and \Eqref{Eq:Currents:Decomposed:2}, the magnetic field reads as
\Equation{Eq:TRSB:MagneticField:2}{
\B=\Curl\left(\frac{\J}{e^2\varrho^2}\right) -\Curl\Grad\varphi_\Sigma
+\sum_a\sum_{b> a}
\Curl\left(\frac{|\psi_a|^2-|\psi_b|^2}{Ne\varrho^2}\Grad\varphi_{ab}\right) \,.
}
Here again, the first term is the contribution of the Meissner currents and the second 
term is the contribution of the total vorticity. Finally, the most interesting new 
contribution is the last term. 

In the absence of phase windings, then the equation \Eqref{Eq:TRSB:MagneticField:2} 
can further be simplified. Indeed, when none of the phase has a singularity then all 
$\Curl\Grad\varphi\equiv0$, and the magnetic 
field reads as
\Equation{Eq:TRSB:MagneticField:3}{
\B=\Curl\left(\frac{\J}{e^2\varrho^2}\right)  +\sum_a\sum_{b> a}
\Grad\left(\frac{|\psi_a|^2-|\psi_b|^2}{Ne\varrho^2}\right)\times\Grad\varphi_{ab}
~~~~~\text{if}~~\oint\Grad\varphi_a\!\cdot d{\bs\ell}=0~~\forall a
\,.
}
It is clear from there, that when the relative density gradients are collinear with the 
relative phase gradients, then there are no new contributions to the magnetic field. 
The most interesting situation is when they are perpendicular. The additional contribution 
can be screened only partially and thus leads to non-trivial signatures of the magnetic 
field \Eqref{Eq:TRSB:MagneticField:3}. As discussed below, there are various situations 
to take advantage of this, to highlight new properties of the $\sis$ state.

As already repeated on various occasions the time-reversal symmetry breaking $\sis$ 
state feature domain wall excitations. It is known that such domain walls do not 
carry magnetic field at constant temperature \cite{Garaud.Babaev:14}. Clearly, the 
relative phase gradients, are the most important, where the domain is located. 
So intuitively this is a good starting point for searching the extra contribution 
to the magnetic field of \Eqref{Eq:TRSB:MagneticField:3}. 
One way to impose relative density variations is to apply a temperature gradient 
along the domain wall. For example, the temperature dependence of the coefficients 
of the quadratic term of the Ginzburg-Landau theory can be modelled as 
$\alpha_{aa}\propto\left[ T(\x)/\Tc{,a} -1\right]$ ($\Tc{,a}$ being a characteristic 
constant). 
There, the thermophase coefficients $\Gamma(T)$ have opposite signs in the $s+is/s-is$ 
domains. Therefore, in the vicinity of the interface between these, there should be a 
net superconducting current and a thermally induced magnetic field $\B$ 
\Eqref{Eq:TRSB:MagneticField:3}.
Such a local modification of the parameters was demonstrated to be responsible for the 
existence of spontaneous magnetic field in different models time-reversal symmetry 
breaking states, in various situations. These include the responses to linear thermal 
gradients \CVcite{Silaev.Garaud.ea:15} and \CVcite{Grinenko.Weston.ea:21}, hotspot 
created by a laser pulse \CVcite{Garaud.Silaev.ea:16}, or \cite{Vadimov.Melnikov:17}, 
but also the effect of impurities \cite{Maiti.Sigrist.ea:15,Lin.Maiti.ea:16}, 
and other inhomogeneous arrays \CVcite{Garaud.Corticelli.ea:18a}, \cite{Vadimov.Silaev:18}.

Multicomponent superconductors are characterized by additional intercomponent 
contributions, not only to the magnetic field, but also to the electric field 
${\bs E}=-\partial_t{\A}-\Grad\A_0$. This can be seen by similar procedure that 
when rewriting $\B$, or by combining Eq.\Eqref{Eq:TRSB:MagneticField:3} to Faraday's 
law. The electric field thus reads as
\Equation{Eq:TRSB:ElectricField}{
{\bs E}=\partial_t\left(\frac{\J}{e^2\varrho^2}\right)
+\sum_a\sum_{b> a}
\partial_t\left(\frac{|\psi_b|^2-|\psi_a|^2}{Ne\varrho^2}\Grad\varphi_{ab}\right)
-\Grad\Phi
~~~~\text{where}~~\Phi=A_0-\partial_t\varphi_\Sigma	\,.
}
Here, the gauge invariant potential field $\Phi$ is determined by the sum of chemical
potential differences between the quasiparticles $\mu_q= e\A_0$ and each of the 
condensates $\mu^{(a)}_{p} = - \partial_t\varphi_a/N$. Each of the partial potential 
differences $\Phi^{(a)} = [\mu_{q}-\mu^{(a)}_{p}]/e$ is proportional to charge imbalance 
in the $a$-th band $Q^*_a = 2e^2\nu_0\Phi^{(a)}$ where $\nu_0$ is the density of states 
\cite{Rieger.Scalapino.ea:71,Tinkham.Clarke:72,Kadin.Smith.ea:80}.

Thus, as for the magnetic field, the electric field feature a contribution from 
the intercomponent interactions. As discussed in details in \CVcite{Garaud.Silaev.ea:16},
this additional contribution can be used to probe the properties of superconducting 
states that break the time-reversal symmetry. In particular by measuring the 
nonequilibrium electric responses generated by nonstationary heating when the local 
temperature evolves, recovering from the initial hot spot created, \eg, 
by a laser pulse \cite{Maniv.Polturak.ea:05,Garaud.Silaev.ea:16,Vadimov.Melnikov:17}.
More precisely, in a multicomponent system, a charge imbalance can be generated by 
the spatial and temporal variations of the intercomponent relative phase. 
This originates in a nonequilibrium redistribution of the Cooper pairs between the 
different components, and thus creates an imbalance of partial charge $Q^*_a$. 
The generated charge imbalance can be measured using the normal metal and superconducting 
potential probes \cite{Clarke:72,Yu.Mercereau:75}.

Similarly to the spontaneous magnetic field, the induced charge imbalances are 
sensitive to the broken time-reversal symmetry. Indeed, for the same heating protocol, 
the degenerate $s\!+\!is$ and $s\!-\!is$ states produce opposite electric fields and 
charge imbalances. It is thus possible to discriminate between the usual thermoelectric 
response occurring in conventional superconductors, and the unconventional response 
that signals states that break the time-reversal symmetry. Moreover, the responses 
are also sensitive to the pairing symmetry, and can for example differentiate the 
$\sis$ from the $\sid$ state \CVcite{Garaud.Silaev.ea:16}.


\doPrint{ \newpage \thispagestyle{empty}\ \newpage }{ }
\chapter*{Overview and Perspectives}		
\addcontentsline{toc}{chapter}{Overview and Perspectives}

\vspace{0.5cm}

\section*{Overview}
\addcontentsline{toc}{section}{Overview}

This report tried to convey the message that multicomponent models, and in particular 
multicomponent superconductors, host a very rich physics that is absent in their 
single-component counterparts. 

As emphasized in the introduction, the topological excitations are ubiquitous 
in physics, as they appear for example in solid state physics, condensed and soft 
matter systems, high-energy physics, and more. Depending on the associated topological 
properties, these kind of objects have different structure. They can be particle-like, 
point-like, "wall"-like, or line-like. In the later case, the topological defects are 
termed vortices, and they have been extensively studied in the context of superfluidity 
and superconductivity. Vortices can determine to a large extent the thermodynamic, 
electric and magnetic properties of the considered materials. 
The choice of the narrative in the introduction tries to emphasize that vortices 
attracted a lot of attention for a long time, and that some old concepts are still 
relevant in modern physics. 

Because of the larger number of degrees of freedom, the multicomponent models of 
superconductivity allow for a rich spectrum of topological defects. The first chapter 
was essentially dedicated to formalize the topological properties of the multicomponent 
superconductors. It was further discussed various contributions of the author, in the 
construction of new kind of topological defects in different models of multicomponent 
superconductivity. These new topological defects may be used to identify properties 
of the underlying models.
Moreover, it was emphasized in the second chapter, that multicomponent superconductors 
not only host new kind of topological excitations, but also that they can interact 
differently than the usual vortices. This new interaction between the vortices is 
essentially different from that of type-1 or type-2 single-component superconductors.
It follows that vortices can form aggregates, and this have important impact on the 
various observable physical processes. 
Finally, there can also exist superconducting states that break the time-reversal 
symmetry, because of the competition between different pairing channels. These states 
are associated with new effects that can be measured, as discussed in the last chapter.

It is important to stress again that all the author's contributions rely  
on an intensive use of numerical techniques. The numerical aspects are quite often 
disregarded, in favour of the discussions of the physical properties. It seemed 
important to take the opportunity of this report, to present in more details these 
numerical aspects.

All the results discussed in this report seek to emphasize the richness of the physics 
of multicomponent system. This is just the tip of the iceberg, and many more can be said. 
Despite the physics of topological defects is quite an old story now, there is still 
a lot to be discovered.

\vspace{0.5cm}
\section*{Perspectives}
\addcontentsline{toc}{section}{Perpsectives}

As it was emphasized in the report, there is a growing number of known 
multiband/multicomponent superconducting materials. Hence this is an always evolving 
playground to look for new relevant theories, and to investigate their topological 
properties. So in some sense, there are always unknown projects that may be worth 
investigating because of their relevance to new materials. In any case, many aspects 
of models of multicomponent superconductors are probably still to be discovered.
Below, we can present three promising directions in relations with the aspects 
discussed in the report.

\paragraph*{Project 1: Anomalous superconducting states that break the 
	time-reversal symmetry.}
Some of the new properties of the $\sis$ state, which spontaneously breaks the 
time-reversal symmetry, were reported in details in this report. These include, 
among other things, the existence of collective modes which includes massless 
\cite{Lin.Hu:12} and mixed phase-density \cite{Carlstrom.Garaud.ea:11a,Stanev:12,
Maiti.Chubukov:13,Marciani.Fanfarillo.ea:13} excitations, unconventional mechanism 
of vortex viscosity \cite{Silaev.Babaev:13}, formation of vortex clusters 
\cite{Carlstrom.Garaud.ea:11a}, unconventional contribution to the thermoelectric 
effect \cite{Silaev.Garaud.ea:15,Garaud.Silaev.ea:16}. The $\sis$ state is also 
predicted to host topological excitations such as skyrmions and domain walls 
\cite{Garaud.Carlstrom.ea:11,Garaud.Carlstrom.ea:13,Garaud.Babaev:14}.

Recently, the specific heat measurement in hole-doped Ba$_{1-x}$K$_x$Fe$_2$As$_2$, 
at doping $x\approx0.8$ showed an intriguing behaviour \cite{Grinenko.Weston.ea:21}. 
Namely, the spontaneous Nernst effect and muon spin rotation experiments indicates 
a state in which the Cooper pairs are incoherent, but which spontaneously breaks 
time-reversal symmetry.
When a multicomponent superconductors breaks the time-reversal symmetry, 
there can be multiple phase transitions. At the level of the mean-field theory, 
the superconducting phase transition $T_c$ always occurs at a temperature equal 
to or higher than the transition temperature of the broken time-reversal symmetry 
$T_{\groupZ{2}}$. The recent results show an opposite behaviour where $T_{\groupZ{2}}>T_c$
\cite{Grinenko.Weston.ea:21}.
All the discussion about the role of the fluctuations, and the implications are beyond 
the discussions here. Yet a few remarks on the structure of the model opens interesting 
perspectives.

As discussed in this report, multi-component superconductor feature an additional 
contribution to the magnetic field because of the inter-component interactions 
\Eqref{Eq:TRSB:MagneticField}. Then, the Ginzburg-Landau free energy expressed 
in terms of charged and neutral modes \Eqref{Eq:Quantization:FreeEnergy:3} can 
further be written as 
\Align{Eq:Perspective:FreeEnergy:1}{
&\F= \frac{1}{2}\left[ \varepsilon_{kij}\left\lbrace 
\nabla_i\left(\frac{J_j}{e^2\varrho^2}\right) +\frac{i}{e\varrho^4}
{\cal Z}_{ij}
 \right\rbrace\right]^2  + \frac{\J^2}{2e^2\varrho^2} 
+\Grad\Psi^\dagger\!\cdot\!\Grad\Psi
+\frac{1}{4\varrho^2}\big(\Psi^\dagger\Grad\Psi-\Grad\Psi^\dagger\Psi\big)^2
+ V(\Psi)	\,, \nonumber \\
&~~~~~\text{where}~~~
{\cal Z}_{ij}= \varrho^2\Grad_i\Psi^\dagger\Grad_j\Psi
+(\Psi^\dagger\Grad_i\Psi)(\Grad_j\Psi^\dagger\Psi)\,.
}

In the anomalous state, the superconducting part of the model is disordered, and the 
part corresponding to London screening is absent, that is $\J=0$. Then an effective 
model for the new state, can be derived from \Eqref{Eq:Perspective:FreeEnergy:1}, 
with requiring that the superconducting current vanish $\J=0$. This amount to 
retain only the degrees of freedom that are related to relative phases. The 
corresponding model thus reads as \cite{Grinenko.Weston.ea:21}
\Equation{Eq:Perspective:FreeEnergy:2}{
\F= \frac{1}{2}\left[  
 \frac{i\varepsilon_{kij}}{e\varrho^4}{\cal Z}_{ij}
 \right]^2   
 +\Grad\Psi^\dagger\!\cdot\!\Grad\Psi
+\frac{1}{4\varrho^2}\big(\Psi^\dagger\Grad\Psi-\Grad\Psi^\dagger\Psi\big)^2
+ V(\Psi)	\,. 
}
As discussed in details in \cite{Grinenko.Weston.ea:21}, the effective theory 
for the anomalous normal state, which breaks the time-reversal symmetry, allows 
for domain walls excitations. These feature magnetic signatures, as those discussed 
in the main body in the Chapter \ref{Chap:TRSB}.

This new effective model offers lot of new opportunity to observe unusual properties 
of multi-component models in an anomalous state. That is, there is an opportunity 
to observe some of the topological properties of multi-component superconductors, 
above the critical temperature. Of course here, the model 
\Eqref{Eq:Perspective:FreeEnergy:2} is obtained heuristically, and a careful derivation 
is required to properly handle how the different terms should be renormalized, 
when the superconducting part of the model is disordered.

\paragraph*{Project 2: Other superconducting states that break the time-reversal symmetry.}
Not only the number of known multiband/multicomponent superconductors is growing, 
but also of those that break the time-reversal symmetry \cite{Ghosh.Smidman.ea:20}.
It was emphasized in this report that the superconducting states which spontaneously 
break the time-reversal symmetry feature new properties. Most of the focus was on the 
$\sis$ state, which is highly relevant for iron based superconductors. Another very 
investigated time-reversal symmetry breaking state, is the $\pip$ in relations with 
spin-triplet superconducting models. There are various other superconducting states,  
which break the time-reversal symmetry, but with other pairing symmetries, \eg 
$\sid$, $\did$. Also for example, it was recently argued that the pairing in \SRO 
could be either $\did$ or $d\!+\!ig$ \cite{Ghosh.Shekhter.ea:20,Agterberg:20}. 
These states that are different from $\pip$ or $\sis$, are much less studied, 
and in particular their topological properties.

Since they break the time-reversal symmetry, all these states should feature 
domain-wall excitations as well. However, the difference is that they break different 
point group symmetries. At the level of the Ginzburg-Landau model, this manifests 
by having different, more rich structure of the kinetic terms in the form of 
anisotropies and mixed gradients. Like for example 
$(D_x\psi_1^*D_x\psi_2-D_y\psi_1^*D_y\psi_2+c.c.)$ for the $\sid$ state which breaks 
the $C_4$ symmetry. Or for example the $d+id$ state, which violates both parity and 
time-reversal symmetries \cite{Balatsky:00,Laughlin:98} .

The peculiarities of these other pairing symmetries, have been much less studied. 
For example because of their different structures, they should also manifest 
thermoelectric responses qualitatively different from those discussed in the 
Chapter \ref{Chap:TRSB}. Moreover, the structure of the topological defect 
should also definitely be sensitive to that.

\paragraph*{Project 3: Knots and vortons in the electroweak theory.} 
The idea here, is to look for topological defects in a theory different than 
that describing multi-component superconductivity. More precisely, the goal is to 
investigate the possibility that the Weinberg-Salam theory of the electroweak 
interactions could host topological defect with a knotted structure.

As emphasized in the introduction, the idea of knotted vortices is an old story 
that received a new breath after knotted topological defects were constructed in 
the Skyrme-Faddeev model \cite{Faddeev.Niemi:97}. Since that, there was a lot of 
activity in tracking similar object in various physical systems as for example in 
spinor Bose-Einstein condensates \cite{Kawaguchi.Nitta.ea:08}, optical beams 
\cite{Dennis.King.ea:10}, nematic colloids \cite{Tkalec.Ravnik.ea:11}, magnetic 
materials \cite{Sutcliffe:17,Sutcliffe:18}, and more; for a review on knots, 
see \cite{Radu.Volkov:08}.

Vortons are objects that, although formally different, are quite alike to knotted 
vortices. These are closed loops of superconducting vortices \cite{Witten:85a}, 
that are expected to be stabilized against contraction by the centrifugal force 
produced by the current \cite{Davis.Shellard:89}. They are expected to occur in a 
model first introduced by Witten \cite{Witten:85a}, which a two-component model but 
with two abelian gauge fields (instead of one for superconductors). The explicit 
construction of vorton, and the demonstration of their potential stability is however 
rather recent \cite{Radu.Volkov:08,Battye.Sutcliffe:08,Battye.Sutcliffe:09,
Garaud.Radu.ea:13}.

The bosonic sector of the Weinberg-Salam theory of the electroweak interactions, 
can be seen, to some extent, as a multi-component theory but more involved than 
those discussed in this report. Indeed, it is a theory of a doublet of complex 
scalars (the Higgs field). But the gauge sector is more complicated, as also contains 
a non-Abelian $\groupSU{2}$ gauge field, in addition to the $\groupU{1}$ gauge field.
It is usually assumed that the electroweak theory admits no solitons, however there 
are indications that it might host some kind of vortons, or knotted vortices.
Not only the theory allows for vortices, but also in some limiting cases, it can 
be very similar to the model of Witten where vortons exist.

Strictly speaking the electroweak theory is different than the models of multicomponent 
superconductivity models discussed in the main body of the report. Yet since they share 
some properties. One could imagine that this theory supports knotted vortex solutions 
similar to those obtained in the framework of two-component superconductors with 
dissipationless Andreev-Bashkin drag interaction \CVcite{Rybakov.Garaud.ea:19}.
If such electroweak vortons or knotted vortices exist, this could be of scientific value.



\doPrint{ \newpage \thispagestyle{empty}\ \newpage }{ }
\addcontentsline{toc}{part}{Appendices}
\part*{Appendices} 
\appendix 

\doPrint{ \newpage \thispagestyle{empty}\ \newpage }{ }
\graphicspath{{Plots/02-Background/}}
\chapter{Single-component Ginzburg-Landau theory}
\label{App:Single-Component}

The main body of this report presents results regarding the properties of the theories 
of superconductivity featuring multiple order parameters or order parameters with 
multiple components. It might be useful, for a better understanding of the peculiarities 
of multicomponent theories, to review the essential properties of  of the conventional, 
single-component, models of superconductivity. Covering all the microscopic aspects of 
conventional superconductivity is well beyond the scope of the present discussions, 
and these will not be discussed here. Both microscopic, and mean field aspects of 
single-component superconductivity are extensively discussed in a great number of 
classical textbooks, see \eg \cite{Saint-James.Thomas.ea,Gennes,Tinkham,Shmidt.Muller.ea,
Chaikin.Lubensky,Huebener,Schmidt,Annett,Fossheim.Sudbo,Svistunov.Babaev.ea}.

Hence the present background review is restricted only to the classical mean-field 
aspects of superconductivity. More precisely, this Appendix presents the general 
theoretical framework, and the textbook properties of the single-component Ginzburg-Landau 
theory.


The Ginzburg-Landau theory \cite{Ginzburg.Landau:50} was introduced in 1950, to account 
for the properties of the superconducting state. This phenomenological theory is based 
on the Landau theory of theh second order phase transitions, where the (macroscopic) 
order parameter $\psi=|\psi|\Exp{i\varphi}$, is a complex scalar field. The order parameter 
$\psi$ is often equivalently termed \emph{superconducting condensate}. The microscopic  
Bardeen-Cooper-Schrieffer theory of superconductivity \cite{Bardeen.Cooper.ea:57} was 
derived later in 1957. Shortly after, in 1959 Gor'kov demonstrated that the Ginzburg-Landau 
theory can be derived as classical approximation of the microscopic theory 
\cite{Gorkov:59}, and the modulus of the order parameters $\psi$ is actually the density 
of Cooper pairs: $n_s=|\psi|^2$. Strictly speaking, the Ginzburg-Landau theory is valid 
only in a close vicinity of the critical temperature $T_c$ where the superconductivity 
is destroyed, and it assumes that $\psi$ is small and is slowly varying (small gradients).

Remark that besides its fundamental applications in solid state physics, the 
Ginzburg-Landau theory attracted a lot of attention in the mathematical community from 
the 1990's, after the report of the well posedness of the problem \cite{Du.Gunzburger.ea:92,
Chen.Hoffmann.ea:93,Du:94,Du:96}. Since these earlier works, there have been a big 
activity in understanding the mathematical properties of that problem, see for example 
\cite{Du:05}. In parallel there also have been continuous efforts in the physics 
community to have optimal formulation for numerical solvers, see for example 
\cite{Gropp.Kaper.ea:96,Sadovskyy.Koshelev.ea:15}.

In the vicinity of the critical temperature, the superconducting state is governed by 
the Ginzburg-Landau free energy, whose density reads as  
\cite{Ginzburg.Landau:50} (see textbook discussions, e.g., \cite{Gennes,Tinkham}):
\Equation{Eq:FreeEnergy:Dim}{
\F=\alpha|\psi|^2+\frac{\beta}{2}|\psi|^4
+\frac{1}{2m_\star}\left|\left(\frac{\hbar}{i}\Grad-\frac{e_\star}{c}\A\right)\psi\right|^2
+\frac{\B^2}{8\pi}.
}
Here $e_\star$ and $m_\star$ are respectively the \emph{effective} charge and mass of 
the Cooper pairs, $\hbar$ is the reduced Planck's constant and $c$ is the speed of light. 
For a consistency with the text of the main body of the report, it is 
convenient to consider the Ginzburg-Landau free energy $F=\int d^3\x\,\F$, whose density 
(in conveniently chosen dimensionless units) reads as
\Equation{Eq:GL}{
\F= 
\frac{\B^2}{2}
+\frac{1}{2}\left|\left(\Grad+ie\A \right)\psi \right|^2
+\alpha|\psi|^2+\frac{\beta}{2}|\psi|^4	\,.
}
Since $\psi$ is a charged scalar field, it is coupled to $\A$, the vector potential of the 
magnetic field $\B=\Curl\A$, via the gauge derivative $\D\equiv\left(\Grad+ie\A\right)$. 
$e$ is a coupling constant sometimes called the \emph{gauge coupling} constant. For the 
energy to be bounded from below, the parameter $\beta$ must be positive, while $\alpha$ 
is either positive or negative. In the presence of an external applied field $\He$, 
this is the Gibbs free energy, 
\Equation{Eq:Gibbs}{
G=F-\int\B\cdot\He\,,
}
that should be considered instead of the Helmholtz free energy $F$. The Ginzburg-Landau 
theory \Eqref{Eq:GL} is thus as classical field theory, where the physical degrees of 
freedom are a complex scalar field $\psi(\x)$ standing for the superconducting condensate 
and the gauge field $\A(\x)$ (the vector potential), a real vector field.

The functional variation of the Ginzburg-Landau functional \Eqref{Eq:GL} with respect to 
$\psi^*$ gives the Ginzburg-Landau equation 
\Equation{Eq:GLeq}{
\D\D\psi=2\left(\alpha+\beta|\psi|^2\right)\psi
~~~~\text{with}~~~\D\equiv\left(\Grad+ie\A\right)	\,,
}
while the variations with respect to the vector potential $\A$ yield the Amp\`ere-Maxwell 
equation 
\Equation{Eq:Ampere}{
\Curl\B+\J=0\,,~~~~\text{with}~~~\J=e\Im(\psi^*\D\psi)=e|\psi|^2(\Grad\varphi+e\A) \,.
}
The right-hand side of the Amp\`ere-Maxwell equation, $\J$, is termed \emph{superconducting 
current} or \emph{supercurrent}. The Ginzburg-Landau equation \Eqref{Eq:GLeq} together 
with the Amp\`ere-Maxwell equation \Eqref{Eq:Ampere} are the Euler-Lagrange equations of 
motion of the Ginzburg-Landau theory \Eqref{Eq:GL}. 

Both the superconducting condensate $\psi$ and the gauge field satisfy the boundary 
conditions that represent the physical properties of the system. The conditions on a 
Superconductor/Insulator interface $\partial\Omega_{SI}$ are
\Equation{Eq:TDGL:BC:SI}{
\D\psi\cdot{\bs n}=0		\,, ~~
(\Curl\A)\times{\bf n}={\bs H}_e\times{\bf n}\,,
}
while the Superconductor/Normal metal interface $\partial\Omega_{SN}$ is described by
\Equation{Eq:TDGL:BC:SN}{
\D\psi\cdot{\bs n}=i\gamma\psi		\,, ~~
(\Curl\A)\times{\bf n}={\bs H}_e\times{\bf n}\,.
}
The real parameter $\gamma$ depends on the details of the materials. 
Here, ${\bf n}$ is the outgoing normal vector to the interface. In all generality, 
the overall boundary $\partial\Omega$ is thus defined as 
$\partial\Omega=\partial\Omega_{SI}\cup\partial\Omega_{SN}$.

\subsubsection*{Gauge invariance}

The theory is well known to be invariant under the local transformations generated by 
the elements having value in the Lie algebra of the $\groupU{1}$ gauge group (see, e.g. 
\cite{Gennes,Tinkham}). The spontaneous breakdown of that symmetry is responsible for 
the longitudinal component of the photon to become massive and then being effectively 
a massive vector (Proca) field. In other words, in the Meissner state the gauge field 
is massive with an exponential decay to due the screening currents. The $\groupU{1}$ 
transformations $\G_\chi$ that are symmetries of the free energy \Eqref{Eq:GL}, of 
the Ginzburg-Landau \Eqref{Eq:GLeq} and of the Amp\`ere-Maxwell \Eqref{Eq:Ampere} 
equations are 
\Equation{Eq:GaugeTR}{
\G_\chi:(\psi,\A)\longmapsto\left(\psi\Exp{i\chi},\A-\frac{1}{e}\Grad\chi\right)\,,
}
for any (sufficiently smooth) real-valued function $\chi:=\chi(t,\x)$. The observable 
physical quantities such as $\B$, $\J$, $|\psi|$, etc are invariant under these gauge 
transformations \Eqref{Eq:GaugeTR}. Obviously any choice of the gauge function 
$\chi(t,\x)$ that preserves the boundary behaviour \Eqref{Eq:TDGL:BC:SI} and 
\Eqref{Eq:TDGL:BC:SN} is physically valid, since it does not affect the physical 
observables. However, it is important to note that different choices of gauge do 
lead to different mathematical structures of the system. This is well known that 
some structures are easier to analyse, for example numerically, than some others 
(see detailed discussions in, \eg , \cite{Du:94,Chen.Hoffmann.ea:93,Tang.Wang:95,
Fleckinger-Pelle.Kaper:96,Fleckinger-Pelle.Kaper.ea:98,Takac:96}.

\subsubsection*{Dynamics}

The Ginzburg-Landau energy describes the magnetostatic properties of superconductors. 
Their dynamics is described by the time-dependent Ginzburg-Landau equations 
\cite{Schmid:66,Gorkov.Eliashberg:68,Gorkov.Kopnin:75}. In the dimensionful units of 
Eq.\Eqref{Eq:FreeEnergy:Dim}, the time-dependent Ginzburg-Landau equations reads as
\SubAlign{Eq:Dim:TDGL}{
\frac{\hbar^2}{2m_\star D}\left(\partial_t+\frac{ie_\star}{\hbar}A_t\right)\psi
&+\frac{1}{2m_\star}\left(\frac{\hbar}{i}\Grad-\frac{e_\star}{c}\A\right)^2\psi 
+\big(\alpha+\beta|\psi|^2\big)\psi=0 \label{Eq:Dim:TDGL:1}
\,, \\
\Curl\Curl\A&=-\frac{4\pi\sigma}{c}\left(\frac{1}{c}\partial_t\A+\Grad A_t\right) 
+\frac{4\pi}{c}\J_s+\Curl{\bs H}_e \label{Eq:Dim:TDGL:2}
\,.
}
Here, $\sigma$ is the normal state conductivity and $D$ is a diffusion constant. 
$A_t$ is the electrostatic potential, ${\bs H}_e$ stands for an externally applied 
field and $\J_s$ is the supercurrent that reads as:
\Equation{Eq:Dim:Supercurrent}{
\J_s=\frac{e_\star\hbar}{m_\star}\left(\Im\left(\psi^*\Grad\psi\right)
-\frac{e_\star}{\hbar c}|\psi|^2\A\right) \,.
}
Up to the $\frac{4\pi}{c}$ factor, the left hand side of Eq.\Eqref{Eq:Dim:TDGL:2} 
is the total current being the superposition of superconducting and normal currents: 
$\J=\J_n+\J_s$. Indeed the normal current satisfies Ohm's law ($\J_n=\sigma{\bs E}$), 
and according to Faraday's law the electric field is 
${\bs E}=-\frac{1}{c}\partial_t\A-\Grad A_t$. As a result Eq.\Eqref{Eq:Dim:TDGL:2}, 
in absence of an external field ${\bs H}_e$, is the Amp\`ere's law: 
$\Curl\B=\frac{4\pi}{c}\J$.

It is important to emphasize the dissipative nature of the time-dependent 
Ginzburg-Landau equations. Indeed, the time-dependent equation \Eqref{Eq:Dim:TDGL}, 
can be understood as a gradient flow of the free energy \Eqref{Eq:FreeEnergy:Dim}.
This implies in particular that the time evolutions leads to stationary solutions, that 
are (local) minima of the free energy \Eqref{Eq:Dim:TDGL}. For rigorous demonstration 
of that statement, see \cite{Du.Gunzburger.ea:92,Chen.Hoffmann.ea:93,Du:94,Du:96,Du:05}.

The next discussions are only about the stationary properties of the Ginzburg-Landau 
theory. Hence it is convenient to re-introduce the dimensionless units of 
Eq.~\Eqref{Eq:GL}.

\subsubsection*{Relativistic version: the Abelian-Higgs Model}

The Ginzburg-Landau model is very similar to the Abelian-Higgs Model that has been 
extensively studied in the framework of high-energy physics. Indeed, this is the theory 
of a complex scalar field charged under the $\groupU{1}$ gauge group of electromagnetism.
In that framework, the scalar field $\psi$ is the Higgs field, and the gauge field is 
the four-potential of the electromagnetic field $(A_t,\A)$ . The energy in the 
Abelian-Higgs model, which reads as 
\Equation{Eq:AH:energy}{
E= \bigintsss d^4{\rm x}\left\lbrace \frac{1}{2}({\bs E}^2+\B^2)
+|D_t\psi|^2+|\D\psi|^2 + \frac{\beta}{8}(|\psi|^2-1)^2 \right\rbrace\,,
}
is very similar to that of the Ginzburg-Landau theory. Again, the gauge derivative is 
$D_\mu\psi=(\partial_\mu+ieA_\mu)\psi $. The constant $\beta$ is the self-interacting 
constant of the scalar field, and it is similar to the Ginzburg-Landau parameter $\kappa$
introduced below.

The close similarity between both models, implies that they share the same static 
solutions. However, it is important to stress that the dynamics is very different. 
Indeed, while the dynamics in the Abelian-Higgs model is relativistic, the dynamics 
of superconductors is determined by the time-dependent Ginzburg-Landau equations 
\Eqref{Eq:Dim:TDGL}. As discussed above, the latter is a dissipative equation, while 
the dynamics of the Abelian-Higgs model is ``wave-like".

\section{Ground state, Length-scales and the Meissner effect}
\label{Sec:Background:GS}
\subsection*{Superconducting ground state}

The ground state is, by definition, the state which gives the minimal value of the energy. 
Since the magnetic and kinetic energy contributions are quadratic, the minimum of the 
energy satisfies 
\Equation{Eq:Minima:Kinetic}{
\left\lbrace \begin{array}{c}\Curl\A=0 \\(\Grad+ie\A)\psi=0\end{array}\right.
~~~\text{whose general solutions are}~~~\left\lbrace \begin{array}{c}
\A=\Grad \chi(\x) \\ \psi=\mathrm{const.}\times\Exp{ie\chi(\x)}\end{array}\right.
\,,
}
for an arbitrary regular function $\chi(\x)$. The solutions \Eqref{Eq:Minima:Kinetic} are 
easily identified with the gauge transformations \Eqref{Eq:GaugeTR}. It is thus 
straightforward to choose the simplest solution $\A=0$ and $\psi =\text{constant}$, that 
simultaneously minimize both magnetic and kinetic contributions in the free energy.
It follows that the minimal energy configuration corresponds to the minimum of the 
potential energy
\Equation{Potential}{
V(\psi)=\alpha|\psi|^2+\frac{\beta}{2}|\psi|^4	\,.
}
Depending on the sign of the parameter $\alpha$, there are two possible minima 
\Equation{Eq:Minima:Potential}{
\argmin V(\psi) :=
\left\lbrace \begin{array}{c}
|\psi|=0 ~,~\text{if}~~ \alpha\geq0 \\
|\psi|=\sqrt{\frac{-\alpha}{\beta}}~,~\text{if}~~ \alpha<0
\end{array}\right.
~~~\text{and}~~~
\min V(\psi)=
\left\lbrace \begin{array}{c}
0 ~,~\text{if}~~ \alpha\geq0 \\
\frac{-\alpha^2}{\beta}~,~\text{if}~~ \alpha<0
\end{array}\right.	\,.
}
Note that the extremality of $V$ only imposes value of the modulus of $\psi$, whereas 
its phase can assume any value. Thus the ground state is degenerate, defined only up to 
the pure gauge transformations \Eqref{Eq:GaugeTR}.
The two different minima \Eqref{Eq:Minima:Potential} termed the \emph{normal state} 
(where $\psi=0$) and the \emph{superconducting ground state} $\psi_0:=\sqrt{-\alpha/\beta}$.

\begin{wrapfigure}{R}{0.35\textwidth}
\hbox to \linewidth{ \hss
\includegraphics[width=.975\linewidth]{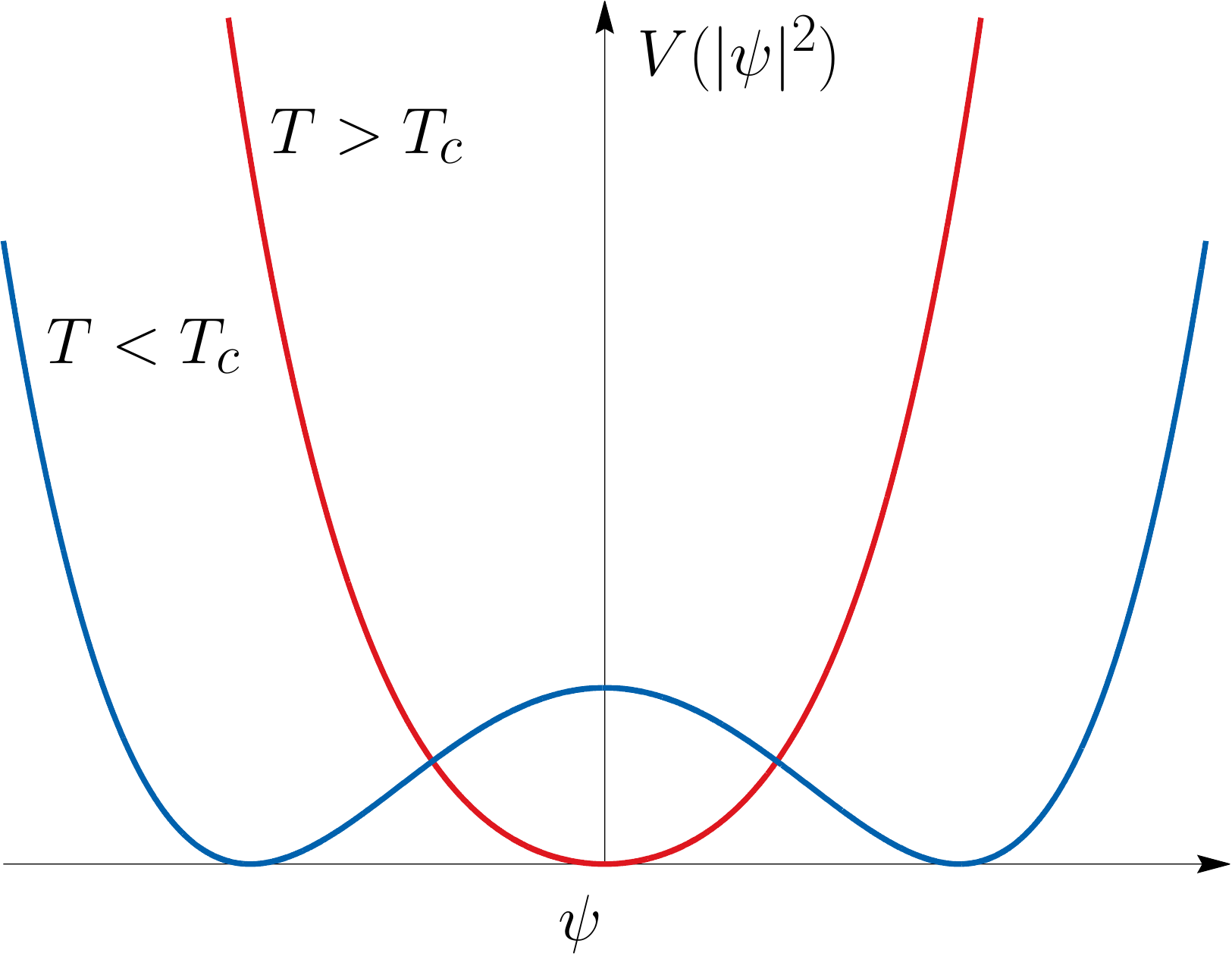}
\hss}
\caption{
The potential energy depending on the sign of the parameter 
$\alpha$. Above $T_c$, $\alpha>0$ and there is a unique minimum, while 
below $T_c$, $\alpha<0$ and the potential looks like the `mexican hat'.
}\label{Fig:Potential}
\end{wrapfigure}

In a first approximation, the parameter $\alpha$ is the only temperature 
dependent parameter 
\Equation{Alpha}{
\alpha(T)=\alpha_0\left(\frac{T}{\Tc{}}-1 \right)\,.
}
This implies that above the critical temperature $\Tc{}$, the parameter $\alpha>0$ and thus 
the normal state is energetically favoured. On the other hand, below the critical 
temperature $\alpha<0$, and the superconducting state is preferred. These two different 
regimes are qualitatively displayed in \Figref{Fig:Potential}.

Finally, the \emph{condensation energy}, is defined as the energy difference between 
the normal state $\psi=0$ and the superconducting state $\psi_0$
\Equation{Eq:Condensation}{
f_n=\F(\psi=0)-\F(\psi_0)=\frac{\alpha^2}{2\beta}
=\frac{H_c^2}{2}
}
The value of magnetic field $H_c$ is called the thermodynamical critical magnetic field. 
While the external magnetic field penetrates without change, into the volume occupied 
by the normal state, it is expelled from the superconducting state. Thus, free energy 
of the normal state in a magnetic field $H>\Hc{t}$ is lower than that of the uniform 
superconducting state, with expelled field. In the simplest case to be discussed below, 
$\Hc{t}$ denotes the value of the external fields which destroys the superconductivity.

\subsection*{Length-scales in the Ginzburg-Landau theory}

Borrowing the terminology of classical field theories, the length-scales are defined 
as the inverses masses of the mass spectrum of the theory. The mass spectrum of the 
theory is defined by the fluctuations of the fields around their ground state values.
Since it is always possible to find a gauge where the ground state is
$(\psi,\A)_{\rm GS}=(\psi_0,0)$, the perturbative expansion around the ground state, 
in terms of the infinitesimal parameter $\eps$ reads as
\Equation{Eq:GL:Expansion}{
\psi = (\psi_0 + \eps f)
\exp\left\{i\eps\frac{\phi}{\psi_0}\right\} 
\,~~~\text{and}~~~
\A=\eps{\bs a}\,.
}      
Here, $f\equiv f(\x)$ is the density amplitude, $\phi\equiv \phi(\x)$ is the normalized 
phase amplitude, and ${\bs a}\equiv {\bs a}(\x)$ are the gauge field fluctuations. 
The fluctuations are thus characterized by a system of Klein-Gordon equations for the 
two condensate fluctuations (one for the density plus one phase), and one Proca equation 
for the gauge field fluctuations. Choosing the gauge $\Div{\bs a}=0$, the system reads as
\Equation{Eq:GL:Fluctuation}{
\frac{1}{2}\Upsilon^T\left(-\Grad^2 + {\cal M}^2\right)\Upsilon
\,,~~~\text{where}~~~
\Upsilon=(f,\phi,{\bs a})^T\,.
}
Here ${\cal M}^2$ is the squared mass matrix that is straightforwardly obtained by 
retaining the quadratic order of the infinitesimal parameter $\eps$, after introducing 
the expansion \Eqref{Eq:GL:Expansion} into the free energy \Eqref{Eq:GL}. The squared 
mass matrix can thus be read from 
\Equation{Eq:GL:KGmass}{
\Upsilon^T{\cal M}^2\Upsilon=2(\alpha+3\beta\psi_0^2)f^2 
+e^2\psi_0^2{\bs a}^2
\,.
}
Note that the fluctuation operator \Eqref{Eq:GL:Fluctuation} and \Eqref{Eq:GL:KGmass} 
can also be obtained by linearizing the equations of motion \Eqref{Eq:GLeq} and 
\Eqref{Eq:Ampere}, in the expansion parameter $\eps$. The eigenspectrum of the matrix 
${\cal M}^2$ determines the squared masses of the excitations and the corresponding 
normal modes. The inverse of each mass defines a characteristic length-scale  of the theory.
Overall, since the equations that define the mass spectrum of the fluctuations are
\Equation{Eq:GL:massEQ}{
\Grad^2f= -4\alpha f
\,,~~~~~~\Grad^2\phi=0
\,,~~~\text{and}~~~\Grad^2{\bs a}= e^2\psi_0^2{\bs a} \,, 
}
the masses are
\Equation{Eq:masses}{
m_{|\psi|}=2\sqrt{-\alpha}
\,,~~~~~~m_\varphi=0
\,,~~~\text{and}~~~m_\A=e\psi_0 	\,.
}
Here $m_\varphi=0$ is the Goldstone boson that gives mass to the longitudinal 
component of the gauge field \cite{Anderson:63}. Having obtained the mass spectrum 
\Eqref{Eq:masses}, the relevant length-scales that characterize the superconducting 
ground state of the single-component Ginzburg-Landau model are
\Equation{Eq:length}{
\xi=\frac{\sqrt{2}}{m_{|\psi|}}=\frac{1}{\sqrt{-2\alpha}}
\,,~~~~\text{and}~~~~\lambda=\frac{1}{m_\A}=\frac{1}{e\psi_0} 	\,,
}
where $\xi$ is coherence length and $\lambda$ is the penetration depth.
The factor $\sqrt{2}$ factor in the definition of coherence length is a matter of 
convention. This convention is that where the non-interacting regime (the Bogomol'nyi 
regime \cite{Bogomolnyi:76}), is $\kappa=1/\sqrt{2}$ for single-component superconductors 
\cite{Tinkham}.

The Ginzburg-Landau functional depends on three parameters, $\alpha$, $\beta$ and $e$. 
These determines the two fundamental length-scales: The coherence length $\xi$ of the 
superconducting condensate and the penetration depth $\lambda$ of the magnetic field.  
Actually the whole theory depends on a unique parameter $\kappa$, the Ginzburg-Landau 
parameter, defined as the ratio of the two length-scales:
\Equation{Eq:Kappa}{
\kappa=\frac{\lambda}{\xi}=\frac{\sqrt{2\beta}}{e}\,.
}

\paragraph{Alternative derivation of the mass spectrum:}
The analysis above contains unphysical degrees of freedom. Indeed, as the gauge is not 
fixed there, the massless Goldstone mode appears. It is possible to find directly the 
physical mass spectrum of the theory, by rewriting the free energy \Eqref{Eq:GL}, only 
in terms of the density and of the magnetic field. 
Given the definition of the supercurrent \Eqref{Eq:Ampere}, the gradient term in the 
free energy can be written as
\Equation{Eq:GradientRelation}{
|\D\psi|^2= \left(\Grad|\psi|\right)^2+|\psi|^2(\Grad\varphi+e\A)^2 
		=\left(\Grad|\psi|\right)^2+\frac{\J^2}{e^2|\psi^2|}	\,.
}
Hence, using the relation \Eqref{Eq:GradientRelation} together with the Amp\`ere-Maxwell 
equation \Eqref{Eq:Ampere}, the free energy \Eqref{Eq:GL} can be rewritten as
\SubAlign{Eq:GLrewritten}{
\F&= 
\frac{1}{2}\left(\frac{\J^2}{e^2|\psi^2|} + \B^2\right)
+\frac{1}{2}\left(\Grad|\psi|\right)^2
+\alpha|\psi|^2+\frac{\beta}{2}|\psi|^4	 \\
&= 
\frac{1}{2}\left(\frac{(\Curl\B)^2}{e^2|\psi^2|} + \B^2\right)
+\frac{1}{2}\left(\Grad|\psi|\right)^2
+\alpha|\psi|^2+\frac{\beta}{2}|\psi|^4	\,.
}
Expressing the free energy in such a way has the advantage to stress the dependence 
on the gauge invariant physical degrees of freedom explicitly. Returning to the 
investigation of the mass spectrum, the perturbative expansion around the ground state, 
in terms of the infinitesinal parameter $\eps$ reads as
\Equation{Eq:GL:Expansion:2}{
|\psi| = \psi_0 + \eps f
\,~~~\text{and}~~~
\B=\eps e\psi_0{\bs b}\,.
}     
Here, $f\equiv f(\x)$ is the density amplitude,  
${\bs b}\equiv {\bs b}(\x)$ are the magnetic field fluctuations. In contrast with 
the expansion \Eqref{Eq:GL:Expansion}, the expansion \Eqref{Eq:GL:Expansion:2} depends 
only on the physical fields. 
The fluctuations are thus characterized by a system of one Klein-Gordon equation for the 
condensate fluctuations, and one Proca equation for the magnetic field fluctuations. 
Introducing the expansion \Eqref{Eq:GL:Expansion:2} into the free energy \Eqref{Eq:GL}, 
and retaining the quadratic order of the infinitesimal parameter $\eps$ yields the 
equation 
\Equation{Eq:GL:Fluctuation:2}{
\frac{1}{2}\Upsilon^T\left(-\Grad^2 + {\cal M}^2\right)\Upsilon
\,,~~~\text{where}~~~
\Upsilon=(f,{\bs b})^T\,.
}
The squared mass matrix ${\cal M}^2$ can be read from
\Equation{Eq:GL:KGmass:2}{
\Upsilon^T{\cal M}^2\Upsilon=2(\alpha+3\beta\psi_0^2)f^2 
+e^2\psi_0^2{\bs b}^2
\,.
}
Again, the eigenspectrum of the matrix ${\cal M}^2$ determines the squared masses of 
the excitations 
\Equation{Eq:masses:2}{
m_{|\psi|}=2\sqrt{-\alpha}
\,,~~~\text{and}~~~m_\B=e\psi_0 	\,,
}
and the associated length-scales $\xi$ and $\lambda$ are obviously the same as in 
\Eqref{Eq:length}.

\subsection*{Meissner effect}
The Meissner effect is the expulsion of a magnetic field from a superconductor, when it 
is in the superconducting state. In a weak external applied field, the superconductor 
expels (almost) all magnetic flux, by setting up surface currents. This relates to the 
above mentioned fact, that the Amp\`ere-Maxwell equation becomes a Proca equation on a 
constant superconducting ground state $|\psi|=\psi_0=\sqrt{\frac{-\alpha}{\beta}}$.
This is more intuitive physically, when considering the London equation for the magnetic 
field. On a constant superconducting state $\psi=\psi_0$, taking the curl of the Amp\`ere's 
equation \Eqref{Eq:Ampere}, yields the London equation
\Equation{Eq:London}{
\Curl\Curl\B+e^2\psi_0^2\B =0 
~~~~\Leftrightarrow~~~~\Grad^2\B=e^2\psi_0^2\B 
~~~~~~(\text{since}~\Curl\Curl\B=\Grad(\Div\B)-\Grad^2\B) \,.
}
The London equation thus turns into a Helmholtz equation for the magnetic field, 
with the eigenvalue of the Laplacian $m_\B=e\psi_0$. In other words, the magnetic 
field is a massive vector field of squared mass $m_\B^2=e^2|\psi_0|^2$. This defines a 
length-scale, as the inverse mass of the magnetic field, called the London penetration 
depth $\lambda=1/m_\B=1/e|\psi_0|$.
The London equation \Eqref{Eq:London} implies that an externally applied field decays 
exponentially inside the superconductor with the decay length given by $\lambda$. 
Alternatively, as demonstrated earlier, the London penetration depth can be understood 
as the length-scale, at which a small fluctuation of the vector potential recovers 
to its ground state value $\A=0$.

\section{Interface energy -- Type-I/type-II dichotomy}

Now, consider the full non-linear problem of the interface between the normal state 
and the superconducting state, in an external field $\He = (0, 0, H)$. The value of 
the external magnetic field is set equal to the thermodynamical critical field $\Hc{}$.
The associated magnetic field reads as $\B = (0, 0, B(x))$. The interface is located at 
$x=0$ and the normal state fills the semi-infinite space $x<0$, while $x > 0$ corresponds 
to the bulk superconductor. As the problem is considered in an external field, the 
relevant energy is the Gibbs energy \Eqref{Eq:Gibbs}.

On the right boundary, far in the bulk superconductor, the superconducting state 
completely recovers. On the left boundary, this is the normal state in an external field 
$\Hc{}$. This sets the boundary conditions 
\SubAlign{InterfaceBC1}{
\psi(x\to+\infty)&=\psi_0~~~~\text{and}~~~~\B(x\to+\infty)=0  \,,\\
\psi(x\to-\infty)&=0~~~~~~\text{and}~~~~\B(x\to-\infty)=\Hc{} \,.
}
As a result, the values of the Gibbs energy on these boundaries are 
$\G(x\to+\infty) = \F(|\psi_0|)$ (the free energy of a superconductor without an external  
field) and  $\G(x\to-\infty) = f_n-\Hc{}^2/2$ (since the free energy of the normal state 
in external field is $\F = f_n + \Hc{}^2/2$).
The interface energy (or boundary surface energy between normal and superconducting state) 
is defined as the difference
\SubAlign{Interface}{
\sigma_{ns}&=\int_{-\infty}^{+\infty}\Big(\G(x)-f_n+\frac{\Hc{}^2}{2}\Big) dx 
 \\
&=\int_{-\infty}^{+\infty}\Big( 
\frac{|\B-H_c|^2}{2}+\frac{1}{2}|\D\psi|^2
+\alpha|\psi|^2+\frac{\beta}{2}|\psi|^4
\Big) dx 
}
The evaluation of the interface energy \Eqref{Interface} has to be done numerically. 
However, in the limiting cases $\kappa\gg1$ and $\kappa\ll1$, it can be estimated 
analytically.

In the $\kappa\gg1$, $|\psi|$ recovers its ground state value $\psi_0$ very quickly. 
Thus, the energy cost associated with the density gradients can be neglected. Now the 
problem reduces to dealing with the screening of the magnetic field in a region of width 
$\lambda$. The associated interface energy is 
\Equation{Interface1}{
\sigma_{ns}(\kappa\gg1)=\int_{-\infty}^{+\infty}\Big(
\frac{\lambda^2}{2}B^{\prime 2}+\frac{1}{2}B^2 -B\Hc{}
\Big) dx \,,
}
where $^\prime$ denotes the differentiation with respect to $x$ and 
$B=B_0\Exp{-x/\lambda}$. This give a negative interface energy
\Equation{Interface2}{
\sigma_{ns}(\kappa\gg1)\approx\frac{\Hc{}^2\lambda}{2}
=-\frac{1}{2e}\frac{(-\alpha)^{3/2}}{\beta^{1/2}}\,.
}

In the limit where $\kappa\ll1$ the magnetic field is much more localized than the 
superconducting condensate. It is screened at the length-scale $\lambda$, which is much 
smaller than the coherence length $\xi$. This means that the dominant contribution to 
the energy comes from the condensate $\psi$. As a result the boundary of width $\xi$ is 
approximated by 
\Align{Interface3}{
\sigma_{ns}(\kappa\ll1)&=\frac{\alpha^2}{2\beta}\int_{-\infty}^{+\infty}\Big(
4\xi^2|\tilde{\psi}|^{\prime 2}+(1-|\tilde{\psi}|^2)^2
\Big) dx \,,
}
where $\tilde{\psi}=\psi/\psi_0$. Substituing, $\tilde{\psi}=\tanh(x/2\xi)$, estimates 
the integral 
\Equation{Interface4}{
\sigma_{ns}(\kappa\ll1)\approx\frac{4\alpha^2}{3\beta}\xi
=\frac{(-2\alpha)^{3/2}}{3\beta}\,,
}
which is positive.

In the limit where $\kappa\gg1$, the interface energy \Eqref{Interface2} is negative, 
while it is positive in the limit $\kappa\ll1$ \Eqref{Interface4}. The crossover 
between positive and negative interface energies is obtained for the critical value 
$\kappa=1/\sqrt{2}$.
The sign of the interface energy determines the division between two classes of 
superconductors. The type-1 superconductors ($\kappa<1/\sqrt{2}$) have a positive 
interface energy ($\sigma_{ns}>0$), while the type-2 superconductors ($\kappa>1/\sqrt{2}$) 
have a negative interface energy ($\sigma_{ns}<0$). 

Physically this means that in type-1 superconductors there is an energy penalty for 
forming interfaces. This results in a preference for forming macroscopically large 
normal domains that minimize the length of the interface.
In type-2 superconductors, this is the opposite. Since the interface energy is negative, 
it is beneficial to have the maximal length of interface. And thus creating large number 
of small domains. These small domains are the vortices which are discussed in the next 
section.

\section{Quantization of the magnetic flux -- Vortices}
\label{Sec:Single-component:Quantization}

The Stokes' theorem implies that the flux of the magnetic field through a given area 
${\cal A}$ can be expressed as the line integral over the contour ${\cal C}$ bounding 
that area $\Phi=\int_{\cal A}\B\!\cdot\!d{\bs S}=\oint_{\cal C} \A\!\cdot\!d{\bs\ell}$. 
Given the definition of the current \Eqref{Eq:Ampere}, the vector potential $\A$
can be written in terms of the phase gradient $\Grad\varphi$ and the current $\J$. 
It follows that the magnetic flux reads as
\Equation{Flux0}{
   \Phi=\frac{1}{e}
    \bigointsss_{\!\!\!\!\!\!{\cal C}} 
    \left(\frac{\J}{e^2|\psi|^2}-\Grad\varphi_a\right)\!\cdot\!d{\bs\ell}\,.
} 
Since in the bulk of the superconductor, $|\psi|$ quickly converges to its ground state 
$\psi_0$ and that $\J$ decays exponentially fast due to Meissner screening, the magnetic 
flux through ${\cal A}$ reads as
\Equation{Flux1}{
\Phi=\frac{-1}{e}\bigointsss_{\!\!\!\!\!\!{\cal C}} 
    \Grad\varphi \cdot {d\bs{\ell}} \,.
}
Since $\psi$ has to be single-valued, the circulation of phase can only take values 
$2\pi n$ where $n$ is an integer. As a result, 
the associated flux is 
\Equation{Flux2}{
\Phi=\frac{-2\pi}{e}n=\Phi_0n \,.
}
The flux is thus quantized in units of $\Phi_0=-2\pi/e$, the flux quantum. 
The integer $n$, called the winding number, counts the number of times the phase 
$\varphi$ winds along the (large) contour ${\cal C}$.
The winding number, which is defined as a line integral over a closed path, is 
related to the maps $\groupS{1}\to\groupS{1}$. It is associated with the 
elements of the first homotopy of the circle: $n\in\pi_1(\groupS{1})=\groupZ{}$.

\begin{wrapfigure}{R}{0.33\textwidth}
\hbox to \linewidth{ \hss
\includegraphics[width=.975\linewidth]{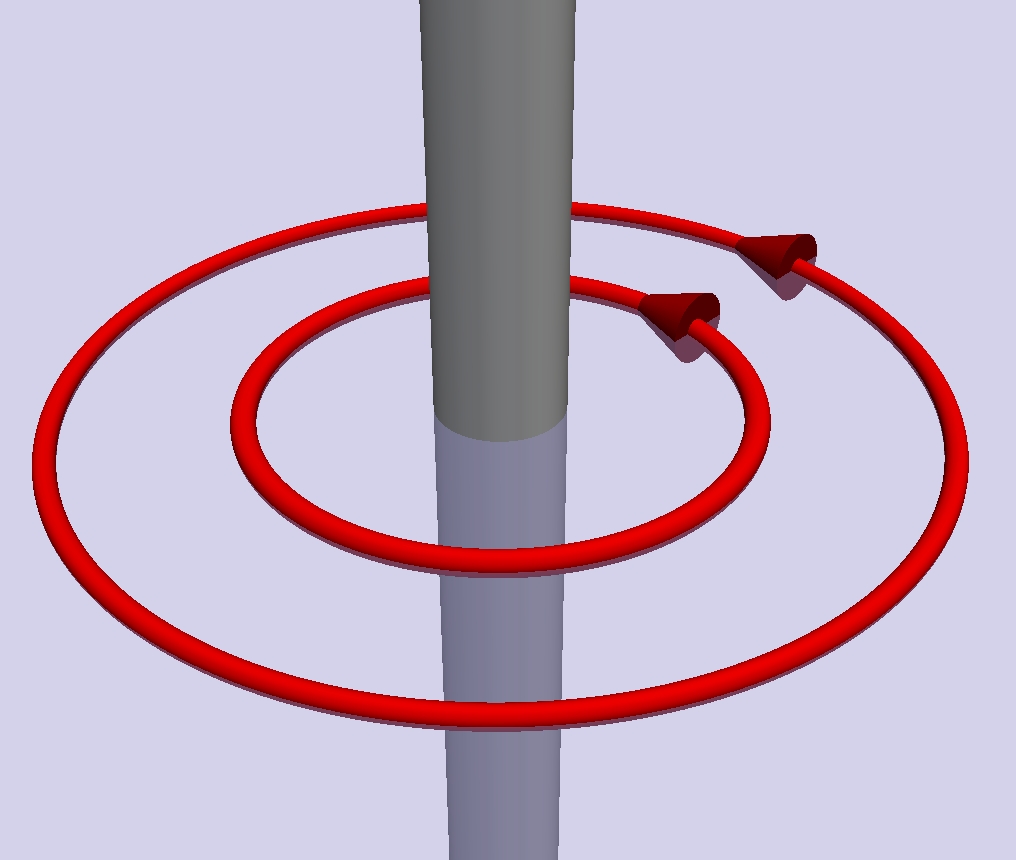}
\hss}
\caption{
Schematic illustration of a singular vortex line surrounded by screening currents.
} \label{Fig:Vortex}
\end{wrapfigure}

If a configuration have a non-zero phase winding, then the density ($|\psi|$) should 
vanish at the point around which the phase winds. The phase winding is compensated by 
having a non-trivial configuration of $\A$ that carries quantized flux. Thus, the
magnetic field cannot penetrate into the bulk of a superconductor without causing 
a singular vortex.

\begin{figure}[!htb]
\hbox to \linewidth{ \hss
\includegraphics[width=.85\linewidth]{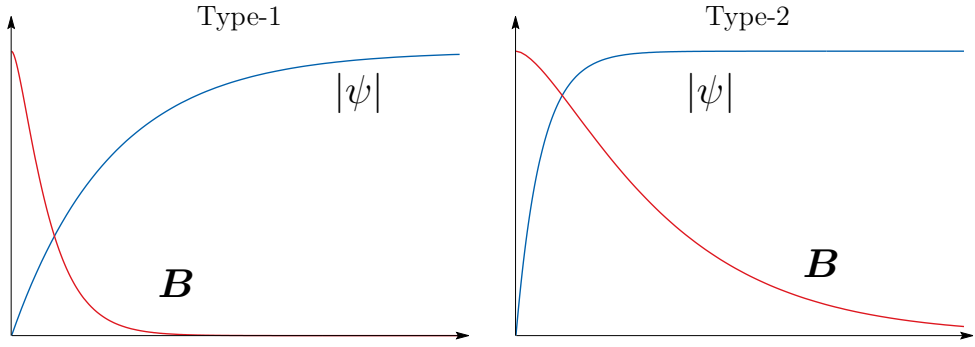}
\hss}
\caption{
The panels illustrate the vortex profiles in single-component type-1 and type-2 
regimes. 
}\label{Fig:Profiles-vortex}
\end{figure}

The finiteness of the energy imposes that at spatial infinity, the condensate has to be 
of constant modulus $\psi_0$ with an integer phase winding, and $\A$ has to be a pure gauge. 
Assuming axial symmetry, in cylindrical coordinates, the fields can be parametrized as
\Equation{Eq:ANO}{
\psi=f(\rho)\Exp{in\theta}\,,
~~~\text{and}~~~
\A=v(\rho)(\sin\theta,-\cos\theta) \,.
}
This is the Abrikosov-Nielsen-Olesen ansatz \cite{Abrikosov:57,Nielsen.Olesen:73} which 
simplifies the Ginzburg-Landau equation \Eqref{Eq:GLeq} and \Eqref{Eq:Ampere} to a set 
of two ordinary differential equations. Yet these equations are nonlinear, so they 
cannot, in general, be solved analytically. A typical vortex configuration is sketched 
on the figure \ref{Fig:Vortex}. A vortex is a singular line around which circulate Meissner 
currents, that screen the magnetic field away from the vortex. Typical vortex profiles 
in the type-1 and type-2 regimes are displayed in \Figref{Fig:Profiles-vortex}.

\subsection*{Vortex interaction: Type-I/type-II dichotomy}

As previously emphasized, depending on the value of the Ginzburg-Landau parameter 
$\kappa$, superconductors can classified into two different classes; type-1 and type-2. 
This classification followed from the different behaviour of interface energy between 
the normal and superconducting state. This dichotomy can also be understood by considering 
the interaction between vortices that should form in an external field.

In the case of type-2 superconductors, \ie for $\kappa > 1/\sqrt{2}$, the long-range 
interaction between vortices is dominated by the the magnetic interaction. It follows 
that vortices repel (as is also the case in the London limit $\kappa\gg1$). 
On the other hand, for type-1 superconductors, \ie for $\kappa < 1/\sqrt{2}$, 
the attractive interaction between the cores dominates the magnetic interactions 
at all separations. As a result, in the regime where $\xi>/\sqrt{2}\lambda$, 
separated vortices will collapse onto each other to form a vortex with a larger winding 
number (a megavortex). This leads to the formation of the macroscopically large domains 
of the normal state. Note that this is obviously consistent with the results from the 
interface energy where, when $\kappa<1/\sqrt{2}$ it is beneficial to minimize the 
interface area.
In the limiting case where $\kappa=1/\sqrt{2}$, \ie when $\xi=\sqrt{2}\lambda$, 
the vortices are non-interacting \cite{Kramer:73,Bogomolnyi:76}. More precisely, 
the repulsion due to the magnetic field exactly compensates the attraction of 
the cores. This non-interacting regime is called the Bogomol'nyi point of the 
phase diagram.

The properties of the vortex matter and their interactions, in the single-component 
Ginzburg-Landau for conventional superconductors, can thus be summarized as follows
\Itemize{
\item {\bf Type-1} : $\sqrt{2}\lambda<\xi$: The vortices attract each other to 
form macroscopically large domains.
\item {\bf BP} : $\sqrt{2}\lambda=\xi$: the vortices are non-interacting and 
all vortex superpositions have the same energy \cite{Kramer:73,Bogomolnyi:76}.
\item {\bf Type-2} : $\sqrt{2}\lambda>\xi$: The vortices repel each other to 
form a vortex lattice \cite{Abrikosov:57}.
}
The qualitative difference between the type-1 and type-2 regimes can be seen in the 
vortex profiles displayed in the \Figref{Fig:Profiles-vortex}.

\section{Phase diagrams and critical fields}

\begin{figure}[!htb]
\hbox to \linewidth{ \hss
\includegraphics[width=.50\linewidth]{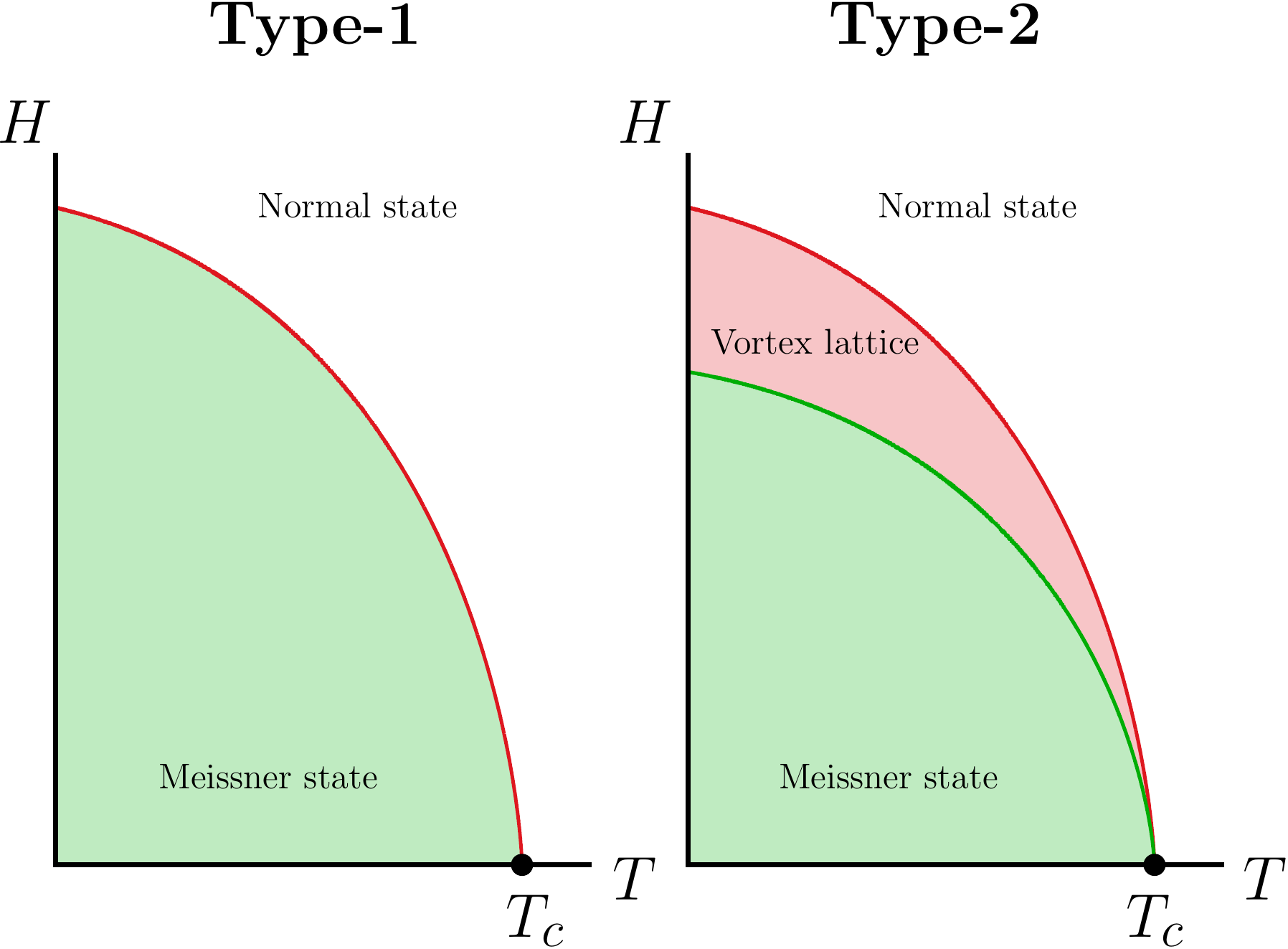}
\hfill
\includegraphics[width=.45\linewidth]{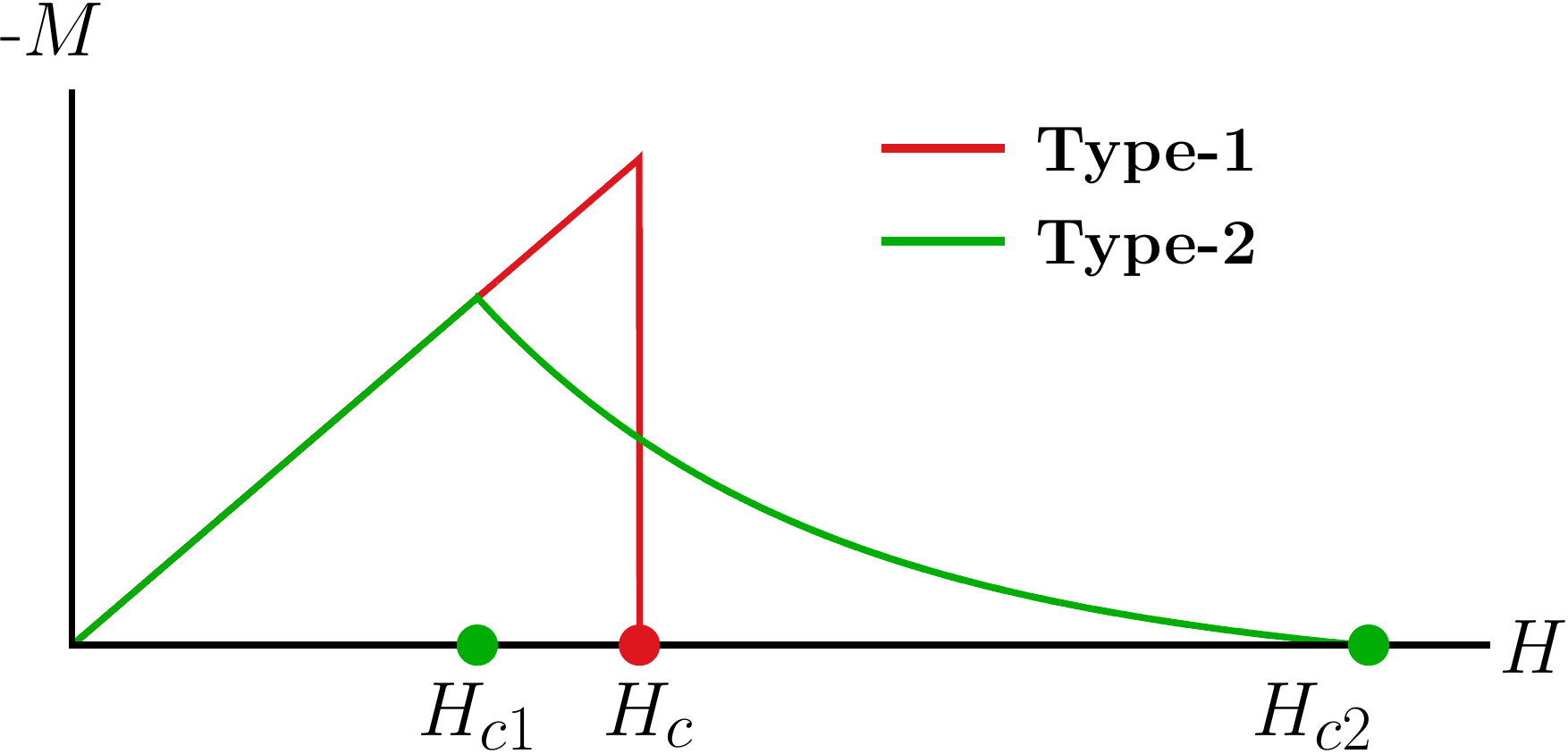}
\hss}
\caption{
Phase diagrams of the different regimes in single-component Ginzburg-Landau theory. 
The two $H-T$ diagrams sketch the difference between the response of type-1 and type-2 
superconductors to an external field $H$. As discussed below, in the type-2 regime, 
since the first critical field is smaller that the thermodynamical critical field, 
there exist an intermediate state between the Meissner state and the normal state. 
This intermediate state is that of a vortex lattice. 
The rightmost panel shows the difference of the magnetization response $M$ between 
type-1 and type-2. The type-1 regime features a first order phase transition between 
the superconducting state and the normal state. In the type-2 regime, at the level of 
mean-field theory, there are second order phase transitions between Meissner and vortex 
states and between vortex and normal states.
}
\label{Fig:Phase-Diagram}
\end{figure}

As stated earlier, in the presence of an applied external field $\He$, this is the 
Gibbs free energy \Eqref{Eq:Gibbs} that should be considered instead of the Helmholtz 
free energy $F$ \Eqref{Eq:GL}.
The response to an applied external magnetic field is summarized in the diagrams 
\Figref{Fig:Phase-Diagram}.

\subsection*{First critical field}

The first critical field $\Hc{1}$ is defined as the applied magnetic field at which the 
formation of a single vortex becomes energetically favourable. In other words, $\Hc{1}$ 
is the thermodynamic field at which the presence of vortices into the sample becomes 
energetically favourable as compared to the Meissner state.
It is defined as $\Hc{1}= E_v/\Phi_v$, where $E_v$ and $\Phi_v$ are
the energy and magnetic flux of the vortex. That is 
\Equation{AppHc1}{
\Hc{1}= \frac{\int (\F(\psi_v,\A_v)-\F(\psi_0,0))}
{\int\Curl\A}\,,
}
where $(\psi_0,0)$ denotes the ground state configuration of the superconducting 
condensates and $(\psi_v,\A_v)$, the configuration of a vortex. Note however that 
the corresponding external field $\He$ should be smaller than the thermodynamical 
critical magnetic field $\Hc{t}$ \Eqref{Eq:Condensation}
\Equation{AppHct}{
\Hc{t}=2\sqrt{\F(0,0) - \F(\psi_0,0) }		\,, 
}
above which it is no longer beneficial to form a superconducting condensate.

If the first critical field is smaller than the thermodynamical critical field 
$\Hc{1}<\Hc{t}$ (this is the case for type-2 and not for type-1), then vortices start 
to enter the superconductor. Thus in the type-2 regime, increasing the applied field, 
results in an intermediate state between the Meissner and normal states.

\subsection*{Second critical field}

When higher field is applied, more and more vortices enter the system. They repel each 
other with length-scale given by $\lambda$. When vortices are more and more packed, and 
vortex cores start to overlap. This results in an overall suppression of the density 
$|\psi|$, so that there is a certain value of the field that completely destroys the 
superconductivity. 
This value of the field is called the second critical field at which superconductivity is 
destroyed. Thus, close to the second critical field $\Hc{2}$, the magnetic field 
is approximately constant, $\B=B\ez=H\ez$ and the density $|\psi|$ is small. As a result, 
the Ginzburg-Landau equation can be linearized 
\Equation{EOMLinearized}{
   \D\D\psi=2\alpha\psi \,.
}
In the Landau gauge, the vector potential reads as $\A=(0,Hx,0)^{-1}$. There, the 
equation \Eqref{EOMLinearized} do not depend on $y$, so a convenient choice (in a 
first approximation) of variable separation is $\psi\equiv\psi(x)$. The linearized 
Ginzburg-Landau equation \Eqref{EOMLinearized} thus becomes the equation of an 
harmonic oscillator: 
\Equation{Harmonic}{
   \psi^{\prime\prime}-(eHx)^2\psi=2\alpha\Psi \,.
}
Thus the solution has the form $\psi(x)=C\exp\left(-\frac{x^2}{2\ell^2}\right)$ with the 
arbitrary constant $C$ and $eH=1/\ell^2$. The equation \Eqref{Harmonic} further simplifies 
\Equation{Hc2a}{
   e\Hc{2}=\frac{1}{\ell^2}\equiv-2\alpha=\frac{1}{\xi^2}	\,.
}
As a result, 
\Equation{Hc2b}{
   \Hc{2}=\frac{1}{e\xi^2}=\frac{\Phi_0}{2\pi\xi^2}	\,.
}
 Note that this means in particular that close 
to $\Hc{2}$, there is one (diverging) length-scale.

The thermodynamical critical field \Eqref{Eq:Condensation} is 
$\Hc{}^2=\alpha^2/\beta=e^{-2}\xi^{-2}\lambda^{-2}/2$. Hence the ratio of the second 
critical field with the thermodynamical critical field is 
\Equation{Hc2c}{
   \frac{\Hc{2}}{\sqrt{2}\Hc{}}=\frac{\lambda}{\xi}=\kappa	\,.
}
In terms of the critical fields, type-1 and type-2 superconductors 
have the phase diagram summarized in \Figref{Fig:Phase-Diagram}.
Type-1 and type-2 superconductors, thus have a different behaviour in an external field, 
summarized in the phase diagrams in \Figref{Fig:Phase-Diagram}. For type-1, the critical 
field simply destroys superconductivity to favour the normal state. In type-2 
superconductors, when the applied field $\He$ exceeds $\Hc{1}$ the magnetic field starts 
to penetrate the bulk superconductor in the form of quantized vortices.


\doPrint{ \newpage \thispagestyle{empty}\ \newpage }{ }
\graphicspath{{Plots/X-Numerics/}}
\chapter{Numerical methods}		\label{App:numerics}

Many of the results discussed in the main body rely on numerical simulations,  
but the focus is made on the physical properties and very few words are said 
about these simulations. Here is a detailed discussion of the numerical methods 
used to investigate the physics discussed in the main part.
This starts, in Section \ref{App:numerics:FEM}, with a general overview of the 
finite element methods used for the spatial discretization. Next, details of the 
algorithm used to solve the (nonlinear) Ginzburg-Landau problems are discussed in 
the Section~\ref{App:numerics:NLCG}. The important aspects of having a suitable choice 
of an initial guess, are detailed in the Section \ref{App:numerics:IG}. Finally, 
the question of the evolution of time-dependent problems is addressed in the Section 
\ref{App:numerics:CN}.

\section{Basic concepts for Finite Element methods}
\label{App:numerics:FEM}

There exists various approaches to address the spatial discretization of partial 
differential equations. The finite difference method, which is based upon local Taylor 
expansions to approximate the derivative, is most historical. This discretization 
technique covers the space with lattices which are topologically square or cuboid 
network. This method is rather intuitive, but it makes it difficult to handle complex 
geometries. This difficulty motivated the approach of the finite element method. 
The finite element methods have often been historically preferred for their 
rigorous provable stability, and because of their natural applicability to complex 
geometries. The relative advantages of both methods have been heavily debated, 
and it is fair to say that nowadays both methods lie more or less on the same ground. 
Both with their own advantages and difficulties. There exist many more different 
approaches, such as the spectral methods, which will not be exhaustively listed here. 
Finite differences or finite elements methods are more or less frequent, depending on 
the different scientific communities. For example, and very roughly speaking, finite 
differences are extensively used for simulations of lattice gauge theories, while 
finite elements are customary for example in engineering and mathematics.

The numerical investigations discussed in the main body, and the corresponding papers, 
extensively used finite element methods for a broad variety of problems including 
direct solving, minimization, constrained optimization, time-evolution, etc. In practice, 
the spatial discretization is handled within a the finite-element framework provided 
by the FreeFEM++ \footnote{ \url{https://freefem.org/}} library \cite{Hecht:12}.
There, the finite element methods are based one the weak formulation (the variational 
formulation) of partial differential equation. Below is presented a brief and non 
exhaustive description of the concepts used in finite element methods. Detailed 
introductions can be found in many textbooks, see for example \cite{Cook.Malkus.ea,
Hutton:03,Reddy:05,Allaire:05}.

\subsection{Finite-element formulation}
\label{Sec:numerics:FEM}

Consider the domain $\Omega$ which a bounded open subset of $\Real^d$ and denote 
$\partial\Omega$ its boundary. $W^{k,p}(\Omega)$ denotes the Sobolev space of order 
$k$ on $\Omega$. That is, the subset of functions $f$ in $L^p(\Omega)$ such that $f$ 
and their weak derivatives have a finite $L^p$-norm up to order $k$. The Hilbert space 
equipped with the $L^2$-norm is denoted $H^{k}(\Omega)=W^{k,2}(\Omega)$, and  
$\mathcal{H}^{k}_{\Omega}=\lbrace u+iv\,|\,u,v\in H^{k}(\Omega)\rbrace$ denotes the
Hilbert spaces of complex-valued functions. The inner products 
are denoted by $\langle\cdot,\cdot\rangle$, as for example:
\Equation{Eq:Inner}{
\ScalarProd{u}{v}=\int_\Omega uv  \,,~\text{for}~u,v\in H^{k}({\Omega}) 
\,,~~~~\text{and}~~~~
\ScalarProd{u}{v}=\int_\Omega u^*v\,,~\text{for}~u,v\in\mathcal{H}^{k}({\Omega}) 
\,.
}
Once the equations are formulated in their weak form, the appropriate Hilbert spaces  
have to be discretized using finite element spaces for a given partition $\mathcal{T}_h$ 
of $\Omega$.

\begin{wrapfigure}{R}{0.4\textwidth}
\hbox to \linewidth{ \hss
\includegraphics[width=.975\linewidth]{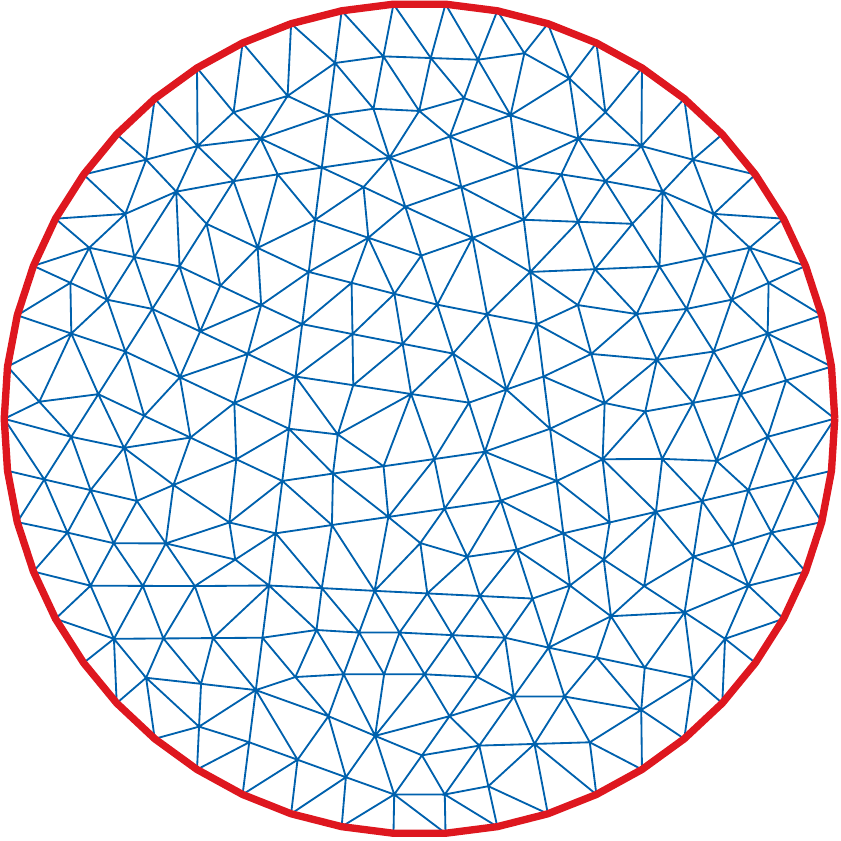}
\hss}
\caption{
Example of the mesh of a disc.
}
\label{Fig:Numerics:Mesh}
\end{wrapfigure}

More precisely, $\mathcal{T}_h$ is a regular partition of the domain $\Omega$ with 
$\Omega=\cup_{k=1}^{n_t}\Omega_k$. The name $\mathcal{T}_h$ refers to the family 
$\lbrace T_k\rbrace_{k=1,\cdots,n_t}$ of the $n_t$ triangles 
\footnote{
In the case of a three dimensional space, the mesh is composed of $n_t$ tetrahedra.
}
that compose the mesh,
as illustrated for example in \Figref{Fig:Numerics:Mesh}. Typically $h$ refers to the 
mesh size $h=\mathrm{max}_{\Omega_k\in\mathcal{T}_h}\lbrace\mathrm{diam}\,\Omega_k\rbrace$.
In practice, for a given boundary of the domain, the mesh can be automatically 
generated using for example the Delaunay-Voronoi algorithm. Next, given a spatial 
discretization, the functions are approximated to belong to a \emph{finite element 
space} whose properties correspond to the details of the Hilbert space to which 
the functions belong. A finite element space associated to a scalar function, say $w$,
is typically, a space of polynomial functions $\mathrm{P}^{(r)}$ of order $r$ on 
the elements (the triangles), with certain matching properties at edges, vertices, etc. 
$\mathrm{P}^{(r)}$ denotes the $r$-th order Lagrange finite-elements.
This describes a linear vector space of finite dimension, for which a basis can be 
found. The canonical basis is made of functions, called the hat functions $\phi_k$ 
and thus 
\Equation{Eq:FEM:space}{
V_h(\mathcal{T}_h,\mathrm{P}^{(r)})=\Big\lbrace w(\x)=\sum_{k=1}^M w_k\phi_k(\x)
,\phi_k(\x)\in \mathrm{P}^{(r)}
\Big\rbrace\,.
}
Here $M$ is the dimension of $V_h$ (the number of vertices), the $w_k$ are called the 
degree of freedom of $w$ and $M$ the number of the degrees of freedom.
Now for a given order of approximation of polynomial functions $\mathrm{P}^{(r)}$, 
the shape functions $\phi_k(\x)$ are constructed from the triangle $T_{k}$. For 
example $V_h(\mathcal{T}_h,\mathrm{P}^{(1)})$ denotes the space of continuous, piecewise 
linear functions of $x$, $y$ on each triangle of $\mathcal{T}_h$.
To summarize a given function is approximated as its decomposition:
\Equation{Eq:FEM:approx}{
w(\x)=\sum_{k=1}^M w_k\phi_k(\x)\,, 
}
on a given basis of $\phi_k(\x)$ that is chosen according to the mathematical 
prescriptions of the problem.

\subsection{A simple example: The Poisson equation}
\label{App:numerics:Poisson}

To illustrate the concept of weak formulation of partial differential equations, 
let consider the simple case of the Poisson equation with Dirichlet condition $u=0$ 
on $\partial\Omega$:
\Equation{Eq:FEM:Poisson:1}{
\nabla^2u=f\,.
}
The weak form, or variational form, is derived by multiplying \Eqref{Eq:FEM:Poisson:1} 
by an arbitrary test function $v$ and integrating over $\Omega$: 
\SubAlign{Eq:FEM:Poisson:2}{
-\int_\Omega v\nabla^2u+\int_\Omega vf =&0\\
\int_\Omega \Grad u\cdot\Grad v+\int_\Omega vf =&
\int_{\partial\Omega}v\Grad u\cdot {\bf n}	\,,
}
where the second line follows from the Green's formula (integration by parts), and 
${\bf n}$ is the outgoing vector normal to $\partial\Omega$. The test functions $v$ 
should satisfy the boundary condition $v=0$ on $\partial\Omega$, and thus the problem 
can be written as 
\Equation{Eq:FEM:Poisson:3}{
a(u,v)+\ell(f,v)=0 	}
with $a(u,v)=\ScalarProd{\Grad u}{\Grad v}$ and $\ell(f,v)=\ScalarProd{f}{v}$.
The weak form \Eqref{Eq:FEM:Poisson:3} defines a linear algebraic system. Indeed inserting 
a finite element decomposition, as in \Eqref{Eq:FEM:approx}, into \Eqref{Eq:FEM:Poisson:3}, 
the problem boils down to 
\Equation{Eq:FEM:Poisson:4}{
\sum_{k=1}^M A_{ik}u_k+L_i=0 \,,\text{with}~~L_i=\ScalarProd{f}{\phi_i(\x)}\,,
}
for all $i=0,\cdots,M-1$. $L_i$ is a vector and $A_{ik}$ is called the stiffness matrix. 
In a matrix notation, the Poisson equation problem \Eqref{Eq:FEM:Poisson:1} reads as 
\Equation{Eq:FEM:Poisson:5}{
\left[{\bs A}\right]\left[u_h\right]+\left[{\bs L}\right]=0\,,
}
whose solution simply requires the inversion of the stiffness matrix
$\left[{\bs A}\right]$:
\Equation{Eq:FEM:Poisson:6}{
\left[u_h\right]= -\left[{\bs A}\right]^{-1}\left[{\bs L}\right]\,.
}
Note that the stiffness matrix $\left[{\bs A}\right]$ is a sparse matrix, which can 
in principle be efficiently preconditioned. Variety of numerical tools and libraries 
are available to perform linear algebra operations such as the matrix inversion.

\subsection{A less simple example: The Amp\`ere-Maxwell equation}
\label{App:numerics:Maxwell}

As a further example, let's consider the Amp\`ere-Maxwell equation with the 
boundary condition $\B=\B_0$ on $\partial\Omega$: 
\Equation{Eq:FEM:Maxwell:1}{
\Curl\B+\J=0\,.
}
Again, the weak form is derived by multiplying 
\Eqref{Eq:FEM:Maxwell:1} by an arbitrary test function $\A^v$ and integrating 
over $\Omega$: 
\SubAlign{Eq:FEM:Maxwell:2}{
\int_\Omega \Curl\A^v\cdot\B+\int_{\partial\Omega}\A^v\times{\bf n}\cdot\B 
+\int_\Omega \A^v\cdot\J=&0 \\
\int_\Omega \Curl\A^v\cdot\Curl\A +\int_\Omega \A^v\cdot\J=&
\int_{\partial\Omega}{\bf n}\times\A^v\cdot\B_0 \,,\label{Eq:FEM:Maxwell:2:b}
}
where the magnetic field on the \rhs of \Eqref{Eq:FEM:Maxwell:2:b} has been replaced 
by the boundary condition $\B=\B_0$ (${\bf n}$ is the outgoing vector normal to 
$\partial\Omega$). The problem can be written as 
\Equation{Eq:FEM:Maxwell:3}{
a(\A,\A^v)+\ell(\J,\A^v)=0 	
}
with $a(\A,\A^v)=\ScalarProd{\Curl\A}{\Curl\A^v}$ and $\ell(\J,\A^v)=\ScalarProd{\J}{\A^v}$. 
Here, the scalar product acts on the vectorial space. For detailed discussions, see 
\eg \cite{Chen:97,Gao.Sun:15}. Again, inserting the finite element decomposition 
for $\A$ defines a linear system that can be solved similarly to the discussion in 
Sec.~\ref{App:numerics:Poisson}. Note that in the case of Ginzburg-Landau models, 
the current $\J$ depends on the gauge field $\A$.

\section{Minimization of the Ginzburg-Landau free energy}
\label{App:numerics:NLCG}

Many of the problems discussed in the main body require to numerically minimized 
the Ginzburg-Landau free energy. This is a non-trivial nonlinear optimization problem 
for a field theory. The choice of the discretization of the fields was usually done 
within the finite element formulation discussed in the previous section, an the choice 
for the optimization algorithm was typically the Non-Linear Conjugate Gradient method.
Below, are briefly introduced the principles of this method, while the actual formulation 
of the Ginzburg-Landau minimization problem is discussed after.

\subsection{The Nonlinear Conjuguate Gradient method} 

The NonLinear Conjugate Gradient method (NLCG) is a numerical method to solve iteratively 
unconstrained optimization problems. It is used to find a local minimum of a nonlinear 
function $f(\x)$ using only its gradient $\nabla_\x f$ (unlike the Newton-Raphson method 
that needs the second variation of the function). This method requires the function $f$ 
to be approximately quadratic close to the minimum. That is, the function should be 
twice differentiable and the second derivative should be non-singular, near the minimum
\cite{Fletcher:87,Gill.Murray.ea:81,Shewchuk:94,Nocedal.Wright:99}.
Given the function of $n$ variables, the conjugate gradient method aims at solving 
the nonlinear, unconstrained optimization problem
 \Equation*{
\min\{ f(\x) | \x\in\Real^n\}	\,,
}
where $f:\Real^n\mapsto\Real$ is continuously differentiable and bounded from below.  
Starting from an initial guess $\x_0\in\Real^n$, the nonlinear conjugate gradient 
method generates a recurrence 
\Equation*{
\x_{k+1}= \x_k + \alpha_k {\bs d}_k	\,.
}
Here, $\alpha_k$ is a (positive) step size obtained by a line search. The conjugate 
direction ${\bs d}_k$ is generated by the rule 
\Equation*{
{\bs d}_k = -{\bs g}_k + \beta_{k-1}{\bs d}_{k-1}	\,.
}
Here ${\bs g}_k$ is a column vector of the gradient of $f$ in all directions, 
${\bs g}_k= \nabla_\x f(\x_k)$. Finally $\beta_k$ is the conjugate gradient parameter 
that can be updated according to different rules \cite{Shewchuk:94}. The most common 
update rules are those of Fletcher-Reeves \cite{Fletcher.Reeves:64} and 
Polak-Ribi\`ere-Polyak \cite{Polak.Ribiere:69,Polyak:69}; these are defined as: 
\Equation{Eq:NLCG:beta}{
\beta_{k-1}^{FR}= \frac{ {\bs g}^T_{k}{\bs g}_{k} }{ {\bs g}^T_{k-1}{\bs g}_{k-1}}
\,,~~~\text{and}~~~
\beta_{k-1}^{PRP}= \frac{ {\bs g}^T_{k}({\bs g}_{k}-{\bs g}_{k-1})}
				   { {\bs g}^T_{k-1}{\bs g}_{k-1}}
				   \,.
} 
For other kind of updates of $\beta_k$ see, \eg, \cite{Shewchuk:94}. 
The whole algorithm can be summarized as 
\begin{algorithm}[H]
 \caption{NonLinear Conjugate Gradient algorithm for the optimization problem: 
	$\min\{ f(\x) | x\in\Real^n\}$ }\label{Algo:NLCG}
   \begin{algorithmic}[1]
	\State Initialize the recurrence  with the initial guess $\x_0$.
	\State Calculate the steepest direction ${\bs d}_0:=-{\bs g}_0=-\nabla_\x f(\x_0)$
	\State Find the adjustable step length $\alpha$ by performing a line 
		search in this direction \newline{}
		$\alpha_0=\argmin_\alpha f(\x_0+\alpha{\bs d}_0)$
		and generate $\x_{1}= \x_0 + \alpha_0 {\bs d}_0$
	\While{unconverged}	
		\State Calculate the steepest direction ${\bs g}_k=\nabla_\x f(\x_k)$
		\State Compute $\beta_k$ according to the Flecther-Reeves or Polak-Ribi\`ere-Polyak  
			formulas \Eqref{Eq:NLCG:beta}
		\State Update the conjugate direction: 
		${\bs d}_k = -{\bs g}_k + \beta_{k-1}{\bs d}_{k-1}$
		\State Perform a line search: optimize 
		$\alpha_k=\argmin_\alpha f(\x_k+\alpha{\bs d}_k)$
		\State Update the degrees of freedom 
		$\x_{x+1}= \x_k + \alpha_k {\bs d}_k$
		\State Check the convergence
	\EndWhile
   \end{algorithmic}
\end{algorithm}
\paragraph{The line search} to find the optimal size of the adjustable step
$\alpha_k=\argmin_\alpha f(\x_k+\alpha{\bs d}_k)$ can be done with different methods. 
Depending on the problem, it can sometimes be done with an exact line search by solving 
for example polynomial equations. Alternatively, the line search can be approximated 
numerically by methods like bisection. The optimal step size can more accurately be 
estimated using Wolfe conditions (see \eg \cite{Nocedal.Wright:99}). 

\vspace{1.5cm}

\paragraph{The convergence criterion} can be chosen in various ways. One can simply 
choose a maximum number of iteration, but this does not guarantee that an actual solution 
has been obtained. One criterion is to iterate until the gradient is smaller than a 
specified tolerance $\epsilon_\mathrm{tol}$: $|{\bs g}_k|<\epsilon_\mathrm{tol}$. 
Another choice is to put a condition on the relative value of the gradient
$|{\bs g}_k|/|{\bs g}_{k-1}|<\epsilon_\mathrm{tol}$. For detailed discussions about 
the appropriate choice of the convergence criterion, see \cite{Shewchuk:94}.

\subsection{Nonlinear Conjugate Gradients for Ginzburg-Landau}

Let $\Omega$ denote the superconductor where both the superconducting condensate $\psi$ 
and the vector potential $\A$ has value \footnote{
Note that in principle, the gauge field $\A$ is valued on the whole $\Real^3$ space. 
Thus in principle the gauge field part should take the infinite space into account, to 
accurately describe effects such as the stray fields. This is rather technical and 
these aspects shall be omitted here. 
}. 
In dimensionless quantities, the theory is described by the functional
\Equation{Eq:App:FreeEnergy}{
\F=\int_{\Omega}
\frac{\B^2}{2}
+\frac{1}{2}\left|\left(\Grad+ie\A \right)\psi \right|^2
+\alpha|\psi|^2+\frac{\beta}{2}|\psi|^4	\,,
}
where $e$ is the gauge coupling and $\alpha$ and $\beta$ are the potential parameters.
The Euler-Lagrange equations of motion obtained by varying the free energy functional with 
respect to the superconducting ($\psi$) and gauge ($\A$) degrees of freedom 
are the Ginzburg-Landau equations

\SubAlign{Eq:App:EOM}{
\D\D\psi&=2\left(\alpha+\beta|\psi|^2\right)\psi
\,,&&\text{with}~~~\D\equiv\left(\Grad+ie\A\right)
	\,, \label{Eq:App:GLeq}\\
\Curl\Curl\A+\J&=0\,,&&\text{with}~~~\J=e\Im(\psi^*\D\psi)=e|\psi|^2(\Grad\varphi+e\A) 
\,.\label{Eq:App:Maxwell}
}
%
It is convenient to write the superconducting degrees of freedom in terms of real and 
imaginary parts: $\psi=\psi_R+i\psi_I$. Now, multiplying by the test functions $\psi_R^v$, 
$\psi_I^v$ and $\A^v$ in $\Omega$ yields, after integration by parts, the weak form for 
the Ginzburg-Landau equations 
\SubAlign{Eq:App:GL:Weak}{
&\Grad\F_\psi:= \int_{\Omega}
	(\Grad\psi_R-e\A\psi_I)\cdot(\Grad\psi^v_R-e\A\psi^v_I)
   +(\Grad\psi_I+e\A\psi_R)\cdot(\Grad\psi^v_I+e\A\psi^v_R) \nonumber  \\
	&~~~~~~~~+\int_{\Omega}2\big(\alpha+\beta(\psi_R^2+\psi_I^2)\big)
					(\psi_R\psi_R^v+\psi_I\psi_I^v)
	\nonumber  \\
	&~~~~~~~~
	-\int_{\partial\Omega}\Big[\psi^v_R(\Grad\psi_R-e\A\psi_I)
	+\psi^v_I(\Grad\psi_I+e\A\psi_R)\Big]\cdot{\bf n}=0
	\, \label{Eq:App:GL:Weak:1}\\
& \Grad\F_\A:=\int_{\Omega}
	\Curl\A\cdot\Curl\A^v
+e\A^v\cdot\Big[ \psi_R(\Grad\psi_I+e\A\psi_R)
-\psi_I(\Grad\psi_R-e\A\psi_I)\Big]
	\nonumber  \\
	&~~~~~~~~
	-\int_{\partial\Omega}\A^v\times {\bf n}\cdot \B_0=0
	\label{Eq:App:GL:Weak:2} \, 
}
where $\B_0$ is an external field. Given the inner products \Eqref{Eq:Inner}
the equations \Eqref{Eq:App:GL:Weak} can be rewritten as
\SubAlign{Eq:App:GL:WeakScal}{
&\Grad\F_\psi:= 
	\ScalarProd{\Grad\psi_R-e\A\psi_I}{\Grad\psi^v_R-e\A\psi^v_I}
   +\ScalarProd{\Grad\psi_I+e\A\psi_R}{\Grad\psi^v_I+e\A\psi^v_R} \nonumber  \\
	&~~~~~~~~
	+\ScalarProd{2\big(\alpha+\beta(\psi_R^2+\psi_I^2)\big)\psi_R}{\psi_R^v}
	+\ScalarProd{2\big(\alpha+\beta(\psi_R^2+\psi_I^2)\big)\psi_I}{\psi_I^v}
	=0\label{Eq:App:GL:WeakScal:1}
\\
& \Grad\F_\A:=
	\ScalarProd{\Curl\A}{\Curl\A^v} + \ScalarProd{e^2|\psi|^2\A}{\A^v} 
+\ScalarProd{e(\psi_R\Grad\psi_I-\psi_I\Grad\psi_R)}{\A^v}
=0	\label{Eq:App:GL:WeakScal:2} \, .
}
The Ginzburg-Landau equation is obviously nonlinear, thus it cannot be directly formulated 
as a linear system as in for example Sec.~\ref{App:numerics:Poisson}. The Amp\`ere-Maxwell 
equation \Eqref{Eq:App:GL:WeakScal:2}, on the other hand, is bilinear in $\A, \A^v$. 
Hence, given a configuration of the superconducting degrees of freedom $\psi$, it can be 
solved as a linear system, as in Sec.~\ref{App:numerics:Maxwell}.
This do not take into account the backreaction of the gauge field, on the superconducting 
degrees of freedom. So this cannot give a solution of the Ginzburg-Landau equations. 
However, this can be used to provide an initial configuration of the gauge field 
that satisfies the Maxwell equation {\it on-shell}.

\vspace{1.5cm}

The two equations \Eqref{Eq:App:GL:WeakScal}, describe the gradient of the free energy 
with respect to all degrees of freedom. The set of degrees of freedom can be cast in 
$\x:=\{\psi_R,\psi_I,\A\}$, and the gradient, which can be seen as the Fr\'echet 
derivative of $\F$, becomes $\nabla_\x\F:=\{\Grad\F_\psi,\Grad\F_\A \}$. From this, 
given an appropriate initial configuration, the nonlinear conjugate gradient 
Algorithm~\ref{Algo:NLCG} can straightforwardly be used to minimize the free energy. 
In a nutshell, the NLCG Algorithm~\ref{Algo:NLCG} for the Ginzburg-Landau energy, 
is summarized in Algorithm~\ref{Algo:NLCG-GL}. 

\begin{algorithm}[H]
 \caption{NonLinear Conjugate Gradient algorithm for Ginzburg-Landau}\label{Algo:NLCG-GL}
   \begin{algorithmic}[1]
	\State Define an initial guess $\psi_0$ for the superconducting degrees of freedom
		   (see discussion in Sec.~\ref{App:numerics:IG})
 		\newline{}
		   Find the solution $\A_0$ of the Amp\`ere-Maxwell equation 
		   \Eqref{Eq:App:GL:WeakScal:2} with a linear solver
 		\newline{}
		Initialize the recurrence with $\x_0=\{\psi_{R,0},\psi_{I,0},\A_0\}$.
	\State Calculate the steepest direction ${\bs d}_0:=-{\bs g}_0=-\nabla_\x \F(\x_0)$, 
		with $\nabla_\x\F:=\{\Grad\F_\psi,\Grad\F_\A \}$
	\State Find the adjustable step length $\alpha$ by performing a line 
		search in this direction \newline{}
		$\alpha_0=\argmin_\alpha \F(\x_0+\alpha{\bs d}_0)$
		and generate $\x_{1}= \x_0 + \alpha_0 {\bs d}_0$
	\While{unconverged}	
		\State Calculate the steepest direction ${\bs g}_k=\nabla_\x \F(\x_k)$
			with $\nabla_\x\F:=\{\Grad\F_\psi,\Grad\F_\A \}$
		\State Compute $\beta_k$ according to Flecther-Reeves or Polak-Ribi\`ere-Polyak  
			formulas \Eqref{Eq:NLCG:beta}
		\State Update the conjugate direction: 
		${\bs d}_k = -{\bs g}_k + \beta_{k-1}{\bs d}_{k-1}$
		\State Perform a line search: optimize 
		$\alpha_k=\argmin_\alpha f(\x_k+\alpha{\bs d}_k)$
		\State Update the degrees of freedom 
		$\x_{x+1}= \x_k + \alpha_k {\bs d}_k$
		\State Check the convergence
	\EndWhile
   \end{algorithmic}
\end{algorithm}

The whole technical discussion here about the finite-element formulation, and the 
numerical minimization of the Ginzburg-Landau free energy has been done for a 
single-component model. The extension of all this machinery to multiple superconducting 
condensates is rather straightforward. The last key ingredient is now the initial guess 
for the superconducting degrees of the freedom.

\subsection{Initial guess for the Ginzburg-Landau energy minimization}
\label{App:numerics:IG}

A crucial step for the minimization is the choice of the initial guess. A wise initial 
guess not only improves the convergence of the algorithm, but also makes it possible 
to imprint some of the desired properties. This is a crucial step when numerically 
studying topological excitations. The key idea is that for the minimization algorithm 
to converge to a configuration that has the desired topological properties, the starting 
configuration itself should exhibit these properties. The rationale is the following: 
different topological sectors are typically separated by big energy barriers, so gradient 
minimization easily converges to a configuration that has the same topological properties 
than the starting configuration
\footnote{
The infinite energy barriers that separate the different topological sectors are usually 
defined in infinite space. Moreover the spatial discretization also limits the arguments 
from the topology. So, strictly speaking, in finite domains, there are only finite energy 
barriers between different topological sectors.
}.

As explained previously the gauge degrees of freedom can be found easily by solving 
the Amp\`ere-Maxwell equation \Eqref{Eq:App:GL:WeakScal:2}, given a configuration of 
the superconducting degrees of freedom $\Psi$. Two qualitatively different excitations 
are present in the various examples discussed in the main body: vortices and domain-walls.
As in the main body, the superconducting degrees of freedom are cast into the $N$-component 
complex field $\Psi$, such that $\Psi^\dagger=(\psi_1^*,\cdots,\psi_N^*)$. 
The starting field configuration can be expressed as the product of two fields that 
carry the information about the qualitatively different topological excitations:
\Equation{Eq:numerics:Guess}{
\Psi=\Psi^{(v)}\Psi^{(dw)}\,.
}
Here $\Psi^{(v)}$ contains the informations about vortices, and $\Psi^{(dw)}$ encodes 
the  informations about domain-walls.

\paragraph{Ground-state:} In the ground-state $\Psi_0$, the superconducting condensates 
are constant, while $\A$ is a pure gauge and can be chosen to be zero. As discussed in 
the main body, for example in Sec.~\ref{Sec:TRSB:GS}, in practice it is always better 
to work with real and imaginary parts $\psi_a=X_a+iY_a$, rather than moduli and phase.
 The real and imaginary parts $X_a,Y_a$ of the ground-state condensate are configurations 
that minimize the potential energy $V$. They should thus be extrema
\Equation{Eq:Extrema}{
 \frac{dV}{dX_a}=0~~\text{and}~~
 \frac{dV}{dY_a}=0\,.
}
The solutions of this system, except under very special conditions, cannot be solved 
analytically. They determine the ground-state densities $u_a$ and phases $\bvarphi_a$ according to
\Equation{Eq:numerics:GS}{
   u_a=X_a^2+Y_a^2~~~\text{and}~~~
   \bvarphi_a=\arctan\left(Y_a/X_a\right)\,.
}
The ground-state is thus the complex vector that reads as 
$\Psi_0^\dagger=(u_1\Exp{-i\bvarphi_1},\cdots,u_N\Exp{-i\bvarphi_N})$.
For the extrema to be a minimum, the Hessian matrix should 
have only positive eigenvalues:
\Equation{Eq:numerics:GS:Stability}{
 H\Psi_n=\lambda_n\Psi_n\,,~\lambda_n>0~~\text{with}~~
 H_{ab}=\left.\frac{d^2V}{df_adf_b}\right|_{\Psi_0}~~,~\text{and}~~f_a=X_a,Y_a\,.
}
Note that the physical length-scales, are determined by the stability condition of the 
obtained ground-state. That is, the mass of the various excitations are determined by 
the eigenvalues of the Hessian matrix.

\paragraph{Starting guess for vortices:}

The initial field configuration with $n^{a}_v$ vortices in a given condensate $\psi_a$ 
(where $a=1,\cdots N\,$) is prepared by using an ansatz which imposes phase windings 
around each of the spatially separated $n^{a}_v$ vortex cores 
\Equation{Eq:numerics:Guess:Vortex1}{
\psi_a^{(v)}= |\psi_a^{(v)}|\mathrm{e}^{ i\Theta_a(x,y)} \, ,~~~\text{with}~~~  
|\psi_a^{(v)}| =\prod_{k=1}^{n_v} 
\sqrt{\frac{1}{2} \left( 1+\tanh\left(\frac{4}{\xi_a}({\cal R}^a_k(x,y)-\xi_a) 
\right)\right)}\, .  
}
Here, $\xi_a$ parametrizes the size of the core while the total phase $\Theta_a(x,y)$ and 
the distance ${\cal R}^a_k(x,y)$ from the position $(x^a_k,y^a_k)$ of the core of the 
$k$-th vortex of the $a$-condensate, are defined as
\Equation{Eq:numerics:Guess:Vortex2}{
\Theta_a(x,y)=\sum_{k=1}^{n_v}
      \tan^{-1}\left(\frac{y-y^a_k}{x-x^a_k}\right)  \,,~~~\text{and}~~~
{\cal R}^a_k(x,y)=\sqrt{(x-x^a_k)^2+(y-y^a_k)^2}\,. 
}
Here, $\psi_a^{(v)}$ is not normalized to its ground-state value $u_a$ 
\Eqref{Eq:numerics:GS}. The normalization to the appropriate ground-state density is 
carried out by $\Psi^{(dw)}$ that carries the informations about domain-walls.

Remark that, in all generality, the numbers $n^{a}_v$ of the vortices in a given 
condensate $\psi_a$ can be different. However, as explained in details in 
Sec.~\ref{Sec:Factional-vortices}, the finite energy considerations dictate that 
the different condensates should have the same number of vortices $n^{a}_v=n_v$.

Note also that many of the new topological defects discussed in the main body are 
characterized by non-overlapping cores. It is thus a good strategy to chose different 
positions $(x^a_k,y^a_k)$ for the core of $k$-th vortex of the $a$-condensate. Indeed, 
it may be difficult for the minimization algorithm to split the cores. On the other hand 
if there are no mechanism to enforce the core splitting, it is easy for the minimization 
algorithm to re-unite the fractional vortices because of the long-range attraction.

\paragraph{Starting guess for domain-walls}
The domain-walls are topological excitations that are associated with the spontaneous 
breakdown of a discrete symmetry. These are field configurations that interpolate 
between inequivalent ground-states that are disconnected.
If there exist two disconnected ground states, say $\Psi_0$ and $\Psi_0^\prime$, the 
configuration that interpolates between the two inequivalent ground-states can be 
parametrized as follow: 
\Equation{Eq:numerics:DW}{
\Psi^{(dw)}=\frac{\Psi_0+\Psi_0^\prime}{2} 
	+\frac{\Psi_0-\Psi_0^\prime}{2}
	\tanh\left( \frac{\x_\perp-\x_0}{\xi^{(dw)}} \right)
\,,
}
where $\xi^{(dw)}$ determines the width of the domain-wall. In \Eqref{Eq:numerics:DW}, 
$\x_0$ is the curvilinear abscissa that determines the position of the domain-wall, 
and $\x_\perp$ is the coordinate perpendicular to the domain-wall.

In the main body, say in Sec.~\ref{Sec:TRSB:Domain-walls} we have been interested in 
domain-walls that interpolate between regions with inequivalent relative phases between 
the condensates. When the ground-state breaks time-reversal symmetry, its complex 
conjugate is not a gauge equivalent. That is, there exists no real number $\chi_0$ such 
that $\Psi_0^*=\Exp{i\chi_0}\Psi_0$. If no such transformation exists, then 
$\Psi_0^*\not\equiv\Psi_0$ and the configurations with $\bvarphi_a$ and $-\bvarphi_a$ 
are disconnected and degenerate in energy. Domain-walls that interpolate between 
$\Psi_0$ and $\Psi_0^*$ are topologically protected as their unwinding would require 
to overcome an infinite energy barrier, see for example textbook discussion in 
\cite{Manton.Sutcliffe,Rajaraman}. The domain-wall that interpolates between $\Psi_0^*$ 
and $\Psi_0$ can thus be parametrized by: 
\Equation{Eq:PhaseDW}{
\psi_a^{(dw)}=u_a\exp\left[i\bar{\varphi}_a\tanh\left( 
\frac{\x_\perp-\x_0}{\xi^{(dw)}_a} \right)\right]\,.
}
If there are no domain-walls, the $\psi_a^{(dw)}=u_a$ is simply the ground-state density.

The knowledge of the ground-state properties $\Psi_0$, together with the domain-walls 
$\Psi^{(dw)}$ and vortex $\Psi^{(v)}$ contents, thus prepares an initial configuration 
of the superconducting degrees of freedom $\Psi$ \Eqref{Eq:numerics:Guess} that has 
the desired topological content. Solving the Amp\`ere-Maxwell equation 
\Eqref{Eq:App:GL:WeakScal:2} for $\Psi$ gives a gauge field $\A$ that satisfies the 
gauge part \emph{on-shell}. This together makes an initial field configuration that is 
suitable for the minimization of the Ginzburg-Landau energy using the Non-Linear Conjugate 
Gradient Algorithm~\ref{Algo:NLCG-GL}. This typically converges rather fast. 

Note that it is often necessary to rely on parallel computing, so that the numerical 
calculations, are efficient. Simple parallelization requires parallel linear algebra 
distributed solvers, and the knowledge of the Message Passing Interface (MPI). There are 
also very involved elegant techniques such as the Domain Decomposition Methods (DDM)
\cite{Toselli.Widlund,Dolean.Jolivet.ea,Jolivet.Dolean.ea:12}. These very technical 
aspects are not discussed further here.

\section{Time evolution: forward extrapolated Crank-Nicolson}
\label{App:numerics:CN}

Most of the problems discussed in the main body require the numerical minimization 
techniques for non-linear problems discussed in Sec.~\ref{App:numerics:NLCG}. 
Sometimes it is also important to know the dynamical (time-evolution) properties of 
the system. For example, the time-dependent Ginzburg-Landau equation (see \eg 
\cite{Schmid:66,Gorkov.Eliashberg:68,Gorkov.Kopnin:75}), was important to investigate 
the thermoelectric properties of non-stationary processes in 
\CVcite{Garaud.Silaev.ea:16}. Here is detailed an algorithm used for time-evolution 
of (non)linear systems, the Crank-Nicolson algorithm \cite{Crank.Nicolson:47}. 
It is a finite difference implicit method in the time dimension that was for example 
originally designed to simulate the time-evolution of the heat equation 
\cite{Crank.Nicolson:47}. Afterwards, it was also used in framework of the non-linear 
Schr\"odinger equation \cite{Delfour.Fortin.ea:81}. It is also used to investigate 
the dynamics of superfluids and Bose-Einstein condensates of ultracold atoms 
\cite{Bao.Cai:18}. Here, the details of this algorithm, are discussed in the context 
of the Gross-Pitaevskii equation. A slightly different 
code is necessary for superconductors in order to account for the gauge field as in 
\CVcite{Garaud.Silaev.ea:16}. Yet it can be straightforwardly adapted to Ginzburg-Landau 
problems. See for example, the related works \cite{Chen:97,Gao.Sun:15,Gao.Li.ea:14}.

The algorithm presented here is a semi-implicit version that accounts efficiently for the 
non-linearities. More precisely, for efficient calculations, the nonlinear part is 
linearized using a forward Richardson extrapolation. 
The time-dependent Gross-Pitaevskii equation reads as:
\Equation{Eq:numerics:TDGP:0}{
i\partial_t\psi=-\frac{1}{2}\nabla^2\psi +\left[V(\x)+g|\psi|^2\right]\psi \,.
}
The weak from is obtained by multiplying by test functions $\psi_w\in\mathcal{H}(\Omega)$ 
and by integrating by parts the Laplace operator. In terms of the inner products 
\Eqref{Eq:Inner}, the weak form of \Eqref{Eq:numerics:TDGP:0} reads as
\Equation{Eq:numerics:TDGP:1}{
\ScalarProd{\psi_w}{i\partial_t\psi} =\frac{1}{2}\ScalarProd{\Grad\psi_w}{\Grad\psi}
+\ScalarProd{\psi_w}{\left[V(\x)+g|\psi|^2\right]\psi}\,.
}
The time is discretized as $t=k\Delta t$ and the wave function at the step $k$ is 
$\psi_k:=\psi(k\Delta t)$. This turns the continuous evolution into a recursion
over the uniform partition $\lbrace t\rbrace_{k=0}^{N_t}$ of the time variable. 
The Crank-Nicolson scheme, uses the Forward-Euler definition of the time derivative, 
while the \emph{r.h.s} of Eq.~\Eqref{Eq:numerics:TDGP:1} is evaluated at the averaged 
times. Using the following notations 
\Equation{Eq:numerics:TDGP:Notations}{
\partial_t\psi :=\delta_t\psi_k = \frac{\psi_{k+1} - \psi_{k}}{\Delta t}
\,,~~\text{and}~~
\bar{\psi}_k = \frac{\psi_{k+1} + \psi_{k}}{2}\,,
}
the Crank-Nicolson scheme for the Gross-Pitaevskii equation \Eqref{Eq:numerics:TDGP:1} 
reads as:
\Equation{Eq:numerics:TDGP:2}{
\ScalarProd{\psi_w}{i\delta_t\psi_k} =
\frac{1}{2}\ScalarProd{\Grad\psi_w}{\Grad\bar{\psi}_k}
+\ScalarProd{\psi_w}{V(\x)\bar{\psi}_k}
+\ScalarProd{\psi_w}{g|\bar{\psi}_k|^2\bar{\psi}_k}	\,.
}
This fully implicit scheme results in a nonlinear algebraic system which is 
difficult to solve. Alternatively the nonlinear part can be approximated in terms 
of the values at previous time steps. Namely, the fields in the nonlinear term are 
approximated by using an extrapolation of the previous time steps. This extrapolation 
should retain the same order of truncation error as the rest of time series. Using the 
forward extrapolation, the averaged wave function in the non-linear term becomes 
$\bar{\psi}_k\approx(3\psi_{k}-\psi_{k-1})/2$. 
By, defining the time-discretized operators: 
\SubAlign{Eq:numerics:TDGP:Operators}{
\Op_1\psi&=\ScalarProd{\psi_w}{\frac{i\psi}{\Delta t}} 
-\ScalarProd{\Grad\psi_w}{\frac{1}{4}\Grad\psi}
-\ScalarProd{\psi_w}{\frac{1}{2}V(\x)\psi}
\,, \\
\Op_2\psi&=\ScalarProd{\psi_w}{\frac{i\psi}{\Delta t}} 
+\ScalarProd{\Grad\psi_w}{\frac{1}{4}\Grad\psi}
+\ScalarProd{\psi_w}{\frac{1}{2}V(\x)\psi}
\,, \\%
\U_k\psi&=\ScalarProd{\psi_w}{\frac{g}{8}|3\psi_{k}-\psi_{k-1}|^2\psi}
\,,
}
the Eq.~\Eqref{Eq:numerics:TDGP:2} can be written in a compact form as
\Equation{Eq:numerics:TDGP:3}{
\Op_1\psi_{k+1}=\Op_2\psi_{k} + \U_n(3\psi_{k}-\psi_{k-1})\,.
}
Hence the time-evolution is formally given by the recursion 
\Equation{Eq:numerics:TDGP:OP}{
\psi_{k+1}=\Op_1^{-1}\big[\Op_2\psi_{k} + \U_k(3\psi_{k}-\psi_{k-1})\big]\,.
}
Next, as discussed in Sec.~\ref{Sec:numerics:FEM}, the spatial discretization is achieved 
by replacing the wave function $\psi$ with its finite-element space representation 
$\psi^{(h)}\in V_h(\mathcal{T}_h,\mathrm{P}^{(2)})$ in the Gross-Pitaevskii equation 
\Eqref{Eq:numerics:TDGP:OP}. The time-discretized evolution operators 
\Eqref{Eq:numerics:TDGP:Operators} thus become matrices: 
\Equation{Eq:numerics:TDGP:Matrices}{
\Op_1\mapsto{\bs M}_{\psi} \,,~~~
\Op_2\mapsto{\bs N}_{\psi} \,,~~~\text{and}~~~~
\U_k(3\psi_{k}-\psi_{k-1})\mapsto{\bs L}_{\psi} \,, 
}
and the recursion Eq. \Eqref{Eq:numerics:TDGP:OP} reduces to a linear algebraic system 
that reads as: 
\Equation{Eq:TDGL:modLCN:Matrix}{
\left[{\bs M}_{\psi}\right]\left[\psi_{k+1}^{(h)}\right] 
-\left[{\bs N}_{\psi}\right]\left[\psi_k^{(h)}\right]
=\left[{\bs L}_\psi\right]	\,.
}
The vector ${\bs L}_\psi$ which is a function of $\psi_k^{(h)}$ and $\psi_{k-1}^{(h)}$,  
has to be recalculated at each time step. On the other hand, the matrices 
${\bs M}_{\psi}$ and ${\bs N}_{\psi}$ are constant over time. They thus need to be 
allocated just once. Finally the recursion is thus 
\Equation{Eq:TDGL:modLCN:Solve}{
\left[\psi_{k+1}^{(h)}\right]=
\left[{\bs M}_{\psi}\right]^{-1}\left(
\left[{\bs N}_{\psi}\right]\left[\psi_k^{(h)}\right]
+\left[{\bs L}_\psi\right]
\right)
\,.
}
The numerical simulations of the time-evolution algorithm \Eqref{Eq:TDGL:modLCN:Solve} 
accurately reproduce the expected properties of the Gross-Pitaevskii equation. For 
example, it preserves the conserved quantities like the energy, the angular momentum, 
and the norm of the wave-function. 
For superconductors, on the other hand, there are no such conserved quantities, as the 
dynamics is dissipative. Yet the algorithm discussed here can easily be upgraded to 
include gauge degrees of freedom.

\doPrint{ \newpage \thispagestyle{empty}\ \newpage }{ }
\chapter{Author's publications discussed in this report}		
\label{Chap:Papers}

\vspace*{-0.75cm}

\newcounter{PaperList} 
\setcounter{PaperList}{1}
\newcommand{\Bibitem}[1]{\item[\namedlabel{#1}{\arabic{PaperList}}.~]
						  \addtocounter{PaperList}{1}}
\begin{itemize}

\Bibitem{CV:Grinenko.Weston.ea:21} 
	V.~Grinenko, D.~Weston, F.~Caglieris, C.~Wuttke, C.~Hess, T.~Gottschall, 
	I.~Maccari, D.~Gorbunov, S.~Zherlitsyn, J.~Wosnitza, A.~Rydh, K.~Kihou, 
	C.-H.~Lee, R.~Sarkar, S.~Dengre,  {\bf{J.~Garaud}}, A.~Charnukha, 
	R.~H\"uhne, K.~Nielsch, B.~B\"uchner, H.-H.~Klauss, and E.~Babaev \\
	\href{http://dx.doi.org/10.1038/s41567-021-01350-9}
	{State with spontaneously broken time-reversal symmetry above superconducting 
	phase transition} \\
	{\it{Nat. Phys.}} {\bf{17}}, 1254 (2021).  \hfill 
	{[cond-mat.supr-con] arXiv:\href{https://arxiv.org/abs/2103.17190}{2103.17190}}

\Bibitem{CV:Garaud.Dai.ea:21} 
	{\bf{J.~Garaud}}, J.~Dai and A.~J.~Niemi\\
	\href{https://doi.org/10.1007/JHEP07(2021)157}
	{Vortex precession and exchange in a Bose-Einstein condensate} \\
	{\it{J.~High~Energ.~Phys.}} {\bf{2021}}, 157 (2021). \hfill 
	{[cond-mat.quant-gas] arXiv:\href{https://arxiv.org/abs/2010.04549}{2010.04549}}

\Bibitem{CV:Krohg.Babaev.ea:21} 
	F.~N.~Krohg, E.~Babaev, {\bf{J.~Garaud}}, H.~H.~Haugen, and A.~Sudb\o \\
	\href{http://dx.doi.org/10.1103/PhysRevB.103.214517}
	{Thermal fluctuations and vortex lattice structures in chiral $p$-wave 
	superconductors: \\ robustness of double-quanta vortices} \\
	{\it{ Phys.~Rev.~B}} {\bf{103}}, 214517 (2021). \hfill 
	{[cond-mat.supr-con] arXiv:\href{https://arxiv.org/abs/2007.09161}{2007.09161}}

\Bibitem{CV:Rybakov.Garaud.ea:19} 
	F.~N.~Rybakov, {\bf{J.~Garaud}} and E.~Babaev\\
	\href{http://dx.doi.org/10.1103/PhysRevB.100.094515}
	{Stable Hopf-Skyrme topological excitations in the superconducting state} \\
	{\it{ Phys.~Rev.~B}} {\bf{100}}, 094515 (2019). \hfill 
	{[cond-mat.supr-con] arXiv:\href{https://arxiv.org/abs/1807.02509}{1807.02509}}

\Bibitem{CV:Garaud.Corticelli.ea:18a} 
	{\bf{J.~Garaud}}, A.~Corticelli, M.~Silaev and E.~Babaev\\
	\href{http://dx.doi.org/10.1103/PhysRevB.98.014520}
	{Properties of dirty two-bands superconductors with repulsive 
	interband interaction: \\
	normal modes, length scales, vortices and magnetic response} \\
	{\it{ Phys.~Rev.~B}} {\bf{98}}, 014520 (2018). \hfill 
	{[cond-mat.supr-con] arXiv:\href{https://arxiv.org/abs/1802.07252}{1802.07252}}

\Bibitem{CV:Garaud.Corticelli.ea:18} 
	{\bf{J.~Garaud}}, A.~Corticelli, M.~Silaev and E.~Babaev\\
	\href{http://dx.doi.org/10.1103/PhysRevB.97.054520}
	{Field-induced coexistence of $s_{++}$ and $s_\pm$ superconducting states \\
	in dirty multiband superconductors} \\
	{\it{ Phys.~Rev.~B}} {\bf{97}}, 054520 (2018). \hfill 
	{[cond-mat.supr-con] arXiv:\href{https://arxiv.org/abs/1712.09273}{1712.09273}}

\Bibitem{CV:Zyuzin.Garaud.ea:17} 
	A.~A.~Zyuzin, {\bf{J.~Garaud}} and E.~Babaev\\
	\href{http://dx.doi.org/10.1103/PhysRevLett.119.167001}
	{Nematic skyrmions in odd-parity superconductors} \\
	{\it{ Phys.~Rev.~Lett.}} {\bf{119}}, 167001 (2017). \hfill 
	{[cond-mat.supr-con] arXiv:\href{https://arxiv.org/abs/1705.01718}{1705.01718}}

\Bibitem{CV:Garaud.Silaev.ea:17a} 
	{\bf{J.~Garaud}}, M.~Silaev and E.~Babaev\\
	\href{https://doi.org/10.1103/PhysRevB.96.140503}
	{Change of the vortex core structure in two-band superconductors \\
	at impurity-scatering-driven $s_\pm/s+is$ crossover} \\
	{\it{ Phys.~Rev.~B}} {\bf{96}}, 140503(R) (2017). \hfill 
	{[cond-mat.supr-con] arXiv:\href{https://arxiv.org/abs/1707.06412}{1707.06412}}

\Bibitem{CV:Silaev.Garaud.ea:17} 
	M.~Silaev, {\bf{J.~Garaud}} and E.~Babaev\\
	\href{https://doi.org/10.1103/PhysRevB.95.024517}
	{Phase diagram of dirty two-band superconductors and observability \\
	of impurity-induced $s+is$ state} \\
	{\it{ Phys.~Rev.~B}} {\bf{95}}, 024517 (2017). \hfill 
	{[cond-mat.supr-con] arXiv:\href{https://arxiv.org/abs/1610.05846}{1610.05846}}

\Bibitem{CV:Garaud.Silaev.ea:17} 
	{\bf{J.~Garaud}}, M.~Silaev and E.~Babaev\\
	\href{http://dx.doi.org/10.1016/j.physc.2016.07.010}
	{Microscopically derived multi-component Ginzburg-Landau theories for $s+is$ \\
	superconducting state} \\
 	{\it{Physica~C}} {\bf{533}}, 63--73 (2017). \hfill 
	{[cond-mat.supr-con] arXiv:\href{http://arxiv.org/abs/1601.02227}{1601.02227}}

\Bibitem{CV:Garaud.Babaev.ea:16} 
	{\bf{J.~Garaud}}, E.~Babaev, T.~A.~Bojesen and A.~Sudb\o  \\
	\href{http://dx.doi.org/10.1103/PhysRevB.94.104509}
	{Lattices of double-quanta vortices and chirality inversion in $p_x+ip_y$ 
	superconductors} \\
	{\it{ Phys.~Rev.~B}} {\bf{94}}, 104509 (2016). \hfill 
	{[cond-mat.supr-con] arXiv:\href{http://arxiv.org/abs/1605.03946}{1605.03946}}
 
\Bibitem{CV:Garaud.Silaev.ea:16} 
	{\bf{J.~Garaud}}, M.~Silaev and E.~Babaev\\
	\href{http://dx.doi.org/10.1103/PhysRevLett.116.097002}
	{Thermoelectric Signatures of Time-Reversal Symmetry Breaking States \\
	in Multiband Superconductors} \\
	{\it{Phys.~Rev.~Lett.}} {\bf{116}}, 097002 (2016). \hfill 
	{[cond-mat.supr-con] arXiv:\href{http://arxiv.org/abs/1507.04712}{1507.04712}}

\Bibitem{CV:Garaud.Babaev:15a} 
	{\bf{J.~Garaud}} and E.~Babaev\\
	\href{http://dx.doi.org/10.1038/srep17540}
	{Properties of skyrmions and multi-quanta vortices in chiral 
	$p$-wave superconductors} \\
	{\it{ Sci.~Rep.}} {\bf{5}}, 17540 (2015). \hfill 
	{[cond-mat.supr-con] arXiv:\href{http://arxiv.org/abs/1507.04634}{1507.04634}}

\Bibitem{CV:Silaev.Garaud.ea:15} 
	M.~Silaev, {\bf{J.~Garaud}} and E.~Babaev\\
	\href{http://dx.doi.org/10.1103/PhysRevB.92.174510}
	{Unconventional thermoelectric effect in superconductors that break 
	time-reversal symmetry} \\
	{\it{ Phys.~Rev.~B}} {\bf{92}}, 174510 (2015). \hfill 
	{[cond-mat.supr-con] arXiv:\href{http://arxiv.org/abs/1503.02024}{1503.02024}}

\Bibitem{CV:Garaud.Babaev:15} 
	{\bf{J.~Garaud}} and E.~Babaev \\
	\href{http://dx.doi.org/10.1103/PhysRevB.91.014510}
	{Vortex chains due to nonpairwise interactions and field-induced phase 
	transitions between \\states with different broken symmetry in superconductors 
	with competing order parameters} \\
	{\it{ Phys.~Rev.~B}} {\bf{91}}, 014510 (2015). 
	~~[\href{http://journals.aps.org/prb/kaleidoscope/prb/91/1/014510}
	{\it{Kaleidoscope}}] 	\hfill 
	{[cond-mat.supr-con] arXiv:\href{http://arxiv.org/abs/1411.6656}{1411.6656}}

\Bibitem{CV:Garaud.Babaev:14b} 
	{\bf{J.~Garaud}} and E.~Babaev \\
	\href{http://dx.doi.org/10.1103/PhysRevB.90.214524}
	{Vortex matter in $\mathrm{U}(1)\times\mathrm{U}(1)\times\mathbb{Z}_2$
	phase-separated superconducting condensates} \\
	{\it{Phys.~Rev.~B}} {\bf{90}}, 214524 (2014). \hfill 
	{[cond-mat.supr-con] arXiv:\href{http://arxiv.org/abs/1410.2985}{1410.2985}}

\Bibitem{CV:Agterberg.Babaev.ea:14} 
	D.~F.~Agterberg, E.~Babaev and {\bf{J.~Garaud}} \\
	\href{http://dx.doi.org/10.1103/PhysRevB.90.064509}
	{Microscopic prediction of skyrmion lattice state in clean interface 
	superconductors} \\
	{\it{Phys.~Rev.~B}} {\bf{90}}, 064509 (2014). 
	~~[\href{http://journals.aps.org/prb/kaleidoscope/prb/90/6/064509}
	{\it{Kaleidoscope}}]	\hfill 
	{[cond-mat.supr-con] arXiv:\href{http://arxiv.org/abs/1403.6655}{1403.6655}}

\Bibitem{CV:Garaud.Babaev:14a} 
	{\bf{J.~Garaud}} and E.~Babaev \\
	\href{http://dx.doi.org/10.1103/PhysRevB.89.214507}
	{Topological defects in mixtures of superconducting condensates 
	with different charges} \\
	{\it{Phys.~Rev.~B}} {\bf{89}}, 214507 (2014). \hfill 
	{[cond-mat.supr-con] arXiv:\href{http://arxiv.org/abs/1403.3373}{1403.3373}}

\Bibitem{CV:Garaud.Babaev:14} 
	{\bf{J.~Garaud}} and E.~Babaev \\
	\href{http://dx.doi.org/10.1103/PhysRevLett.112.017003 }
	{Domain walls and their experimental signatures in $s+is$ superconductors} \\
	{\it{Phys.~Rev.~Lett.}} {\bf{112}}, 017003 (2014). \hfill 
	{[cond-mat.supr-con] arXiv:\href{http://arxiv.org/abs/1308.3220}{1308.3220}}

\Bibitem{CV:Garaud.Sellin.ea:14} 
	{\bf{J.~Garaud}},  K.~Sellin, J.~J\"aykk\"a and E.~Babaev \\
	\href{http://dx.doi.org/10.1103/PhysRevB.89.104508}
	{Skyrmions induced by dissipationless drag in 
	$\mathrm{U}(1)\times\mathrm{U}(1)$ superconductors} \\
	{\it{Phys.~Rev.~B}} {\bf{89}}, 104508 (2014). \hfill 
	{[cond-mat.supr-con] arXiv:\href{http://arxiv.org/abs/1307.3211}{1307.3211}}

\Bibitem{CV:Garaud.Carlstrom.ea:13} 
	{\bf{J.~Garaud}}, J.~Carlstr\"om, E.~Babaev and M. Speight \\
	\href{http://dx.doi.org/10.1103/PhysRevB.87.014507}
	{Chiral $\ {\mathbbm{C}}{{P}}^2\ $ skyrmions in three-band superconductors} \\
	{\it{Phys.~Rev.~B}} {\bf{87}}, 014507 (2013).~~[{\it{Editors' Suggestion}}] \hfill 
	{[cond-mat.supr-con] arXiv:\href{http://arxiv.org/abs/1211.4342}{1211.4342}}

\Bibitem{CV:Garaud.Agterberg.ea:12} 
	{\bf{J.~Garaud}}, D.~F.~Agterberg and E.~Babaev \\
	\href{http://dx.doi.org/10.1103/PhysRevB.86.060513}
	{Vortex coalescence and type-1.5 superconductivity in Sr$_2$RuO$_4$} \\	
	{\it{Phys.~Rev.~B}} {\bf{86}}, 060513(R) (2012).~~[{\it{Rapid Comm.}}] \hfill	
	{[cond-mat.supr-con] arXiv:\href{http://arxiv.org/abs/1207.6395}{1207.6395}}

\Bibitem{CV:Garaud.Babaev:12} 
	{\bf{J.~Garaud}} and E.~Babaev \\
	\href{http://dx.doi.org/10.1103/PhysRevB.86.060514}
	{Skyrmionic state and stable half-quantum vortices in chiral 
	$p$-wave superconductors} \\
	{\it{Phys.~Rev.~B}} {\bf{86}}, 060514(R) (2012). ~~[{\it{Rapid Comm.}}] \hfill	
	{[cond-mat.supr-con] arXiv:\href{http://arxiv.org/abs/1201.2946}{1201.2946}}

\Bibitem{CV:Babaev.Carlstrom.ea:12} 
	E.~Babaev, J.~Carlstr\"om, {\bf{J.~Garaud}}, M.~Silaev and J.~M.~Speight \\
	\href{http://dx.doi.org/10.1016/j.physc.2012.01.002}
	{Type-1.5 superconductivity in multiband systems: magnetic response, \\
	broken symmetries and microscopic theory. A brief overview} \\
 	{\it{Physica~C}} {\bf{479}}, 2--14 (2012). \hfill 
	{[cond-mat.supr-con] arXiv:\href{http://arxiv.org/abs/1110.2744}{1110.2744}}

\Bibitem{CV:Carlstrom.Garaud.ea:11a} 
	J.~Carlstr\"om, {\bf{J.~Garaud}} and E.~Babaev \\
	\href{http://dx.doi.org/10.1103/PhysRevB.84.134518}
	{Length scales, collective modes, and type-1.5 regimes in three-band 
	superconductors} \\
 	{\it{Phys.~Rev.~B}} {\bf{84}}, 134518 (2011). 
 	~~[see \href{https://journals.aps.org/prb/abstract/10.1103/PhysRevB.87.219904}
	{\it{Erratum}}]	\hfill 
	{[cond-mat.supr-con] arXiv:\href{http://arxiv.org/abs/1107.4279}{1107.4279}}

\Bibitem{CV:Garaud.Carlstrom.ea:11} 
	{\bf{J.~Garaud}}, J.~Carlstr\"om and E.~Babaev \\
	\href{http://dx.doi.org/10.1103/PhysRevLett.107.197001}
	{Topological solitons in three-band superconductors with broken 
	time reversal symmetry} \\
 	{\it{Phys.~Rev.~Lett.}} {\bf{107}}, 197001 (2011). \hfill 
	{[cond-mat.supr-con] arXiv:\href{http://arxiv.org/abs/1107.0995}{1107.0995}}

\Bibitem{CV:Carlstrom.Garaud.ea:11} 
	J.~Carlstr\"om, {\bf{J.~Garaud}} and E.~Babaev \\
	\href{http://dx.doi.org/10.1103/PhysRevB.84.134515}
	{Semi-Meissner state and nonpairwise intervortex interactions 
	in type-1.5 superconductors} \\
 	{\it{Phys.~Rev.~B}} {\bf{84}}, 134515 (2011). \hfill 
	{[cond-mat.supr-con] arXiv:\href{http://arxiv.org/abs/1101.4599}{1101.4599}}

\end{itemize}


\newpage \thispagestyle{empty}\ \newpage
\chapter{Synthèse en fran\c cais du mémoire d'HDR}		
\graphicspath{{Plots/01-Introduction/}}
\label{Chap:Synthese}

\vspace{1cm}

\textbf{
Le corps principal du mémoire, les différents chapitres et les annexes sont rédigés 
en anglais. Ce chapitre de synthèse reprend dans les grandes lignes, en fran\c cais, 
les éléments et les résultats principaux qui y sont discutés. En particulier, 
on reprend ici les éléments de l'introduction, ainsi que les motivations et résumés 
des résultats présentés dans chaque chapitre. On reprend  également  les résumés 
décrivant les différentes annexes.
}

\vspace{1.25cm}

\begin{wrapfigure}{R}{0.35\textwidth}
\hbox to \linewidth{ \hss
\includegraphics[width=.975\linewidth,angle=0]{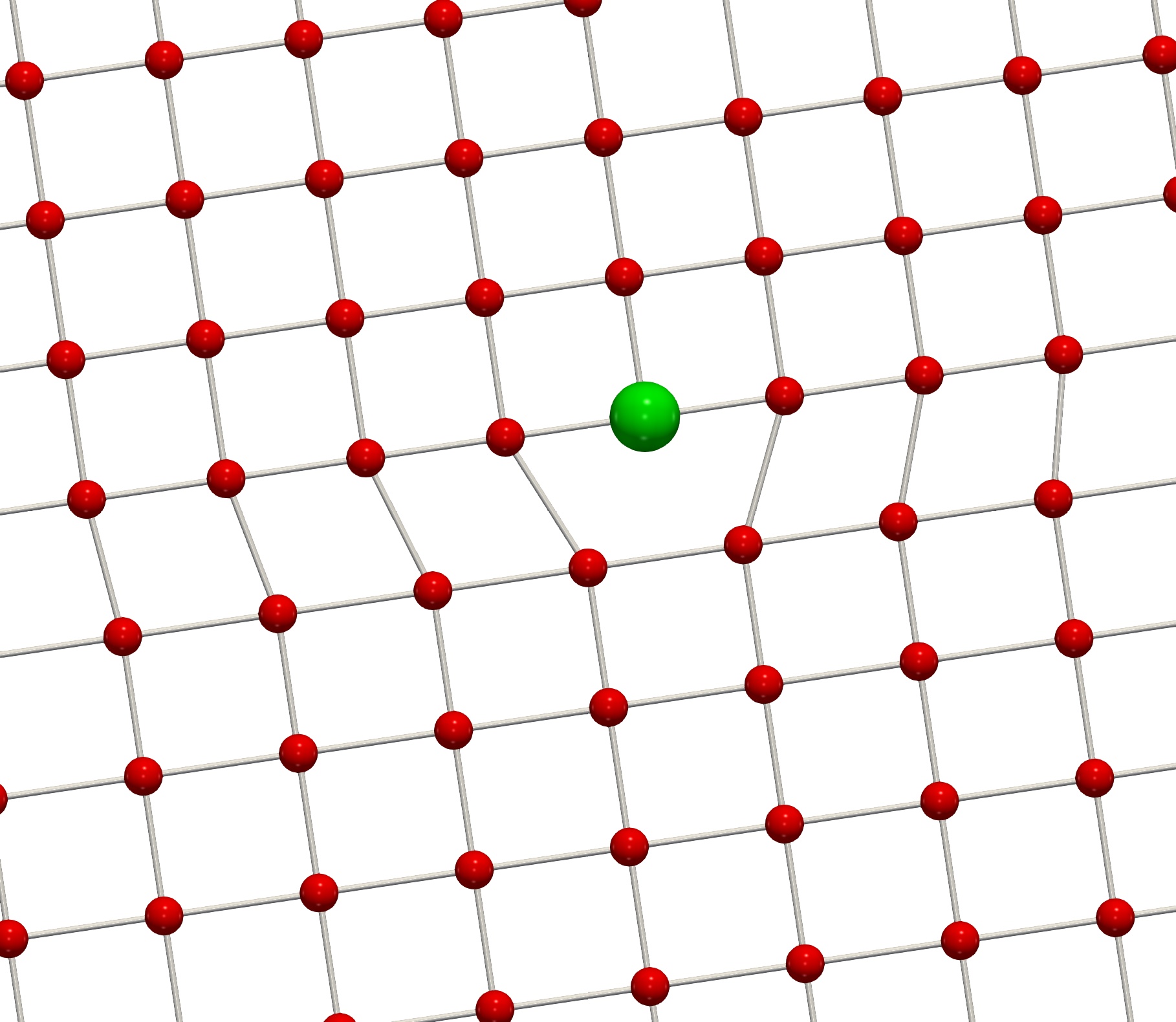}
\hss}
\caption{
Un défaut topologique dans un cristal. Une des rangées d'atomes en bas disparaît 
à mi-chemin de l'échantillon. L'endroit où la rangée disparaît est un défaut, 
car il ne ressemble pas localement à un morceau du cristal parfait. 
}\label{FigFR:Dislocation}
\end{wrapfigure}
Les défauts topologiques et leur compréhension sont au coeur de la physique moderne. 
La formalisation de leurs propriétés et la compréhension de leur rôle dans de nombreux 
processus physiques est relativement récente. Cependant, ils sont connus heuristiquement 
par l'humanité depuis probablement plus de trois mille ans. C'est en effet 
approximativement aussi loin que l'on puisse retracer les processus de trempe de métaux 
par les forgerons \cite{Mackenzie:08}. La trempe, c'est-à-dire le refroidissement rapide, 
utilisée dans les procédés de durcissement des métaux crée des dislocations de la 
structure cristalline, qui s'apparentent à des défauts topologiques. Ce procédé est 
similaire à la prolifération de défauts topologiques qui se produit lors des transitions 
de phase \cite{Kibble:76,Zurek:85}. 
Comme illustré sur la \Figref{FigFR:Dislocation}, une dislocation dans un cristal 
est un défaut topologique, car elle ne peut être supprimée par aucun réarrangement local.

Associés aux symétries brisées, les défauts topologiques sont omniprésents en physique. 
Ils apparaissent en effet dans un contexte très vaste, allant de la cosmologie de 
l'univers primordial et de la physique des particules \cite{Rajaraman,Manton.Sutcliffe,
Vachaspati,Vilenkin.Shellard,Shnir,Volovik,Shnir:18}, à la physique du solide et de la 
matière condensée \cite{Mermin:79,Mineev,Volovik}.
En fonction la théorie sous-jacente, les défauts topologiques peuvent être de 
différentes natures, comme par exemple les dislocations dans les cristaux liquides, 
les monopôles, les murs de domaine, les vortex, les skyrmions, les hopfions, et bien 
plus encore. Ils sont intimement liés aux transitions de phase \cite{Kosterlitz:17,
Kibble:76,Zurek:85}, et leur simple existence peut avoir un grand impact. Par exemple, 
la possible formation de défauts topologiques dans l'univers primordial aurait pu 
avoir un impact considérable sur la formation des structures de l'univers aux grandes 
échelles\cite{Vilenkin.Shellard,Kibble:76,Brandenberger:94}. De même, on pense qu'ils 
sont à l'origine de certaines transitions de phase dans divers systèmes physiques, 
comme par exemple la prolifération de vortex dans les superfluides et les 
supraconducteurs \cite{Kosterlitz:17}. 
Les vortex, qui sont des objets linéiques avec des propriétés topologiques spécifiques, 
sont probablement les défauts qui ont été le plus étudiés.

\section*{Défauts topologiques -- Supraconducteurs et superfluides}
\addcontentsline{toc}{section}{Défauts topologiques -- Supraconducteurs et superfluides}

La physique des vortex fait l'objet d'une intense activité scientifique depuis la 
seconde moitié du XIXe siècle. Peu de temps après les premiers travaux de Helmholtz 
\cite{Helmholtz:58} sur la dynamique des fluides, les vortex sont devenus la clé de voûte 
de la théorie de la matière de l'« atome vortex » conjecturée par Kelvin \cite{Thomson:69}. 
Cette tentative de classer les éléments chimiques en tant qu'excitations constituées 
de boucles de vortex dans l'éther luminifère
\footnote{
L'éther luminifère était un fluide idéal postulé. Sensé remplir l'univers, il 
devait servir de support à la propagation des ondes lumineuses. 
},
liées et nouées se solda par un échec. Cependant elle conduit à des progrès 
considérable en topologie car elle a motivé la création des premières tables des noeuds 
par Tait \cite{Tait:78,Tait:84,Tait:86} et les travaux sur la théorie des noeuds 
peu de temps après.

\subsection*{La théorie de « l'atome vortex »}

Fait intéressant, la théorie de « l'atome vortex » de Kelvin et Tait résonne encore 
avec certains concepts de la physique moderne \cite{Kragh:02,Laan:12}, 
et elle a inspiré plusieurs travaux au fil des ans. Ainsi, c'est une histoire 
qui mérite d'être racontée.

Dans son travail de 1858 sur la dynamique des fluides, Helmholtz \cite{Helmholtz:58} 
démontra que dans un fluide parfait (c'est-à-dire caractérisé pas un écoulement 
incompressible et non visqueux), la circulation d'un filament vortex est invariante 
dans le temps. 
Il a par ailleurs démontré qu'un vortex ne peut pas se terminer à l'intérieur d'un 
fluide, mais qu'il doit soit s'étendre jusqu'aux limites du fluide, soit former des 
boucles fermées. Enfin qu'en l'absence de forces de rotation externes, un écoulement 
initialement irrotationnel reste irrotationnel.

\begin{wrapfigure}{R}{0.4\textwidth}
\hbox to \linewidth{ \hss
\includegraphics[width=.975\linewidth]{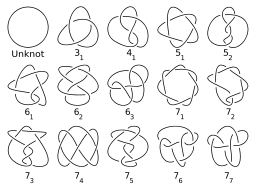}
\hss}
\caption{
Une des noeuds inéquivalents.
}\label{FigFR:Kelvin-Knots}
\end{wrapfigure}
Connaissant les théorèmes de Helmholtz, Kelvin remarqua en 1867 \cite{Thomson:69} que 
\Quote{(...) this discovery inevitably suggests the idea that Helmholtz's rings 
are the only true atoms.} L'idée générale étant que, puisque les lignes de vortex sont 
figées dans l'écoulement d'un fluide idéal, alors leur topologie doit être invariante 
dans le temps : 
\Quote{
It is to be remarked that two ring atoms linked together, or one knotted in any manner 
with its ends meeting, constitute a system which, however it may be altered in shape, 
can never deviate from its own peculiarity of multiple continuity (...)
}. Ce fluide idéal serait l'éther luminifère dont on pensait à l'époque qu'il emplissait 
l'univers. Kelvin a ensuite attribué les propriétés spectroscopiques de la matière à la 
topologie des lignes de vortex : 
\Quote{It seems, therefore, probable that the sodium atom may not consist of a single
vortex line; but it may very probably consist of two approximately equal vortex rings 
passing through one another, like two links of a chain.}. Il remarqua de plus que 
les modèles de \Quote{(...) knotted or knitted vortex atoms, the endless variety of
which is infinitely more than sufficient to explain the varieties and allotropies of 
known simple bodies and their mutual affinities.}. En bref, Kelvin a conjecturé que les 
différents éléments chimiques sont constitués de boucles fermées de vortex dans l'éther, 
liées et nouées entre elles de fa\c con topologiquement inéquivalentes, illustrées dans 
\Figref{FigFR:Kelvin-Knots}. 

Par la suite, Tait a commencé à classifier les différentes manières inéquivalentes de 
réaliser de tels nœuds \cite{Tait:78,Tait:84,Tait:86}. Ces travaux ont été les pionniers 
du domaine de la théorie des nœuds en topologie algébrique.
La théorie de Kelvin a finalement été falsifiée, lorsque l'expérience de Michelson 
et Morley a exclu l'existence de l'éther \cite{Michelson.Morley:87}. Cependant, 
le paradigme d'associer des vortex d'un champ sous-jacent à des ``particules 
élémentaires'' est réapparu à plusieurs reprises.
D'une certaine manière, ces vortex noués peuvent être considérés comme les premiers 
exemples théoriques de défauts topologiques
\footnote{
On ne manquera pas d'y voir une certaine analogie avec la théories des cordes.
}.

\subsection*{Vortex et défauts topologiques dans la physique contemporaine}

Environ 80 ans après les travaux de Kelvin, il a été réalisé par Onsager 
\cite{Onsager:49}, et plus tard formalisé sur des bases théoriques solides 
par Feynman \cite{Feynman:55}, que les vortex occupent une part importante 
dans les processus physiques modernes. En particulier dans son travail sur l'$^4$He 
superfluide, Onsager \cite{Onsager:49} a observé que la circulation de la vélocité 
du superfluide est quantifiée. De plus il compris que la matière-vortex contrôle 
essentiellement la plupart des réponses clés des superfluides. Par exemple, que 
la transition de phase de l'état superfluide à l'état normal est une en fait génération 
thermique et une prolifération de boucles et de noeuds de vortex \cite{Onsager:49}. 
Également que les vortex apparaissent comme la réponse rotationnelle des superfluides.

Ces idées résonnent en quelque sorte (en partie) avec la théorie de Kelvin de l'atome 
vortex. En effet, puisque la vitesse de circulation du superfluide est quantifiée, 
alors les vortex sont des défauts topologiques dans le superfluide. De plus, 
la rotation d'un superfluide se traduit par la formation d'un réseau ou d'un liquide 
de ces vortex quantiques. Ces réseaux peuvent être considérés comme la réalisation 
de cristaux et de liquides de la matière-vortex.
Plus tard, Abrikosov a prédit que les supraconducteurs de type-2 doivent former 
des vortex magnétiques en réponse à un champ magnétique appliqué \cite{Abrikosov:57a}, 
par analogie avec les superfluides ou les vortex sont formés en réponse à une rotation. 
Plus tard, il a également été réalisé qu'en trois dimensions, les transitions de phases 
superfluides et supraconductrices sont une génération thermique et une prolifération de 
boucles de vortex \cite{Peskin:78,Dasgupta.Halperin:81}.

\begin{wrapfigure}{R}{0.4\textwidth}
\hbox to \linewidth{ \hss
\includegraphics[width=.975\linewidth]{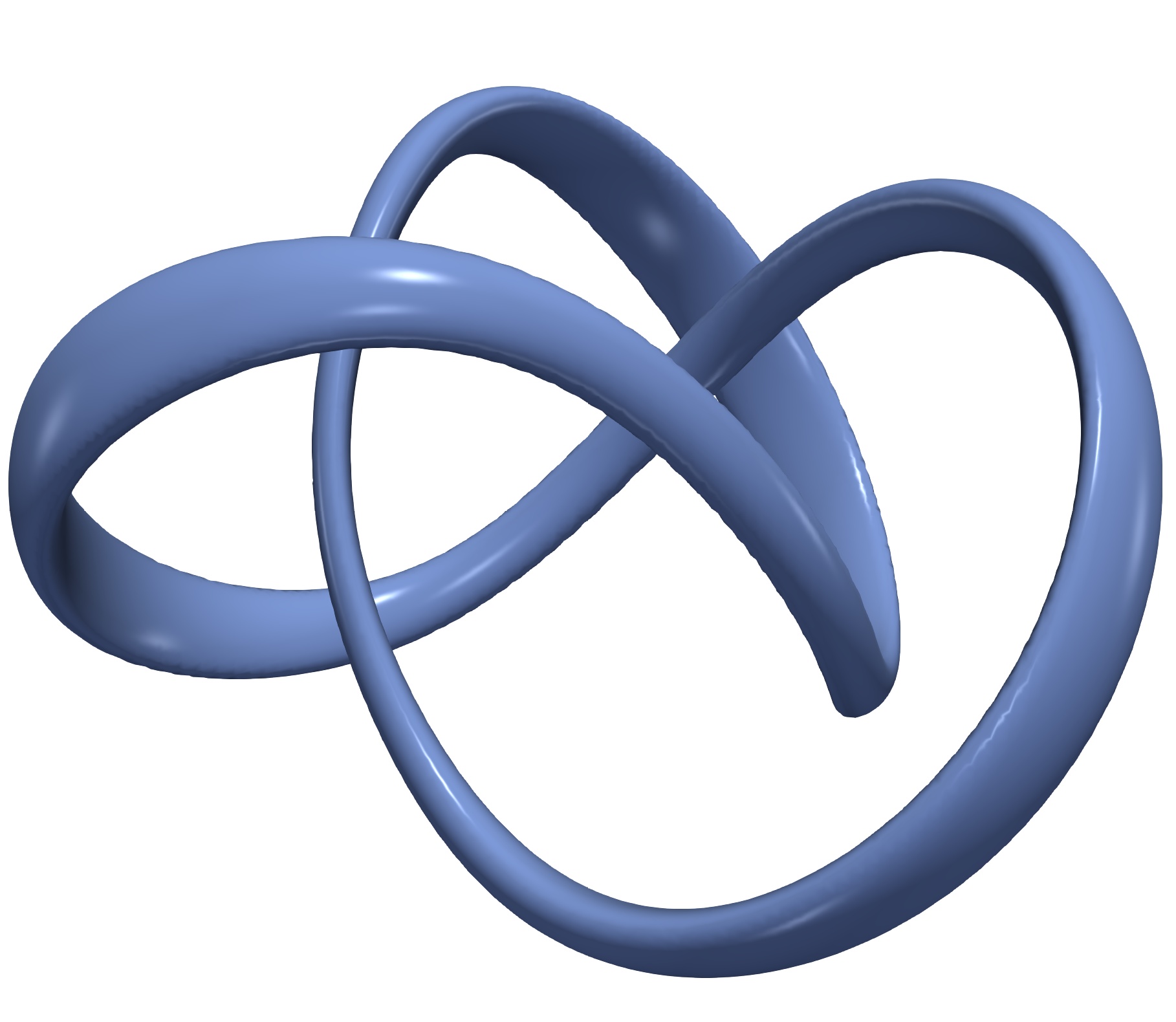}
\hss}
\caption{
Un noeud \emph{trèfle} dans le modèle de Skyrme-Faddev.
}\label{FigFR:Knots-SF}
\end{wrapfigure}

A noter que d'importants progrès dans la physique moderne, où les vortex jouent un 
rôle clé, ont reçu un prix Nobel. Comme par exemple à Abrikosov en 2003 \cite{Abrikosov:04} 
pour ses travaux sur la compréhension de leur rôle dans les supraconducteurs, ou plus 
récemment en 2016 à Haldane, Kosterlitz et Thouless pour avoir déterminé leur rôle dans 
les transitions de phase dans les systèmes bidimensionnels \cite{Haldane:17,Kosterlitz:17}.
Le concept de vortex quantiques a ensuite été généralisé à des théories relativistes, 
comme par exemple dans le modèle Higgs-abélien \cite{Nielsen.Olesen:73}; des théories 
qui auraient pu être pertinentes dans le contexte de l'univers primordial 
\cite{Vilenkin.Shellard,Witten:85,Witten:85a}, ou encore au secteur bosonique de la 
théorie de Weinberg-Salam des interactions électrofaibles \cite{Achucarro.Vachaspati:00}. 
D'après le scénario de Kibble-Zurek \cite{Kibble:76,Zurek:85}, divers types de défauts 
topologiques auraient dû être produit lors des possibles transitions de phase de l'univers 
primordial. Cela impliquerait, entre autres, que si de tels défauts topologiques étaient 
créés, ils pourraient contribuer substantiellement au contenu en matière de l'univers et 
avoir un impact non négligeable sur son histoire \cite{Vilenkin.Shellard,Brandenberger:94}. 
Ces idées sont à l'origine d'un grand intérêt pour les défauts topologiques. Cela a abouti 
à de nombreux travaux fondateurs et à une meilleure compréhension des propriétés 
mathématiques des défauts topologiques.

Ainsi, comme déjà mentionné, il existe une pléthore de différents types de défauts 
topologiques, et dans un large éventail de systèmes physiques. C'est un peu dénué 
de sens de les lister de manière exhaustive. Mentionnons plutôt deux types particuliers 
de défauts topologiques qui résonnent particulièrement avec la théorie de Kelvin, 
car ils sont en quelque sorte identifiés en lien avec des états de la matière.
Un premier exemple est celui des défauts topologiques dans le modèle Skyrme 
\cite{Skyrme:61,Skyrme:62}. Les défauts topologiques y sont appelés \emph{skyrmions}
\footnote{
Dans le texte principal, la terminologie \emph{skyrmions} est utilisée pour caractériser 
des types de défauts topologiques légèrement différents. Ils sont davantage liés aux 
``baby-skyrmions''. Comme ils en partagent de nombreuses propriétés topologiques, 
ils sont souvent également appelés skyrmions, par abus de langage. 
}, 
et l'invariant topologique associé est interprété comme le nombre de baryons 
\cite{Brown.Rho}.
De même, la recherche sur les modèles supportant des défauts topologiques, noués, 
stables a été d'un grand intérêt en mathématiques et en physique, après que la stabilité 
de ces objets nommés \emph{hopfions} ait été démontrée dans le modèle Skyrme-Faddeev 
\cite{Faddeev:75,Faddeev.Niemi:97,Gladikowski.Hellmund:97,Hietarinta.Salo:99,
Battye.Sutcliffe:99,Sutcliffe:07} (pour une revue, voir \cite{Radu.Volkov:08}). 
Les hopfions du modèle de Skyrme-Faddeev ressemblent aux noeuds illustrés dans 
\Figref{FigFR:Knots-SF}.

Après cette introduction générale sur les défauts topologiques, l'essentiel de 
l'attention sera porté sur les vortex, avec un focus particulier sur ceux qui 
apparaissent dans les modèles de supraconductivité avec de multiples composantes du 
paramètre d'ordre. 

\section*{Supraconducteurs à multiple composantes}
\addcontentsline{toc}{section}{Supraconducteurs à multiple composantes}

La supraconductivité et la superfluidité sont des états de la matière caractérisés 
par la cohérence macroscopique des excitations quantiques sous-jacentes. La physique 
qui décrit de tels systèmes est celle des théories quantiques des champs, et ce sont 
les propriétés du problème à N-corps, au sein de ces théories qui déterminent 
la cohérence macroscopique des excitations quantiques. La propriété intéressante 
est que ces problèmes quantiques à N-corps peuvent être réduits, dans l'approximation 
de champ moyen, à des théories classiques des champs, non linéaires, décrivant les 
propriétés macroscopiques de l'état cohérent représenté par un seul champ scalaire 
complexe (le \emph{paramètre d'ordre}).
Ces approximations de champ moyen sont les équations de Gross-Pitaevskii pour les 
superfluides et les équations de Ginzburg-Landau pour les supraconducteurs. A noter 
que dans le cas des supraconducteurs, le champ scalaire complexe est complété par 
un champ vectoriel abélien réel, décrivant le potentiel du champ électromagnétique. 
Ce champ de jauge devient massif par le mécanisme d'Anderson-Higgs\cite{Anderson:58,
Anderson:63}, qui est responsable de l'effet Meissner \cite{Meissner.Ochsenfeld:33}. 
Alors que dans les cas les plus simples, les paramètres d'ordre sont des singulets, 
il peuvent être des multiplets scalaires dans des situations plus compliquées.

Dans les systèmes de la matière condensée comme les superfluides ou les condensats 
de Bose-Einstein d'atomes ultra-froids, les théories avec des paramètres d'ordre à 
plusieurs composantes (c'est-à-dire décrites par des multiplets ou des matrices de 
champs scalaires complexes) ont été envisagées depuis longtemps. Elles sont connus 
pour admettre un éventail extrêmement riche de défauts topologiques, comme par exemple 
dans l'hélium superfluide \cite{Volovik,Volovik:92}, ou dans les superfluides 
neutroniques $^3P_2$ \cite{Fujita.Tsuneto:72,Masuda.Nitta:20}.
Dans le contexte de la supraconductivité, les théories avec de multiples gaps 
supraconducteurs ont été considérées depuis les débuts de la théorie 
Bardeen-Cooper-Schrieffer \cite{Moskalenko:59,Suhl.Matthias.ea:59,Tilley:64}. 
Pourtant, ces théories multi-bandes/multi-composantes ont longtemps été considérées 
comme décrivant des matériaux exotiques.

Cependant, il y a eu relativement récemment un intérêt accru pour de tels matériaux, 
car le nombre de supraconducteurs multi-bandes/multi-composantes connus augmente 
rapidement. En effet, dans de nombreux supraconducteurs, l'appariement des électrons 
est supposé se produire sur plusieurs feuillets de la surface de Fermi formée par le 
chevauchement des bandes électroniques. Pour n'en citer que quelques-uns, c'est par 
exemple le cas du MgB$_2$ \cite{Nagamatsu.Nakagawa.ea:01,Mazin.Antropov:03}, du 
supraconducteur pérovskite Sr$_2$RuO$_4$ \cite{Maeno.Hashimoto.ea:94,Mackenzie.Maeno:03,
Damascelli.Lu.ea:00}, ou de la famille des supraconducteurs à base de fer 
\cite{Kamihara.Watanabe.ea:08,Mazin.Singh.ea:08,Kuroki.Onari.ea:08,Chubukov.Efremov.ea:08}. 
Au-delà de la physique du solide, les théories à multiples composantes s'appliquent 
également à des systèmes plus exotiques, comme certains modèles de supraconductivité 
nucléaire à l'intérieur des étoiles à neutrons \cite{Jones:06}, ou aux états 
supraconducteurs de l'hydrogène liquide métallique \cite{Ashcroft:68,Ashcroft:00}, 
du deutérium liquide métallique \cite{Oliva.Ashcroft:84,Oliva.Ashcroft:84a} 
et d'autres types de superfluides métalliques \cite{Ashcroft:05}.
Cela ouvre la possibilité de modèles de théorie des champs plus complexes où, 
généralement en raison de l'existence de plusieurs symétries brisées, la physique 
des vortex (et des autres défauts topologiques) est extrêmement riche et sans équivalent 
dans des modèles plus simples.

Notons que les modèles où les vortex apparaissent, dans le contexte de la physique 
des hautes énergies, sont généralement très symétriques en raison des propriétés 
sous-jacentes de la théorie. Par exemple, dans le cas de la théorie de Weinberg-Salam, 
la théorie est invariante (entre autres symétries) sous les rotations locales 
$\groupSU{2}$ au sein du doublet scalaire (le champ de Higgs). En matière condensée
les modèles sont généralement beaucoup moins contraints sur pour des raisons de symétrie.
Ils autorisent ainsi plus de termes d'interaction qui brisent explicitement diverses 
symétries. Par exemple, dans les supraconducteurs à deux composantes (décrits par un 
doublet scalaire), l'invariance globale $\groupSU{2}$ est explicitement brisée vers 
un sous-groupe plus petit (comme par exemple $\groupU{1}\times\groupU{1}$ ). 
L'absence de contraintes fortes sur les symétries, et donc l'existence de divers termes 
de brisure explicite de symétrie sont à l'origine de nombreuses nouvelles propriétés. 
Il en résulte notamment que les vortex peuvent acquérir de nouvelles caractéristiques, 
et sont associés à un large éventail de nouveaux phénomènes physiques.
Ces nouvelles propriétés exotiques peuvent être comprises comme provenant des nouvelles 
symétries brisées. Par conséquent, ces nouveaux phénomènes peuvent être utilisés comme 
signatures pour obtenir des informations sur les symétries d'un état inconnu.

L'importance cruciale des excitations topologiques dans la physique de la 
supraconductivité a fait des vortex de Ginzburg-Landau l'un des exemples les 
plus étudié de défauts topologiques. En effet, les propriétés de transport des 
supraconducteurs dépendent de manière cruciale du comportement des vortex magnétiques 
dans ces matériaux. Par exemple, les courants critiques élevés dans les lignes de 
transmission supraconductrices commerciales actuelles ne sont atteints qu'en 
contrôlant soigneusement le mouvement des vortex dans ces matériaux.
Les théories pour les supraconducteurs multi-bandes/multi-composantes étendent la 
théorie de Ginzburg-Landau en considérant plus d'un paramètre d'ordre supraconducteur. 
En raison des champs supplémentaires et des nouvelles symétries brisées, le spectre 
des excitations topologiques et les signatures associées sont beaucoup plus riches 
dans les systèmes multicomposantes que dans leurs homologues monocomposantes. 
Par exemple, les supraconducteurs multicomposantes peuvent présenter des vortex 
fractionnaires, des vortex singuliers/sans-coeur (coreless), des skyrmions, des hopfions, 
des murs de domaine, etc. Toutes ces excitations topologiques peuvent être utilisées 
comme signatures expérimentales de la nature multicomposante d'un système supraconducteur. 
Leur observabilité peut par exemple fournir des informations précieuses sur 
la nature du paramètre d'ordre et de la symétrie d'appariement sous-jacente.

Les travaux présentés dans ce rapport traitent de divers aspects de la supraconductivité 
des théories avec plus d'un condensat. Aussi bien pour des modèles génériques de 
supraconducteurs multi-composantes que pour les modèles de matériaux particuliers.
Notamment par l'analyse des propriétés inhabituelles des défauts topologiques qui 
y apparaissent.

\subsection*{Théorie de champ moyen de Ginzburg-Landau}

En considérant l'approximation du champ moyen et à couplage faible, dans la théorie 
de Bardeen-Cooper-Schrieffer \cite{Bardeen.Cooper.ea:57} à une composante, l'état 
supraconducteur est décrit par un champ classique complexe qui est proportionnel 
à la fonction de gap. 
C'est-à-dire, la théorie de Ginzburg-Landau \cite{Ginzburg.Landau:50a}, introduite 
phénoménologiquement peut être dérivée comme l'approximation classique du champ moyen 
de la théorie microscopique \cite{Gorkov:59}, et le module du paramètre d'ordre est 
la densité de paires de Cooper.
Il existe différentes approches pour caractériser les propriétés des matériaux 
supraconducteurs, qui sont différentes/complémentaires à la théorie de Ginzburg-Landau. 
Par exemple, des méthodes telles que le formalisme de Bogoliubov-de Gennes \cite{Gennes,
Zhu:16}, les équations d'Eilenberger \cite{Eilenberger:68} et d'Usadel \cite{Usadel:70} 
pour le transport, etc.
Cependant, le reste de ce rapport se limite uniquement aux aspects classiques et en 
champ moyen, de la supraconductivité des systèmes à plusieurs composantes. C'est-à-dire 
à la théorie de Ginzburg-Landau à plusieurs composantes et aux excitations topologiques 
qui s'y apparaissent.

Il est intéressant de noter que la théorie de Ginzburg-Landau de la supraconductivité 
n'a attiré beaucoup d'attention de la part de la communauté des analystes numériques 
que depuis les années 1990, après qu'il ait été rapporté que le problème est bien posé 
\cite{Du.Gunzburger.ea:92,Du:94}. Depuis, il y a eu une grande activité pour comprendre 
les propriétés mathématiques de ce problème, voir par exemple \cite{Du:05}.

\section*{Théorie générale à multiples composantes}
\addcontentsline{toc}{section}{Théorie générale à multiples composantes}

Les détails des modèles à multiples composantes peuvent varier, selon le contexte 
du problème physique sous-jacent considéré. On exposera ici brièvement la structure 
mathématique des modèles génériques qui décrivent les supraconducteurs à plusieurs 
composantes (dans l'approximation de champ moyen).
Les propriétés macroscopiques de tels systèmes physiques sont généralement décrites 
par la fonctionnelle d'énergie (libre) de Ginzburg-Landau de la forme: 
\Equation{EqFR:General:FreeEnergy}{
\F/\F_0=\int_{\mathbb{R}^3} \frac{1}{2}\big|\Curl\A\big|^2 
+\frac{\kappa_{ab}}{2}(\D\psi_a)^*\D\psi_b^{} 
+ \alpha_{ab} \psi_a^*\psi_b^{}
+ \beta_{abcd} \psi_a^*\psi_b^*\psi_c^{}\psi_d^{}
}
où $\psi_a$ sont les composantes du multiplet scalaire $\Psi\in\mathbb{C}^N$. 
Le multiplet scalaire est $\Psi^\dagger=(\psi_1^*,\psi_2^*,\cdots,\psi_N^*)$, où $
a,b,c,d=1,\cdots, N $ ; et les indices répétés sont implicitement sommés. Les champs 
scalaires sont couplés au champ de jauge (abélien) $\A$ via la dérivée de jauge 
$\D=\Grad+ie\A$, avec $e$ la constante couplage de jauge (les caractères gras désignent 
les quantités vectorielles).
Tous les coefficients tensoriels $\kappa$, $\alpha$, $\beta$ obéissent à des relations 
de symétrie, de sorte que l'énergie est une quantité réelle, définie positive 
\footnote{
Le modèle de Ginzburg-Landau \Eqref{EqFR:General:FreeEnergy} est isotrope. 
Les anisotropies peuvent être incorporées en utilisant un terme cinétique 
plus général : $\kappa_{ab;\mu\nu}(D_\mu\psi_a)^*D_\nu\psi_b^{}$. 
}.

Il peut être pratique de collecter tous les termes potentiels dans 
\Eqref{EqFR:General:FreeEnergy} en un seul terme potentiel $V(\Psi,\Psi^\dagger)$ 
comme 
\Equation{EqFR:General:Potential}{
V(\Psi,\Psi^\dagger)=\alpha_{ab} \psi_a^*\psi_b^{}
		+ \beta_{abcd} \psi_a^*\psi_b^*\psi_c^{}\psi_d^{} \,.
}
Parfois, la structure spécifique du potentiel $V(\Psi,\Psi^\dagger)$ sera sans 
importance. À d'autres occasions, le potentiel d'interaction aura un rôle central 
pour définir des nouvelles propriétés physiques. Ainsi, la restriction pertinente 
du potentiel le plus générique \Eqref{EqFR:General:Potential}, sera spécifiée lorsque  
nécessaire.

L'état fondamental est l'état qui minimise l'énergie potentielle 
\Eqref{EqFR:General:Potential}, et qui est spatialement constant: 
$\Psi_0:=\argmin V(\Psi,\Psi^\dagger)$. De plus l'état fondamental \emph{supraconducteur} 
est l'état qui minimise l'énergie et qui a $\Psi^\dagger\Psi=const.\neq0$. L'état 
fondamental est dégénéré en énergie et cela définit une variété appelée 
\emph{variété du vide}. En gros, c'est la topologie de cette variété du vide qui 
spécifie la nature des défauts topologiques qui peuvent apparaître dans la théorie. 
Par exemple, l'énergie de l'état fondamental est invariante sous les rotations globales 
de la phase du multiplet, cela définit ainsi variété du vide qui est isomorphe au cercle. 
Les configurations des champs sont ainsi classées par un nombre d'enroulement (le winding 
number) qui est un élément du premier groupe d'homotopie du cercle $\pi_1(\groupS{1})$ 
(cela peut également être compris comme une conséquence du fait que $\Psi$ doit être 
univalué). Ce nombre d'enroulement détermine le contenu en vortex de la théorie.

La variation fonctionnelle de l'énergie libre par rapport aux condensats supraconducteurs 
$\psi_a^*$ donne les équations du mouvement d'Euler-Lagrange. Dans le cadre de la 
supraconductivité, ce sont les équations de Ginzburg-Landau 
\Equation{EqFR:General:GL}{
\kappa_{ab}\D\D\psi_b=2\frac{\delta V}{\delta\psi_a^*}\,.
}
De même, la variation par rapport au champ de jauge $\A$ donne l'équation 
d'Ampère-Maxwell \Equation{EqFR:General:Maxwell}{
\Curl\B +e\sum_{a,b}\kappa_{ab}\Im\big( \psi_a^*\D\psi_b \big)=0 	\,.
}
Cette équation permet d'introduire les supercourants
\Equation{EqFR:General:Current}{
\J:=\sum_a \J^\oa \,,~~~~~\text{where}~~~~~
\J^\oa=e\sum_{b}\kappa_{ab}\Im\big( \psi_a^*\D\psi_b \big) 	\,.
}
Ici $\J$ est le supercourant total, et $\J^\oa$ désigne le supercourant partiel 
associé à un condensat donné $\psi_a$.

En fonction des propriétés du modèle microscopique considéré, il peut y avoir diverses 
exigences supplémentaires contraignant davantage la structure des paramètres tensoriels 
$\kappa$, $\alpha$, $\beta$. Cela peut donner lieu à de différentes situations qu'il 
est inutile d'énumérer ici.
Comme mentionné plus haut, la teneur en vortex est spécifiée par le nombre d'enroulement 
de la configuration du champ (plus précisément par l'enroulement à l'infini). L'étape 
suivante consiste à construire explicitement les solutions de vortex dans un secteur 
topologique donné, spécifié par ce nombre d'enroulement. La théorie étant clairement 
non linéaire, la construction explicite d'une configuration du champ pour un 
nombre d'enroulements donné doit donc être abordée numériquement.
Dans les travaux qui sont discutés ici, cela est fait en utilisant des algorithmes 
de minimisation sur l'énergie, conjointement à une formulation du problème selon la 
méthode des éléments finis. Ces aspects techniques sont détaillés dans 
la seconde annexe.

\section*{Plan du rapport}
\addcontentsline{toc}{section}{Plan du rapport}

Il est difficilement concevable de séparer tous les aspects liés à la nouvelle physique 
qui apparaissent dans les systèmes multi-composantes. Il y aura ainsi sûrement des 
chevauchements ou bien des redites de temps à autre. Quoi qu'il en soit, le corps 
principal de ce rapport est organisé comme suit : D'abord, le chapitre 
1 met en lumière les nouvelles propriétés associées à la 
topologie des modèles phénoménologiques à multiples composantes. Ensuite, le chapitre 
2 présente certaines nouvelles propriétés physiques qui apparaissent 
en raison de l'existence d'échelles de longueur supplémentaires. Enfin, les propriétés 
des états supraconducteurs multi-composantes qui brisent spontanément la symétrie 
d'inversion temporelle sont discutées dans le chapitre 3.

Plus précisément, le premier chapitre est dédié à la nature 
des excitations topologiques qui apparaissent dans les supraconducteurs à multiples 
composants. Il est tout d'abord démontré que la condition de quantification du flux 
magnétique implique que les excitations topologiques élémentaires y sont des 
\emph{vortex fractionnaires}. Ce sont des configurations des champs qui portent une 
fraction arbitraire du quantum de flux, mais qui ont une énergie par unité de longueur 
divergente. Cependant, lorsque des vortex fractionnaires se combinent pour former 
un objet qui portent une quantité entière de flux, ils forment un défaut topologique 
d'énergie finie. Selon la position relative des vortex fractionnaires, le défaut 
topologique qui en résulte est soit \emph{singulier} soit \emph{sans-coeur} (coreless). 
Dans ce dernier cas, on peut alors démontrer qu'il existe un invariant topologique 
supplémentaire, dont la nature est différente de celle du nombre d'enroulement 
le plus courant. Néanmoins, l'analyse la plus simple montre que typiquement les 
vortex fractionnaires s'attirent pour former un défaut singulier. Il s'ensuit qu'un 
mécanisme de stabilisation est nécessaire pour l'existence de défauts sans-coeur. 
Diverses occurrences de tels défauts topologiques sans-coeur et stables, appelés 
\emph{skyrmions}, sont présentées tout au long de ce chapitre. Comme ils ont une 
structure du coeur différente, les skyrmions peuvent interagir différemment en 
comparaison des vortex singuliers (d'Abrikosov). Ainsi il ont des propriétés 
observables qui sont significativement différentes.

Alors que les défauts sans-coeur présentent de nouvelles propriétés intéressantes, 
les défauts singuliers présentent également une nouvelle physique riche. Cette 
nouvelle physique des défauts singuliers est discutée dans le second chapitre. 
Les propriétés de la réponse magnétique des supraconducteurs 
peuvent, dans une certaine mesure, être considérées comme la conséquence de 
l'interaction entre les vortex. Plus précisément, la dichotomie classique qui classe 
les supraconducteurs conventionnels en type-1 ou type-2, peut être comprise selon que 
les vortex s'attirent ou s'ils se repoussent. Les interactions entre vortex peuvent 
être déterminées, en partie, par l'analyse des échelles de longueur de la théorie. 
Les vortex s'attirent lorsque la longueur de cohérence du condensat est supérieure 
à la longueur de pénétration du champ magnétique (c'est le régime de type-1). 
En revanche, si la longueur de pénétration est la plus grande échelle de longueur, 
les vortex se repoussent (c'est le régime de type-2). Dans les supraconducteurs à 
plusieurs composantes, une telle dichotomie n'est pas toujours possible. En effet, 
puisqu'ils comportent plusieurs condensats supraconducteurs, les supraconducteurs 
à multiples composantes présentent généralement des échelles de longueur supplémentaires. 
Il peut donc arriver que la longueur de pénétration (qui est unique) soit une échelle 
de longueur intermédiaire, et que l'interaction entre vortex soit attractive à longue 
portée (comme dans le type-1) et répulsive à courte portée (comme dans le type-2). 
Un tel régime avec des forces inter-vortex non monotones est appelé type-1.5. 
Dans ce régime, les vortex ont tendance à s'agréger pour former des clusters entourés 
de régions de l'état de Meissner (donc sans vortex). La possible formation de tels 
agrégats impacte fortement les propriétés de magnétisation, par rapport aux régimes 
conventionnels de type-1 ou de type-2.

Les interactions non monotones entre les vortex sont en partie déterminées par 
les échelles de longueur, et celles-ci sont déterminées par les perturbations autour 
de l'état fondamental de la théorie. Certains supraconducteurs à multiples composantes 
présentent des états fondamentaux inhabituels qui brisent spontanément la symétrie 
d'inversion temporelle. Ces aspects sont discutés dans le troisième chapitre.
Ces états, qui brisent la symétrie d'inversion temporelle, sont caractérisés par un 
état fondamental dont les phases relatives entre les condensats ne sont ni 0 ni $\pi$; 
par phase relative, on entend la différence entre les phases de différents condensats.
Il existe divers états supraconducteurs de ce type, par exemple appelés $\pip$, $\sis$, 
$\sid$, $\did$, etc. Cependant, l'accent sera mis sur l'état $\sis$, qui est l'extension 
de l'état $s$-wave le plus commun, et qui brise la symétrie d'inversion temporelle. 
La brisure spontanée de la symétrie d'inversion du temps dans l'état $\sis$ se produit 
typiquement à cause de la compétition entre différents termes de verrouillage des 
phases (phase-locking). La transition de phase vers les états brisant la symétrie 
d'inversion temporelle est du second ordre, et est donc associée à une échelle de 
longueur divergente. Notamment, cette transition peut se produire dans l'état 
supraconducteur où la longueur de pénétration est finie. Il s'ensuit qu'au voisinage 
de la transition de cette brisure de la symétrie d'inversion temporelle, la longueur 
de pénétration peut être une échelle de longueur intermédiaire, menant ainsi à de 
interactions non monotones entre vortex évoquées plus haut. De plus, la symétrie 
d'inversion temporelle est une opération discrète, donc si elle est spontanément brisée, 
alors l'état fondamental possède une dégénérescence discrète $\groupZ{2}$ en plus de la 
dégénérescence $\groupU{1}$ habituelle. Ceci implique qu'en plus des vortex, la théorie 
admet des excitations du type mur de domaine. Ceux-ci interagissent de manière non 
triviale avec les vortex, ce qui mène un nouveau type d'excitations topologiques avec 
différentes propriétés magnétiques. Enfin, les états supraconducteurs qui brisent la 
symétrie d'inversion du temps présentent des propriétés thermoélectriques inhabituelles. 
Celles-ci peuvent être utilisés pour induire des réponses électriques et magnétiques 
spécifiques, lorsque le matériau est exposé à des variations locales inhomogènes de la 
température.

Comme expliqué ci-dessus, l'essentiel de ces effets qui apparaissent dans les 
supraconducteurs à multiples composantes, sont discutés ici dans le cadre de la théorie 
de Ginzburg-Landau. La première annexe, présente le cadre théorique 
et les propriétés de la théorie de Ginzburg-Landau à une seule composante, qu'on 
retrouvera dans la plupart des manuels. Il est en effet parfois utile, de comparer 
les propriétés des théories de Ginzburg-Landau multi-composantes avec les résultats 
classiques. Cela peut fournir un meilleur apper\c cu afin de mieux appréhender les 
nouvelles caractéristiques qui apparaissent dans les théories à multiples composantes.

Il est à plusieurs reprises souligné que la théorie de Ginzburg-Landau est une théorie 
des champs classique qui est non linéaire. Être non linéaire implique que, sauf dans 
des circonstances très particulières, il n'y a pas de solutions analytiques, et que 
le problème doit être traité numériquement. C'est notamment le cas des résultats 
affichés dans cet ouvrage, et des résultat des travaux qui y sont discutés. 
Les aspects techniques des méthodes numériques sont présentés dans la seconde annexe. 
Cela comprend une présentation des méthodes d'éléments finis 
utilisées pour gérer la discrétisation spatiale des équations aux dérivées partielles; 
également l'algorithme d'optimisation pour gérer la non linéarité du problème. 
La construction numérique des défauts topologiques repose également sur l'implémentation  
appropriée des propriétés topologiques pour l'algorithme numérique; c'est également 
discuté dans cet annexe.

\section*{Synthèse du Chapitre 1~-~Les défauts topologiques dans les systèmes à plusieurs
		condensats supraconducteurs}
\addcontentsline{toc}{section}{Résumé du chapitre 1}

Contrairement à la théorie de Ginzburg-Landau à une seule composante, où les excitations 
topologiques consistent uniquement en des vortex quantiques, les théories dont le
paramètre d'ordre a plusieurs composantes présentent un spectre d'excitations topologiques 
beaucoup plus riche. Les supraconducteurs et les superfluides à plusieurs composantes sont 
décrits par des paramètres d'ordre duquel chacune des composantes est décrite par un 
champ scalaire complexe. Ainsi, tous les degrés de liberté supraconducteurs/
superfluides peuvent être assemblés en un multiplet de champs scalaires complexes.

Ce chapitre présente des résultats relatifs aux propriétés topologiques des théories 
de la supraconductivité comportant des paramètres d'ordre multiples ou des paramètres 
d'ordre à plusieurs composantes. Ces propriétés topologiques sont étudiées ici dans le 
cadre des théories de Ginzburg-Landau. Les excitations topologiques les plus élémentaires 
y sont des \emph{vortex fractionnaires} qui portent une fraction du quantum de flux 
\cite{Babaev:02,Babaev:04b,Babaev.Ashcroft:07}. En bref, il s'agit d'une configuration 
de champ pour laquelle une \emph{seule} des composantes a un nombre d'enroulement non nul.

Dans certains modèles spécifiques de supraconductivité, la fraction portée par ces 
vortex fractionnaires est d'un demi quantum de flux. Là, les vortex fractionnaires sont 
plutôt appelés des vortex à \emph{un demi-quantum} (half-quantum vortices). L'existence 
des vortex à un demi-quantum a été à l'origine prédite dans la phase $A$ de l'$^3$He 
superfluide \cite{Volovik.Mineev:76a,Mineev:13}. Leur existence a été étudiée sans 
relâche et leur observation a finalement été rapportée dans la phase polaire de 
l'$^3$He superfluide \cite{Autti.Dmitriev.ea:16}.

La recherche de vortex porteurs d'un demi-quantum de flux a également été très active 
en physique du solide. En particulier pour les supraconducteurs dont on a soutenu 
qu'ils avaient un appariement de type $p$-wave, comme \SRO \cite{Chung.Bluhm.ea:07,
Chung.Agterberg.ea:09,Chung.Kivelson:10}. L'observation de pas d'un demi-quantum de 
flux dans les courbes de magnétisation d'échantillons mésoscopiques de \SRO, a été 
revendiquée comme la marque des vortex portant un demi-quantum \cite{Jang.Ferguson.ea:11}.
L'intérêt pour la réalisation de tels vortex vient du fait que leur spectre 
d'excitation contient des fermions de Majorana d'énergie nulle \cite{Ivanov:01}. 
La statistique de ces vortex est non abélienne \cite{Ivanov:01}, ce qui pourrait 
potentiellement être utilisé comme éléments de base pour l'informatique quantique 
(des qbits) \cite{Kitaev:03}. 

L'énergie par unité de longueur des vortex fractionnaires n'est malheureusement pas 
finie. Ainsi, dans des conditions habituelles, les vortex fractionnaires sont 
thermodynamiquement instables dans le volume. Notons cependant que des échantillons 
mésoscopiques, peuvent permettre de favoriser énergétiquement les vortex fractionnaires 
\cite{Chibotaru.Dao:10,Silaev:11}.

Les vortex fractionnaires sont des objets assez insaisissables qui ne peuvent donc, 
en général, pas être observés individuellement. Cependant, leurs états liés portant 
un flux entiers ont une énergie finie et sont donc observables.
Les vortex fractionnaires sont importants non seulement parce qu'ils sont les éléments 
constituants des excitations topologiques plus complexes, ils sont également la pierre 
angulaire des propriétés thermodynamiques des systèmes à plusieurs composantes. 
Dans les supraconducteurs à une seule composante, la transition de phase est entraînée 
par la prolifération de boucles de vortex excitées thermiquement \cite{Peskin:78,
Dasgupta.Halperin:81}. De même, dans les supraconducteurs à plusieurs composantes, 
c'est la prolifération de vortex fractionnaires qui entraîne les transitions de phase, 
comme démontré dans la limite de London \cite{Smiseth.Smorgrav.ea:05,Smorgrav.Babaev.ea:05,
Herland.Babaev.ea:10}, ou dans Ginzburg-Landau \cite{Smorgrav.Smiseth.ea:05a}. 
En présence d'un champ externe, les vortex fractionnaires jouent également un rôle 
dans la fusion des réseaux de vortex \cite{Smorgrav.Smiseth.ea:05a}.
De même, les propriétés thermodynamiques des superfluides à plusieurs composantes 
reposent fortement sur le rôle des vortex fractionnaires \cite{Dahl.Babaev.ea:08a,
Dahl.Babaev.ea:08b,Dahl.Babaev.ea:08}.

\subsection*{Plan détaillé du chapitre 1}

Pour commencer, une première Section 1.1 aborde la question de la 
quantification du flux du champ magnétique dans les supraconducteurs à multiples 
composantes. On y montre que la quantification du flux autorise formellement l'existence 
de vortex fractionnaires. Leurs propriétés élémentaires sont également discutées. 
Dans les systèmes à plusieurs composantes, les vortex fractionnaires sont des 
configurations des champs, pour lesquelles un seul condensat possède un nombre 
d'enroulement de la phase non nul, tandis que les autres non. Considérés individuellement, 
l'énergie des vortex fractionnaires est divergente. Cependant, cette divergence de 
l'énergie disparaît s'ils forment des états liés. Plus précisément, si tous les 
condensats possèdent le même nombre d'enroulement.

Cela implique que, dans le volume de ces systèmes, seuls les objets composites ont 
une énergie finie. La section 1.1.3 développe davantage 
les propriétés topologiques des défauts composites pour les supraconducteurs à multiple 
composantes. En particulier, il existe un invariant topologique caché associé 
à la topologie de l'espace projectif complexe, qui caractérise les défauts 
topologiques sans-coeur (core-less). Cet invariant, qui classe les applications 
$\Real^2\to\groupCP{N-1}$, permet de différencier les vortex sans-coeur des vortex 
singuliers. Dans le cas d'un système à deux composantes, l'espace cible $\groupCP{1}$ 
peut être identifié avec la 2-sphère unité $\groupS{2}$. L'invariant topologique 
peut alors être interprété comme l'indice de Hopf, et peut être utilisé pour caractériser 
les vortex noués dans les supraconducteurs à deux composantes.

La quantification du flux implique que ces invariants supplémentaires sont non nuls, 
tant que tous les condensats supraconducteurs ne disparaissent pas simultanément. 
C'est-à-dire tant que les vortex fractionnaires dans les différents composantes ne 
se superposent pas. L'interaction entre les vortex fractionnaires est analysée dans 
la section 1.1.4. Étant donné que l'interaction 
entre les vortex fractionnaires est attractive, l'observation des défauts topologiques 
sans-coeur est plutôt difficile. Comme expliqué en détails après, divers mécanismes 
peuvent compenser l'attraction entre les vortex fractionnaires et ainsi mener à la 
formation de défauts sans-coeur. Ces défauts sans-coeur qui sont un état lié de 
vortex fractionnaires sont souvent appelés \emph{skyrmions}. Cette terminologie 
trouve son origine dans l'existence d'une relation formelle entre les modèles de 
Ginzburg-Landau à deux composantes et le modèle de Skyrme-Faddeev. La relation 
entre ces deux modèles est présentée explicitement dans la section 1.2.

La section 1.3 présente des diverses situations où différents mécanismes 
physiques permettent de stabiliser des défauts sans-coeur au dépens des vortex 
singuliers.
Tout d'abord, dans la Section 1.3.1, dans un modèle d'une mixture 
de condensats possédant différentes charges électriques (commensurables), introduit dans 
\CVcite{Garaud.Babaev:14a}. Dans ce modèle assez exotique, les différents condensats 
supraconducteurs peuvent avoir des charges électriques qui sont différentes (c'est-à-dire 
qu'ils ont un couplage au champ de jauge qui est différent). Là, les vortex fractionnaires 
sont naturellement divisés et forment ainsi un état lié sans-coeur. Ce modèle peut être 
appliqué pour décrire phénoménologiquement l'état supraconducteur du deutérium liquide 
métallique, où des paires de Cooper électroniques coexistent avec un condensat 
de Bose-Einstein de deutérons.

Ensuite, dans la section 1.3.2, il est démontré que l'entraînement  
inter-composantes non dissipatif, connu sous le nom d'effet Andreev-Bashkin, 
est responsable de l'existence des skyrmions \CVcite{Garaud.Sellin.ea:14}. De plus, 
cet entraînement non dissipatif peut également stabiliser des états liés des vortex 
fractionnaires noués \CVcite{Rybakov.Garaud.ea:19}. Ces noeuds, caractérisés par 
l'indice de Hopf, sont ainsi appelés \emph{hopfions}. Fait intéressant, ces hopfions 
rappellent l'idée antérieure de Kelvin de vortex noués d'éther luminifère  pour expliquer 
la classification des atomes.

Après, les propriétés des défauts topologiques qui se produisent dans les états 
supraconducteurs $\sis$ sont analysées dans la section 1.3.3. 
Cet état $\sis$ sera considéré plus en détail dans le chapitre 3. 
L'état $\sis$ brise une symétrie discrète $\groupZ{2}$ associée à la symétrie 
d'inversion temporelle, en plus de l'habituelle symétrie de jauge $\groupU{1}$.
La brisure spontanée d'une symétrie discrète est associée à la formation de murs 
de domaine. Suivant \CVcite{Garaud.Babaev:14}, ces murs de domaine peuvent être se 
former lors de refroidissements rapides du matériau supraconducteur et peuvent être 
stabilisées géométriquement contre leur expulsion. Comme démontré dans 
\CVcite{Garaud.Carlstrom.ea:11} et dans \CVcite{Garaud.Carlstrom.ea:13}, 
l'interaction complexe entre les mur de domaine et les vortex fractionnaires conduit 
à l'existence de nouveaux états skyrmioniques.

\subsection*{Résumé des résultats qui seront présentés dans le chapitre 1}

\begin{itemize}
\setlength\itemsep{0.025em}

\item Dans \CVcite{Garaud.Babaev:12} et \CVcite{Garaud.Babaev:15a}, nous avons démontré 
que l'état supraconducteur $p_x\!+ip_y$ permet des états de type skyrmion qui sont 
caractérisées par les invariants d'homotopie des applications $\groupS{2}\!\to\!\groupS{2}$. 
Ils peuvent être alternativement compris comme des vortex portant deux quanta de flux 
du champ magnétique. De plus ils sont favorisés énergétiquement par rapport aux vortex 
portant un seul quantum de flux \CVcite{Garaud.Babaev:15a}. Dans un champ magnétique 
appliqué, ces vortex à deux quanta se forment spontanément et s'organisent en réseaux 
hexagonaux \CVcite{Garaud.Babaev.ea:16}. Au voisinage du second champ critique $\Hc{2}$ 
le réseau hexagonal des vortex à deux quanta se dissocient en un réseau carré de vortex 
quantiques simples \CVcite{Garaud.Babaev.ea:16}. Ce scenario persiste au-delà de 
l'approximation du champ moyen lorsque les fluctuations sont prisent en compte 
\CVcite{Krohg.Babaev.ea:21}.

\item Démonstration que les supraconducteurs interfaciaux avec un fort couplage 
spin-orbite de Rashba ont une réponse magnétique inhabituelle 
\CVcite{Agterberg.Babaev.ea:14}. Nous démontrons microscopiquement que dans la limite 
pure, les supraconducteurs interfaciaux, tels que SrTiO$_3$/LaAlO$_3$, 
sont des candidats idéaux pour observer les défauts caractérisés par le groupe d’homotopie 
des applications $\groupS{2}\!\to\!\groupS{2}$ (en supplément des invariant des 
applications $\groupS{1}\!\to\!\groupS{1}$). Des états skyrmioniques similaires existent 
également dans les supraconducteurs nématiques tels que Cu$_x$Bi$_2$Se$_3$ 
\CVcite{Zyuzin.Garaud.ea:17}.

\item Identification des propriétés topologiques des mixtures de condensats chargés 
ayant des charges électriques différentes (commensurables) \CVcite{Garaud.Babaev:14a}. 
Une telle situation devrait apparaître par exemple dans le deutérium liquide métallique.

\item Prédiction d'une nouvelle phase dans les supraconducteurs 
$\groupU{1}\!\times\!\groupU{1}$ avec un entraînement inter-composantes non dissipatif  
\CVcite{Garaud.Sellin.ea:14}. L'interaction non dissipative rend les vortex singuliers 
instables au profit des skyrmions, dont l'interaction à longue portée modifie 
sensiblement les processus d'aimantation. Ces modèles de supraconductivité avec un 
entraînement non dissipatif stabilise également des vortex noués stables 
\CVcite{Rybakov.Garaud.ea:19}.

\item Découverte d'un nouveau type de solitons topologiques stables dans des 
supraconducteurs à trois composantes qui brisent spontanément la symétrie d'inversion 
temporelle \CVcite{Garaud.Carlstrom.ea:11} et \CVcite{Garaud.Carlstrom.ea:13}. 
Ces défauts topologiques porteurs de flux, qui sont caractérisés par des invariants 
topologiques $\groupCP{2}$, sont des skyrmions. Leur observation pourrait signaler 
des états supraconducteurs qui brisent la symétrie d'inversion temporelle, par exemple 
dans certains supraconducteurs à base de fer, ainsi que dans des multi-couches entre 
$s_\pm$ et $s$-wave ordinaire, couplées par l'interaction Josephson.

\end{itemize}

\section*{Synthèse du Chapitre 2~-~Le régime supraconducteur de type-1.5}
\addcontentsline{toc}{section}{Résumé du chapitre 2}

Les supraconducteurs avec des paramètres d'ordre multi-composantes ont non seulement 
une grande variété de défauts topologiques, sans équivalent dans les systèmes 
mono-composantes (vortex fractionnaires, skyrmions, hopfions, etc.), 
mais ils permettent également des interactions plus riches entre eux.
Comme introduit dans le chapitre précédent, les excitations topologiques élémentaires 
sont des vortex portant une fraction du quantum de flux. Ceux-ci se combinent pour 
former des défauts composites qui portent un flux entier. Indépendamment de leur 
structure centrale, les interactions entre les défauts topologiques sont régies, 
dans une large mesure, par les échelles de longueur caractéristiques de la théorie.
Les condensats supraconducteurs à une seule composante sont caractérisés par la longueur 
de cohérence $\xi$ associée aux variations de densité (mode d'Anderson-Higgs). 
Les paramètres d'ordre multi-composantes, d'autre part, comportent généralement plus 
d'échelles de longueur. Alors que les modes scalaires associés sont typiquement attractifs, 
les modes chargés, associés à la longueur de pénétration $\lambda$ du champ de jauge, 
médient la répulsion entre les défauts porteurs de flux.

La classification usuelle divise les supraconducteurs en deux classes, 
en fonction de leur comportement dans un champ externe. Cette classification est 
quantifiée par le paramètre de Ginzburg-Landau sans dimension $\kappa$ défini comme 
le rapport des deux échelles de longueur fondamentales $\kappa=\lambda/\xi$. Lorsque 
$\sqrt{2}\lambda<\xi$ (type-1), les supraconducteurs expulsent un champ magnétique 
faible (l'état de Meissner), tandis que des domaines normaux macroscopiques se forment 
lorsque les champs appliqués sont importants \cite{Ginzburg.Landau:50,Gennes}. 
D'autre part, les supraconducteurs de type-2, pour lesquels $\xi<\sqrt{2}\lambda$, 
présentent des excitations de type vortex qui sont thermodynamiquement stables 
\cite{Abrikosov:57}. Le champ magnétique est expulsé en dessous d'une certaine valeur 
critique $\Hc{1}$. Au-dessus de cette valeur, et jusqu'à la destruction de la 
supraconductivité au second champ critique $\Hc{2}$, les supraconducteurs de type-2 
forment des réseaux ou des liquides de vortex portant un quantum de flux. 
Le coût énergétique de l'interface entre l'état normal et l'état supraconducteur est 
positif dans le régime de type-1. L'absence de vortex thermodynamiquement stables et la 
formation de domaines normaux macroscopiques découlent donc de la minimisation de cette 
énergie d'interface. Il en résulte également que les forces inter-vortex sont purement 
attractives. Ainsi les vortex s'effondrent en un vortex géant, minimisant ainsi 
l'interface. D'autre part, les supraconducteurs de type-2 supportent des uniquement 
des vortex à un seul quantum, thermodynamiquement stables, car l'énergie de l'interface 
entre l'état normal et l'état supraconducteur est négative. L'interaction entre les 
vortex est purement répulsive, ils forment des réseaux d'Abrikosov (triangulaires) 
\cite{Abrikosov:57}. Cet ensemble de vortex maximise l'interface. Dans la théorie de 
Ginzburg-Landau, à la valeur critique $\kappa=1/\sqrt{2}$ (appelée le point de 
Bogomol'nyi), les vortex n'interagissent pas \cite{Kramer:73,Bogomolnyi:76}. Là, 
la répulsion courant-courant compense exactement l'attraction coeur-coeur à toutes 
les distances.

Contrairement aux supraconducteurs à une seule composante, il n'est pas possible de 
construire un seul paramètre sans dimension pour les supraconducteurs à plusieurs 
composantes, présentant plusieurs longueurs de cohérence $\xi_a$. Par conséquent, 
la dichotomie habituelle type-1/type-2 est insuffisante pour capturer l'ensemble de 
la physique et pour classer les supraconducteurs à plusieurs composantes. En effet, 
comme les longueurs de cohérence $\xi_a$ associées aux condensats supraconducteurs 
sont typiquement différentes, la longueur de pénétration $\lambda$ peut être 
formellement une échelle de longueur \emph{intermédiaire} : 
$\xi_1\!<\!\cdots\! <\!\sqrt{2}\lambda\!<\!\cdots\!<\!\xi_N$. Pour une telle hiérarchie 
des échelles de longueur, les modes associés aux échelles de longueur supérieures à 
$\sqrt{2}\lambda$ fournissent une attraction à longue portée comme dans les 
supraconducteurs de type-1. En revanche, les modes avec des échelles de longueur plus 
courtes (que $\sqrt{2}\lambda$) permettent une répulsion à courte portée, 
comme dans le type-2. 

Considérons par exemple un supraconducteur à deux composantes satisfaisant cette 
hiérarchie des échelles de longueur : $\xi_1<\sqrt{2}\lambda<\xi_2$. Là, le mode associé 
à la plus grande échelle de longueur $\xi_2$ devrait fournir une attraction à longue 
portée entre vortex (en gros en raison du chevauchement des coeurs). D'autre part, 
les interactions courant-courant et électromagnétique associées à $\lambda$ fournissent 
une interaction répulsive à courte portée. La compétition entre ces comportements ouvre 
la possibilité d'un potentiel d'interaction inter-vortex non monotone, qui est attractif 
à longue portée (comme dans le régime de type-1) et répulsif à courte portée (comme 
dans régime de type-2). Ce compromis entre les comportements de type-1 et de type-2 
a motivé la terminologie \emph{type-1.5} de tels états \cite{Moshchalkov.Menghini.ea:09}.
La hiérarchie où $\lambda$ est une échelle de longueur intermédiaire est une condition 
nécessaire, mais pas suffisante, pour réaliser des interactions inter-vortex non monotones. 
Pourtant, si elles sont réalisées, les forces non monotones entraînent une séparation 
préférentielle entre vortex telle que deux vortex forment un état lié. Ainsi, de nombreux 
vortex fusionnent pour former un agrégat de vortex (cluster), coexistant avec les 
domaines de l'état de Meissner (sans vortex): l'\emph{état semi-Meissner} 
\cite{Babaev.Speight:05}.

Dans l'ensemble, c'est l'interaction inter-vortex non monotone qui définit les propriétés 
essentielles du régime de type-1.5, mais ce n'est pas suffisant pour définir cet état. 
En effet, l'attraction entre les vortex peut également survenir dans certaines 
circonstances dans les matériaux à une seule composante. Cependant, dans le cas du 
type 1.5, l'attraction à longue portée est une conséquence de plusieurs longueurs de 
cohérence et s'accompagne de plusieurs nouveaux effets physiques discutés ci-dessous.

\subsection*{Plan détaillé du chapitre 2}

L'ingrédient essentiel pour la réalisation du régime de type-1.5 dans les supraconducteurs 
à multiples composantes est d'avoir une hiérarchie des longueurs caractéristiques 
telle que la longueur de pénétration soit une échelle intermédiaire. 
Ainsi, comme point de départ, la Section 2.1 présentera 
le cadre général pour l'analyse des échelles de longueur. Ceci résulte de l'analyse du 
spectre propre de l'opérateur (linéaire) des perturbations autour de l'état fondamental. 
Le cadre général est ensuite complété par un exemple particulier qui peut être traité 
analytiquement. 

Les échelles de longueur, définies à partir du spectre propre des perturbations, 
déterminent le comportement asymptotique des vortex. En particulier, comme détaillé dans 
la section 2.1.3, cela contrôle l'interaction à longue 
portée entre les vortex. Si la longueur de pénétration est une échelle de longueur 
intermédiaire, alors les forces inter-vortex sont attractives à longue portée et répulsives 
à courte portée. Cela suggère fortement que les vortex devraient former des états liés 
avec une séparation préférentielle.

Lorsque la hiérarchie des échelles de longueur autorise des forces inter-vortex 
non monotones, alors les vortex peuvent s'agréger ensemble formant ainsi des 
\emph{clusters de vortex} entourés de régions sans vortex, de l'état de Meissner. 
Quelques exemples de tels clusters sont considérés dans la section 2.2. 

Ensuite, la section 2.3 examine les possibles mécanismes qui 
devrait conduire à la formation d’agrégats de vortex; ainsi que les différents modèles 
où cela a été observé. Les signatures expérimentales possibles de ces amas de vortex, et 
leur pertinence expérimentale y sont également discutés.

\subsection*{Résumé des résultats qui seront présentés dans le chapitre 2}

\begin{itemize}

\item Découverte d'un nouveau type de forces inter-vortex à plusieurs corps dans les 
supraconducteurs multibandes \CVcite{Carlstrom.Garaud.ea:11}. Les interactions 
inter-vortex sont non monotones et mènent à la formation d'amas de vortex entourés de 
domaines de Meissner macroscopiques (c'est-à-dire des état sans vortex). La formation 
des structures peut être fortement impactée par les interactions entre vortex 
non-par-paires (non pairwise), provenant de la superposition non linéaire des vortex. 
Les forces inter-vortex non monotones entraînent également la formation de clusters 
dans les supraconducteurs à trois bandes \CVcite{Carlstrom.Garaud.ea:11a}. Pour une 
revue de ces phénomènes, voir \CVcite{Babaev.Carlstrom.ea:12}.
Des forces inter-vortex non monotones peuvent également se produire dans des systèmes 
supraconducteurs avec des paramètres d'ordre en compétition. C'est-à-dire lorsque, 
dans l'état fondamental, les interactions inter-composantes interdisent la coexistence 
des deux condensats \CVcite{Garaud.Babaev:14b} et \CVcite{Garaud.Babaev:15}.

\item Explication de la coalescence de vortex, dans un modèle putatif à deux bandes pour 
le matériau supraconducteur \SRO \CVcite{Garaud.Agterberg.ea:12}. Nous avons soutenu que 
la coalescence de vortex parfois observée dans \SRO peut s'expliquer par des interactions 
non monotones provenant de la nature multibande de \SRO. Ce scenario de la coalescence 
des vortex dans \SRO a reçu un soutien expérimental des mesures de $\mu$SR dans 
\href{http://dx.doi.org/10.1103/PhysRevB.89.094504} {Phys. Rev. B 89, 094504 (2014)} 
\cite{Ray.Gibbs.ea:14}.

\item Prédiction de la réponse magnétique inhabituelle dans les supraconducteurs 
interfaciaux avec un fort couplage spin-orbite (du type Rashba) 
\CVcite{Agterberg.Babaev.ea:14}. Nous démontrons microscopiquement, et au travers de 
diverses simulations, que les supraconducteurs interfaciaux tels que SrTiO$_3$/LaAlO$_3$, 
peuvent présenter la formation de clusters de vortex.

\item Dans une série de travaux sur les propriétés microscopiques des supraconducteurs 
à deux bandes, avec des impuretés, nous avons démontré qu'il existe des régions du diagramme 
des phase où la hiérarchie des échelles de longueur permet en principe la formation d'amas 
de vortex. L'origine d'une telle hiérarchie est la proximité d'une transition de phase 
du second ordre, cachée au sein de l'état supraconducteur \CVcite{Silaev.Garaud.ea:17}. 
Cela devrait se produire de la même manière dans les systèmes pures à trois bandes  
\CVcite{Garaud.Silaev.ea:17}. Des simulations numériques montrent qu'en effet cela se 
produit, et que cela se traduit par des signaux particuliers qui peuvent être distinguer 
la coalescence par rapport à d'autres scénarios, via des mesures globales de la réponse 
d'expériences de $\mu$SR \CVcite{Garaud.Corticelli.ea:18a}.

\end{itemize}

\section*{Synthèse du Chapitrer 3~-~Les états supraconducteurs qui brisent la symétrie d'inversion temporelle}
\addcontentsline{toc}{section}{Résumé du chapitre 3}

Les théories qui décrivent la physique des supraconducteurs (ou des superfluides) sont 
invariantes sous conjugaison complexe. Cette invariance est généralement appelée 
symétrie d'inversion du temps. Dans les systèmes à plusieurs composantes, la symétrie 
d'inversion temporelle peut être spontanément brisée par l'état fondamental. 
C'est-à-dire que l'état fondamental n'est \emph{pas} invariant, aux rotations globales 
de phase près, sous conjugaison complexe.
De tels états peuvent apparaître si la phase relative entre les fonctions de gap 
supraconducteur dans les différentes composantes diffère de $0$ ou $\pi$ 
\cite{Balatsky:00,Lee.Zhang.ea:09,Platt.Thomale.ea:12,Stanev.Tesanovic:10,
Fernandes.Millis:13,Agterberg.Barzykin.ea:99,Ng.Nagaosa:09,Lin.Hu:12,
Carlstrom.Garaud.ea:11a,Bobkov.Bobkova:11,Maiti.Chubukov:13,Maiti.Sigrist.ea:15}.
Il en résulte qu'en plus de la brisure habituelle de la symétrie de jauge $\groupU{1}$, 
ces états supraconducteurs présentent une dégénérescence discrète associée à la 
brisure spontanée de la symétrie d'inversion temporelle.

Les états brisant spontanément la symétrie d'inversion temporelle ont suscité beaucoup 
d'intérêt dans le contexte des modèles supraconducteurs non conventionnels à triplet 
de spin, en particulier l'état $p_x+ip_y$ qui a été intensivement étudié en relation 
avec le supraconducteur \SRO. L'état supraconducteur $\sis$ est un autre état brisant 
la symétrie d'inversion du temps, qui a attiré beaucoup d'attention. En effet, 
il a reçu un fort soutien théorique en relation avec certains supraconducteurs à base 
de fer, et en particulier Ba$_{1-x}$K$_x$Fe$_2$As$_2$ avec un dopage de trous 
\cite{Hirschfeld.Korshunov.ea:11,Maiti.Korshunov.ea:11,Maiti.Korshunov.ea:11a,
Maiti.Korshunov.ea:12,Maiti.Chubukov:13}.
L'état $\sis$ est un mélange complexe d'états supraconducteurs distincts, ayant la même 
symétrie, et qui sont en compétition au travers de termes de verrouillage des phases. 
Dans les pnictides, on pense qu'il provient de la compétition entre différents canaux 
d'appariement \cite{Maiti.Chubukov:13}. Cet état pourrait également être conçu 
artificiellement sur des interfaces de bicouches supraconductrices \cite{Ng.Nagaosa:09}.

La brisure spontanée de la symétrie d'inversion temporelle a diverses conséquences 
physiques intéressantes. Certaines, comme l'existence de murs de domaine, ont été discutés 
précédemment au chapitre 1. Les supraconducteurs à base 
de fer \cite{Kamihara.Watanabe.ea:08} sont parmi les candidats les plus prometteurs 
pour l'observation des états $\sis$.
En effet, les données expérimentales montrent que dans les composés 122 
Ba$_{1-x}$K$_x$Fe$_2$As$_2$ dopés en trous, la symétrie de l'état supraconducteur 
change en fonction du niveau de dopage $x$. Une structure de bande typique de 
Ba$_{1-x}$K$_{x}$Fe$_2$As$_2$ se compose de deux poches de trous au point $\Gamma$ 
et de deux poches d'électrons à $(0, \pi)$ et $(\pi,0)$. A des niveaux modérés de 
dopage $x\sim 0.4$ diverses mesures, incluant ARPES \cite{Ding.Richard.ea:08,
Khasanov.Evtushinsky.ea:09,Nakayama.Sato.ea:11}, de conductivité thermique 
\cite{Luo.Tanatar.ea:09} et des expériences de diffusion de neutrons 
\cite{Christianson.Goremychkin.ea:08}, sont cohérentes avec l'hypothèse d'un état 
$s_\pm$ où l'etat supraconducteur change de signe entre les poches d'électrons 
et de trous.
D'autre part, la symétrie de l'état supraconducteur à fort dopage $x\rightarrow 1$ 
n'est pas aussi claire quant à la question de savoir si le canal $d$ domine, ou si 
le gap conserve la symétrie $s_\pm$ qui change de signe entre le bandes de trous 
internes au point $\Gamma$ \cite{Maiti.Korshunov.ea:11,Maiti.Korshunov.ea:11a}. 
En effet, il existe des preuves que canal $d$-wave  domine \cite{Reid.Juneau-Fecteau.ea:12,
Reid.Tanatar.ea:12,Tafti.Juneau-Fecteau.ea:13,Tafti.Clancy.ea:14} tandis que d'autres 
données ARPES ont été interprétées en faveur d'une symétrie $s$-wave \cite{Okazaki.Ota.ea:12,
Watanabe.Yamashita.ea:14}.
Dans les deux situations, cela implique l'existence possible d'un état complexe 
intermédiaire qui compromet entre les comportements à dopage modéré et élevé. Selon que le 
canal $d$ ou $s$ domine à fort dopage, un tel état complexe est nommé $\sis$ ou $\sid$.

L'état $\sis$ est isotrope et préserve les symétries cristallines \cite{Maiti.Chubukov:13}. 
D'autre part, l'état $\sid$ brise la symétrie $C_4$, alors qu'il reste invariant sous 
la combinaison d'une opération de symétrie par inversion du temps et des rotations $C_4$. 
Étant anisotrope, l'état $\sid$ est donc qualitativement différent de l'état $\sis$. 
Notons que l'état supraconducteur $\sid$ est aussi qualitativement différent des états 
$s\!+\!d$ (préservant l'inversion temporelle), qui ont suscité l'intérêt dans le contexte 
des supraconducteurs cuprates à haute température (voir par exemple \cite{Joynt:90,
Li.Koltenbah.ea:93,Berlinsky.Fetter.ea:95}).
Il contraste également avec l'état $d+id$, qui peut apparaître en présence d'un champ 
magnétique externe, et qui viole à la fois les symétries de parité et d'inversion du 
temps \cite{Balatsky:00, Laughlin:98} .
Bien qu'il s'agisse d'un scénario intéressant, peut-être pertinent pour les pnictides, 
les propriétés de l'état $\sid$ ne seront pas davantage examinées ici. L'accent sera mis 
sur l'analyse de l'état supraconducteur de $\sis$. Cet état devrait résulter de diverses 
physiques microscopiques \cite{Stanev.Tesanovic:10,Platt.Thomale.ea:12,Suzuki.Usui.ea:11,
Chubukov.Efremov.ea:08,Maiti.Chubukov:13,Ahn.Eremin.ea:14}.

L'observation expérimentale des états $\sis$ ou $\sid$, qui brisent la symétrie 
d'inversion temporelle, est un défi. En effet, 
cela nécessite de sonder les phases relatives entre les composantes du paramètre d'ordre, 
ce qui est une tâche difficile.
Par exemple, l'état $\sis$ ne brise pas les symétries des groupes ponctuels et n'est 
donc pas associé à un moment cinétique intrinsèque des paires de Cooper. Par conséquent, 
il ne peut pas produire de champ magnétique local et est donc {\it a priori} invisible 
pour les méthodes conventionnelles telles que la relaxation du spin du muon et les 
mesures de l'effet Kerr polaire qui ont par exemple été utilisées pour sonder l'état 
supraconducteur $\pip$ qui brise l'inversion temporelle.
Plusieurs propositions ont été émises, chacune avec diverses limitations, pour 
observer indirectement des signatures de la brisure de la symétrie d'inversion du temps 
dans les pnictides. Celles-ci, par exemple, incluent l'étude du spectre des modes 
collectifs, qui comprend des excitations à masse nulle \cite{Lin.Hu:12}, et qui mélangent 
des modes de densité et des modes de phases \cite{Carlstrom.Garaud.ea:11a,Stanev:12,
Maiti.Chubukov:13,Marciani.Fanfarillo.ea:13} .
En outre, il a été proposé de considérer les propriétés des excitations topologiques 
exotiques telles que les skyrmions et les murs de domaine \cite{Garaud.Carlstrom.ea:11,
Garaud.Carlstrom.ea:13,Garaud.Babaev:14}, des mécanismes non conventionnels de la 
viscosité des vortex \cite{Silaev.Babaev:13}, la formation d'amas de vortex 
\cite{Carlstrom.Garaud.ea:11a}, des  phases exotiques réentrantes et précurseures induites 
par les fluctuations \cite{Bojesen.Babaev.ea:13,Bojesen.Babaev.ea:14,Carlstrom.Babaev:15,
Hinojosa.Fernandes.ea:14}.
Il a été prédit que des courants spontanés existent à proximité d'impuretés non magnétiques 
dans des états $\sid$ supraconducteurs anisotropes \cite{Lee.Zhang.ea:09,Maiti.Sigrist.ea:15} 
ou dans des échantillons soumis à une contrainte \cite{Maiti.Sigrist.ea:15}. Cependant, 
cette dernière proposition implique en fait un changement de symétrie des états $\sis$ 
et repose sur la présence d'impuretés qui peuvent avoir une distribution incontrôlable.
Il a également été souligné que l'état $\sis$ brisant la symétrie d'inversion temporelle  
présente une contribution non conventionnelle à l'effet thermoélectrique 
\cite{Silaev.Garaud.ea:15}. En lien avec cela, un montage expérimental, basé sur un 
chauffage local a été proposé \cite{Garaud.Silaev.ea:16}. L'idée clé étant que le 
chauffage local induit des variations locales de phases relatives qui donnent 
une réponse électromagnétique directement observable.

\subsection*{Plan détaillé du chapitre 3}

L'existence de l'état $\sis$ peut provenir de divers mécanismes, y compris la 
compétition entre différents canaux d'appariement, ou la diffusion d'impuretés. 
Ces aspects microscopiques ne seront pas abordés ici. Au lieu de cela, on présentera 
la manière dont cet état $\sis$ apparaît dans les théories phénoménologiques de 
Ginzburg-Landau.

Comme point de départ, la Section 3.1 présente le mécanisme 
de \emph{frustration de phases}, responsable de la brisure spontanée de la symétrie 
d'inversion temporelle, dans les supraconducteurs à trois composantes. Il s'agit ici 
de la compétition entre les différents termes de verrouillage des phases qui peut 
entraîner l'état $\sis$. Puisque la symétrie d'inversion du temps est une opération 
discrète, sa brisure spontanée implique que l'état fondamental présente une dégénérescence 
discrète. Les propriétés d'un tel état fondamental seront analysées dans la section 
3.1.1, tandis que les échelles de longueur correspondantes seront dérivées 
dans la section 3.1.2. Le cas d'un état $\sis$ à deux composantes sera 
également traité dans cette section.

Ensuite, les propriétés des défauts topologiques qui peuvent apparaître dans les 
états supraconducteurs qui brisent la symétrie d'inversion du temps sont abordées 
dans la section 3.2. Il y aura un chevauchement partiel avec les discussions de la 
section 1.3.3 du chapitre 1. Le fait que l'état fondamental brise une symétrie 
discrète implique que la théorie admet des murs de domaine. Ces murs de domaine entre 
différents états brisant la symétrie d'inversion temporelle seront construits 
explicitement dans la section 3.2.1. Les mur de domaine 
interagissent de manière non triviale avec les vortex. Comme détaillé dans la section 
3.2.1, des murs de domaine fermés peuvent former des états liés 
avec des vortex. Comme discuté d'abord dans le chapitre 1, 
puisque les défauts composites qui en résultent sont sans-coeur, ils ont des propriétés 
topologiques supplémentaires: ce sont les skyrmions chiraux $\groupCP{2}$. 

Enfin, les propriétés thermoélectriques de l'état $\sis$ seront discutées dans la 
section 3.3.

\subsection*{Résumé des résultats qui seront présentés dans ce chapitre}

\begin{itemize}
\setlength\itemsep{0.025em}

\item Dans \CVcite{Garaud.Silaev.ea:17} nous avons démontré que les théories du champ 
moyen pour les états supraconducteurs $\sis$, qui brisent la symétrie d'inversion 
temporelle, sont quantitativement cohérentes avec les modèles microscopiques à plusieurs 
bandes. Nous avons par ailleurs démontré que l'état $\sis$ peut également apparaître 
dans les systèmes à deux bandes en raison de la diffusion due à des impuretés  
\CVcite{Silaev.Garaud.ea:17}. Dans l'approximation quasi-classique, nous montrons qu'en 
fonction du niveau des impuretés, l'état $\sis$ est une phase intermédiaire entre 
les états $\spm$ et $\spp$. Nous avons en outre établi dans \CVcite{Garaud.Silaev.ea:17} 
et dans \CVcite{Garaud.Corticelli.ea:18a} que la phase $\sis$ est entouré d'une ligne de 
transition de phase du second ordre. Cela implique l'existence d'un soft-mode avec une 
échelle de longueur divergente. Les autres longueurs de cohérence restent finies à cette 
transition, et il existe donc une disparité infinie des longueurs de cohérence, ce qui 
peut conduire à une physique des vortex inhabituelle avec des forces non monotones 
\CVcite{Garaud.Corticelli.ea:18a} et \CVcite{Garaud.Corticelli.ea:18}. Ces forces 
inter-vortex, attractives à longue portée et répulsive à courte portée, permettent 
la formation d'agrégats de vortex.

\item Dans \CVcite{Silaev.Garaud.ea:15} et \CVcite{Garaud.Silaev.ea:16} nous avons 
démontré que l'existence d'états brisant la symétrie d'inversion du temps a un impact 
mesurable sur la réponse thermoélectrique des supraconducteurs. Dans 
\CVcite{Silaev.Garaud.ea:15}, nous avons prédit que les supraconducteurs qui brisent 
la symétrie d'inversion temporelle présentent un effet thermoélectrique géant, dont la 
nature est essentiellement différente de celui des supraconducteurs à une seul composante. 
Cet effet provient des contre-courants inter-composantes induits thermiquement, 
contrairement aux contre-courants entre les courants normaux et supraconducteurs 
dans le mécanisme de Ginzburg traditionnel.
Nous avons par ailleurs démontré dans \CVcite{Garaud.Silaev.ea:16}, que ces propriétés 
thermoélectriques non conventionnelles peuvent être utilisées pour induire des champs 
magnétiques et électriques, mesurables expérimentalement, en réponse à un chauffage 
local des matériaux candidats. Les champs induits sont sensibles à la présence de murs 
de domaine, ainsi qu'à l'anisotropie cristalline. De plus, un processus de chauffage 
non stationnaire produit un champ électrique et un déséquilibre de charge dans les 
différentes bandes \CVcite{Garaud.Silaev.ea:16}, qui est également mesurable.

\item Description des signatures expérimentales des murs de domaine se formant lors 
de quenchs, via le mécanisme de Kibble-Zurek, dans des supraconducteurs à brisant 
spontanément la symétrie d'inversion temporelle, ayant la structure du gap $\sis$ 
\CVcite{Garaud.Babaev:14}. Comme il s'agit d'une symétrie discrète, la brisure spontanée 
de la symétrie d'inversion temporelle dans l'état $\sis$, dicte qu'elle possède des 
excitations de type mur de domaine. Nous discutons également de l'influence des murs 
de domaine, stabilisées géométriquement, sur les processus de magnétisation.

\item Découverte d'un nouveau type de solitons topologiques stables, dans les 
supraconducteurs à trois composantes, qui brisent spontanément la symétrie d'inversion 
du temps \CVcite{Garaud.Carlstrom.ea:11} et \CVcite{Garaud.Carlstrom.ea:13}. Ces 
défauts topologiques portant un flux magnétique, sont caractérisés par un ivariant  
topologique cachée, associée à la topologie de l'espace projectif complexe $\groupCP{N-1}$. 
Ces skyrmions $\groupCP{2}$ peuvent se former spontanément lors d'expérience de 
refroidissement sous champs  \CVcite{Garaud.Babaev:14}, lorsque le processus de 
refroidissement passe par la transition de phase vers l'état brisant la symétrie 
d'inversion du temps.

\item Découvertes d'un nouveau type de mode collectif, dans les supraconducteurs à trois 
composantes brisant la symétrie d'inversion temporelle \CVcite{Carlstrom.Garaud.ea:11a}. 
Ce mode est associé à des excitations collectives mélangeant densités et phases. 
Il est donc différent du mode Leggett.

\end{itemize}

\section*{Synthèse de l'Annexe A~-~Théorie de Ginzburg-Landau à une composante}
\addcontentsline{toc}{section}{Résumé de l'annexe A}

Le corps principal de ce rapport présente les résultats concernant les propriétés 
des théories de la supraconductivité comportant plusieurs paramètres d'ordre, ou bien 
avec un paramètre d'ordre à plusieurs composantes. Il est utile, pour une meilleure 
compréhension des particularités des théories multicomposantes, de passer en revue 
les propriétés essentielles des modèles conventionnels de supraconductivité à une 
seule composante. Couvrir tous les aspects microscopiques de la supraconductivité 
conventionnelle est au-delà des présentes discussions, et ceux-ci ne seront pas 
discutés ici. Les aspects microscopiques et de champ moyen de la supraconductivité 
à une seule composante sont largement discutés dans un grand nombre de manuels 
classiques, voir par exemple \cite{Saint-James.Thomas.ea,Gennes,Tinkham,Shmidt.Muller.ea,
Chaikin.Lubensky,Huebener,Schmidt,Annett,Fossheim.Sudbo,Svistunov.Babaev.ea}.

Par conséquent, la présente revue se limite uniquement aux aspects classiques de la 
théorie de champ moyen de la supraconductivité. Plus précisément, cette annexe présente 
le cadre théorique général et les propriétés de la théorie de Ginzburg-Landau à une 
seule composante. 

La théorie de Ginzburg-Landau \cite{Ginzburg.Landau:50} a été introduite en 1950, 
pour rendre compte des propriétés macroscopiques de l'état supraconducteur. Cette 
théorie phénoménologique est basée sur la théorie de Landau de transition de phase 
du second ordre, où le paramètre d'ordre macroscopique, noté $\psi=|\psi|\Exp{i\varphi}$, 
est un champ scalaire complexe. Le paramètre d'ordre $\psi$ est souvent appelé de 
manière équivalente \emph{condensat supraconducteur}. La théorie microscopique de la 
supraconductivité de Bardeen-Cooper-Schrieffer \cite{Bardeen.Cooper.ea:57} a été 
dérivée plus tard en 1957. Peu de temps après, en 1959, Gor'kov a démontré que la 
théorie de Ginzburg-Landau peut être dérivée comme une approximation classique de 
la théorie microscopique \cite{Gorkov:59}, et le module du paramètre d'ordre $\psi$ 
est en fait la densité de paires de Cooper : $n_s=|\psi|^2$. Au sens strict, la théorie 
de Ginzburg-Landau n'est valable que dans un proche voisinage de la température critique 
$T_c$ où la supraconductivité est détruite, et suppose que $\psi$ est petit et varie 
lentement (que les gradients sont petits). 

Notons qu'en plus de ses applications fondamentales en physique du solide, la théorie 
de Ginzburg-Landau a attiré beaucoup d'attention dans la communauté des mathématiques 
à partir des années 1990, après qu'il ait été démontré qu'il s'agit d'un problème bien 
posé \cite{Du.Gunzburger.ea:92,Chen.Hoffmann.ea:93,Du:94,Du:96}. Depuis ces travaux, 
il y a eu une activité important pour comprendre les propriétés mathématiques de ce 
problème, voir par exemple \cite{Du:05}. Parallèlement, des efforts continus ont 
également été déployés dans la communauté des physiciens pour avoir une formulation 
optimale pour les solveurs numériques, voir par exemple \cite{Gropp.Kaper.ea:96,
Sadovskyy.Koshelev.ea:15}.

\subsection*{Plan détaillé de l'annexe A}

Dans ce chapitre on introduit la théorie de Ginzburg-Landau ainsi que les équations 
décrivant la dynamique du condensat supraconducteur. Ensuite, les propriétés de l'état 
fondamental sont discutées, ainsi que la détermination des échelles de longueur 
caractéristiques et de l'effet Meissner. En particulier, les échelles de longueur sont 
déterminées à partir du spectre de masse de la théorie, obtenu en étudiant l'opérateur 
des perturbations linéaires autour de l'état fondamental supraconducteur.

Ensuite, nous considérons la condition de quantification du flux magnétique, et il 
est démontré que cela implique l'existence de vortex magnétiques. Une dichotomie peut 
être établie pour classer les supraconducteurs en deux types selon les propriétés 
d'interaction des vortex. En particulier que dans le régime de type-1 les vortex 
s'attirent, alors qu'ils se repoussent dans le régime de type-2. Cette image est 
complétée par l'analyse des champs critiques, ce qui permet d'établir qualitativement 
les diagrammes de phases des différents types de supraconducteurs.

\section*{Synthèse de l'Annexe B~-~Méthodes numériques }
\addcontentsline{toc}{section}{Résumé de l'annexe B}


La plupart des résultats présentés dans le corps principal du rapport reposent sur des 
simulations numériques. L'accent à été mis sur les propriétés physiques et très peu de 
mots ont été dits sur les détails de ces simulations. On présente dans cette annexe, 
une discussion détaillée des méthodes numériques utilisées pour étudier la physique 
des différents systèmes introduits dans la partie principale.
Cela commence, dans la section B.1, par un aperçu général des méthodes 
d'éléments finis utilisées pour la discrétisation spatiale. Ensuite, les détails de 
l'algorithme utilisé pour résoudre les problèmes (non linéaires) de Ginzburg-Landau 
sont présentés dans la section B.2. Les aspects importants du choix 
approprié de la configuration initiale sont détaillés dans la section B.2.3. 
Enfin, la question de l'évolution des problèmes dépendant du temps est abordée dans la 
section B.3.

\subsection*{Plan détaillé de l'annexe B}

Il existe différentes méthodes pour aborder la discrétisation spatiale des équations 
aux dérivées partielles. La méthode des différences finies, qui est basée sur des 
développements en séries de Taylor pour approximer la dérivée, est la plus ancienne. 
Cette technique de discrétisation représente l'espace en réseaux topologiquement 
carrés ou en réseau de cuboïdes. Cette méthode est plutôt intuitive, mais elle rend 
difficile la manipulation de géométries complexes. Cette difficulté a motivé l'approche 
de la méthode des éléments finis. Historiquement, les méthodes des éléments finis ont 
souvent été préférées pour la démontrabilité rigoureuse de leur stabilité, ainsi que pour 
leur applicabilité naturelle à des géométries complexes. Les avantages relatifs des deux 
méthodes ont longtemps été fortement débattus, et il est juste de dire qu'aujourd'hui 
les deux méthodes sont sensiblement équivalentes. Chacune avec ses avantages et ses 
inconvénients. Il existe de nombreuses approches différentes comme les méthodes 
spectrales, qui ne sont pas exhaustivement listées ici.
Les méthodes aux différences finies ou aux éléments finis sont plus ou moins fréquentes 
selon les différentes communautés scientifiques. Par exemple, et très grossièrement, 
les différences finies sont largement utilisées pour les simulations des théories de 
jauge sur réseau, tandis que les éléments finis sont plus communs, par exemple en 
ingénierie et en mathématiques.

Les simulations numériques discutées dans le corps principal, et les articles 
correspondants, ont largement utilisé les méthodes d'éléments finis, pour une grande 
variété de problèmes incluant la résolution directe, la minimisation, l'optimisation 
contrainte, l'évolution temporelle, etc. En pratique, la discrétisation spatiale est 
gérée dans un cadre fourni par la bibliothèque FreeFEM++ 
\footnote{ \url{https://freefem.org/}} \cite{Hecht:12}.
Les méthodes des éléments finis sont basées sur la formulation faible (la formulation 
variationnelle) d'équation aux dérivées partielles. On présentera une description brève 
et non exhaustive des concepts utilisés dans les méthodes des éléments finis. Des 
introductions détaillées peuvent être trouvées dans de nombreux manuels, voir par 
exemple \cite{Cook.Malkus.ea,Hutton:03,Reddy:05,Allaire:05}.

Après cette introduction aux principes des méthodes d'éléments finis, on se focalisera 
sur les aspects algorithmiques qui permettent la résolution des problèmes non linéaires. 
En effet, la plupart des problèmes discutés dans le corps principal nécessitent de 
minimiser numériquement l'énergie libre de Ginzburg-Landau. Il s'agit d'un problème 
d'optimisation non linéaire, pour la théorie des champs considérée, et le choix 
l'algorithme était généralement la méthode du gradient conjugué non linéaire. 
Il s'agit d'une méthode numérique pour résoudre des problèmes d'optimisation non 
contraints, de manière itérative. 
On présentera brièvement les principes de cette méthode, puis on en montrera la 
formulation explicite pour le problème de minimisation dans la théorie de Ginzburg-Landau.

Ensuite on présentera en détails les aspects relatif au choix de la configuration 
initiale pour l'algorithme de minimisation. En effet, le choix de la configuration 
initiale pour la minimisation est une étape cruciale dans la construction de nouvelles 
solutions. Un choix initial judicieux améliore non seulement la convergence de l'algorithme, 
mais permet également d'initier certaines des propriétés souhaitées. L'idée clé est 
que pour que l'algorithme de minimisation converge vers une configuration qui possède 
les propriétés topologiques souhaitées, la configuration de départ doit elle-même 
avoir ces propriétés. Le raisonnement est le suivant : différents secteurs topologiques 
sont généralement séparés de grandes barrières énergétiques, ainsi la minimisation 
converge facilement vers une configuration qui a les mêmes propriétés topologiques que la 
configuration de départ. Remarquons que les barrières énergétiques infinies, qui séparent 
les différents secteurs topologiques, sont généralement définies dans un espace infini. 
De plus la discrétisation spatiale limite également les arguments issus de la topologie. 
Donc, à proprement parler, dans les domaines finis, il n'y a que des barrières d'énergie 
finies entre les différents secteurs topologiques.

On discutera enfin des algorithmes pertinents pour l'analyse de problèmes dépendants 
du temps. En effet, la plupart des problèmes discutés dans le corps principal nécessitent 
les techniques de minimisation pour les problèmes non linéaires. Cependant, il est parfois 
important de connaître les propriétés dynamiques d'évolution temporelle du système. 
Par exemple, avec l'équation de Ginzburg-Landau dépendante du temps, voir par exemple 
\cite{Schmid:66,Gorkov.Eliashberg:68,Gorkov.Kopnin:75}. On présentera en détail un 
algorithme utilisé pour l'évolution temporelle des systèmes (non)linéaires, l'algorithme 
de Crank-Nicolson \cite{Crank.Nicolson:47}. Il s'agit d'une méthode implicite aux 
différences finies dans la dimension temporelle qui a été conçue à l'origine pour 
simuler l'évolution temporelle de l'équation de la chaleur \cite{Crank.Nicolson:47}. 
Par la suite, cette méthode a également été utilisé dans le cadre de l'équation de 
Schr\"odinger non linéaire \cite{Delfour.Fortin.ea:81}, et pour étudier la dynamique 
des superfluides et des condensats de Bose-Einstein d'atomes ultra-froids \cite{Bao.Cai:18}. 
On présentera les détails de cet algorithme dans le contexte de l'équation de 
Gross-Pitaevskii. Une version légèrement différente est nécessaire pour les 
supraconducteurs, afin de prendre en compte le champ de jauge, voir par exemple 
les travaux connexes \cite{Chen:97,Gao.Sun:15,Gao.Li.ea:14}.


\section*{Conclusion et perspectives}		
\addcontentsline{toc}{section}{Conclusion et perspectives}

\vspace{0.5cm}

\subsection*{Conclusion}
\addcontentsline{toc}{subsection}{Conclusion}

Ce mémoire tente de faire passer le message que les modèles à plusieurs composantes, 
et en particulier les supraconducteurs à multiples composantes, hébergent une physique 
très riche, qui est absente de leurs homologues à une seule composante. 

Comme souligné dans l'introduction, les excitations topologiques sont omniprésentes 
en physique. Ainsi, elles apparaissent par exemple dans la physique du solide, dans 
les systèmes de la matière condensée, la physique des hautes énergies, etc. Selon les 
propriétés topologiques associées, ces objets ont des structures qui sont différentes. 
Ils peuvent être semblables à des particules, à des singularités ponctuelles, 
à des parois ou encore à des lignes. Dans ce dernier cas, les défauts topologiques 
sont les vortex, et ils ont été largement étudiés dans le contexte de la superfluidité 
et de la supraconductivité. Les vortex déterminent, dans une large mesure, les propriétés 
thermodynamiques, électriques et magnétiques des matériaux considérés.
Le choix du récit dans l'introduction tente de souligner que les vortex ont attiré 
l'attention depuis longtemps, et que certains concepts relativement anciens sont 
toujours d'actualité dans la physique moderne.

En raison du plus grand nombre de degrés de liberté, les modèles multicomposantes 
de supraconductivité permettent un riche spectre de défauts topologiques. 
Le premier chapitre était essentiellement consacré à la formalisation des propriétés 
topologiques des supraconducteurs à multiples composantes. Il a ensuite été discuté 
diverses contributions de l'auteur, dans la construction de nouveaux types de défauts 
topologiques, dans différents modèles de supraconductivité multicomposantes. 
Ces nouveaux défauts topologiques peuvent être utilisés pour identifier les propriétés 
des modèles sous-jacents.
De plus, il a été souligné dans le deuxième chapitre, que les supraconducteurs 
multicomposantes hébergent non seulement de nouvelles d'excitations topologiques, 
mais aussi qu'ils peuvent interagir différemment des vortex habituels. Cette nouvelle 
interaction entre les vortex est essentiellement différente de celle des supraconducteurs 
à une seule composante, de type 1 ou de type-2. Il s'ensuit que les vortex peuvent former 
des agrégats, ce qui a un impact important sur divers processus physiques observables.
Enfin, il peut également exister des états supraconducteurs qui brisent la symétrie 
d'inversion temporelle, à cause de la compétition entre différents canaux d'appariement. 
Ces états sont associés à de nouveaux effets mesurables, comme on l'avons vu dans le 
dernier chapitre. 

Il est important de souligner à nouveau, que toutes les contributions de l'auteur 
reposent sur une utilisation intensive des méthodes numériques. Les aspects numériques 
sont bien souvent négligés, au profit des discussions sur les propriétés physiques. 
Il à semble opportun de profiter de ce mémoire, pour présenter plus en détail ces 
aspects des méthodes numériques.

Les résultats présentés dans ce rapport visent à souligner la richesse de la physique 
des systèmes à plusieurs composantes. Ce n'est que la pointe de l'iceberg, et beaucoup 
d'autres choses peuvent encore être dites. Bien que la physique des défauts topologiques 
soit une histoire assez ancienne maintenant, il reste encore beaucoup à découvrir.

\vspace{0.5cm}
\subsection*{Perspectives}
\addcontentsline{toc}{subsection}{Perpsectives}

Comme cela a été souligné dans le mémoire, il existe un nombre croissant de matériaux 
supraconducteurs multibandes/multicomposantes connus. C'est donc un terrain de jeu en 
constante évolution pour rechercher de nouvelles théories pertinentes, et en étudier 
les propriétés topologiques. Ainsi, dans un certain sens, il y a toujours des projets 
encore inconnus qui méritent d'être étudiés, en raison de leur pertinence par rapport 
à certains nouveaux matériaux. En tout cas, de nombreux aspects des modèles de 
supraconducteurs multi-composants sont probablement encore à découvrir. On présentera 
trois directions intéressantes, en relation avec les aspects abordés dans le rapport.

\paragraph*{Projet 1 : États supraconducteurs anormaux qui brisent la symétrie 
d'inversion temporelle. }
Certaines des nouvelles propriétés de l'état $\sis$, qui brise spontanément la symétrie 
d'inversion du temps, ont été décrites en détail dans ce mémoire. Elles sont, entre 
autres, l'existence de modes collectifs qui incluent des excitations de masse nulle
\cite{Lin.Hu:12}, des modes mélangeant phase et densité \cite{Carlstrom.Garaud.ea:11a,
Stanev:12,Maiti.Chubukov:13,Marciani.Fanfarillo.ea:13}, un mécanisme non 
conventionnel de la viscosité des vortex \cite{Silaev.Babaev:13}, la formation d'agrégats 
de vortex \cite{Carlstrom.Garaud.ea:11a}, une contribution inhabituelle à l'effet 
thermoélectrique \cite{Silaev.Garaud.ea:15,Garaud.Silaev.ea:16}. L'état $\sis$ doit 
également héberger des excitations topologiques telles que des skyrmions et des murs 
de domaine \cite{Garaud.Carlstrom.ea:11,Garaud.Carlstrom.ea:13,Garaud.Babaev:14}.

Récemment, la mesure de chaleur spécifique du composé Ba$_{1-x}$K$_x$Fe$_2$As$_2$ 
dopé en trous, au dopage $x\approx0.8$, a montré un comportement intrigant 
\cite{Grinenko.Weston.ea:21}. À savoir, les expériences de l'effet Nernst spontané 
et  de rotation de spin du muon (muon spin rotation), indiquent un état qui brise 
spontanément la symétrie d'inversion du temps, mais dans lequel les paires de Cooper 
ne sont pas cohérentes.
Lorsqu'un supraconducteur à plusieurs composantes brise la symétrie d'inversion 
temporelle, il peut y avoir plusieurs transitions de phase. Au niveau de la théorie 
du champ moyen, la transition de phase supraconductrice $T_c$ se produit toujours
à une température supérieure ou égale à la température de transition de la brisure 
de la symétrie d'inversion de temps $T_{\groupZ{2}}$. Les résultats récents montrent 
un comportement opposé où $T_{\groupZ{2}}>T_c$ \cite{Grinenko.Weston.ea:21}.
Toutes les discussions sur le rôle des fluctuations et leurs implications dépassent 
les discussions ici. Cependant, quelques remarques sur la structure du modèle ouvrent 
des perspectives intéressantes.

Comme indiqué dans ce mémoire, les supraconducteurs à plusieurs composantes possèdent 
une contribution supplémentaire au champ magnétique, en raison des interactions entre 
composantes. En reprenant l'expression de l'énergie libre de Ginzburg-Landau, en termes 
de modes chargés et neutres, on peut réécrire le modèle comme 
\Align{EqFR:Perspective:FreeEnergy:1}{
&\F= \frac{1}{2}\left[ \varepsilon_{kij}\left\lbrace 
\nabla_i\left(\frac{J_j}{e^2\varrho^2}\right) +\frac{i}{e\varrho^4}
{\cal Z}_{ij}
 \right\rbrace\right]^2  + \frac{\J^2}{2e^2\varrho^2} 
+\Grad\Psi^\dagger\!\cdot\!\Grad\Psi
+\frac{1}{4\varrho^2}\big(\Psi^\dagger\Grad\Psi-\Grad\Psi^\dagger\Psi\big)^2
+ V(\Psi)	\,, \nonumber \\
&~~~~~\text{où}~~~
{\cal Z}_{ij}= \varrho^2\Grad_i\Psi^\dagger\Grad_j\Psi
+(\Psi^\dagger\Grad_i\Psi)(\Grad_j\Psi^\dagger\Psi)\,.
}

Dans l'état anormal, la partie supraconductrice du modèle est désordonnée, 
et la partie correspondant l'écrantage de London est absente, c'est-à-dire $\J=0$. 
Alors, un modèle effectif décrivant le nouvel état anormal, peut être dérivé à partir 
\Eqref{EqFR:Perspective:FreeEnergy:1}, en exigeant que le courant supraconducteur 
disparaisse $\J=0$. Cela revient à ne retenir que les degrés de liberté qui sont 
liés aux phases relatives. Ce modèle s'écrit donc \cite{Grinenko.Weston.ea:21} 
\Equation{EqFR:Perspective:FreeEnergy:2}{
\F= \frac{1}{2}\left[  
 \frac{i\varepsilon_{kij}}{e\varrho^4}{\cal Z}_{ij}
 \right]^2   
 +\Grad\Psi^\dagger\!\cdot\!\Grad\Psi
+\frac{1}{4\varrho^2}\big(\Psi^\dagger\Grad\Psi-\Grad\Psi^\dagger\Psi\big)^2
+ V(\Psi)	\,. 
}
Comme discuté en détail dans \cite{Grinenko.Weston.ea:21}, la théorie effective de 
cet état anormal, qui brise la symétrie d'inversion du temps, permet des excitations 
de murs de domaine. Ceux-ci comportent des signatures magnétiques, analogues à celles 
discutées dans le chapitre 3.

Ce nouveau modèle effectif offre de nombreuses opportunités pour observer des propriétés 
inhabituelles des modèles multicomposantes dans un état anormal. C'est-à-dire qu'il 
est possible d'observer certaines des propriétés topologiques des supraconducteurs à 
plusieurs composantes, au-dessus de la température critique. Bien entendu ici, le modèle 
\Eqref{EqFR:Perspective:FreeEnergy:2} est obtenu de manière heuristique, et une dérivation 
rigoureuse est nécessaire pour gérer correctement la manière dont les différents termes 
doivent être renormalisés, lorsque la partie supraconductrice du modèle est désordonnée.

\paragraph*{Projet 2 : Autres états supraconducteurs qui brisent la symétrie 
d'inversion temporelle.}
Non seulement le nombre de supraconducteurs multibandes/multicomposants connus augmente, 
mais aussi de ceux qui brisent la symétrie d'inversion du temps \cite{Ghosh.Smidman.ea:20}. 
Il a été souligné dans ce rapport que les états supraconducteurs qui brisent spontanément 
la symétrie d'inversion temporelle présentent de nouvelles propriétés. L'accent a été 
mis principalement sur l'état $\sis$, qui est pertinent pour la famille des 
supraconducteurs à base de fer. Un autre état brisant la symétrie d'inversion du temps, 
qui a été très étudié en lien avec les modèles supraconducteurs à triplets de spin, 
est l'état $\pip$ . Il existe divers autres états supraconducteurs, qui brisent la 
symétrie d'inversion temporelle, mais avec d'autres symétries d'appariement, comme 
$\sid$, $\did$. Par exemple, il a été récemment avancé que l'appariement dans \SRO 
pourrait être soit $\did$ soit $d\!+\!ig$ \cite{Ghosh.Shekhter.ea:20,Agterberg:20}. 
Ces états, différents de $\pip$ ou $\sis$, sont beaucoup moins étudiés, et notamment 
leurs propriétés topologiques. 

Puisqu'ils brisent la symétrie d'inversion du temps, tous ces états doivent également 
comporter des excitations du type mur de domaine. Cependant, la différence est ces 
états brisent différentes symétries de groupes ponctuel. Au niveau du modèle de 
Ginzburg-Landau, cela se manifeste par une structure différente et plus riche des 
termes cinétiques, sous forme d'anisotropies et de mélange de gradients. Comme par 
exemple $(D_x\psi_1^*D_x\psi_2-D_y\psi_1^*D_y\psi_2+c.c.)$, pour l'état $\sid$ qui brise 
la symétrie $C_4$. Ou par exemple l'état $d+id$, qui viole à la fois les symétries 
de parité et d'inversion temporelle \cite{Balatsky:00,Laughlin:98} .

Les particularités de ces autres symétries d'appariement ont été beaucoup moins étudiées. 
Par exemple, en raison de leurs structures différentes, ils devraient également manifester 
des réponses thermoélectriques qualitativement différentes de celles discutées dans le 
chapitre 3. De plus, la structure des défauts topologiques doit aussi certainement 
y être sensible.

\paragraph*{Projet 3 : Noeuds et vortons dans la théorie électrofaible. } 
L'idée ici, est de rechercher des défauts topologiques dans une théorie différente 
de celle décrivant la supraconductivité à plusieurs composantes. Plus précisément, 
l'objectif est d'étudier la possibilité que la théorie de Weinberg-Salam, des 
interactions électrofaibles, puisse héberger des défauts topologiques avec une 
structure nouée. 

Comme souligné dans l'introduction, l'idée des vortex noués est une histoire ancienne 
qui a été ravivé, après la construction de défauts topologiques noués dans le modèle 
de Skyrme-Faddeev \cite{Faddeev.Niemi:97}. Depuis, il y a eu beaucoup d'activité dans 
la recherche d'objets similaires, dans divers systèmes physiques comme par exemple dans 
les condensats de Bose-Einstein spinoriels \cite{Kawaguchi.Nitta.ea:08}, les faisceaux 
optiques \cite{Dennis.King.ea:10}, les colloïdes nématiques \cite{Tkalec.Ravnik.ea:11}, 
des matériaux magnétiques \cite{Sutcliffe:17,Sutcliffe:18}, et plus encore. 
Pour une revue sur les noeuds, voir \cite{Radu.Volkov:08}. 

Les vortons sont des objets qui, bien que formellement différents, sont assez 
semblables aux vortex noués. Ce sont des boucles fermées de vortex supraconducteurs 
\cite{Witten:85a} qui devraient être stabilisées contre la contraction, par la force 
centrifuge produite par le courant \cite{Davis.Shellard:89}. Ces vortons doivent se 
produire dans un modèle introduit pour la première fois par Witten \cite{Witten:85a}. 
Il s'agit d'un modèle à deux composantes, mais avec deux champs de jauge abéliens 
(au lieu d'un seul pour les supraconducteurs). La construction explicite des vortons, 
et la démonstration de leur stabilité potentielle est cependant assez récente 
\cite{Radu.Volkov:08,Battye.Sutcliffe:08,Battye.Sutcliffe:09,Garaud.Radu.ea:13}.

Le secteur bosonique de la théorie de Weinberg-Salam des interactions électrofaibles, 
peut être vu, dans une certaine mesure, comme une théorie à plusieurs composantes mais 
plus complexe que celles discutées dans ce mémoire. En effet, il s'agit d'une théorie 
d'un doublet de champs scalaires complexes (le champ de Higgs). Par contre le secteur 
de jauge est plus compliqué, car il contient également un champ de jauge non-abélian 
$\groupSU{2}$, en plus du champ de jauge $\groupU{1}$. On suppose généralement que la 
théorie électrofaible n'admet pas de solitons, cependant, il y a des indications qu'elle  
pourrait héberger une sorte de vortons, ou des vortex noués. En effet, d'une part 
la théorie héberge des vortex, mais aussi dans certains cas limites, elle est très 
similaire au modèle de Witten où l'on sait que les vortons existent.

À proprement parler, la théorie électrofaible est différente des modèles de 
supraconductivité à plusieurs composants discutés dans le corps principal du rapport. 
Pourtant, puisqu'ils partagent certaines propriétés, on peut imaginer que cette théorie 
possède des solutions de type vortex noués, similaires à ceux obtenus dans le cadre 
des supraconducteurs à deux composantes avec l'interaction d'entrainement non dissipatif  
d'Andreev-Bashkin \CVcite{Rybakov.Garaud.ea:19}. La potentielle existence de tels 
vortons électrofaibles, ou de vortex noués, pourrait être d'une grande valeur scientifique.




%

\end{document}